\documentclass[11pt]{article}
\usepackage{epsf} 
\makeatletter
\renewcommand{\@biblabel}[1]{#1.}
\let\jnl@style=\it

\def\vec#1{{\mathbf{#1}}}
\newcommand{\cbeg}[1]{\relax}
\newcommand{\cend}[1]{\relax}
\newcommand{\cdel}[1]{\if\relax}
\newcommand{\cbbeg}[1]{\relax}
\newcommand{\cbend}[1]{\relax}

\def\M{{\cal M}}
\def\R{{\cal R}}
\def\D{{\cal D}}

\def\beq#1{\begin{equation}\label{#1}}
\def\eeq{\end{equation}}
\def\beqa#1{\begin{eqnarray}\label{#1}}
\def\eeqa{\end{eqnarray}}

\def\myfrac#1#2{\left(\frac{#1}{#2}\right)}
\def\comment#1{\relax}
\def\const{\hbox{\rm const}}

\def\spose#1{\hbox to 0pt{#1\hss}}
\def\simlt{\mathrel{\spose{\lower 3pt\hbox{$\mathchar"218$}}
     \raise 2.0pt\hbox{$\mathchar"13C$}}}
\def\simgt{\mathrel{\spose{\lower 3pt\hbox{$\mathchar"218$}}
     \raise 2.0pt\hbox{$\mathchar"13E$}}}
\def\simpropto{\mathrel{\spose{\lower 3pt\hbox{$\mathchar"218$}}
     \raise 2.0pt\hbox{$\propto$}}}

\def\eqalign#1{\null\,\vcenter{\openup\jot\m@th
  \ialign{\strut\hfil$\displaystyle{##}$&$\displaystyle{{}##}$\hfil
      \crcr#1\crcr}}\,}
\def\eqalignleft#1{\null\,\vcenter{\openup\jot\m@th
  \ialign{\strut$\displaystyle{##}$\hfil&$\displaystyle{{}##}$\hfil
      \crcr#1\crcr}}\,}
\def\ref@jnl#1{{\jnl@style#1\/}}

\def\aj#1#2{\ref@jnl{Astron. J.}}			
\def\araa#1#2{\ref@jnl{Ann. Rev. Astron. Astrophys.}}		
\def\jaa#1#2{\ref@jnl{J. Astron. Astrophys.}}			
\def\apj#1#2{\ref@jnl{Astrophys. J.}}			
\def\apjl#1#2{\ref@jnl{Astrophys. J. Lett.}}		
\def\apjs#1#2{\ref@jnl{Astrophys. J. Suppl.}}		
\def\ao#1#2{\ref@jnl{Appl. Optics}}		
\def\apss#1#2{\ref@jnl{Astrophys. Space Sci.}}		
\def\aap#1#2{\ref@jnl{Astron. Astrophys.}}		
\def\aapr#1#2{\ref@jnl{Astron. Astrophys. Rev.}}		
\def\aaps#1#2{\ref@jnl{Astron. Astrophys. Suppl.}}	
\def\baas#1#2{\ref@jnl{BAAS}}		
\def\jrasc#1#2{\ref@jnl{JRASC}}		
\def\memras#1#2{\ref@jnl{MmRAS}}		
\def\mnras#1#2{\ref@jnl{Montly Not. RAS}}		
\def\pra#1#2{\ref@jnl{Phys. Rev. A}}		
\def\prb#1#2{\ref@jnl{Phys. Rev. B}}		
\def\prc#1#2{\ref@jnl{Phys. Rev. C}}		
\def\prd#1#2{\ref@jnl{Phys. Rev. D}}		
\def\prl#1#2{\ref@jnl{Phys. Rev. Lett.}}	
\def\pasp#1#2{\ref@jnl{Publ. Astron. Soc. Pacif.}}		
\def\pasj#1#2{\ref@jnl{Publ. Astron. Soc. Jap.}}		
\def\qjras#1#2{\ref@jnl{QJRAS}}		
\def\skytel#1#2{\ref@jnl{S\&T}}		
\def\solphys#1#2{\ref@jnl{Solar~Phys.}}	
\def\ssr#1#2{\ref@jnl{Space~Sci. Rev.}}	
\def\zap#1#2{\ref@jnl{ZAp}}			
\def\aspr#1#2{\ref@jnl{Astrophys. Scape Phys. Rev.}}

\def\nat#1#2{\ref@jnl{Nature}}			
\def\sci#1#2{\ref@jnl{Science}}			
\def\bain#1#2{\ref@jnl{Bull. Astron. Inst. Netherland}}			
\def\cqg#1#2{\ref@jnl{Class. Quant. Gravity}}
\def\nyasa#1#2{\ref@jnl{Ann. NY Acad. Sci.}}
\def\iaus#1#2{\ref@jnl{IAU Symp.}}
\def\newa#1#2{\ref@jnl{New Astron.}}
\def\nuphs#1#2{\ref@jnl{Nucl. Phys. B Proc. Suppl.}}
\def\phlb#1#2{\ref@jnl{Phys. Lett. B}}
\def\rvma#1#2{\ref@jnl{Rev. Mod. Astr.}}
\def\rvmph#1#2{\ref@jnl{Rev. Mod. Phys.}}
\def\aas#1#2{\ref@jnl{Amer. Astr. Soc. Meeting}}
\def\aatr#1#2{\ref@jnl{Astron. Astroph. Trans.}}
\def\aph#1#2{\ref@jnl{Astropart. Phys.}}
\def\nucima#1#2{\ref@jnl{Nucl. Instr. Meth. A}}
\def\osajb#1#2{\ref@jnl{Opt. Soc. Amer. J. B}}
\def\pramjp#1#2{\ref@jnl{Pramana J. Phys.}}
\def\pthph#1#2{\ref@jnl{Progress Theot. Phys.}}
\def\azh#1#2{\ref@jnl{\if#2r{áö}\else\if#2t{Astron. Zhurn.}\else\ifnum#1<1993 Sov. Astron.\else Astron. Rep.\fi\fi\fi}}
\def\pazh#1#2{\ref@jnl{\if#2r{ðáö}\else\if#2t{Pis'ma Astron. Zhurn.}\else\ifnum#1<1993 Sov. Astron. Lett.\else Astron. Lett.\fi\fi\fi}}
\def\ufn#1#2{\ref@jnl{\if#2r{õæî}\else\if#2t{Usp. Fiz. Nauk}\else\ifnum#1<1993 Sov. Phys. Usp.\else Phys. Usp.\fi\fi\fi}}
\def\jetp#1#2{\ref@jnl{\if#2r{öüôæ}\else\if#2t{Zh. Eksp. Teor. Fiz.}\else Sov. Phys. JETP\fi\fi}}
\def\pjetp#1#2{\ref@jnl{\if#2r{ðÉÓØÍÁ öüôæ}\else\if#2t{Pis'ma Zh. Eksp. Teor. Fiz.}\else Sov. Phys. JETP Lett.\fi\fi}}

\makeatother
\def\figplace{t} 

\def\la{\mathrel{\mathchoice {\vcenter{\offinterlineskip\halign{\hfil
$\displaystyle##$\hfil\cr<\cr\noalign{\vskip1.5pt}\sim\cr}}}
{\vcenter{\offinterlineskip\halign{\hfil$\textstyle##$\hfil\cr<\cr
\noalign{\vskip1.0pt}\sim\cr}}}
{\vcenter{\offinterlineskip\halign{\hfil$\scriptstyle##$\hfil\cr<\cr
\noalign{\vskip0.5pt}\sim\cr}}}
{\vcenter{\offinterlineskip\halign{\hfil$\scriptscriptstyle##$\hfil
\cr<\cr\noalign{\vskip0.5pt}\sim\cr}}}}}
\def\ga{\mathrel{\mathchoice {\vcenter{\offinterlineskip\halign{\hfil
$\displaystyle##$\hfil\cr>\cr\noalign{\vskip1.5pt}\sim\cr}}}
{\vcenter{\offinterlineskip\halign{\hfil$\textstyle##$\hfil\cr>\cr
\noalign{\vskip1.0pt}\sim\cr}}}
{\vcenter{\offinterlineskip\halign{\hfil$\scriptstyle##$\hfil\cr>\cr
\noalign{\vskip0.5pt}\sim\cr}}}
{\vcenter{\offinterlineskip\halign{\hfil$\scriptscriptstyle##$\hfil
\cr>\cr\noalign{\vskip0.5pt}\sim\cr}}}}}
\begin{document}

\thispagestyle{empty}

\begin{center}
{\LARGE\bf Gravitational Wave Astronomy: \\
in Anticipation of First Sources to be Detected \\ 
\bigskip
}
{\large
L.\,P.\,Grishchuk$^{1,3}$,
V.\,M.\,Lipunov\,$^{2,3}$,
K.\,A.\,Postnov\,$^{2,3}$, \\
M.\,E.\,Prokhorov\,$^3$ and
B.\,S.\,Sathyaprakash$^1$ \\ }
\end{center}

\vfill

{\large
\begin{abstract}
The first generation of long-baseline laser 
interferometric detectors of gravitational waves
will start collecting data in 2001--2003. We carefully analyse their
planned performance and compare it with the expected strengths of 
astrophysical sources. The scientific importance of the anticipated discovery 
of various gravitatinal wave signals and the reliability of theoretical 
predictions are taken into account in our analysis. We try to be 
conservative both in evaluating the theoretical uncertainties about a 
source and the prospects of its detection. After having considered 
many possible sources, we place our emphasis on (1) inspiraling binaries 
consisting of stellar mass black holes and (2) relic gravitational waves.
We draw the conclusion that inspiraling binary black holes are likely
to be detected first by the initial ground-based interferometers. We
estimate that the initial interferometers will see 2--3 events per year 
from black hole binaries with component masses 10--15 M$_{\odot}$, 
with a signal-to-noise ratio of around 2--3, in each of a network of 
detectors consisting of GEO, VIRGO and the two LIGOs. It appears that 
other possible sources, including coalescing neutron stars, are unlikely 
to be detected by the initial instruments. We also argue that relic 
gravitational waves may be discovered by the space-based interferometers 
in the frequency interval $2\times 10^{-3}$~Hz--$10^{-2}$~Hz, at the 
signal-to-noise ratio level around~3. 
\end{abstract}
}
\vspace{1cm}

{\it
\noindent
$^1$~Cardiff University, P.O. Box 913, Cardiff, CF2 3YB, U.K. \\
$^2$~Physics Department, Moscow University, 117234 Moscow, Russia \\
$^3$~Sternberg Astronomical Institute, Moscow University,  
	119899 Moscow, Russia \\[\baselineskip]
}
e--mail:\\[1mm]
\begin{tabular}{ll}
L.\,P.\,Grishchuk:		& grishchuk@astro.cf.ac.uk \\
V.\,M.\,Lipunov:		& lipunov@sai.msu.ru \\
K.\,A.\,Postnov			& pk@sai.msu.ru \\
M.\,E.\,Prokhorov		& mike@sai.msu.ru \\
B.\,S.\,Sathyaprakash		& B.Sathyaprakash@astro.cf.ac.uk \\
\end{tabular}

\newpage

\setcounter{tocdepth}{3}
\tableofcontents
\newpage

\section{Introduction}

	The goal of this review article is quite ambitious. We want to
foretell the first gravitational wave signals that will be seen by sensitive 
detectors, several of which are currently in the final stage of construction. 
The detectors will start collecting data in a couple of years from now. 
Obviously, we present a subjective point of view. It is based on our 
evaluation of what we consider the best theoretical knowledge available 
today, in conjunction with the expected sensitivity of the instruments.
Possibly, other authors would regard other sources more promising, and
would place their bet on something else. It is also possible that our view
is biased, because it is partially guided by the work that we personally 
were involved in. We will not be very disappointed if we are proved wrong. Nature
may have many surprises in store for us. It is important, however, that for 
the first time in the long history of gravitational wave research, the 
conservative astrophysical estimates overlap with the detecting capabilities 
of real instruments. It is an appropriate time to prepare strategies for the 
search and analysis of signals that appear to be more probable than others.
	
	The general theory of gravitational radiation is well understood 
and is described in textbooks 
\cite{Weber61,L&L_v2,MTW73}.
The status of the 
gravitational wave astronomy has been regularly reviewed 
\cite{Thorne87_300yr, Thorne95_Next_Mill,LPG:sch},
including papers in Uspekhi 
\cite{Braginsky&Rudenko70:UFN,Grishchuk77:UFN,LPG:g1:c,Will94:UFN}. 
Here, we will only remind the reader that the gravitational waves are an 
inescapable consequence of Einstein's general relativity 
and, indeed, of any gravitational theory
which respects special relativity. Gravitational waves are similar to 
electromagnetic waves in several aspects. They propagate with the velocity 
of light $c$, have two independent transverse polarisation states, 
and in their action on masses have analogs of electric and magnetic 
components. Gravitational waves carry away from the radiating
system its energy, angular momentum, and linear momentum. The 
gravitational--wave field is dimensionless, and its strength is qualitatively 
characterized by a single quantity --- the gravitational wave amplitude $h$. 
The amplitude falls off in course of propagation from a localized source,
in proportion to the inverse power of the traveled distance: $h \propto 1/r$. 
The difficulty of direct detection of gravitational waves can be seen 
from the fact that the expected amplitude $h$ on Earth from realistic 
astronomical sources is exceedingly small, of the order or smaller
than $10^{-21}$. The conceivable amplitudes from laboratory sources are
even smaller than that. This small number $h$ enters any possible scheme of 
detection of gravitational waves and makes the detection difficult to 
achieve. For instance, gravitational waves cause a tiny variation $\Delta l$ 
of the distance $l$ between two free masses: $\Delta l = hl$. In an 
interferometer with a 1~km arm-length the variation of the distance
between the two end-mirrors would be of the order $\Delta l = 10^{-16}~$cm.  
This tiny variation is
supposed to be measured and distinguished against background noise.
However, in the cosmos, gravitational waves are an important factor of cosmic
evolution. Gravitational waves are routinely taken into account in the 
study of orbital evolution of close pairs of compact 
stars \cite{King_LesHoushes}. The measured secular change of orbital 
parameters in the binary system of neutron stars, which includes the 
pulsar PSR~1913+16, agrees with the gravitational wave prediction of 
general relativity to within 1\% accuracy \cite{S:psr1913}.
For the study of pulsars and this discovery, Hulse and Taylor were
awarded a Nobel prize in 1993.

	Like any other observational science, gravitational wave astronomy
operates with sources, detectors, data analysis, and interpretation.
In what follows, we devote some discussion to each of these notions. However,
we are not aimed at reviewing all interesting astrophysical theories and 
all possible signals and detection techniques. We concentrate on 
sources, which, we believe, rest on the most solid theoretical 
foundation, are scientifically important, and involve minimal number of 
additional hypotheses. To be interesting from the point of view of its 
detection, the source should be sufficiently powerful, should fall in 
the frequency band of the
detector, and occur reasonably often during the life-time of the instrument.
The frequency range of the discussed signals is determined by the frequency
intervals of the detectors's sensitivity. The currently operating bar 
detectors
are sensitive at frequencies around $10^3$~Hz. The ground-based laser
interferometers are sensitive in the interval 10~Hz -- $10^4$~Hz. The 
space-based laser antennas will be sensitive in the 
interval $10^{-4}$~Hz -- 1~Hz. The great expectations are related with
the forthcoming sensitive instruments. The Japanese scientists have
already built a 300~m laser intergerometer called TAMA. The British-German 
collaboration is in
the phase of completion of a 600~m laser interferometer 
called GEO600 \cite{S:geo600}. 
The French-Italian collaboration 
is building a 3~km interferometer called VIRGO 
\cite{S:virgo}.
The American project LIGO is building two 4~km arm-length 
interferometers \cite{S:ligo}. 
It is expected that 
these instruments will become operational in 1--2 years. The proposal 
to build a Laser Interferometer Space Antenna (LISA) \cite{LPG:lisa} has 
been tentatively approved by the European Space Agency and NASA, and the 
launch may occur around the year 2010. There exists also plans for 
advanced ground-based interferometers, such as LIGO-II \cite{LPG:ligoII}. 

	The ability of a given instrument to detect a
signal depends on the nature of the signal. The burst sources, which
accompany cosmic catastrophes, emit gravitational radiation at some 
characteristic frequency during just a few cycles. They have tendency to
be inherently powerful, but their event rate is very low. It is very unlikely
to expect such an event to happen in our own Galaxy during, say, a 1--year 
observational run. To see a few events per year, one needs to survey 
a large (cosmological) volume of space and, hence, to possess a sensitive 
instrument capable of detecting the sources from the edges of this volume.   
The quasi-periodic astrophysical sources are expected to be more frequent 
than the burst
sources, but they produce much weaker signals in terms of $h$. However,  
the amount of the radiated energy during some long time $T$ may be
not much smaller than that of a burst source. If one knows, or can model, the 
temporal structure of the signal, one can monitor the detector's output 
during many cycles that are covered by the observation time $T$. This can 
make a weak periodic signal not much more diffucult to detect than a 
burst signal. Some rare but reliable astrophysical sources, such as binary 
neutron stars and black holes at their latest stage of evolution, exhibit a 
kind of 
quasi-periodic gravitational wave signal
at the inspiral phase, and more like a burst signal in the last moments
of their coalescence and merging. The stochastic backgrounds of gravitational 
waves are typically weak and difficult to distinguish from the instrumental 
noise. However, if one can cross-correlate the outputs of two or more 
instruments, and can do this during a long integration time, the stochastic
background can also be measured. The fundamentally important 
relic gravitational waves form a sort of a stochastic background. 
They are the only direct probe of the evolution of the very
early Universe, up to the limits of the Planck era and Big Bang. It would
be extremely valuable, even if difficult, to detect relic gravitational 
waves. 

	The balance between the expected scientific payoff and theoretical 
likelihood of various astrophysical
sources versus their detectability by the forthcoming and planned instruments
is the major thrust of this paper. After having analysed many possible
sources of gravitational waves, and taking all the factors into account, we
place our emphasis on compact binaries (neutron stars and black holes) and
relic gravitational waves. In fact, we argue that inspiraling black holes,
formed as a result of stellar evolution, are the most likely sources to be
seen first by the forthcoming sensitive instruments. Also, we think that relic 
gravitational waves are likely to be detected by the advanced ground-based
and space-based laser interferometers. To justify our point we go into 
a great detail in describing compact binary stars and relic gravitons. 

	Section \ref{sec:secI} is devoted to the formation 
and evolution of binary 
systems. Binary stars are as numerous as single stars. Binaries 
emit gravitational radiation at twice their orbital frequency. To radiate
gravitational waves with large intensity and at frequencies accessible 
to ground-based interferometers, the objects forming a pair should
be massive and should orbit each other at very small separations --- a few
hundred kilometers. According
to the existing views, these massive objects can only be the end-products of 
stellar evolution --- neutron stars and black holes. Because of the loss of
the angular momentum due to gravitational waves, these binary objects are
in the late thousands of cycles at their inspiral phase. They are 
only tens of minutes away from the final coalescence and merging, or, 
possibly, from another spectacular event, a gamma--ray burst. 
The central question is how many
such close systems exist in our Galaxy and at cosmological
distances. This determines the event rate --- the number of coalescence events 
that can occur in a given volume of space during, say, 1 year. A detector, 
sensitive enough to see the most distant objects in this volume, will detect
all of them. A detector of lower sensitivity is capable of seeing the
coalescing systems at shorter distances and, hence, will register a 
smaller number of such systems, or will not be expected to see them 
at all during a 1--year interval of observation. 

	In sub-section \ref{sec:secI:observ} we review all the observational data on binary
neutron stars. Even these data alone, allow one to derive some estimates 
on the rate of neutron star coalescences. So far, there is no observational 
evidence of binaries consisting of a neutron star and a black hole 
or two black holes. However, we certainly do not see all the products 
of stellar evolution in binary systems. We need to
take into account the predicitons of a theory which successfuly 
explains the formation and 
relative abundance of various populations of observed binaries consisting of 
normal stars and neutron stars. Such a theory predicts the existence of 
close binaries involving neutron stars and black holes, as the outcomes of 
processes along certain channels of the binary evolution. It is these 
channels of evolution that are most important for gravitational 
wave astronomy.   

The sub-section \ref{sec:secI:population} 
is devoted to the population synthesis
method of describing the continuing birth and future fate of binary 
stars. The purpose of this analysis is to find the statistically
expected number of massive and sufficiently close binaries, which could
be in their final stage of inspiral at the present cosmological time. 
This means that we are interested only in those binaries whose expected 
total life-time, from formation to coalescence, is shorter than the
Hubble time. As usual, the results of evolution 
depend on initial conditions and on physical processes along the 
evolutionary path. We combine the well-established observational facts
with reasonable theoretical assumptions. Two parameters are especially
important --- the kick velocity $\vec{w}$ imparted to a newly born neutron
star during a supernova explosion, and the fraction $k_{BH}$ of a
pre-collapse massive star that goes into a resulting black hole. The
formation of a black hole can also be accompanied by the impartation of 
some kick velocity. A large 
kick velocity can either disrupt a binary system --- a would-be powerful 
source of gravitational waves, or, on the contrary, to make the binary
orbit more eccentric, thus increasing the gravitational wave luminosity.    
In our evolutionary calculations we vary $\vec{w}$ and $k_{BH}$ in the
observationally allowed limits. We also take into account the stellar
wind and the loss of mass as factors of binary evolution. The kick 
velocity is so important a factor of binary evolution
that we devote to its analysis a separate sub-section \ref{sec:secI:kick}.  

	The results of the population synthesis are summarised in
Section \ref{sec:secI:DetectRate}. 
These results are at the same time our predictions
for the detection rate of various compact binary inspiral signals. 
In a given cosmological volume of space, the estimated event rate 
for coalescing black holes is about 10 times lower than 
that for coalescing neutron stars and neutron star -- black hole systems.
However, since the masses of black holes are significantly larger than the 
masses of neutron stars, they are more luminous gravitational wave sources
than pairs of neutron stars. 
Hence, a given detector can observe inspiralling black holes at greater
distances than pairs of inspiralling neutron stars.  We conclude
that a network of initial laser interferometers are likely to see black hole 
inspirals more often than the neutron star inspirals, and as often as 
2--3 events per year. 

	Section \ref{sec:secII} is devoted to transient and periodic 
sources. They
include supernovae explosions, various unstable modes in rapidly rotating
neutron stars, and quasi-normal modes of black hole perturbations.
All these sources are interesting and potentially detectable. However, we
do not place them at the beginning of our priority list.
The asymmetric supernovae explosions, as well as the merging event of
binaries can produce powerful bursts of gravitational radiation, but the
estimates of their performance rely on factors which are not well understood
theoretically and do not have much of observational evidence. The merging
event will be probably seen as a confirming signature of the inspiral 
phase, but one cannot rely on this event alone. However, if we err in our
priorities, a special kind of hypernovae explosions can top the list.     
As for the unstable modes in rotating neutron stars, they require quite
sophisticated mechanisms of their excitation and can be hampered by viscosity
and other physical processes. The collision of black holes and 
quasi-normal modes of newly born 
black holes is an intriguing possibility, but should probably be treated 
as something to be discovered by gravity wave observations, rather 
than reliably calculated on purely theoretical grounds.  

	In Section \ref{sec:secIII} we review stochastic gravitational wave 
backgrounds of astrophysical origin. These are the overlaping signals from
many individual sources. The populations of sources that we consider
include unresolved binary white dwarfs in our Galaxy and at cosmological 
distances, and the population of rotating neutron stars.
The detection of these backgrounds would carry some scientific information
on its own, but it is also necessary to study these sources for another
reason. These stochastic backgrounds set a confusion limit for detection
of more interesting signals by space-based and ground-based interferometers.
We conclude that the LISA will be free of the gravitational wave noise
from unresolved binaries at frequencies near and higher 
than $2\times 10^{-3}$~Hz. This noise is mostly from unresolved binaries
in our Galaxy, the extra-Galactic binaries contribute only 
about 10\% to this noise. Any detected stochastic background at 
frequencies above $2\times 10^{-3}$~Hz in the LISA window of sensitivity
is expected to be of primordial origin. 
The population of non-axisymmetric rotating neutron stars could potentially
blur the view of the ground-based interferometers. However, we find that
this background is below the instrumental noise of initial interferometers,  
and can possibly present a problem only for the advanced LIGO. We estimate
that other stochastic backgrounds of astrophysical origin are weaker than
those that we have considered.  

	Section \ref{sec:secIV} is devoted to relic gravitational waves. 
In contrast
to all other sources, which are based on classical physics, the generation
mechanism of relic gravitons includes some elements of quantum physics.
It is the inevitable zero-point quantum oscillations of gravitational 
waves amplified by the strong, variable gravitational field of
the early Universe that ends up in a stochastic background of 
relic gravitational waves measurable today. Despite the fact that the
existence of this gravitational wave signal involves an extra 
element --- quantum physics, it is not less reliable than many other 
sources. The generation of relic gravitons relies essentially only on 
the validity of general relativity and basic principles of quantum 
field theory. Since the same mechanism is thought to be responsible for
the generation of primordial density perturbation seeding the formation
of galaxies, we present a qualitative picture of this mechanism in
sub-section \ref{sec:secIV:intro}. The calculation of the expected relic gravitational
wave background is given for a class of cosmological models supported by 
other observations. In particular, we use the data on the measured 
microwave background anisotropies. The results of this analysis are
presented in sub-section \ref{sec:secIV:detectability}. 
We find that in the most favorable
case, the detection of relic gravitational waves can be achieved by
the cross-correlation of outputs in the initial ground-based laser 
interferometers. In the more realistic case, the sensitivity of the 
advanced ground-based
and space-based instruments will be needed. We also discuss a specific
statistical signature of relic gravitons, associated with the phenomenon
of squeezing. This phenomenon is also known in formal quantum mechanics 
and quantum optics. The signature of squeezing could potentially help in 
further improving the signal-to-noise ratio (SNR).  

	The problems of detectability are systematically referred to 
throughout the paper. However a rigorous discussion of detectors and data 
analysis is concentrated in Section~\ref{sec:detectors} 
and Section~\ref{sec:DataAn}. Whenever we qualify a source as detectable
or undetectable, we base our conclusions on the more detailed treatment
of these Sections. Section~\ref{sec:detectors} gives a  
general description of detectors and their sensitivity curves. The important
notion in the detection of a signal with a known or suspected temporal
structure, is the notion of a template. A template allows one to use the
mathched filtering technique 
(sub-section~\ref{sec:matched filtering}) 
in order to increase 
the SNR. This method will be indispensable in the search 
for signals from inspiraling binaries. The related issues are the 
practically accesible number of templates, their overlap in the parameter  
space, the computational cost, etc. These issues are important not only
for a confident detection of a signal, but also for the extraction of 
astrophysical information from a signal 
(sub-sections \ref{sec:compute costs}--\ref{sec:estimation}) 
--- the ultimate purpose of the gravitational wave astronomy.

	Some mathematical details on the Keplerian motion of a binary
system and gravitational reaction force are described 
in Appendix \ref{sec:appA}.
Appendix~\ref{sec:appB} contains some technical 
issues of the mass transfer modes and mass loss in binary stars.
Appendix~\ref{sec:appC} gives post-Newtonian expressions for energy and
gravitational wave flux. The main conclusions of the review are
formulated in the Abstract.

\section{Astrophysical sources. Close binary neutron stars and black holes.}
\label{sec:secI}

In this Section we discuss observational and theoretical
estimates for the coalescence rate of close binary neutron stars and
black holes. We start from a review of observational limits on the
coalescence rate of binary neutron stars. Then, we describe the basics of
the population synthesis of binary evolution which allows one to predict
theoretically the event rates for systems involving neutron stars and 
black holes. The role of the kick velocity in the binary evolution 
is discussed in subsection \ref{sec:secI:kick}. 
The expected detection rates in the
forthcoming sensitive gravitational wave detectors are summarised
in Section~\ref{sec:secI:DetectRate}. %

\subsection{Observational limits on the binary neutron star
coalescence rate}
\label{sec:secI:observ}

What do we know about compact binary stars and the rate of their
mergings on {\it observational} grounds?
More than a thousand of single neutron stars
(NS) are currently 
(\protect\ref{bin:LL})observed as radio-pulsars (see \cite{PSR_catalog_93}; new 
data are being continuously added at 
{\bf http://puppsr.princeton.edu}). In addition,
about 30 NS are
seen as X-ray pulsars and yet more 100 NS are seen
as burst and transient X-ray
sources. These NS enter binary systems with
non-degenerate companions, that is, their companions are normal stars 
rather than neutron stars or black holes. 
Only six NS are known to enter binary
systems with another NS as a secondary component
\footnote{New binary pulsars
are found in recent pulsar surveys (see e.g \cite{D'Amico&99,Camilo&2000}). 
However,
a reliable determination of the component masses 
is only possible after sufficiently long-term observations.}.  
All these six systems
belong to binary radio-pulsars. The systems and some of their
parameters are listed in
Table~\ref{t:psr}. Orbital periods are given in days and masses
in units of the solar mass $M_\odot$. Three of these systems
(namely, B1913+16, B1534+12 and B2127+11c)
are close enough to merge due to GW emission in a time interval
shorter than the Hubble time $t_H$. 
We loosely refer to binaries as coalescing
or merging binaries if their expected life-time up to coalescence,
$t_{coal}$, is shorter than $t_H$. For numerical estimates we 
use the value $t_H\simeq 12 {\rm Gyr} = 12\times 10^9 {\rm yr}$. 

Much less is known about black holes (BH).
A dozen of BH candidates participate in binary systems with
non-degenerate companions.
They are observed as persistent X-ray sources (like Cyg~X-1) or 
X-ray transients (mostly X-ray Novae) (see \cite{Cher96_BH} for a review).
Neither single BH nor BH forming a binary with radio-pulsar
or another BH have been found so far.
Parameters of the BH candidates in binary systems are listed 
in Table~\ref{t:bh}. Note that according to
these data, the mean BH mass is $M_{\rm BH}\simeq 8.5M_\odot$,
i.e. notably higher than a typical NS mass $M_{\rm
NS}\simeq 1.4M_\odot$. (Of course, we mean a black
hole in astrophysical sense, i.e. as a highly compact
gravitating object of certain mass. The presence or absence of 
the event horizon is irrelevant for our
discussion.) For a recent summary of NS mass determination see 
\cite{Thorsett&Chakrabarty99}.

\begin{table}[h]
\begin{center}
\caption{Binary PSR with NS secondaries (data from \protect\cite{Nice&96}) }
\label{t:psr}
\begin{tabular}{lcccccr}
\noalign{\smallskip}
\hline\hline
                     &       &       &&&\\[-2mm]
\quad PSR            &$P_b(d)$& $e$  &$M_1+M_2$& $M_1$&$M_2$      & $t_{coales}$, yr\\[2mm]
\hline
                     &       &       &&&\\[-3mm]
J1518+4904           & 8.634 & 0.249 &2.62     &$\ldots$&$\ldots$ & $\ge$$3.6\,10^{12}$ \\
B1913+16$^1$         & 0.323 & 0.617 &2.8284   &1.44    &1.39     & $1.0\,10^{8\ }$    \\
B1534+12$^1$         & 0.420 & 0.274 &2.6784   &1.34    &1.34     & $1.0\,10^{9\ }$    \\
B2127+11c$^{1,\,2}$  & 0.335 & 0.681 &2.712    &1.35    &1.36     & $8.0\,10^{7\ }$    \\
B2303+46             &12.340 & 0.658 &2.60     &$\ldots$&$\ldots$ & $\ge$$1.6\,10^{12}$ \\
B1820-11$^3$         &357.762& 0.795 &         &        &         & $\ga$$2.4\,10^{15}$ \\[1mm]
\hline\hline
\noalign{\smallskip}
\multicolumn{6}{l}{$^1$ Coalescing binary pulsars} \\
\multicolumn{6}{l}{$^2$ Binary pulsar in a globular cluster} \\
\multicolumn{6}{l}{$^3$ The secondary companion may not be a NS} \\
\end{tabular}
\end{center}
\end{table}

\begin{table}[t]
\caption{BH Candidates \protect\cite{Cher96_BH}.}
\label{t:bh}
\smallskip
\centerline{%
\begin{tabular}{llcccc}
\hline\hline
&&&&&\\[-2mm]
\multicolumn{1}{c}{System} & Spectral class & $P_{orb}$, d & $f_v(m), M_\odot$
    & $m_x, M_\odot$ & $m_v, M_\odot$ \\[2mm]
\hline
&&&&&\\[-3mm]
Cyg X-1     & O9,7 Iab       & 5.6 & 0.23 & 7--18  & 20--30    \\
LMC X-3     & B(3--6)II--III & 1.7 & 2.3  & 7--11  & 3--6      \\
LMC X-1     & O(7--9)III     & 4.2 & 0.14 & 4--10  & 18--25    \\
A0620-00    & K(5--7)V       & 0.3 & 3.1  & 5--17  & $\sim$0.7 \\
GS2023+338  & K0IV           & 6.5 & 6.3  & 10--15 & 0.5--1.0  \\
GSR1121-68  & K(3--5)V       & 0.4 & 3.01 & 9--16  & 0.7--0.8  \\
GS2000+25   & K(3--7)V       & 0.3 & 5.0  &5.3--8.2& $\sim$0.7 \\
GRO J0422+32& M(0--4)V       & 0.2 & 0.9  &2.5--5.0& $\sim$0.4 \\
GRO J1655-40& F5IV           & 2.6 & 3.2  & 4--6   & $\sim$2.3 \\
XN Oph 1977 & K3             & 0.7 & 4.0  & 5--7   & $\sim$0.8 \\[1mm]
Cyg X-3 ?\\[1mm]
\hline
&&&&&\\[-3mm]
\multicolumn{3}{l}{Mean Value of the BH mass}
                                   &      &$\sim$8.5M$_\odot$ &  \\[1mm]
\hline\hline
\end{tabular}}
\end{table}

There are two types of estimates of the 
binary NS coalescence rate. The estimates of the first type are derived 
directly from observations (see Table~\ref{t:cls-observ}), while the 
estimates of the second type are inferred from theory
of binary stellar evolution (see Table~\ref{t:cls-theor}).
We will consider each of these estimates in turn.

\begin{table}[h]
\begin{center}
\caption{Observational estimates of binary NS coalescence rate.}
\label{t:cls-observ}
\begin{tabular}{lc}
\hline\hline
\multicolumn{1}{c}{Author(s)} & Coalescence rate (yr$^{-1}$)\\
\hline
Phinney 1991~\cite{Phinney91} & $1/10^6$\\
Narayan et al 1991~\cite{Narayan&91} & $1/10^6$\\
Curran, Lorimer 1995~\cite{Curran&Lorimer95} & $3/10^6$\\
van den Heuvel, Lorimer 1996~\cite{Heuvel&Lorimer96} & $8/10^6$\\
``Bailes limit'' 1996~\cite{Bailes96_limit} & $<1/10^5$\\
Arzoumanian et al. 1999~\cite{Arzoumanian&99} & $<1/10^4$\\
\hline\hline
\end{tabular}
\end{center}
\end{table}

\cbeg{1}
The estimates of the first type are based on the data
on three binary radio-pulsars which should merge within the Hubble
time (Table~\ref{t:psr}). These estimates
use the following argumentation.
The average coalescence time for these pulsars is approximately
$3 \times 10^8$~years. So the binary NS merging rate based on
these 3 pulsars would be approximately once per
100 million years. As we observe only about 1\% of the Galactic
volume, a {\it lower} limit for the binary NS merging rate becomes
one every million years \cite{Phinney91}. In fact, this estimate was
formulated at the time when only two of the presently known three
coalescing binary radio-pulsars were known. Taking into account the spatial 
distribution of pulsars inside the Galaxy and the fact that a typical
radiopulsar switches-off long before the coalescence, the lower limit
for NS merging rate
can be increased by almost an order of magnitude
\cite{Heuvel&Lorimer96}, thus reaching $10^{-5}$ per year.

An interesting {\it upper} limit, the so-called ``Bailes limit'', was
derived from independent arguments \cite{Bailes96_limit}.
It was noted that the properties of the pulsars in the
three coalescing binary radio pulsars (most of all, their surface
magnetic fields) are
quite different from those found in ordinary single radio pulsars.
Since the number of single radio pulsars is about 1000,
it is estimated that the radio pulsars similar to those residing 
in merging binary NS should be formed al least $\sim 1000$
times rarer than single radio pulsars.
Taking the birth rate of single pulsars from a large sample of
known pulsars, Bailes proposed to put
an upper bound on the birth
rate of binary NS as the (formation rate of single pulsars)$\times$
(number of pulsars with ordinary properties among binary radio-pulsars) = 
(1/60 yr)$\times$(1/1000)$\simeq 2\times 10^{-5}\,$yr$^{-1}$.
%
%

One should note, however, that in
both the estimates -- the one based on statistics of coalescing binary
radio pulsars, and the other based on the Bailes' limit --
suffer from selection effects. They depend
on the pulsar distances (in some cases known not better
than up to a factor of 2) \footnote{For example, 
recent observations of PSR~1534+12
\cite{Stairs&98}
suggest a distance which is two times larger than
was previously thought, so the ``observational estimate''
of binary NS mergings should be decreased by a factor $\simgt 2$.}, 
on the
characteristic pulsar life-time (known not better than up to an order of
magnitude), and on the differences in the properties of single and binary
pulsars. So, one cannot infer from observations an absolutely reliable 
estimate of the binary NS merging rate. 
Indeed, a recent re-assessment of the Bailes limit
\cite{Arzoumanian&99} taking into account the current pulsar numbers and
the reduction in search sensitivity to short orbital period binaries
gave $10^{-4}$ yr$^{-1}$ for the upper limit of binary NS Galactic
merging rate. An alternative way of deriving the upper limit
based on empirical pulsar birth rate and theoretical understanding
of binary NS formation was used in \cite{Kalogera&Lorimer99} to yield
a few mergers per $10^5$ years. The cited papers clearly demonstrate that
(1) there is a steady tendency to
{\it increase} the empirical upper limit of the binary NS coalescence rate
and (2) various selection effects generic to radio pulsar surveys
and lack of detailed knowledge of Galactic pulsar population properties still
prevent us from the derivation of a fully reliable estimate.

\subsection{Population synthesis of coalescing binary NS and BH}
\label{Scenario}
\label{sec:secI:population}

Now we turn to estimates partially based on theoretical grounds. 
The merging rates of binary compact stars have been calculated by
different independent research groups, 
mostly with the help of the population synthesis numerical simulations
(see Table \ref{t:cls-theor}).
The reliability of these results depend on whether
the binary evolution scenarios properly reflect different aspects
of the real, observed populations. It is encouraging that 
these independent calculations yield similar results.

\begin{table}[b]
\begin{center}
\caption{Theoretical estimates of binary NS coalescence rate.}
\label{t:cls-theor}
\begin{tabular}{lc}
\hline\hline
\multicolumn{1}{c}{Author(s)} & Coalescence rate (yr$^{-1}$)\\
\hline
Clark et al 1979~\cite{Clark&79} & $1/10^4$--$1/10^6$\\
Lipunov et al 1987~\cite{LPP87_gw} & $1/10^4$\\
Hills et al 1990~\cite{Hils&90} & $1/10^4$\\
Tutukov, Yungelson 1993~\cite{Tut&Yung93} & $3/10^4$--$1/10^4$\\
Lipunov et al 1995~\cite{L&95_gwsky} & $<3/10^4$\\
Portegies Zwart, Spreeuw 1996~\cite{PortZw&Spreeuw96} & $3/10^5$\\
Lipunov et al 1996~\cite{LPP96_SM} & $3/10^4$--$3/10^5$\\
Portegies Zwart, Yungelson 1998~\cite{PortegiesZwart&Yungelson98}
& $\sim 1/10^4-3/10^{5}$\\
\hline\hline
\end{tabular}
\end{center}
\end{table}

Theoretical estimates of double NS
coalescence rate are systematically higher than
the observational ones by, on average, an order of magnitude. 
This does not mean that the estimates are in conflict with each other. 
The main reason for the discrepancy is that the observational estimates 
directly refer only to the merging rates of binary NS, in which one
of the components is a radio pulsar.
The participation of a pulsar in a binary system is of course 
not a {\it necessary}
condition for the system to be interesting from the point
of view of gravitational wave astronomy. 
For example, the neutron stars could be born 
with weak magnetic fields and/or slow rotational periods, and,
hence, they would never manifest themselves as radio pulsars.
Theoretical calculations provide a broader range of estimates because
they depend on several evolutionary parameters which are not well
known. However, the population synthesis calculations are the
only way to estimate the coalescence rates of, so far unobserved, 
compact binaries consisting of two BH or a NS and a BH.   
These systems are of a great importance for gravitational wave astronomy. We will consider below the population synthesis for all possible pairs: NS+NS, NS+BH, and
BH+BH.

\subsubsection{Basics of population synthesis}

Coalescence rates
of different types of binary compact stars are calculated using
modern theory of binary star evolution
(e.g. \cite{Heuvel94_int_bin} 
and references therein).
A full description of the method can also be found in 
\cite{LPP96_SM}. Some key formulas for binary
system evolution are summarized in Appendix~\ref{A:mass_transf}.

Binary stars are formed with different initial 
masses, semi-major axes, eccentricities, etc.. These initial 
parameters are drawn
from certain distribution laws. Also, there are some other physical 
parameters important for binary evolution,
such as the efficiency $\alpha_{CE}$ of the angular momentum 
removal in the common envelope
stage, or kick velocity distribution $f(\vec{w})$
for newly born neutron stars.
This means that in order  to calculate
the expected rate of binary star mergers, we need to derive 
the number of binaries formed in all appropriate regions of the
parameter space, and then integrate over these regions.
Some distributions can be more or less accurately deduced from astronomical 
observations. To these belong the initial stellar mass function
and the distribution over binary semi-major axis. Distributions 
of other physical parameters are being adopted on theoretical grounds.
We perform evolutionary calculations using the ``Scenario Machine'' 
code --  a version of the Monte-Carlo method (see \cite{LPP96_SM} for review). 
In a typical numerical experiment, some 
$\sim$$10^6$ binary evolutionary tracks with different initial conditions
are calculated. Similar approaches in the binary evolution studies 
are being used by other groups
(e.g. \cite{PortegiesZwart&Yungelson98}) and are commonly named ``population
synthesis" methods.

\subsubsection{Initial binary parameters}

The initial components of a binary system are taken  
as zero age main-sequence stars.
The initial  parameters which determine the subsequent binary evolution are:
the mass of the primary component 
$M_1$, the binary mass ratio
$q=M_2/M_1<1$, and the orbital separation $a$.
For sufficiently close binaries, which are capable of producing a
merging NS or BH, we assume that the orbits are circular. This assumption is 
justifiable since the tidal interaction between the components
is effective enough to circularize the orbit.  
           
The distribution of binaries
over initial orbital separations is partially known 
from observations~\cite{Abt83}:
\begin{equation}\label{init_distr_a}
\eqalign{
&f(\log a) =\mbox{\rm const,}~ \mbox{\rm for $a$ such that}  \cr
&\max~\{10~ {\rm R}_\odot,~\hbox{Roche Lobe}~(M_1)\} < a < 10^7~{\rm R}_\odot.
\cr}
\end{equation}

We assume that the distribution of masses of the primary (more massive)
components obeys the Salpeter mass
function found for the birth rates of main-sequence stars in the solar
vicinity~\cite{Salpeter_law}: 
\beq{e:Salpeter_law}
f\myfrac{M_1}{M_\odot}=\myfrac{M_1}{M_\odot}^{-2.35}.  
\eeq
The observed star formation rate in the Galactic disk 
relates to this distribution as 
\beq{init_distr}
\label{init_distr_m}
\frac{dN}{dM_1dt}=0.9~\hbox{yr}^{-1} f\myfrac{M_1}{M_\odot}\,,\quad 0.1~{\rm M}_\odot < M_1< 120~{\rm M}_\odot\,.
\eeq
Assuming that $50\%$ of the total number of
stars in the Galaxy reside in binaries,
this distribution law predicts one massive star
($M_1>10$M$_\odot$ to be able to produce a compact remnant)
to form in a binary system, approximately every 60 years.
This estimate agrees with the binary birth rates
derived from observations of binary stars~\cite{Popova&82}.

The initial mass ratio $q$ in a binary is crucial for
its subsequent evolution \cite{Trimble83}
because it determines the mode of the first mass transfer 
between the components. The initial distribution over $q$
has not been reliably derived from observations due to various selection
effects. Usually, one makes the `zero-order assumption', according to which 
the mass ratio distribution is flat, i.e. low mass ratio binaries
are formed as frequently as those with equal masses
(e.g.~\cite{Heuvel94_int_bin}):  
\beq{init_distr_q}
 \frac{dN}{dq} = const\,,\quad q\equiv M_2/M_1<1 \,.
\eeq
In the calculations presented below we used this prescription.
The effect of the initial $q$ distribution 
on binary pulsar statistics was studied in
ref.~\cite{LPP96_restrict}.

\subsubsection{Neutron star kick velocities}

In the course of evolution of massive binary stars, one or two neutron stars
can form. 
One of the most important parameters affecting 
the eventual binary NS coalescence rate is the kick velocity $\bf w$
imparted to a NS at its birth. There exists a plenty of observational 
evidence for the kick velocities. 
The impact of a kick velocity $\sim$ 100 km/s explains the 
precessing binary pulsar orbit in 
PSR~J0045-7319 \cite{Kaspi&96_kick}. The evidence of the kick velocity
is seen in the inclined, with respect to the orbital plane, 
circumstellar disk around the Be star SS~2883 --- an optical component 
to the binary pulsar PSR~B1259-63 
\cite{PP97_kick}. One more direct evidence comes from observations of the
geodetic precession in the famous binary pulsar
PSR~1913+16 \cite{Wex&00a,Wex&00b}.
All these observations indicate that in order to produce
the observed mis-alignment between the orbital angular 
momentum and the neutron star spin, 
a component of the kick velocity perpendicular to the orbital
plane is required. 
A non-zero kick velocity is also required in order to properly explain 
the observed pulsar velocity distribution
\footnote{For an alternative point of view see~\protect\cite{Iben&Tut96}
and for its criticism \cite{PortegiesZwart&vdH99}.}.

Most likely, the origin of the kick velocity should be
attributed to the asymmetry in the supernova explosion. 
Astrophysical evidence for the existence of a substantial
kick velocity during supernova explosions was discussed
in \cite{vdH&vP97} and recently summarized in
\cite{Tauris&vdH00a,Tauris&vdH00b}. 
However, the nature of concrete physical mechanisms 
giving rise to the kick velocity is still unclear
(see \cite{Lai99} for recent review). A promising possibility 
capable of producing small and moderate kick velocities (up to 100 km/s)
involves asymmetric neutrino emission in strong magnetic field of
a newly born NS \cite{Dorofeev&85,BK93}.

Spatial velocities of NS are usually derived from direct
observations of proper motions of single radio pulsars
\cite{Lyne&Lorimer94,Lyne&Lorimer94,Blaauw&Ramachandran98}. 
Or, with more uncertainty,
from the observed offsets in the positions of young pulsars 
relative to the centers of their associated supernova remnants 
(e.g. \cite{Frail&94}). Both methods determine, 
however, only the component of the space
velocity which is transversal to the line of sight. 
One should also bear in mind 
the existing uncertainty in pulsar distances, which affects the 
evaluation of the kick velocity. On average, the uncertainty in
the distance scale is $\sim 30\%$ \cite{Lorimer97_priv},
and can be as much as a factor 2 in individual cases. 
In general, the observed distribution of transversal pulsar velocities 
is recognized to have, (1) a high mean
value ($\sim$200-350 km/s) and (2) a broad shape with
a high-velocity tail up to 1500 km/s. 

It is a difficult problem to derive the intrinsic kick
velocity distribution from these data.
If all the pulsars, presently seen as single pulsars, took their origin from 
single massive stars,
their velocity distribution would have exactly reflected the initial
kick velocity distribution.  This is because single massive stars have very small 
(of order 10 km/s) spatial velocities. However, when a supernova explosion
occurs in a binary system, which gets disrupted as a result of the 
explosion, the neutron star can acquire a substantial
space velocity, equal to the orbital velocity of the progenitor,
even without any additional kick velocity. If a non-zero kick is present, 
it becomes practically impossible to solve analytically the inverse problem
for the intrinsic kick velocity distribution from the 
observed pulsar velocities.

So, the only way to check
the very assumption of the non-zero kick is to 
find numerically the pulsar velocity distributions 
arising from various theoretical kick distributions, 
and compare the results with observations.  
This is usually done by the Monte-Carlo
simulations of binary evolution.
Presently, there is no general agreement
with regard to the form of the kick distribution.
A Maxwellian distribution for $w = |\vec{w}|$  
\begin{equation}
f_m( w)\propto w^2 \exp(-w^2/w_0^2)\,,
\label{bin:Mxw}
\end{equation}
with $w_0=190$ km/s, was used by Hansen and 
Phinney \cite{Hansen&Phinney97} in order to fit
the observed pulsar velocities.
A different form of the kick velocity
distribution was suggested in \cite{LPP96_restrict}.
It was found that this latter distribution
fits well with the observed 2D pulsar velocity distribution
given by Lyne and Lorimer~\cite{Lyne&Lorimer94}.
In contrast to the Maxwell-like distribution, the proposed 
distribution function has a power-law shape:
\beq{bin:LL}
f_{LL}(|\vec{w}|)\propto  \frac{(|\vec{w}|/w_0)^{0.19}}
{(1+(|\vec{w}|/w_0)^{6.72})^{0.5}}\,
\eeq
and assumes $w_0 \approx 400$ km/s. 
In our calculations described below, we have used this form of the NS
kick velocity distribution, where $w_0$ is treated as a free parameter.
A study of the kick velocity effects 
on the binary neutron star merging rate can be found in \cite{LPP97_mnras}
and \cite{Belczynski&Bulik99}. We give a detailed analysis of these
effects in subsection \ref{sec:secI:kick}.

\subsubsection{Binary neutron star formation and merging}

We are interested in evolutionary tracks 
which lead to the formation of a pair of coalescing NS.
Detailed studies of possible evolutionary channels which produce 
merging binary NS can be found in the literature
(e.g. \cite{Tut&Yung93,LPP97_mnras,PortegiesZwart&Yungelson98,Bagot97,Wettig&Brown96}).
Usually, the evolutionary analysis is being done in the following order: 
one starts from the observed parameters of the binary and tries to
deduce the parameters of the supernova progenitor
and then,  the initial binary masses and orbital separation.
In contrast, the Monte--Carlo population synthesis method,
which we apply, evolves a trial binary
and looks for appropriate results by changing 
the initial parameters within their distributions.
One of such typical tracks calculated by us is sketched 
in Fig.~\ref{f:track}, and we will explain it in detail.

\if t\figplace
\begin{figure}
\begin{center}
\epsfxsize=0.5\hsize
\fbox{\epsfbox{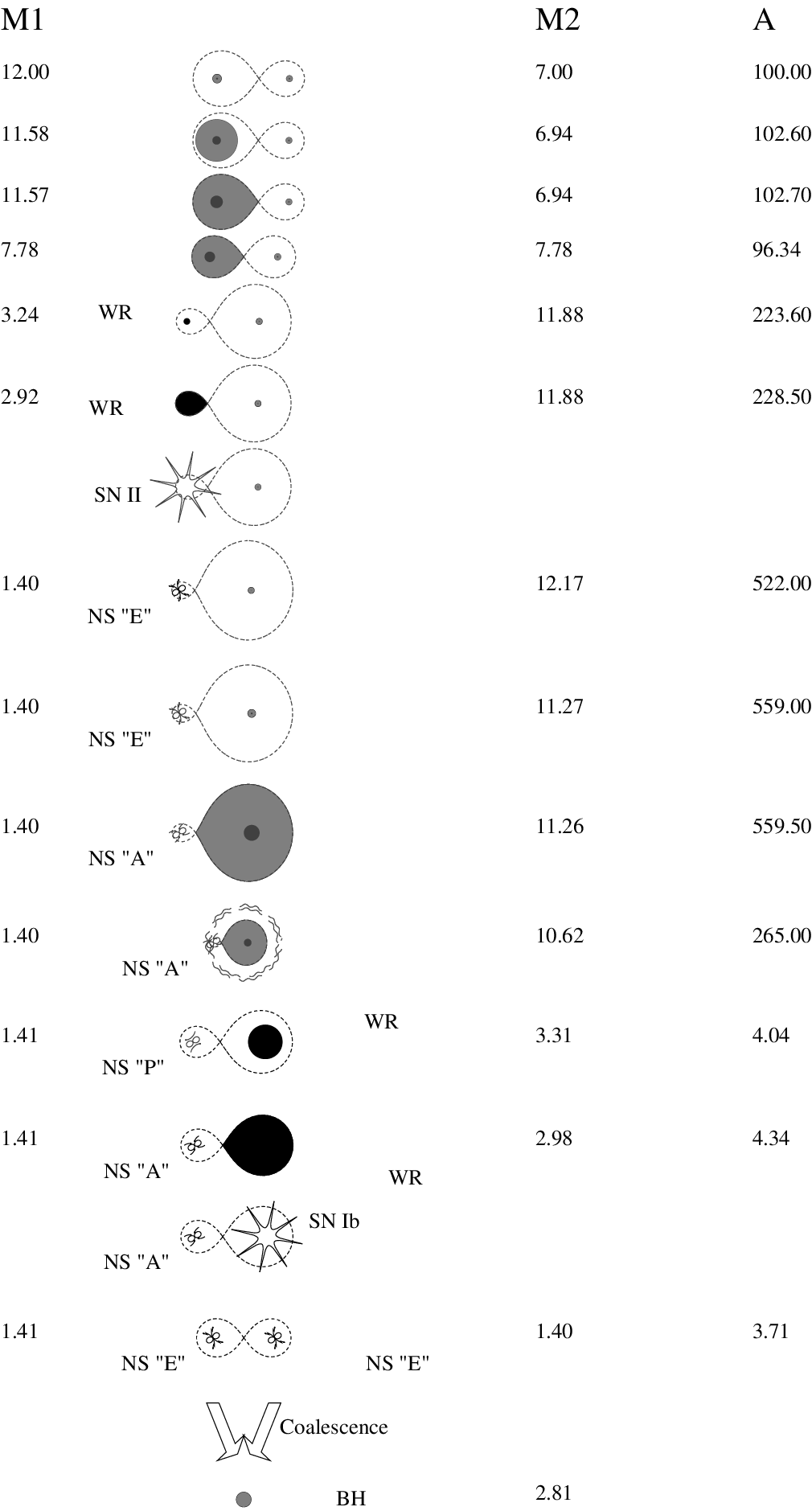}} 
\caption{Evolutionary track of a massive binary star leading
to the formation and coalescence of two NS.}
\label{f:track}
\end{center}
\end{figure}
\fi

Close NS binaries originate from two sufficiently massive 
main-sequence stars separated by a moderate distance of
order 100 solar radii (1-st row in Fig.~\ref{f:track}).
To get a NS in the course of evolution, the mass of the 
progenitor star must be larger than $\approx$10M$_\odot$ 
initially or, in any case, taking into account a possible
mass transfer in close binary, the mass should be $\approx$10M$_\odot$ 
during the stage of nuclear burning.
The more massive the primary star is, the faster it evolves.
For main-sequence stars, the time of core hydrogen burning
is $t_{nucl}\propto M^{-2}$.
The star burns out its hydrogen in its central parts, so that
a dense central helium core with a mass
$M_{He}\sim 0.1 (M/M_\odot)^{1.4}$
forms by the time when the star leaves the main-sequence. 
The outer shell expands and
the star moves towards the red 
super-giant region in the Hertzsprung--Russell
diagram. At some stage of its evolution
the star fills out its Roche lobe [Eq.~(\ref{A:Roche})]
(3-rd row in Fig.~\ref{f:track}).
The hydrogen envelope starts outflowing onto the secondary,
less massive star, which still resides on the main-sequence.
The primary star is being continuously stripped off of its hydrogen 
envelope, until a naked helium core emerges.
This core can be observed as a hot compact
helium star, or, for more massive stars, 
as a Wolf--Rayet star with intensive stellar wind (5-th row). 

While the mass of the primary star reduces,
the mass of the secondary star increases,  
since the mass transfer at this stage is thought 
to be quasi-conservative.
For not too massive main-sequence stars, $M \la 20$~M$_\odot$,
no significant stellar wind mass loss occurs
which could, otherwise, remove matter from the binary.
The secondary star acquires a large angular momentum
due to the infalling material, so that its outer envelope can be spun up
to an angular velocity close to the limiting (Kepler orbit) value.
Such massive rapidly rotating stars are observed as Be-stars.   
During the conservative stage of mass transfer, 
the semi-major axis of the orbit first decreases, reaches a minimum
when the masses of the binary components become equal to each other, and 
then increases. This behavior is dictated by  
the angular momentum conservation law [Eq. (\ref{A:conserv})].
After the completion of the conservative mass transfer, 
the initially more massive star
becomes less massive than its initially lighter companion. The parameter
$q= M_2/M_1$ becomes larger than 1.
In a short time, typically $\sim 10\%$ of the hydrogen
burning time, the nuclear evolution of the helium star is completed
and, provided its mass is larger than 2-3 $M_\odot$,
it explodes as a core-collapse supernova type-II
leaving a neutron star as its remnant.

Even for asymmetric supernova explosions, 
most of such binaries do not get disrupted. 
This is because the mass ratio $q=M_2/M_1$ of the pre-supernova
binary becomes generally high,  $q\approx3$--5.
After the first SN explosion, the binary system consists of 
a Be-star and a NS in an elliptical orbit  (8-th row).
Orbital evolution following the SN explosion
is described in more detail in an Appendix 
[Eqs. (\ref{A:SN:afai}--\ref{A:SN:symm-ecc})].

Be-stars have very rapidly rotating envelopes but
in other respects they do not differ from ordinary main-sequence stars.
After the completion of hydrogen burning in the core, a Be-star
starts expanding until it fills out its Roche lobe while passing through 
the periastron of an elliptical orbit (10-th row).
This initiates the second episode of the mass transfer, which takes
place on the thermal scale of the Be-star, typically 
$\sim 10^{-6} M_\odot$/yr. However, this mass transfer is
qualitatively different from the first one, since 
the mass transfer is now on to a compact star.
Once the accretion rate exceeds the value which provides the
luminosity equal to the Eddington luminosity limit near the NS surface
($\sim 10^{-8} M_\odot$/yr), the NS
cannot accrete all the infalling matter. The so-called common envelope
stage arises (11th row) during which the neutron star finds itself 
inside quite dense outer layers of the companion star.
Numerical hydrodynamic calculations 
\cite{Terman&96,Rasio&Liv96} 
show that the dynamical friction  
of the orbiting NS leads to an efficient transfer of
the orbital angular momentum to 
the common envelope, thus dispersing it on a very short 
timescale (typically, $10^3-10^4$ years). The semi-major 
axis of the binary system reduces dramatically [Eq. (\ref{A:CE:eq})], 
which results in the formation of a close binary system consisting
of a NS and a WR star (12th row). Alternatively, the NS can sink 
into the center of its red giant companion (the so-called Thorne--Zytkow 
object; not shown in this Figure).

In a short time ($\la 10^5$ years), the companion WR star
explodes as a supernova type~Ib, thus producing a second neutron star.
During this explosion, the system is more likely to be disrupted 
than during the first SN explosion, since the exploding star is 
now more massive than its companion. Surviving systems form 
close high-eccentric NS binaries, similar to the NS binary 
PSR~1913+16. Orbital parameters of
such binaries change exclusively due to the emission of
gravitational waves (see section \ref{A:GW_evol}). If the neutron
stars are close enough, they coalesce in a time shorter than $t_H$.

\subsubsection{Black hole formation parameters}

So far, we have considered the formation of individual NS
and their binaries. It is believed that very massive stars 
end up their evolution with the formation of stellar mass black holes.
We will discuss now the formation of an individual black hole.

In the analysis of BH formation, new important 
parameters appear. The first one is the threshold mass $M_{cr}$
beginning from which a main-sequence star,
after the completion of its nuclear evolution, can collapse into a BH.
This mass is not well known, different authors assume
different values: van den Heuvel and Habets~\cite{Heuvel&Habets84} ---
40M$_\odot$; Woosley et~al.~\cite{Woosley&95} --- 60M$_\odot$;
Portegies Zwart, Verbunt, Ergma~\cite{PortZw&97}
--- more than $20M_\odot$. A simple physical argument usually cited
in the literature
is that the mantle of the main-sequence star with 
$M>M_{cr}\approx$30M$_\odot$
is bound before the collapse with the binding energy well above $10^{51}$
ergs (typical supernova energy observed),
so that the supernova shock is not strong enough to expel the mantle.  

The second parameter is the mass $M_{BH}$ of the formed BH.
There are various studies as for what the mass of the BH should be
(e.g. \cite{Timmes&96,Bethe&Brown98,Fryer99,Fryer&Kalogera99}).
In some papers a typical BH mass was found to be not much higher than
the upper limit for the NS mass (Oppenheimer--Volkoff limit
$\sim$1.6--2.5M$_\odot$, depending on the unknown equation
of state for the neutron star matter) even if the fallback 
accretion onto the supernova remnant is allowed \cite{Timmes&96}.
However, observations strongly indicate much higher masses 
of BH candidates, of the order of 6--10M$_\odot$ (see Table~\ref{t:bh}).
To obtain such BH masses, it is sometimes assumed \cite{Bethe&Brown98} 
that $M_{cr}\sim$80M$_\odot$. Recently, a continuous range of BH masses 
up to 10-15 M$_\odot$ was derived in calculations \cite{Fryer&Kalogera99}.
Since the present day calculations are still unable
to reproduce self-consistently even the supernova explosion, 
we have parameterized the BH mass $M_{BH}$ 
by the fraction of the pre-supernova mass $M_*$
that collapses into BH: $k_{BH}=M_{BH}/M_*$. 
In fact, the pre-supernova mass $M_*$ is directly related with 
$M_{cr}$, but the form of this relationship 
is somewhat different in different scenarios for massive star evolution. 
According to our parameterization, the minimal BH mass can be 
$M_{BH}^{min}=k_{BH}M_*$, where $M_*$ itself depends on $M_{cr}$.
We have varied $k_{BH}$ in a wide range from 0.1 to 1.

The third parameter, similar to the case of NS formation,
is a possible kick velocity ${\bf w}_{BH}$ attributed to a newly 
formed BH. In general, one expects that a BH should acquire a smaller kick 
velocity than a NS, as black holes are more massive than neutron stars.
In our calculations we have adopted the relation
\beq{BH_kick}
    \frac{w_{BH}}{w_{NS}}
        = \frac{M_*-M_{BH}}{M_*-M_{OV}}
        = \frac{1-k_{BH}}{1-M_{OV}/M_*}\,,
\eeq
where $M_{OV}=2.5$M$_\odot$ is the maximum NS mass.
When $M_{BH}$ is close to $M_{OV}$, the ratio $w_{BH}/w_{NS}$ 
approaches 1, and
the low-mass black holes acquire kick velocities similar to those of 
neutron stars. When $M_{BH}$ is significantly larger than $M_{OV}$,
the parameter $k_{BH}=1$, and the BH kick velocity becomes vanishingly
small \footnote{Other possible
relationships between $w_{BH}/w_{NS}$ have also been 
checked, but their different forms do not affect
the results significantly.}. As we show below,
the allowance for a quite moderate $w_{BH}$ 
strongly increases the coalescence rate of binary BH. A recent analysis of
space velocities of some BH candidates did not reveal the need for
a non-zero $w_{BH}$ \cite{Nelemans&99}. However,
other studies show that some kick velocity can arise during the 
BH formation, and its presence does not contradict the observational data
\cite{Kuranov&Postnov&Prokhorov00}. From a theoretical point of view, 
the presence of a moderate kick velocity imparted to a BH during 
its formation seems very plausible \cite{Fryer&Kalogera99}.

\subsubsection{Binary black hole merging with $w_{BH} = 0$: typical example}

We begin from the simplest assumption $w_{BH} = 0$. The more realistic 
cases $w_{BH} \neq 0$ will be considered 
in subsection~\ref{sec:secI:kick:coales}.
In contrast to NS+NS binaries, the BH+BH and BH+NS binary systems have not
been observed so far. There is no way of recovering from observations
the range of progenitors for such binary systems.  
We can only apply the population synthesis method and derive theoretically 
the parameters of all the binaries, including the BH+BH and BH+NS pairs, 
that should be produced at the end of the evolution of very massive 
binary stars. 

In addition to the evolutionary uncertainties
existing for stars evolving to binary NS systems, new uncertainties
arise for very massive stars with initial masses $M\simgt 40 M_\odot$. 
First of all,
a large mass-loss via stellar wind is observed for
such stars. According to current views, a massive single star
can lose more than a half of its initial mass already on the 
main-sequence. Further rapid mass decrease is expected during a 
helium star stage. There is no general agreement as to how to
describe the mass loss of a massive star. Yet, one can consider
two extreme cases for the mass loss via stellar wind: a slow 
mass-loss and very fast one. 
Since the exact description of the stellar wind mass loss
is not known, we have considered both
options in our numerical simulations 
(for more details, see \cite{LPP97_mnras}).
A typical evolutionary track that leads to the formation of coalescing
binary BH system, assuming a low mass-loss scenario, is shown
in Fig.~\ref{f:bhbh}. 

\if t\figplace
\begin{figure}
\begin{center}
\epsfxsize=0.5\hsize
\fbox{\epsfbox{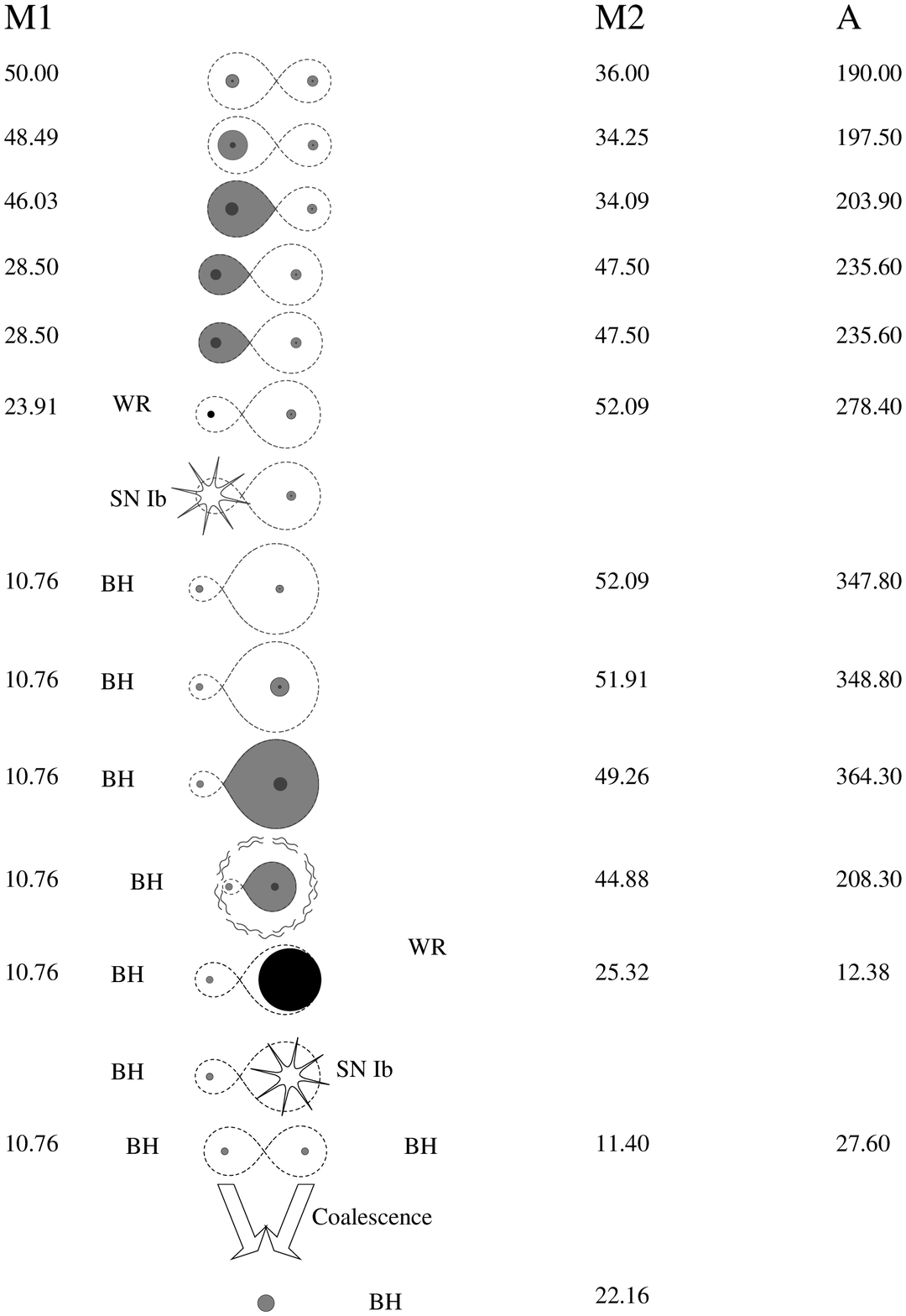}} 
\caption{Evolutionary track of a massive binary star leading
to the formation and coalescence of two BH.
The low stellar wind mass loss scenario is used.}
\label{f:bhbh}
\end{center}
\end{figure}
\fi

Before proceeding further with evolutionary calculations,
we want to explain qualitatively 
how a merging BH binary can form even in the framework of
an extremely high mass loss. The high mass loss makes the 
binary system wider, according to  Eq. (\ref{A:Jeans}), so it should
have been quite tight before the phase of an active mass-loss. 
Consider two already existing black holes, each of 10 $M_\odot$, 
in a circular orbit.  
The orbital separation should be $<20 R_\odot$ for the binary
to merge within the $t_{H}$ (see Fig.~\ref{A:GW:a-e}).
This means that the radius of a Wolf-Rayet (helium) star that collapsed 
second, in the course of a binary evolution, to form a BH  
should have been less than 10 $R_\odot$. 
The mass of the pre-collapse star could not be smaller than  
10 $M_\odot$, since $k_{BH} \le 1$. Such massive helium stars
have very small radii ($\sim 1 R_\odot$) 
and do not expand too much before the collapse,
so the requirement $R<10R_\odot$
is fulfilled. The life-time of a massive helium star is
about $10^{5}$ years, since it loses mass at a high rate of
$10^{-5}$ $M_\odot$/yr. The star can lose a sizable fraction, maybe
a half, of its mass before the collapse. Thus, we will be 
dealing with a 20 $M_\odot$ helium star
in pair with the first 10 $M_\odot$ BH in a circular orbit of
radius $a\approx 13 R_\odot.$ (We applied Eq. (\ref{A:Jeans}) to calculate
the radius of the resulting orbit.) Note that the 8 $R_\odot$ Roche lobe 
of a 20 $M_\odot$ helium star is still quite large. To form such
a close WR+BH binary, a common envelope stage is needed. The 20 $M_\odot$
He core corresponds to at least 55 $M_\odot$ main-sequence star, as follows 
from Eq. (\ref{A:MHe}). According to the models by
Schaller et~al. \cite{Schaller&92_tracks},
a massive star loses about a half of its initial mass
on the main-sequence, so to form a common envelope with 10 $M_\odot$ BH
the star has to lose $\sim 25 M_\odot$ while on the main-sequence. This means
that the common envelope stage should have started with 
a 30 $M_\odot$ red super-giant filling its
Roche lobe and having a 10 $M_\odot$ BH as a companion. The mass ratio 
in such a system is high enough for the common
envelope to develop. The orbital separation at the common envelope stage
should have decreased by 6-12 times, according to Eq. (\ref{A:CE:afai}),
depending on the parameter $\alpha_{CE}$ and the exact value of the
red giant mass. So, before the common envelope stage,
the orbital separation should have been $\sim 130 R_\odot$.
The orbit should be somewhat smaller than this
(i.e., about 120 $R_\odot$ or less) when the first
BH forms, because of the strong wind from the red giant and loss of
the total mass. And in order to collapse first, the
mass of the primary star must have been at least 60 $M_\odot$. 
Assuming isotropic stellar wind and using again Eq. (\ref{A:Jeans}),
we conclude that the initial
system could have widened at most $(60+55)/(10+30)\approx 3$ times since
the time of its formation, i.e. the initial separation of the progenitor 
binary should be larger than 40 $R_\odot$.
The initial separation of 50 $R_\odot$ is sufficient enough  
to harbour two 60 $M_\odot$
stars since their radii are less than 20 $R_\odot$ on the main-sequence.
Even though such initially close massive binaries are rare, they should 
exist. Thus, we see that some fraction of massive binary stars should 
have ended up as sufficiently close pairs of black holes.

\subsection{Effects of the kick velocity} 
\label{sec:secI:kick}

The picture outlined above changes if a non-zero kick velocity is 
present in the process of formation of a NS or a BH.
This, in turn, has a significant effect on the expected rate 
of compact binary mergings, which is of primary interest for our
study. In general, the formation of a compact object (NS or BH) is 
accompanied, both, by a mass loss from the system and by a kick velocity.
The effects of kick velocity during supernova explosions were considered in
many papers (see e.g. \cite{Yamaoka&93,Brandt&Podsiadlowski95}
and also~\ref{sec:appB}). The general formulae for the condition
of system's disruption and for parameters of the resulting elliptical 
orbit, if the system remains bound, are derived in~\ref{sec:appB},
see Eqs. (\ref{A:SN:afai}), (\ref{A:SN:ecc}), (\ref{A:disrupt}). 
Here we will present qualitative arguments enabling the reader 
to see the main consequences of a non-zero kick velocity.
We restrict our attention to circular orbits and assume equal 
probabilities for all possible orientations of the kick 
velocity vector $\vec w$. We argue
that a moderate (not too large) kick velocity increases the
rate of binary mergings. This happens because a moderate 
kick velocity does not change too much the likelihood of the
system's disruption, but, at the same time, always makes the 
periastron of the resulting elliptical orbit smaller than it would
have been without a kick. As a result, some of the binaries, 
whose coalescence time without a kick 
would be longer than the Hubble time, now get a chance to merge in a time 
shorter than $t_H$. This increases the number of detectable 
gravitational wave sources.

\subsubsection{Effect of the kick velocity on
the disruption of a binary system}

The collapse of a star to a BH, or its explosion leading to the
formation of a NS, are normally considered as instantaneous.  
This assumption is well justified in binary systems, 
since typical orbital velocities before the explosion 
do not exceed a few hundred km/s, while most of the mass 
is expelled with velocities about several thousand km/s. 
The exploding star $M_1$ leaves
the remnant $M_c$, and the binary loses a portion of its mass: 
$\Delta M = M_1 - M_c$. The
relative velocity of stars before the event is  
\beq{}
V_i=\sqrt{G(M_1+M_2)/a_i}.
\eeq
Right after the event, the relative velocity is
\beq{V+W}
\vec V_f=\vec V_i+\vec w\,.
\eeq
Depending on the direction of the kick velocity vector $\vec w$, the
absolute value of $\vec {V}_f$ varies in the interval from
the smallest $V_f = |V_i - w|$ to the largest $V_f = V_i + w$. 
The system gets disrupted
if $V_f$ satisfies the condition (see~\ref{sec:appB}):
\beq{I:disrupt}
V_f \ge V_i \sqrt{\frac{2}{\chi}}
\eeq
where $\chi \equiv (M_1 + M_2)/(M_c+M_2)$.

Let us start from the limiting case when the mass loss is 
practically zero ($\Delta M = 0$, $\chi =1$), while a non-zero kick 
velocity can still be present. This is a model for a BH formation with
$k_{BH} =1$. It follows from Eq. (\ref{I:disrupt})  
that, for relatively small kicks, $w< (\sqrt{2}-1)V_i$, 
the system always (independently of the direction of $\vec w$) 
remains bound, while for $w> (\sqrt{2}+1)V_i$ 
the system always unbinds. By averaging over equally probable
orientations of $\vec w$ with a fixed amplitude $w$, one can show
that in the particular case $w= V_i$ the system disrupts or 
survives with equal probabilities. If $V_f < V_i$, 
the semi-major axis of the system becomes smaller than the
original binary separation, $a_f < a_i$ 
(see  Eq. (\ref{A:SN:afai})). This means that the system becomes 
more bound than before, i.e. it has a greater negative
total energy than the original binary. If $V_i <V_f <\sqrt{2}V_i$, 
the system remains bound, but $a_f > a_i$. For small and moderate 
kicks $w \la V_i$, the probabilities for the system to become more or
less bound are approximately equal.

In general, the binary system loses some fraction of its mass
$\Delta M$. For a BH formation this corresponds to $k_{BH}<1$. 
In the absence of the kick
velocity, the system remains bound if $\Delta M < M/2$ and gets
disrupted if $\Delta M \ge M/2$ (see~\ref{sec:appB}). Clearly,
a properly oriented kick velocity (directed against the vector
$\vec {V}_i$) can keep the system bound, even if it would have been 
disrupted without the kick. And, on the other hand, an unfortunate 
direction of $\vec w$ can disrupt the system, which otherwise would 
stay bound. 

Consider, first, the case $\Delta M < M/2$. The parameter $\chi$ varies
in the interval from 1 to 2, and the escape velocity $V_e$ varies in
the interval from $\sqrt{2} V_i$ to $V_i$ (see~\ref{sec:appB}).  
It follows from Eq. (\ref{A:disrupt2}) that the binary always remains
bound if $w< V_e - V_i$, and always unbinds if $w> V_e + V_i$.
This is a generalization of the formulae 
derived above for the limiting case $\Delta M = 0$. Obviously, 
for a given $w$, the probability for the system to disrupt or become 
less bound increases when $\Delta M$ becomes larger. Now turn  
to the case $\Delta M > M/2$. The escape velocity of the compact 
star becomes $V_e<V_i$. The binary is always disrupted if the
kick velocity is too large or too small: $w > V_i + V_e$ or
$w < V_i - V_e$. However, for all intermediate values of $w$,
the system can remain bound, and sometimes even more bound than before,
if the direction of $\vec w$ happened to be approximately
opposite to $\vec {V}_i$. A detailed calculation of probabilities
for the binary's survival or disruption requires integration
over the kick velocity distribution function $f (\vec w)$  
(see e.g. \cite{Brandt&Podsiadlowski95}).

\subsubsection{Effect of the kick velocity on
coalescence rate of compact binary systems}
\label{sec:secI:kick:coales}

Here we consider binary systems that were not disrupted during the 
formation of a compact object. The parameters $a_f$ and $e$ of the 
resulting elliptical
orbit are defined by Eqs. (\ref{A:SN:afai}), (\ref{A:SN:ecc}). The
distance of the closest approach between the stars is given by
the orbit's periastron $a_p=a_f(1-e)$. It follows from 
Eqs. (\ref{A:SN:afai}), (\ref{A:SN:ecc}) that $a_p = a_i$ in the 
absence of kick velocity. The importance of the kick
velocity $w \ne 0$ lies in the fact 
that, although the semimajor axis $a_f$ can increase or decrease 
under the action of the kick, the periastron distance always becomes
smaller: $a_p < a_i$. This relationship follows from the combination of 
Eqs. (\ref{A:SN:afai}), (\ref{A:SN:ecc}) plus the requirement that
the system remains bound, i.e. the quantities participating
in Eq. (\ref{A:disrupt}) satisfy the opposite inequality. 
The decrease of the periastron distance plays an 
important role in the subsequent evolution of the binary, which consists 
now of a newly born compact star and its companion.   

Consider, first, a normal star as the companion. Since the kick has 
diminished the periastron distance, as compared with the no-kick
case, the normal star, while passing through the periastron, 
will fill out its Roche lobe in a shorter time, than it would do 
in the absence of the kick. After the tidal circularization of 
the orbit, a tighter binary is formed. Accordingly,
the subsequent common envelope stage makes the binary 
tighter than it would otherwise do (see Eq. (\ref{A:CE:afai})). 
As a result, the final binary system, consisting of two compact objects, 
will coalesce due to GW radiation in a shorter time (see
Eq. (\ref{A:GW:t_0})). In other words, some of the binaries,
which would be too broad to coalesce in $t_H$, become detectable
sources of GW with the help of a moderate kick velocity.   
If the companion is already a compact star, the orbital
evolution is driven exclusively by GW emission (Section \ref{A:GW_evol}).
Unless the kick velocity is so big that it makes the semimajor
axis $a_f$ very large, these binaries 
will also merge in a time interval 
shorter than the one following from the evolution without a kick. 

These qualitative considerations explain the outcomes of numerical 
simulations with many trial systems. We are interested in results
averaged over many systems with different input parameters. 
These results are presented below. As expected, 
a moderate kick velocity increases, on average, 
the rate of compact star mergings.

\subsubsection{Coalescence rates of compact binaries}

We can now present the results of our numerical calculations for
the coalescence rate of compact binaries in a typical 
galaxy \cite{S:lipunov.etal.a}. 
The total mass of a model 
galaxy is assumed to be $10^{11} M_\odot$. 
We adopt a constant star formation rate defined
by Eq. (\ref{init_distr}). It is believed that Eq. (\ref{init_distr})
reflects well the situation in a galaxy like our own Milky Way.

\if t\figplace
\begin{figure}
\epsfxsize=0.5\hsize
\centerline{\epsfbox{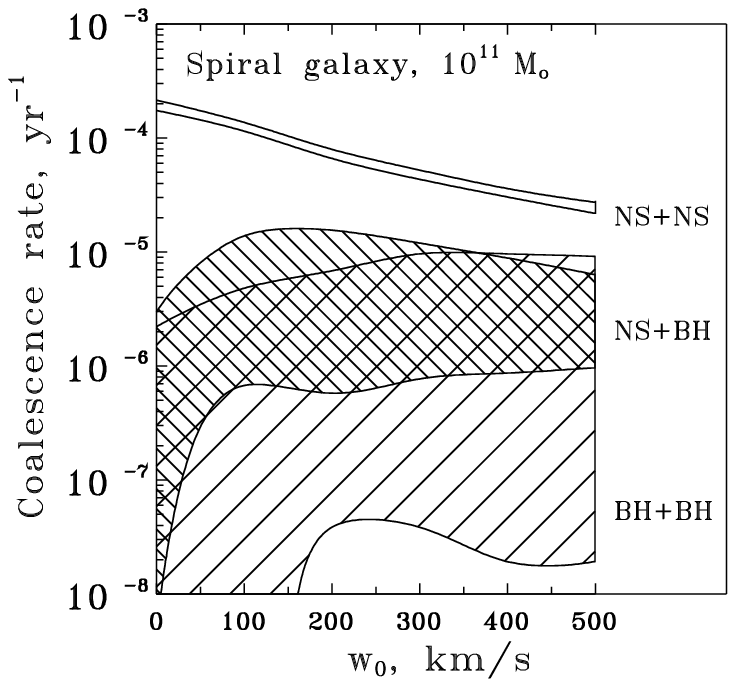}} 
\caption{NS+NS, BH+NS, and BH+BH merging rates in a
$10^{11}$~M$_\odot$ galaxy 
as functions of the kick velocity parameter $w_0$ for
Lyne-Lorimer kick velocity distribution (\protect\ref{bin:LL}).
Star formation rate in the galaxy is assumed constant.
BH formation parameters are $M_*=15$--50~M$_\odot$, $k_{BH}=0.25$.}
\label{f:grate}
\end{figure}
\fi

In Fig. \ref{f:grate} we plot the NS+NS,
BH+NS, and BH+BH merging rates as functions of the kick velocity
parameter $w_0$ in the distribution (\ref{bin:LL}). The calculations
were performed for discrete values of $w_0$, but the resulting
points are joined by smooth curves. The BH formation parameters 
were taken from the range
$M_*=15-50 M_\odot$ with $k_{BH}=0.25$. Both the high mass loss 
and the low mass loss stellar winds were considered. The broad range 
of $M_*$ and the uncertainty in the stellar winds have contributed to the
spread of the results for BH+NS and BH+BH systems. The NS+NS
systems arise from relatively low mass stars, so they are less
sensitive to the uncertainty in the stellar wind.  
It is seen from Fig. \ref{f:grate} that the NS+NS rate lies in the range
$\sim 3\times 10^{-4}$--$\sim 3\times 10^{-5}$ per year.
The rates of BH+NS and BH+BH mergings are 10-100 times lower. 
For the limiting case of zero kick velocity ($w_0 = 0$) our rates 
agree with the independent estimates of Tutukov and Yungelson
\cite{Tut&Yung93_mnras}. In the same limit $w_0 = 0$, our rate for 
NS+BH binaries ($\sim 10^{-6}$ per year) is
smaller than the estimate by Bethe and Brown \cite{Bethe&Brown98}, who
obtained the rate  $\sim 2\times 10^{-5}$ per year. However, we believe  
that their estimate was derived from a somewhat  
simplified picture of binary evolution.

As expected, the BH+NS and BH+BH rates have a tendency to grow with the 
increase of kick velocity from zero. This is seen on the graph as the rise
of the NS+BH and BH+BH curves for small and moderate
$w_0$ (up to $w_0 \sim 100$~km/s). For much larger values of $w_0$,
the kick velocity contributes mostly to the disruption of binary
systems, and this is why the curves have a tendency to turn down.
Generally speaking, the NS+NS rate should also grow for small
deviations of $w_0$ from zero. However, since the NS mass is 
smaller than the BH mass, the increase of the NS+NS rate takes place 
for only a small value of $w_0$, not resolvable on the graph. For larger values 
of $w_0$, the kicks mostly disrupt the binaries, and the NS+NS 
curve goes down. The value of $w_0$ preferred by the radio-pulsar observations
lies in the range 200--400~km/s.        

For a broad range of used parameters and despite all the remaining
uncertainties, the results of evolutionary calculations show that the
number of coalescing BH+BH pairs is only a factor 10-100 smaller
than the number of coalescing NS+NS pairs. This relationship may
have a simple explanation and can be traced back to the initial conditions 
of star formation. The line of argument is as follows. Let us take 
the NS mass at 1.4$M_\odot$ (a typical mass well confirmed
by existing observations), and the BH mass at 8.5M$_\odot$ (the mean
value for BH candidates from Table~\ref{t:bh}). Assume  
that the lower initial mass of NS progenitors is 
$M(NS)\approx 10$M$_\odot$, while the threshold for a BH
formation is at the maximum of the estimates quoted above:
$M(BH)>M_{cr}=80$M$_\odot\,$. Applying the Salpeter initial mass 
function for the formation rate of stars in the Galaxy (see
Eq. (\ref{e:Salpeter_law})): 
$$
\frac{dN}{dt dM}\simeq 1 M_\odot (M/M_\odot)^{-2.35}  
$$
and using the lower limits of integration, one finds  
$$
    \frac{N(M>80M_\odot)}{N(M>10M_\odot)}
    = \left(\frac{80M_\odot}{10M_\odot}\right)^{-1.35} \simeq 0.06. 
$$
This ratio should be valid for binary stars too. It is reasonable
to expect then that despite differences and complexities of binary
evolution, the ratio of coalescence rates will be given, approximately,
by the same number:  
\beq{BH/NS}
\frac{\R_{BH}}{\R_{NS}} 
    = \left(\frac{80M_\odot}{10M_\odot}\right)^{-1.35} \simeq 0.06. 
\eeq
This expectation turns out to be in rough agreement with 
the results of detailed evolutionary calculations presented above. 

The derived rates $\R_G$ for a single galaxy can 
be extrapolated to larger volumes. For the purposes of 
GW detection it is important to know the rate of events from
distances accessible to the instruments in  
LIGO, VIRO,  GEO--600. These are large distances up to and above
100 Mpc (see Section~\ref{sec:detectors}). In such a large volume one can regard 
galaxies as being distributed homogeneously, and at the same time,
one can neglect effects of cosmological evolution on star formation
initial conditions, etc.   
To derive the average density of Galactic events in a large
volume one can use different approaches. One possibility is
to use the luminosity of galaxies per Mpc$^3$ (as in ~\cite{Phinney91}).
Alternatively, one can rely on the estimate of 
density of baryons bound in stars. The baryon density $\rho_b$ is
often expressed in terms of the dimensionless 
parameter $\Omega_b\equiv \rho_b/\rho_{cr}$,
where $H_0$ is the present value of the Hubble parameter and 
$\rho_{cr}=3H_0^2/8\pi G$ is the critical density. 
Then, one can relate the Galactic rate $\R_G$ per a $10^{11} M_{\odot}$
galaxy with the volume rate $\R_V$ per 1 $\hbox{Mpc}^3$:  
\beq{RV}
\R_V=3\times 10^{-3} \R_G \frac{\epsilon}{(0.5)}\frac{\Omega_b h_{70}^2}{(0.0045)} 
\hbox{Mpc}^{-3}
\eeq
where $\epsilon$ is the fraction of binary stars and 
$h_{70}=H_0$/(70 km/s~Mpc). This estimate agrees  
with that of \cite{Phinney91} assuming $\epsilon=1$
(all stars are binaries).
The available astronomical measurements of the total baryon budget 
give $\Omega_b\approx 0.0015 h_{70}^{-1}$ in galactic disks and  
$\Omega_b\approx 0.003h_{70}^{-1}$ in bulges of spirals and 
ellipticals \cite{Fukugita&98} (as well as the somewhat larger 
values~\cite{Edmunds99}). On the other hand, estimates 
of $\Omega_b$ based on the primordial nucleo-synthesis considerations
give as much as $\Omega_b h^2 = 0.016$, but this number can also be a 
factor of 2 smaller \cite{Olive2000}. 
Formula (\ref{RV}) can be rewritten as
\beq{RV2}
\R_V=0.1 \R_G \frac{\epsilon}{0.5}\frac{\Omega_b h^2}{0.016} 
\left(\frac{r}{\hbox{Mpc}}\right)^{3}. 
\eeq
When comparing our numerical simulations, described below, with
qualitative estimates, we rely on the relationship
\beq{RV3}
\R_V=0.1 \R_G \left(\frac{r}{\hbox{Mpc}}\right)^{3}. 
\eeq

This result for $\R_V$ is based on $\R_G$ for spiral galaxies. For elliptical
galaxies the star formation process is more like an instantaneous event
rather than a continuing process described by  
(\ref{init_distr}). The coalescence rates have been calculated for
elliptical galaxies too. However, it was shown \cite{Jorgensen_ea95} that 
the contribution of elliptical galaxies to the coalescence
rates from the discussed distances is only about 10-20 \%.

\section{Detection Rates} 
\label{sec:secI:DetectRate}

Having found the coalescence rates $\R_V$ for binaries of different
nature, one can now predict the detection rates of these binaries
in a given GW detector. We argue that binary black holes have
a better SNR than of binary neutron stars, 
and, despite
their lower abundance, the BH+BH and BH+NS pairs should be seen 
more often than NS+NS pairs. In the 
first subsection, we derive the detection rates that are based on 
the $\R_G$ described above. In the second subsection, we discuss 
possible modifications to our conclusion in connection with the 
recently proposed scenario \cite{Vanbeveren&98}, which applies to 
very massive stars. Since the proposed scenario can affect only 
the BH+BH detection rates, we concentrate on these systems emphasizing
the important role of kick velocities.

\subsection{Detection rates in the usual picture}
\label{sec:secI:DetectRate:usual}

The rate of NS+NS coalescences is higher than the rate of NS+BH and
BH+BH coalescences. However, the BH mass is significantly larger than
the NS mass. A binary involving one or two black holes, placed at the 
same distance as a NS+NS binary, produces a significantly larger 
amplitude of gravitational
waves (see Section~\ref{sec:data analysis} and~\ref{sec:appA}). 
With a given sensitivity of the detector (fixed
SNR), a BH+BH binary can be seen at a greater 
distance than a NS+NS binary. Hence, the registration volume for 
such bright binaries is significantly larger than the registration
volume for relatively weak binaries. 
The detection rate of a given detector depends on the 
interplay between the coalescence rate (spatial density of
sources) and the detector's response to sources of one or 
another kind.  

Coalescing binaries emit gravitational wave signals with a well
known time-dependence (waveform). This allows one to use the 
technique of matched
filtering \cite{Thorne87_300yr}. The signal-to-noise ratio $S/N$ 
depends mostly on the ``chirp'' mass of the 
binary system $\M=(M_1+M_2)^{-1/5}(M_1M_2)^{3/5}$
and its distance $r$. The accurate formula for $S/N$ is presented 
in Section~\ref{sec:DataAn} (formula (\ref{eq:snr3})). 
Here, we will use its simplified
version which is sufficient for our purposes
(\cite{Thorne87_300yr}, see also \cite{Flanagan&Hughes98}): 
\beq{D:S/N}
    \frac{S}{N} = 3^{-1/2}\pi^{-2/3}
	\frac{G^{5/6}}{c^{3/2}} \frac{\M^{5/6}}{r} f^{-1/6}
	/h_{\rm rms}(f)\,.
\eeq
At a fixed
level of $S/N$, the detection volume is proportional to $r^3$ and
therefore it is proportional to $\M^{5/2}\,$.
The detection rate $\D$ for binaries of a given class is the
product of their coalescence rate $\R_V$ with the detector's
registration volume $\propto \M^{5/2}\,$ for these binaries. 

Let us start from a qualitative discussion of the expected ratio
\beq{DBH/DNS}  
    \frac{\D_{BH}}{\D_{NS}} = \frac{\R_{BH}}{\R_{NS}}
        \left( \frac{\M_{BH}}{\M_{NS}} \right)^{5/2}\,
    \label{eq:2nd-situation}
\eeq
where $\D_{BH}$ and $\D_{NS}$ refer to BH+BH and NS+NS pairs,
respectively. Here, we discuss the ratio of the detection rates,
rather than their absolute values. The derivation of absolute 
values require detailed evolutionary calculations
which will be discussed later. As a rough estimate 
for $\R_{BH}/\R_{NS}$ one can take
Eq. (\ref{BH/NS}). Then, Eq. (\ref{DBH/DNS}) gives a remarkable
result:
\beq{DBH/DNS.2}
   \frac{\D_{BH}}{\D_{NS}} =
    \left(\frac{80M_\odot}{10M_\odot}\right)^{-1.35}
    \left(\frac{8.5M_\odot}{1.40M_\odot}\right)^{5/2}
    \simeq 5.5\,.
\eeq
This ratio becomes even larger than $5.5$, if one takes 
$M_{cr}< 80 M_\odot$ as usually
assumed. Thus, the registration rate of 
BH mergers is expected to be {\it higher} than that of NS mergers.
This estimate is, of course, very rough, but it can serve as an indication
of what one can expect from detailed calculations. 

In Fig.~\ref{f:sheja} 
we display the results
of numerical calculations for the absolute registration rates of
various binaries. The detector sensitivity is taken 
as $h_{rms}=10^{-21}$ at $f=100 Hz$, as expected for
initial instruments in LIGO, VIRGO, GEO600. It is assumed that
$S/N=1$. Since the most interesting results refer to systems with 
black holes, we vary the black hole formation  
parameter $k_{BH}$. The calculations were performed
assuming the Lyne-Lorimer kick velocity distribution with
$w_0=400$~km/s. The vertical dispersion of the results is due to 
the uncertainty in the parameter $M_{cr}$. 

\if t\figplace
\begin{figure}
\epsfxsize=0.7\hsize
\centerline{\epsfbox{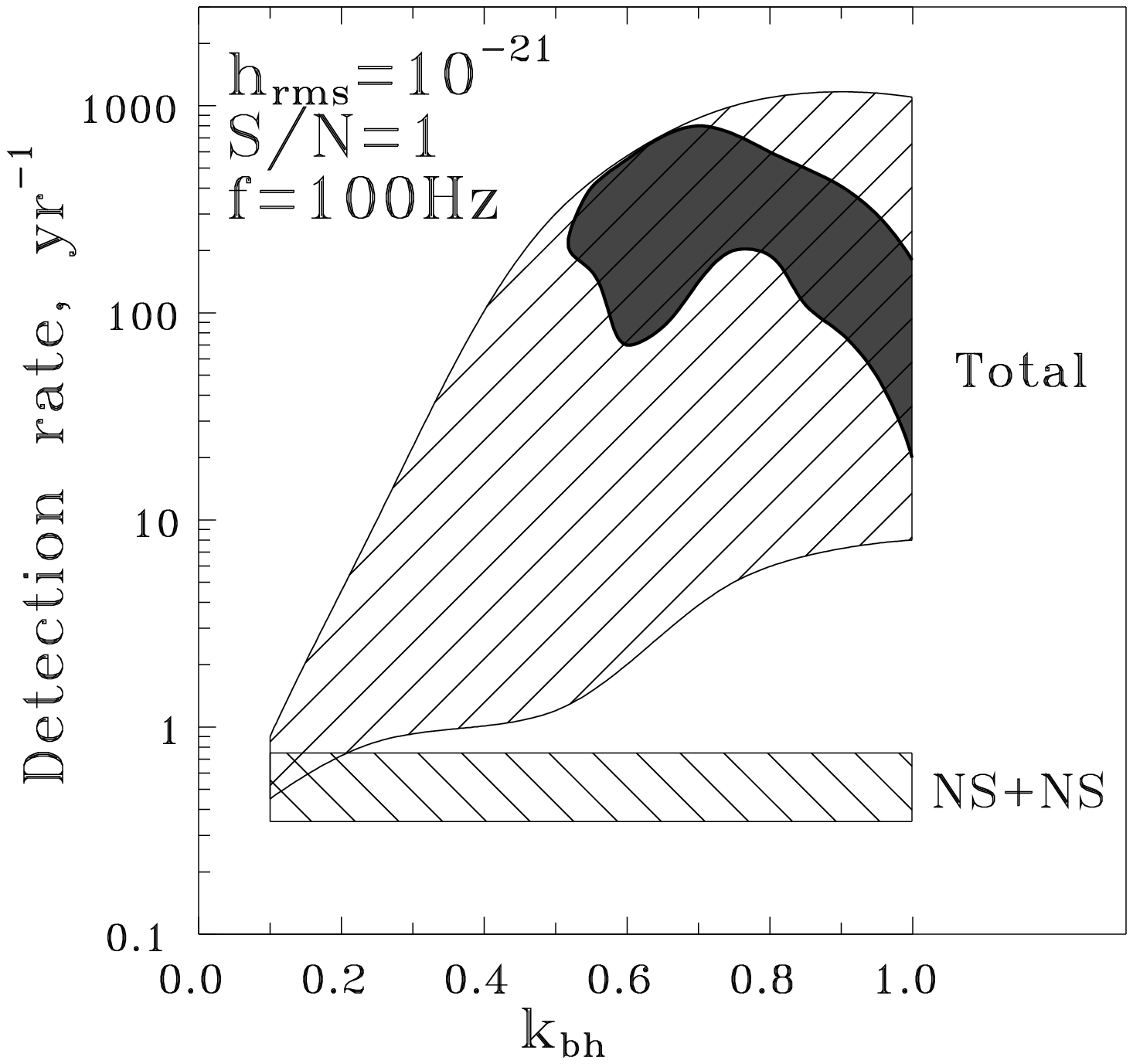}} 
\caption{The detection rate $\D$ of GW events in a detector
with the sensitivity $h_{\rm rms}=10^{-21}$ at frequency 100 Hz and the
signal-to-noise level $S/N=1$, as a function of BH formation parameter
$k_{BH}$. The calculations were performed for the Lyne--Lorimer kick velocity
distribution with $w_0=400$~km/s. The spread of $\D$ at fixed $k_{BH}$
is due to variation of the parameter $M_{cr}$ 
from 15 $M_\odot$ to 50 $M_\odot$. The bottom rectangular area is drawn
for binary NS coalescences.
Their rate is independent 
of $k_{BH}$ (as it should be) and predicts a couple of events per 1--3 years
at this level.   
The total detection rate can be 2--3 orders of magnitude higher 
then the NS+NS rate
at the expense of 
BH+BH and BH+NS coalescences. The hatched area
shows the region of the most probable parameters for the low stellar wind
mass loss scenario. Inside this region
the outcomes of calculations are in agreement with 
the upper limit on the galactic number of binary 
BH with radiopulsars (less than 1 per 700 single pulsars)
and the galactic number of BH candidates similar to Cyg~X--1 (from 1 to 10).}
\label{f:sheja}
\end{figure}
\fi

It is seen from the graph that under the formulated conditions one
can expect to see a couple of NS+NS coalescences in 1-3 years of 
observations, at a $S/N = 1$. These systems
are located, roughly, at a distance of 100 Mpc. The SNR is higher
for closer systems, but the expected event rate would be lower; for more
distant systems, the event rate increases, but the SNR becomes smaller
than 1. So, it is unlikely that NS+NS coalescences will be
detected by the initial instruments.  

The situation is significantly better for systems involving 
black holes. As is seen from Fig.~\ref{f:sheja}, 
the total registration rate of
all binaries, including BH+BH and BH+NS pairs,
can be 2--3 orders of magnitude higher than the registration rate
of NS+NS systems alone, mostly at the expense of
massive BH pairs. This is true unless the $k_{BH}$ parameter is
very small, $k_{BH}< 0.4$.  The hatched area shows the results of
calculations with the stellar wind parameters taken from
the ``most probable" region. This means that under this choice of
the parameters, the
outcomes of other evolutionary tracks are in agreement with
observations, namely, in agreement with the upper limit on the number 
of binary pulsars with BH (less than 1 per 700
single radio-pulsars) and with the number of Cyg X-1-like BH candidates
(from 1 to 10 per Galaxy). Inside this region, one should count on 
100 registrations (at the level $S/N = 1$), mostly from BH mergers.
The mean total mass of the BH pairs in the hatched area is around
$M = 30 M_{\odot}$. The simplified formula (\ref{D:S/N}), used in the 
construction of Fig.~\ref{f:sheja}, 
overestimates the $S/N$ for pairs heavier
than $30 M_{\odot}$, as shown in Section~\ref{sec:DataAn}. 
However, a correction for
more massive binaries is not expected to change significantly the
derived total registration rate.    

For a reliable detection, the $S/N$ ratio should be at least 2
in each of a network of four or more antennas.
Then, the calculated detection rates should be 
decreased by at least a factor $(S/N)^3 = 8$. This is because of the
scaling  $S/N\propto 1/r$ and $\D\propto r^3\propto (S/N)^{-3}$. 
Then, the expected detection rate of merging BH+BH pairs is up to 10 
events per year. As will be explained in Section\ref{sec:DataAn}, 
the $S/N$ ratio
is somewhat different for the three different instruments: LIGO, 
VIRGO and GEO. For a coalescing pair with a total mass $M =
30 M_{\odot}$ the SNR is roughly 4 for sources at
a distance of 100~Mpc. (The
performance of VIRGO is expected to be better than that of other
instruments, since VIRGO will be more sensitive at lower frequencies
and can track the binary for a larger number of cycles.) If one
is satisfied with $S/N = 2$, the accessible radius increases to
$r =$~200~Mpc. Then, the calculated detection rate (several per year) is
in agreement with formula (\ref{RV3}) if one takes for coalescing
black holes a reasonable galactic rate $\R_G = 3\times10^{-6}$ and
$r=$~200~Mpc. In its turn, this value for $\R_G$ fits well the event 
rate derived from numerical simulations, as displayed in Fig.~\ref{f:grate}.

Thus, taking into account all the remaining uncertainties, we 
conclude that the initial network is likely to see
each year 2-3 coalescing black hole binaries with the total mass around
$30 M_{\odot}$, at an SNR level of about 2--3.

\subsection{Non-standard scenarios and effects of kick velocities\\
on BH+BH detection rate}

Some of the recent evolutionary calculations 
\cite{Vanbeveren&98} assume that the primary stars 
with initial masses $M_1>40 M_\odot$
{\it never fill} their Roche lobes, so that the components of the
binaries evolve like single stars. 
As a result, the binary BH systems would be too
wide to merge in $t_H$. 
Although we think the scenario \cite{Vanbeveren&98} 
will face observational difficulties,
since it will lead to the too small 
number of binaries involving a BH and a massive blue star (Cyg~X--1-like
systems), we consider it necessary to follow in detail the possible 
fate of binary BH systems. 
We argue that the kick velocity accompanying the BH formation
increases the eccentricity of the binary, decreases its coalescence
time, and thus keeps the detection  
rate at almost the same level as discussed 
in section~\ref{sec:secI:DetectRate:usual}. 
In addition, the kick velocity leads to interesting modifications in the
relative orientations of the black hole spins with respect to each
other and with respect to the orbital angular momentum.

We have adopted the proposed scenario \cite{Vanbeveren&98} and 
have carried out population synthesis
calculations by varying the kick velocity parameter. The binary BH merging rate
was derived for a model galaxy of $10^{11} M_\odot$ (assuming that all 
stars are formed in binaries) with a constant star formation rate. For
simplicity, the kick velocity distribution was taken as a delta-function.
The more complicated distributions do not change the results
significantly and are not commented upon here.  
The results of these calculations are shown in Fig.~\ref{f:cls-dlt}.
The left panel shows the merging rate, while the right panel shows the
detection rate. 
The detection rate of binary BH coalescences is given for initial laser
interferometers ($h_{\rm rms}=10^{-21}$ at $f =100$ Hz) 
as a function of BH kick velocity. It is seen  from the plot that 
the 
merging rate and the detection rates increase rapidly with
the kick. The merging and detection rates reach the maxima  $\R\sim 2.5\times
10^{-5}$ yr$^{-1}$ and $\D\sim 20$ detections per year for 
$w\simeq 120$ km/s. Since 
$\D\sim \M^{5/2}\R$, the $\R(w)$ and $\D(w)$ functions have similar shapes.  

\if t\figplace
\begin{figure}
\hbox to \textwidth{ 
\hbox to 0.5\textwidth{
\epsfxsize=0.5\textwidth
\epsfbox{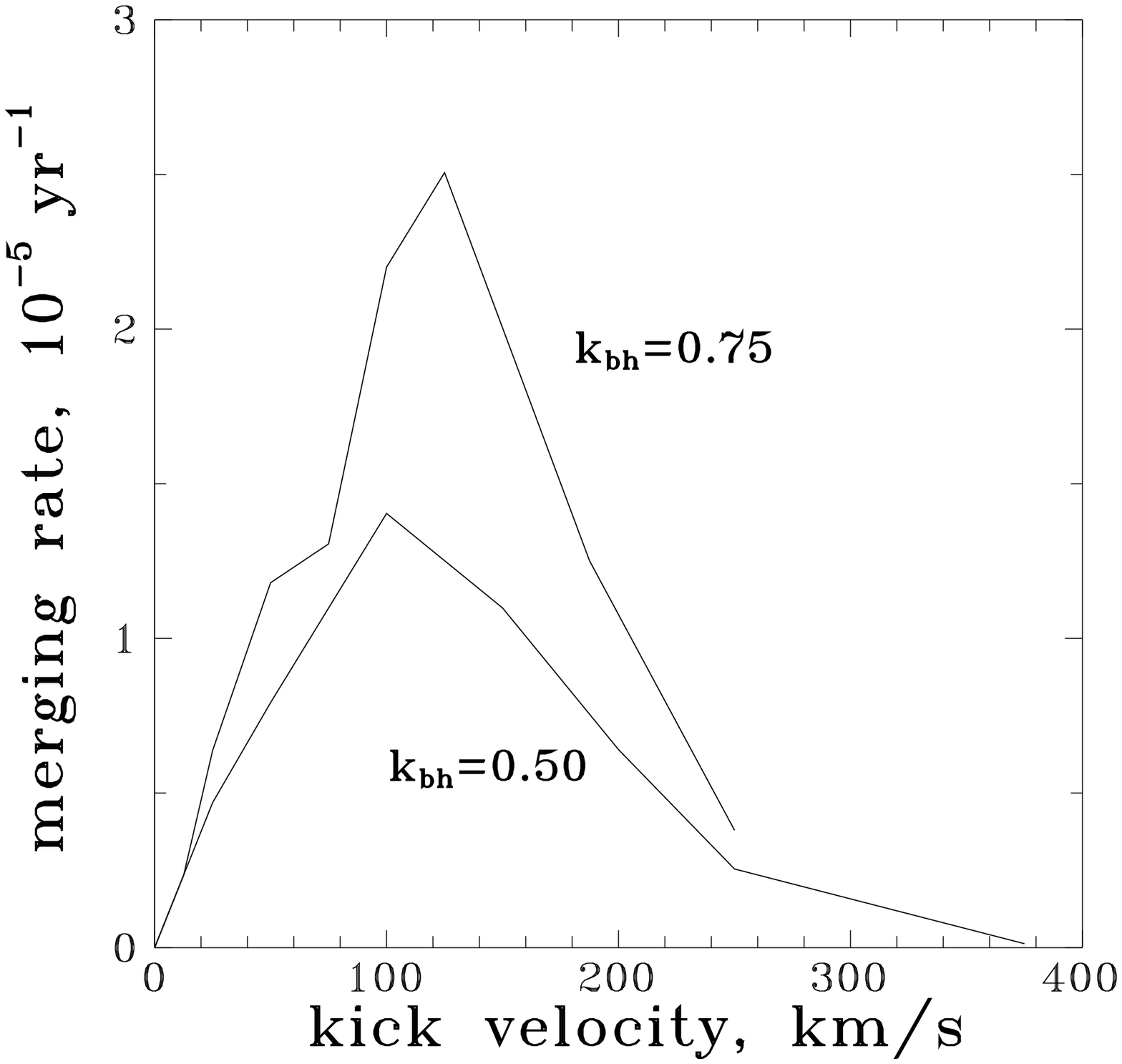} 
}
\hss
\hbox to 0.5\textwidth{
\epsfxsize=0.5\textwidth
\epsfbox{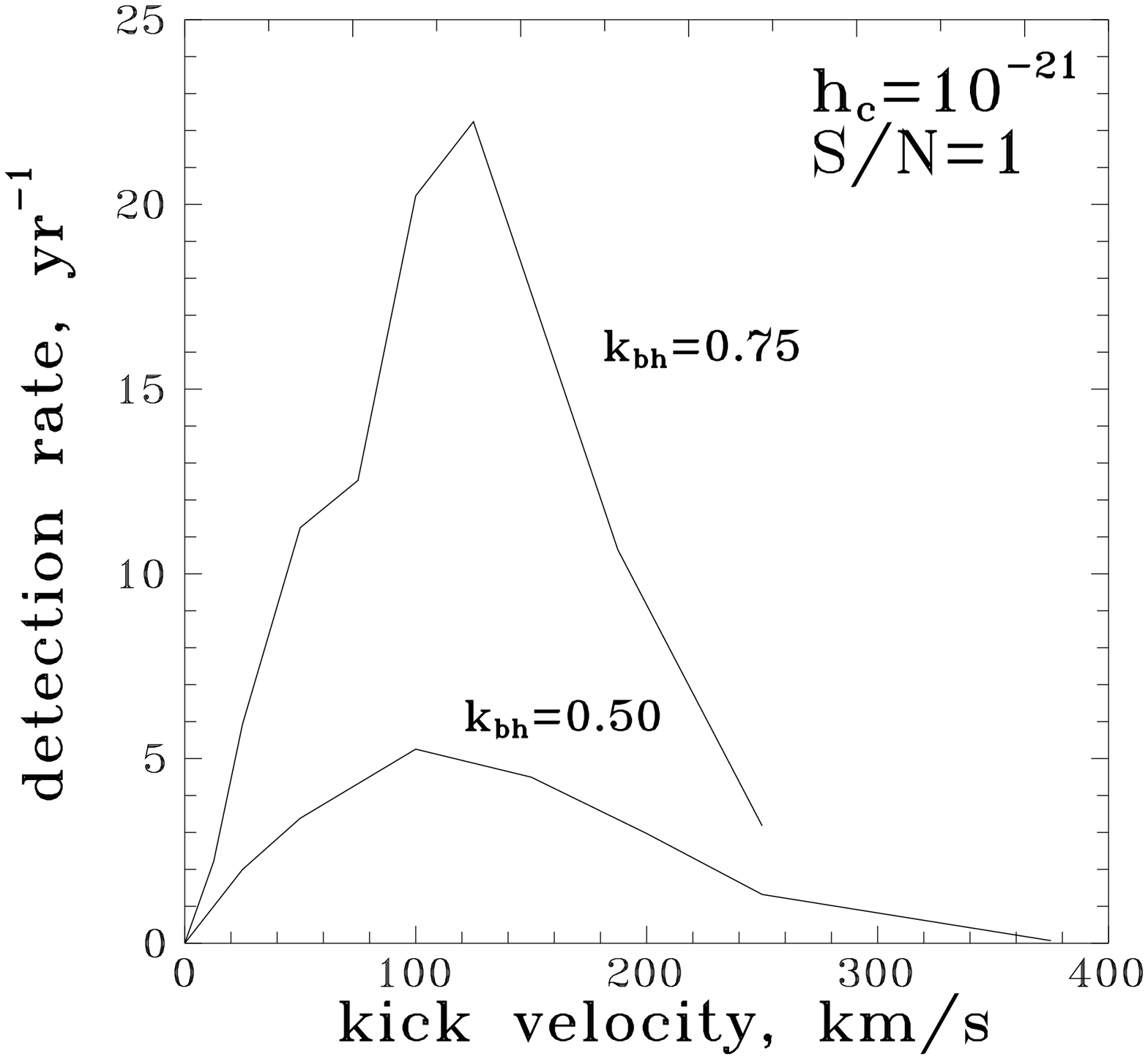} 
}
}
\caption{New scenario --- superhigh wind.
Left: BH+BH merging rate calculated for a $10^{11} M_\odot$
galaxy with a constant star formation rate, as a function of the kick 
velocity during BH formation with $M_{cr}=35 M_\odot$,
for $k_{BH}=0.5$ and $0.75$. 
Right: Detection rate of 
BH+BH mergings by the initial laser
interferometers ($h_{\rm rms}=10^{-21}$ at $f=100$~Hz), as a function of 
the kick velocity during BH formation.}
\label{f:cls-dlt}
\label{f:det-dlt}
\end{figure}
\fi
   
Obviously, the kick velocity imparted to newly born black holes makes
the orbits of survived systems highly eccentric. 
It is important to
stress that some fraction of binary BH retain their large eccentricities 
up to the late stages of their coalescence. This signature should be 
reflected in their emitted waveforms and should be modeled in templates. 

The asymmetric explosions accompanied by a kick
velocity change the space orientation of the orbital angular momentum.
On the other hand, the star's spin axis remains fixed (unless the kick 
was non-central). As a result, some distribution 
of the angle between 
the BH spins and the orbital angular momentum (denoted by $J$) 
will be established \cite{PP99}. 
It is interesting that even for small kicks of a few tens of km/s 
an appreciable fraction (30--50\%) of the merging binary BH should have 
$\cos J<0$. This means that in these
binaries the orbital angular
momentum vector is oriented almost oppositely to the black hole spins. This is one more signature
of imparted kicks that can be tested observationally.
These effects are also discussed in a recent paper \cite{Kalogera99}.

Thus, to conclude this analysis, we stress again that 
binary black hole coalescences remain the most likely sources to be
detected first by the initial network of laser interferometers.


\section{Transients and Continuous Gravitational Waves}
\label{sec:secII}
\label{sec:misc sources}

In this Section we will discuss two distinct types of signals:
(1) transient events, that last few to several
milliseconds, which, on astronomical grounds, are expected
to occur but emit waves of unknown phase evolution as in the case
of supernovae and (2) continuous radiation, that last for several
days or longer, from either newly born neutron stars or old 
recycled neutron stars. The strengths, duration and shapes of these
signals is rather speculative and highly uncertain. On general 
physical grounds we should expect such sources to exist and every
effort should be made in searching for these sources by taking
the best advantage of current knowledge. However, since the
astrophysical uncertainties are so large,
we shall keep the discussion of this topic quite qualitative.

\subsection {Transients}

\subsubsection{Supernovae and asymmetric explosions}
\label{sec:supernovae}


Supernovae (of type II) are associated with violent mass 
ejection with velocities
of order $0.001 c$ and the formation of a compact remnant ---
a neutron star or black hole. The event has at its disposal the difference
in the gravitational binding energy of the pre-collapse star and the
newly formed compact star which, neglecting the former, is:
\begin {equation}
|E| \sim 3 \times 10^{53} \left ( \frac {M}{M_\odot} \right )^2
\left ( \frac {R}{10~{\rm km}} \right )^{-1}~\rm erg.
\end{equation}
99\% of this energy is carried away by neutrinos, about 1\% is
transferred as kinetic energy of ejecta, a fraction $10^{-4}$ of
the total energy is emitted in the form of electromagnetic radiation.
Depending on how asymmetric the collapse is, some fraction of the 
total energy should be deposited into gravitational waves; 
spherically symmetric collapse, of course, cannot emit any radiation.
According to numerical simulations 
(see \cite{S:muller97} for a review) one might expect up to
$10^{-7}$ of the total energy to be emitted in gravitational waves.  
Together with uncertainties in the event rate, this is not a very
encouraging prognosis for the initial instruments
\cite{Thorne87_300yr,Thorne95_Next_Mill}.  It appears that a star collapsing to form a black 
hole is also not particularly well suited for detection 
by the existing resonant detectors and forthcoming 
interferometers \cite{astro-ph/0003321}.
However, second generation interferometers should be able to see a supernova
event as far as the Virgo super-cluster which contains about 200 bright 
galaxies and at least twice as many faint galaxies. 
In addition, there are a few other smaller clusters within that 
distance as also a large number of field galaxies. Therefore, the supernova 
event rate for these instruments could be as large as tens per year. Such
observations would undoubtedly be of great interest and would
shed light on hitherto un-understood processes that occur when a star
collapses to form a compact object.


	An asymmetric collapse generates gravitational waves but is
not necessarily accompanied by a change of linear momentum of the
exploding star. However,
as we emphasized above, the observations of single and binary radio-pulsars
require the presence of a significant kick velocity imparted
to a newly born neutron star. The kick velocity can be as large as
200--400 km/s. This testifies for a non-axisymmetric explosion leading
to an additional linear momentum acquired by the neutron star. A possible
reason for this asymmetry can be related with the asymmetry of the
neutrino emission during the collapse \cite{Burrows&Hayes95}.   
Using the observed
kick velocities one can evaluate asymmetry of the explosion and
calculate the amplitude of emitted gravitational waves.    
Nazin and Postnov \cite{NazinPostnov97} estimate that the
mean energy carried away by the gravitational wave burst
can reach the value $E_{GW}=5\times 10^{-6} M_\odot c^2$.
It is interesting to note that the radiated signal belongs to
the category of bursts ``with memory''
\cite{BraginskyGrishchuk85,BraginskyThorne87}. However, the
estimated strength of the asymmetric explosions is still too
low to count on them as reliable sources for first detection.

\subsubsection{Bar-mode and convective instabilities}
\label{sec:bar-mode}


	Some amount of gravitational waves may be emitted during 
two special stages of the collapse: (1) rotation-induced bars 
and (2) convective 
instabilities set up in the core of the newly born neutron star.  

	If the core's rotation is high enough it may cause the core to 
flatten before it reaches nuclear density, leading to an 
instability that transforms
the flattened core into a bar-like configuration which spins about its
transverse axis. Some of these instabilities could also fragment the
core into two or more pieces which then rotate about each other.
Both are efficient ways of losing energy in the form
of gravitational waves. It is estimated \cite{S:bonnell.pringle}
that the waves could carry up to $10^{-3} M_\odot c^2$
in a few ms.  LIGO and VIRGO detectors can see such an event at a distance 
up to 50 Mpc, or about 5--10 Mpc if the waves come off at 1 kHz.  
GEO will also be able to see these events provided the signal 
comes off around 200~Hz~\cite{S:schutz.98}.

Convective instabilities in the core of a
newly born neutron star, which last for about a second after the collapse,
are likely to produce gravitational radiation due to anisotropic mass
distribution and motion~\cite{Burrows&Hayes96}. 
M\"uller and Janka \cite{S:muller.janka} 
find that the gravitational wave amplitude of a source at 100 kpc is 
$h \sim 10^{-23}$ and that the waves would come off at about 100 Hz.
Since there would be about 100 cycles, one can enhance the amplitude
to about $10^{-22}$ provided we know the development of the signal
but this is still far too weak to be detected beyond about 10 kpc with
high confidence.

\subsubsection {Merger waves}
\label{sec:coalescence}

%
The physics of the merger phase that follows the adiabatic inspiral
regime of a compact binary coalescence is not known. Presently,
it is a subject of active research and contains a lot of uncertainties. 
Some authors expect that this phase could result in detectable amounts 
of radiation \cite{Flanagan&Hughes98} while others conclude that the 
merger phase is just
a continuation of the inspiral phase adding on the order of 1 more
cycle to the inspiral \cite{BuonannoDamour00}.  There is a lot
of effort to solve this important problem by using semi-analytical
and numerical techniques. It is likely that a solution will be in
place by the time the first detectors will begin to operate. 
The fact that the coalescence waves will be
preceded by inspiral waves makes the search easier, though it is not
inconceivable that while the former may be observable with the aid of
accurate search templates, the latter may not be. Flanagan \& Hughes
\cite{Flanagan&Hughes98} 
estimate that for binary systems of total mass in excess of 25M$_\odot$,
coalescence waves are likely to be significantly stronger than
the inspiral waves. The span of a detector is larger for heavier
binaries and therefore, these authors conclude, it is likely that the
first gravitational wave events will be the merger phases of massive 
binaries.

\subsubsection {Sub-stellar mass black hole binaries}
\label{sec:machos}

The results of recent micro-lensing experiments \cite{S:alcock.96}
have revealed massive compact halo objects (MACHOs) of mass
$0.5^{+0.3}_{-0.2}\,$M$_\odot.$ Nakamura et al.,
\cite{S:nakamura.etal} argue that if MACHOs are black holes then
they must have formed in the early Universe and they estimate that
our Galaxy may contain about $10^8$ black hole binaries with
inspiral time scales less than the Hubble time. If this is the case then
the rate of MACHO coalescences in our Galaxy is 
$\sim 5\times 10 ^{-2}$~yr$^{-1},$
implying an event rate of  few per year within 15 Mpc. As we shall
see in the Data Analysis Sections,  
first generation interferometers should be able to detect the
final inspiral phase of these systems.

\subsubsection {Quasi-normal modes}
\label{sec:ring down}

A compact binary coalescing as a result of gravitational radiation reaction 
would most likely result in the formation of a single black hole.
The newly formed hole will initially be somewhat non-spherical, and 
this dynamical non-sphericity will be radiated away in the form of 
gravitational waves.
The late time behaviour of this radiation is well studied in the
black hole literature and there are detailed calculations of the
(quasi) normal modes for both static, i.e.
Schwarzschild, and stationary, i.e. Kerr, black holes. In all cases
the time-evolution of the emitted radiation is well-modeled
by a quasi-periodic signal of the form
\begin{equation}
h(t; \tau, \omega) = A e^{-t/\tau} \cos(\omega t)
\end{equation}
where $\tau$ is the decay time-scale of the mode in question and
$\omega$ is the angular frequency of the mode, both of which depend
on the black hole mass and angular momentum. In all but the
extreme Kerr black holes (extreme Kerr black holes are those
that are spinning at the maximum possible rate) the only dominant
mode, i.e. the mode for which the decay time is the longest and
the amplitude is the highest, is the fundamental mode  whose
frequency is related to the mass $M$ and spin $a$ of the black hole
via $\omega=[1-0.63 (1-a)^{0.3}]/M$ where
$M$ is the mass of the black hole in units $G=c=1$ and $a=J/M$ is the
spin angular momentum of the hole in units of black hole mass. The decay
time $\tau$ is given by $\tau = 4/[\omega (1-a)^{0.45}].$ 
(See Ref.~\cite{S2:echeverria} and references therein for details.)

It is estimated  \cite{Flanagan&Hughes98} that 
during the quasi-normal mode ringing
of a black hole the energy emitted might be as large as 3\% of the 
system's total mass.  By matched filtering 
(cf. Sec.~\ref{sec:matched filtering}) it should be possible to
detect quasi-normal modes, in initial interferometers, from black holes
of mass in the range 60--$10^3$M$_\odot$ and at a distance of 
200~Mpc.  Binary black hole mergers should result in the emission of
such a ring down signal during the late stages. Thus, inspiral signals
emitted before the merger might aid in identifying the quasi-normal modes.

\subsection {Continuous waves} 
\label {sec:periodic.sources}

Our Galaxy is expected to have at least $10^{8}$ spinning neutron
stars that form roughly at a rate of one every 30 years. Some 
population of neutron stars is in binaries. There are a number of 
ways in which a single spinning neutron star could radiate away 
gravitational waves (if the neutron star is axisymmetric, of course, 
then there will be no gravitational wave emission): (1) Neutron stars 
normally spin at high rates (several to 500 Hz) and this must induce
some equatorial bulge and flattening of the poles. The
presence of a  magnetic field may cause the star to spin
about an axis that is different from the symmetry axis
leading to a time-varying quadrupole moment. 
(2) The star may have some density inhomogeneities in the
core/crust set up during its formation and/or subsequent convectively 
unstable motions of the core. (3) The presence of an accretion disc, 
with its angular momentum not necessarily aligned with that of the 
neutron star, can potentially alter axisymmetry. That and electromagnetic 
radiation reaction torques can induce and sustain wobble. 
(4) The normal modes of the neutron star fluid  (radial and other
oscillations) can extract rotational energy and re-emit in the form
of gravitational waves. (5) There are certain classical and relativistic
instabilities in the neutron star fluid which may cause the star
to radiate away energy in the form of gravitational radiation. In what
follows we will only discuss a sample of recent work on the radiation
from spinning neutron stars.

\paragraph{GW amplitude from spinning asymmetric neutron stars}

If $I_{zz}$ is the moment of inertia about the spin axis of a neutron
star emitting gravitational waves at a frequency $f$ then the 
gravitational amplitude at a distance $r$ is:
\begin {equation}
h = 3 \times 10^{-27} \left ( \frac {10~{\rm kpc}}{r} \right )
\left ( \frac {I_{zz}} {10^{45}~g~cm^2} \right )
\left ( \frac {f}{200~{\rm Hz}} \right )^2
\left ( \frac {\epsilon}{10^{-6}} \right ),
\label{eq:amplitude}
\end {equation}
where $\epsilon$ is the ellipticity of the star. 
In a simple model of an equatorial plane of elliptical cross section of
semi-major axis $a_1$ and semi-minor axis $a_2,$ the ellipticity
is $\epsilon \equiv 1-a_2/a_1$. The ellipticity is an unknown but
one can obtain an upper limit on it by attributing the observed 
spin-down of pulsars $\dot P$ to gravitational radiation back reaction, 
namely that the change in the rotational energy $E=I\Omega^2/2$ 
is equal to gravitational wave luminosity. Then, the ellipticity  
is related to the spin-down rate of a pulsar via
\begin {equation}
\epsilon = 5.7 \times 10^{-6} 
\left ( \frac {P}{10^{-2}~{\rm s}} \right )^{3/2}
\left ( \frac {\dot P}{10^{-15}} \right )^{1/2}.  
\end {equation}
Since one knows the observed values of $P$ and $\dot P$ one can obtain
an upper limit on $\epsilon$ using the above equation. Following this
method one can find that for the Crab pulsar 
$\epsilon \le 7 \times 10^{-4}$ and the gravitational amplitude is 
$h \le 10^{-24}.$ One concludes that Crab will have a sufficiently large
amplitude to be observable with the aid of GEO detector \cite{LPG:sch} in
1 year of continuous observing, if all of its spin down can be attributed
to the emission of GW. It is unlikely that the ellipticity is so large.
Yet, the prospect
of seeing Crab event at a 10th or a hundredth of this ellipticity is
quite good with first/second generation of interferometers.

In the next two paragraphs we will discuss some new developments
in relativistic astrophysics that could lead to potential 
GW sources.

\paragraph{Relativistic instabilities in young neutrons stars}

Chandrasekhar \cite{S:chandra} and Friedman \& Schutz 
\cite{S:friedman.schutz} discovered an instability (now called the
CFS instability) in the fundamental,
or ``f'', mode of a neutron star fluid, arising as a result of GW emission.
The mode goes unstable above a critical spin frequency of the star
and progressively grows, instead of decaying, by emitting gravitational waves.

The physics behind this instability can be understood in the following 
manner: 
Imagine exciting a mass-quadrupole mode -- that is a non-uniform distribution
of mass -- in a non-spinning star. The mass inhomogeneity will travel on the
surface of the star and this dynamical asymmetry will cause the star to 
radiate gravitational waves. After a while, all the energy in the mode 
will be radiated away and the mode will decay. Now consider a spinning
neutron star in which a co-rotating and counter-rotating modes are 
excited. These
fluid modes have a certain pattern speed on the surface of the star. 
For low spin rates both these modes will decay in course
of time by emitting gravitational waves. But as the neutron star is
spun up above a critical rate, to an external inertial observer 
both modes will appear to be traveling in the 
same sense as the rotation of the star. 
Therefore, the mode counter-rotating relative to the star will
emit positive angular momentum, causing the angular momentum associated
with the mode to enhance, or for the amplitude of the mode to increase.
In other words, a mode counter-rotating relative to the star, but seen
co-rotating relative to the inertial observer, can only emit negative
angular momentum which causes its own angular momentum to increase.
The energy for this enhancement is supplied by the spin angular momentum
of the neutron star. Thus, while the mode co-rotating 
with the star's spin will decay, the mode 
counter-rotating with the spin will grow in amplitude and emit
more and more radiation. This will go on until the mode has sucked
out enough angular momentum of the star to make the counter-rotating
mode appear to be counter-rotating with respect to an inertial 
observer too. 

It is suspected that the CFS instability will not work in the presence of
viscosity and hence it may be unimportant in old neutron stars. However,
newly born neutron stars will be very hot and viscous forces may be 
insignificant in them.  Recently, Andersson \cite{S:andersson} discovered 
another 
class of modes called $r$--modes, which --- unlike CFS modes that are
mass-quadrupole moments --- are current-quadrupole moments, that are unstable
at all spin frequencies. The role of these modes may be significant in
young neutron stars. It is proposed that $r$--modes
are responsible for the limit on the spin frequencies of newly born
neutron stars \cite{S:Andersson&99}. 
Owen {\it et al.} \cite{S:owen.etal} have
computed the efficiency with which these modes extract energy out
of the system and the expected gravitational wave amplitude from
isolated neutron stars, as well as from the ensemble of all sources up to
cosmological distances. They conclude that the second generation of
interferometric antennas will be able to distinguish such a 
background by coincident observations with a nearby bar detector.
Such observations should provide independent knowledge of the distribution
of galaxies in the high-red shift Universe as also on the star formation
history via the observation of $r$--modes associated with the formation of
neutron stars and pulsars.

\paragraph {Neutron stars in X-ray binaries}

In the recent years, Rossi satellite observations of the X-ray emitting
binaries have shown high-frequency quasi-periodic oscillations (QPO) 
in their X-ray
power spectra. Some authors believe that these QPO could be 
a result of beating
of two frequencies one of which is that of the neutron star. The
neutron star spins inferred in this way seem to lie in a
narrow range of 250--350 Hz and are all within 20 \% of 300 Hz. 
A neutron star may be born with a high spin rate (several 100 Hz)
but quickly spins down to moderate rates (several 10 Hz). 
In a binary system, when the companion becomes a red giant the neutron
star starts accreting mass and angular momentum. Though the mass 
accretion rate is very low ($\dot M \sim 10^{-10}$~M$_\odot$~yr$^{-1}$)
the accretion of angular momentum can spin up a neutron star and heat its
crust substantially. Indeed, the millisecond pulsars are believed to be old
pulsars in binaries recycled in this way. It is puzzling as to why
the spin frequencies of neutron stars in X-ray binaries are all in
a narrow range. 
Bildsten suggests \cite{S:bildsten.98} that absence of efficient 
heat transport 
processes make it possible to set up temperature gradients in 
accreting neutron stars.
Provided that the temperature distribution has large scale asymmetry
then the temperature sensitive electron
captures in the deep crust can build up the mass 
quadrupole ($\sim 10^{-7}MR^2$)
needed to radiate away accreted angular momentum and limit the spin frequency.
This mechanism is present only during accretion and decreases rapidly once
the accretion halts. The frequency of GW radiation will be known in
advance since one knows the spin frequency of the pulsar via X-ray 
observations. It is argued \cite{S:bildsten.98} that the gravitational 
wave strength
will be $h \sim (0.5$--$3) \times 10^{-26}$ for many of these sources and
that the LIGO/VIRGO and signal recycled GEO600, can detect the strongest 
of these
sources, Sco X-1, at an SNR of 5 with a few years of integration.

\section{Astrophysical stochastic backgrounds of gravitational waves}
\label{sec:ApStB}
\label{sec:secIII}

Coalescing binaries of compact stars are at the center of our attention.
A gravitational wave signal will be monitored by a detector as long as 
the changing frequency of the source sweeps through the detector's window
of sensitivity. The detection of a useful gravitational wave source is limited 
by the instrumental noise and by a possible gravitational wave noise
produced by other sources. If a large population of astrophysical sources emit 
overlapping gravitational waves, the resulting signal will be perceived 
by the detector as
a gravitational wave noise. In principle, this noise could be larger than the
instrumental noise. The astrophysical backgrounds of gravitational
waves are important in their own right, but we are discussing them
mostly as possible noises preventing the detection of a signal of a greater
scientific importance. The aim of our discussion is the derivation of
the gravitational mean square amplitude in a given frequency 
interval $\Delta f$ and
comparison of this number with the instrumental noise in the same interval. If
the background is below the instrumental noise, it will not  prevent
the detection of the useful signal. In subsection \ref{sec:Unres}
we derive general
formulas for the performance of a large number of unresolved sources. 
In subsection \ref{sec:NoigeOldNsLIGO} we discuss the
population of rotating neutron stars in our Galaxy. 
They could constitute a major danger for the ground-based interferometers.
The analysis shows that the danger arises only under quite
unrealistic assumptions about the parameters of this 
population. In subsection \ref{sec:NoigeGalWdLISA} we discuss the
stochastic gravitational wave background produced by binary  
white dwarfs in the Galaxy. This background  dominates the LISA instrumental 
noise from $\sim 10^{-4}$~Hz up to $\sim  10^{-3}$~Hz, but 
leaves the gravitational wave sky transparent at lower and higher frequencies. 
Finally, subsection \ref{sec:NoigeExtraGalBin} comments on 
astrophysical backgrounds of
extra-Galactic origin. Generically, these backgrounds are one
order of magnitude lower than those produced by the Galactic sources.

\subsection{Unresolved sources in our Galaxy}
\label{sec:Unres}

A large collection of independent sources produces signals whose 
intensities add. Consider $N$ identical sources located at 
approximately the same distance 
$r$ from the observer. The resulting gravitational wave field is
characterized by the r.m.s. amplitude $h_N$: 
\beq{h_N}
h_N= h_1\sqrt{N},   
\eeq 
where $h_1$ is the averaged amplitude of a single source. 
We are interested in the narrow frequency interval from $f$ 
to $f + \Delta f$. The radiating sources
gradually change their frequency and pass through the window of interest.
The crossing of the window can occur either on the way from lower to 
higher frequencies, as in the case of binaries, or in the opposite 
direction, as in the case of rotating neutron stars. 
To find $h_N$ we need to know the number $N(f, \Delta f)$ of the 
radiating systems in the discussed frequency interval. 
Denote by $\R$ the rate at which
the sources appear in the window. A source is present in the window
during the time $\Delta t$, where $\Delta t = \Delta f/ {\dot f}$. Thus,
\beq{N}
N(f, \Delta f) = \R \Delta t = \R \frac{\Delta f}{\dot f},
\eeq
and
\beq{h_N2}
h_N= h_1\sqrt{\R \frac{\Delta f}{\dot f}}.
\eeq 
If the frequency evolution is driven by gravitational waves only,
the quantity $\dot f$ is determined by the gravitational radiation 
damping. For example, in the case of a binary system, the $\dot f$ is 
given by Eq. (\ref{A:dotf}). Alternatively, the quantity $\dot f$ can be 
determined by the electromagnetic radiation damping. This takes place 
in the case of highly magnetized rotating neutron stars whose
$\dot f$ is determined by electromagnetic, rather than gravitational,
losses. The appearance rate $\R$ of sources of a given population in the 
discussed window $\Delta f$ is equal to the birthrate or 
to the coalescence rate of the sources as such 
(see, for example, $\R_{\rm coales}$ 
in Sec.~\ref{sec:secII}). This is true as long as the number of sources remains 
constant during the frequency evolution throughout the window, what we 
always assume.

For sources driven by gravitational radiation only, 
formula (\ref{h_N2}) can be expressed in
terms of the energy of a single radiating system and the rate $\R$.
One needs to use Eq. (\ref{A:loss2}) in order to express $h_1^2$ in 
terms of $dE/dt$, and to take into account the relationship $dE/dt =
(dE/df) {\dot f}$. Then, Eq. (\ref{h_N2}) takes the form
\beq{h_N^2}
h_N^2 = \frac{G}{c^3} \frac{\R}{r^2 (\pi f)^2} \frac{dE}{df} \Delta f.
\eeq
Usually, the energy of the radiating system is a power-law function of
the frequency $f$: $E(f)\sim f^\alpha$. For example, in the case of 
a binary star in circular orbit one derives from Eq. (\ref{B:E}):
\beq{E(f)}  
E(f)= \frac{G^{2/3}}{2} \M^{5/3} (\pi f)^{2/3},
\eeq
where $\M$ is the chirp mass. So, in the case of binaries, $\alpha = 2/3$.
For a non-axisymmetric rotating star $E(f) \sim f^2$, so that $\alpha = 2$.
Thus, Eq. (\ref{h_N^2}) takes the universal form
\beq{h_N^2.2}
h_N^2 = \frac{G}{c^3} \frac{\R}{r^2 (\pi f)^2} \alpha E(f) \frac{\Delta f}{f}.
\eeq

The quantity $h_N^2/\Delta f$ is the mean square noise amplitude
$h_f^2$ (with dimensionality Hz$^{-1}$) which participates in the expression
\beq{}
\langle h^2 \rangle = \int\limits_{f_{\min}}^{f_{\max}} h_f^2 df\,, 
\eeq
and can now be compared with the frequency-dependent instrumental
noise. We will work with the dimensionless spectral amplitude
\beq{h(f)}
h_{N}(f) \equiv h_N \sqrt{\frac{f}{\Delta f}} =  
\frac{1}{r} \sqrt{\frac{G \alpha E(f) \R}{c^3 (\pi f)^2}}. 
\eeq

Obviously, the independent sources in a frequency bin $\Delta f$ form a
kind of stochastic background
if the number $N(f, \Delta f)$ is much larger than 1. A source becomes
resolvable, if this number is of order 1. For a collection of evolving 
sources one can find the limiting frequency $f_{\lim}$ at which this
happens. For fixed $\Delta f$ and $\R$, one uses the concrete 
function ${\dot f}(f)$ (arising due to the gravitational reaction 
force or by some other reasons) and finds $f_{\lim}$ from the requirement
\beq{flim}
\R \frac{\Delta f}{\strut\dot f} = 1.  
\eeq
For a collection of binary stars one uses Eq. (\ref{A:dotf}) and finds
\beq{}
\eqalignleft{
f_{\lim} \approx &(1.2\times 10^{-3}\hbox{Hz})
\R_{300}^{3/11}\myfrac{\Delta f}{3\times 10^{-8} \hbox{Hz}}^{3/11}
\myfrac{\M}{0.52 M_\odot}^{-5/11}
\cr}\,.
\eeq 
The appearance (coalescence) rate $\R_{300}$ is chosen for compact
white dwarfs, which are expected to coalesce in our Galaxy once per
300 years. The chirp mass is normalized to 
0.52~M$_\odot$ which is true for two CO white dwarfs with masses 
0.6~M$_\odot$. 
This estimate will be needed in subsection \ref{sec:NoigeGalWdLISA} 
which discusses the LISA noise.

\subsubsection{Noise from old neutron stars at frequencies of ground-based
interferometers} 
\label{sec:NoigeOldNsLIGO}

Rotating neutron stars as sources of gravitational radiation can be roughly
divided into two populations: One consists of old neutron stars with relatively
weak magnetic fields and small electromagnetic losses. Their rotational
frequency slowly decreases due to gravitational wave damping. Another 
population consists of young highly magnetized neutron stars. Their
rotational frequency decreases much faster due to electromagnetic damping.
In course of their frequency evolution, members of both populations cross
the window of sensitivity of ground-based interferometers, descending 
from $10^3$ Hz to $10$ Hz. The number of sources simultaneously
radiating in a given frequency interval is proportional to the birthrate
$\R$ of the population and inversely proportional to the velocity of the 
population flow $\dot f$ through the window. We start from old neutron 
stars and then discuss young neutron stars. 

The frequency evolution of old, rotating, deformed neutron stars is governed by 
gravitational radiation damping. Formula for $\dot f$, analogous to Eq. (\ref{A:dotf})
for double stars, is 
\beq{fdot}
\dot f = \frac{32 \pi^4 G}{c^5} I \epsilon^2 f^5,
\eeq
where $I$ is the relevant moment of inertia and $\epsilon$ is the 
ellipticity (deformation)  
parameter. The birthrate $\R$ of neutron stars in this population 
can be estimated using the observed fraction of millisecond pulsars
(which are thought to be old neutron stars with low magnetic fields
that have been spun up by accretion in a binary system)
among radio pulsars: $N_{ms}/N_{PSR}\approx 20/2000= 1/100$. Remembering
that the life-time of a millisecond pulsar is $t_{ms}=10^8$~yr
and that the life-time of an ordinary radio pulsar is $t_{PSR}=10^6$~yr, 
and adopting the Galactic birth rate of radio pulsars 1 per 30 years, 
as for the core collapse supernovae, one can estimate 
the birth rate of millisecond pulsars: 
$$
\R_{ms}=\R_{PSR}\frac{N_{ms}}{N_{PSR}}\frac{t_{PSR}}{t_{ms}}\approx
3\times 10^{-6}\hbox{yr}^{-1}
$$
This estimate is in agreement with 
the one derived in Ref. \cite{Kulkarni_ea} from the observed 
space density of millisecond pulsars.      

From Eq. (\ref{flim}) one finds the limiting frequency
\beq{flimold}
f_{\lim} \approx 53~\hbox{Hz}\, \myfrac{\R}{3\times 10^5\,\hbox{yr}}^{1/5}
\myfrac{\Delta f}{3\times 10^{-8}\hbox{Hz}} 
\myfrac{I}{10^{45}\hbox{g~cm}^2}^{-1/5}
\myfrac{\epsilon}{10^{-9}}^{-2/5}.
\eeq
Thus, at frequencies below $53$~Hz the population of old neutron stars
is likely to produce a stochastic background. Taking 
$E(f) = \pi^2 I f^2 / 2$ one derives from Eq. (\ref{h(f)}):  
\begin{equation}
h_N (f) = \frac{1}{r} \sqrt{\frac{G}{c^3} I\R}\,
\approx 2\times 10^{-26}\left(\frac{10\hbox{kpc}}{r}\right)
\left(\frac{\R}{3\times 10^5\,\hbox{yr}}\right)^{1/2}
\left(\frac{I}{10^{45}\hbox{g~cm}^2}\right)^{1/2}. 
\label{h_lim}
\end{equation}
It is interesting that this quantity does not depend on the
deformation parameter $\epsilon$, as soon as $\epsilon \neq 0$. The $h_1^2$
and $\dot f$ are both proportional to $\epsilon^2$, so that $\epsilon^2$ 
cancels out in the expression for $h_N^2$. The quantity $h_{N}(f)$ is
also independent of frequency $f$. The numerical level 
of $h_N (f)$ is much lower
than the instrumental noise of initial and advanced ground-based 
interferometers.
The $h_N (f)$ can be increased by two orders of magnitude, and hence the
gravitational wave noise from old neutron stars becomes marginally 
detectable, only under the condition that one postulates a significantly
larger (and, we believe, unrealistic) birthrate $\R$ for old neutron stars 
(compare with \cite{&Giazotto&97}). 

At frequencies higher than the limiting
frequency (\ref{flimold}), the  
sources are resolvable during a 1 year interval of observations. However,
to monitor a single neutron star one needs to know its exact location
on the sky and to take care of the Doppler frequency modulation due to the
Earth's motion around Sun.

The young neutron stars differ from old neutron stars in that their
electromagnetic energy loss
\beq{Eem}
{\dot E}_{em} = \frac{2 \pi^4}{3 c^3} \mu^2 f^4,  
\eeq
where $\mu$ is the NS magnetic moment,
is significantly larger than the gravitational wave loss
\beq{Egw}
{\dot E}_{gw} = \frac{32 \pi^6 G}{c^5} I^2 \epsilon^2 f^6. 
\eeq
The ratio $x = {\dot E}_{em}/{\dot E}_{gw}$ is   
\begin{equation}
x \approx 4\times 10^3  (\mu_{30})^2 (\epsilon_{-6})^{-2}(I_{45})^{-2}
\left(\frac{100 \hbox{Hz}}{f}\right)^2, 
\end{equation}
where $\mu_{30}=\mu/(10^{30}\,\hbox{G cm}^3)$,  
$\epsilon_{-6} = \epsilon/10^{-6}$ and $I_{45}=I/10^{45}$\,g~cm$^2$. 
For typical parameters of
young neutron stars one has $x \gg 1$. The ratio $x$ becomes comparable
with 1 only for relatively weak magnetic fields, such that the
magnetic moment $\mu$ satisfies the condition 
$
\mu < 1.5\times 10^{26} (\hbox{G cm}^3) \epsilon_{-6} (f/100\,\hbox{Hz}).  
$
The frequency change $\dot f$ is  
determined by the electromagnetic loss and reads  
\beq{fdotem}
{\dot f} = \frac {2 \pi^2}{3c^3} \frac {\mu^2 f^3}{I}.  
\eeq 
For $f_{lim}$ one derives 
\beq{flimyoung}
f_{\lim} = 0.5~\hbox{Hz}~(\R_{30})^{1/3}
\myfrac{\Delta f}{3\times 10^{-8}\hbox{Hz}}^{1/3} (\mu_{30})^{-2/3}
(I_{45})^{1/3}.
\eeq
The averaged amplitude $h_1$ of a single neutron star amounts to
\beq{h_1}
h_1 = \frac{G \pi^2 \sqrt{32}}{c^4} \frac {1}{r} I f^2 \epsilon.
\eeq
Then, the $h_N (f)$ found from Eq. (\ref{h_N2}) with the help 
of Eq. (\ref{fdotem}) becomes
\begin{equation}
h_{N} (f) =\frac{4 \sqrt{3} \pi G}{c^{5/2}}\frac{1}{r} I^{3/2} \R^{1/2}
\epsilon \mu^{-1} f \approx 3\times 10^{-26}
\left(\frac{10\hbox{kpc}}{r}\right)
\R_{30}^{1/2}I^{3/2}_{45}\epsilon_{-6}\mu_{30}^{-1}{f/100 Hz}. 
\label{h(nu)_em}
\end{equation}
Thus, the gravitational wave noise from young neutron stars is at the same
numerical level as the noise from old neutron stars. The appearance
rate of young neutron stars is much higher than that of old neutron
stars, but they flow through the window of sensitivity much faster, so that
there aren't a sufficiently large number of sources to produce a background 
of a high enough level.  We conclude, that the populations of 
neither the old nor new neutron stars do present   
any danger for the sensitivity curves of initial and advanced ground-based
instruments. It is important that the calculated noise level is below the
expected level of the relic gravitational waves (see Section 6).

\subsubsection{Noise from Galactic binary white dwarfs in LISA}

\label{sec:NoigeGalWdLISA}

The LISA frequency range, $10^{-4}$--$10^{-1}$~Hz, can be contaminated
by the gravitational wave noise from coalescing binary 
white dwarfs (WD) and binary neutron stars (NS).
The appearance rates $\R$ of the two populations in the sensitivity 
window are numerically equal to their coalescence rates. The binary
WD systems are much more numerous than the binary NS systems, and the
coalescence rate of the former population is significantly higher than
that of the latter. The binary WD coalescence rate is about 
1 per 300 yrs, while the coalescence rate of binary NS is
about 1 per several 10000 years. At the same time, the chirp masses, 
$\M \approx 0.52 M_{\odot}$ for a WD binary and 
$\M \approx 1.22 M_{\odot}$ for a NS binary, are not so significantly 
different. This is why the WD background is more important than the NS background,
and we consider only the former.   

For a collection of sources consisting of binary stars, one can use
Eqs. (\ref{A:h1}), (\ref{A:dotf}) in Eq. (\ref{h_N2}), or, alternatively,
Eq. (\ref{E(f)}) and $\alpha = 2/3$ in Eq. (\ref{h_N^2.2}).
By either way one obtains  
\begin{equation}
\label{h_N.bs}
\eqalign{
h_{N}(f)&=\frac{G^{5/6}}{\sqrt{3} \pi^{2/3} c^{3/2}}\frac{1}{r} 
\R^{1/2} \M^{5/6} f^{-2/3} \cr
&= 10^{-20} 
\myfrac {10\,\hbox{kpc}}{r}
\myfrac {\R}{300\,\hbox{yr}}^{1/2} 
\myfrac{f}{10^{-3}\,\hbox{Hz}}^{-2/3}
\myfrac {\M}{0.52 M_\odot}^{5/6}. \cr
}
\end{equation} 
Roughly, this is a result of performance of $\approx 10^6$
binaries in the frequency bin $\Delta f = f = 10^{-3}$ Hz with the
averaged amplitude (see Eq.~(\ref{A:h1})) 
\beq{h_1.individ}
h_1 = \frac {\sqrt{32} \pi^2 G^{5/3}}{\sqrt{5} c^4} \frac{1}{r} 
\M^{5/3} f^{2/3} = 2.5 \times 10^{-23} \myfrac{10\hbox{kpc}}{r} 
\myfrac{\M}{0.52~ M_\odot}^{5/3} \myfrac{f}{10^{-3}\hbox{Hz}}^{2/3}.
\eeq
It is necessary to note that Eq.(\ref{h_N.bs}) gives the estimate 
for the amplitude 
averaged over the whole sky, while the real background is strongly
concentrated toward the Galactic plane~\cite{L&95_gwsky}.
The response of a space-based interferometer should be modulated while
the instrument is turning in its orbit. In principle, this distinctive 
feature of the Galactic background can be used in order to distinguish
it from the backgrounds of cosmological origin~\cite{Giam&Poln97}.

In formula (\ref{h_N.bs}), the quantity $\R$ replaces  
all the astrophysical uncertainties in the binary WD evolution. 
At frequencies higher than $3\times 10^{-4}$~Hz the evolution of
the vast majority of binary white dwarfs is totally controlled by
GW emission. So, at frequencies of interest,  
the GW noise is fully determined by the Galactic rate of binary
WD mergers and is independent of complicated details of binary
evolution at lower frequencies. (For  examples of calculated spectra at
all frequencies see  \cite{LP87_gw,LPP87_gw,Hils&90,Schneider&00}.) 
The coalescence rate of close binary WD is known only up 
to a factor of few. One way to estimate $\R$ is based 
on the search for nearby WD  
binaries. A recent study \cite{Marsh&95} revealed a larger
number of such systems than had been previously believed to exist. However, the
statistics of such binaries in the Galaxy remains very poor.
If coalescing binary WD are associated with SN Ia explosions, as
proposed by \cite{Iben&Tut84} and further investigated by many
authors (for a recent review of SN Ia progenitors see \cite{Branch&95}),
their coalescence rate can be constrained using the much more
representative SN Ia statistics. The authors in Ref. \cite{Branch&95} have concluded 
that the coalescing CO--CO binary WD remain the most plausible candidates 
responsible for the SN Ia events. The Galactic rate of SN Ia is
estimated to be $4\times 10^{-3}$ per year \cite{Tamm&94,vdBergh&mClu94},
which is close to the calculated rate of
CO--CO coalescences $\sim (1$--$3)\times 10^{-3}$. The coalescence rate
of the He--CO and He--He WD pairs (other possible progenitors of SN Ia) 
falls ten times short of that for CO--CO WD \cite{Branch&95}.  As SN Ia
explosions may well be triggered by other mechanisms as well, we conclude 
that the observed SN Ia rate provides an upper limit to the
double WD merger rate, regardless of the evolutionary considerations.

In Fig. \ref{LPG:fig8} we plot the LISA sensitivity curve (thick
solid line) calculated for the frequency bins 
$\Delta f = 3\times 10^{-8}$~Hz as a function of frequency.
The binary confusion limit is shown with the dash-dotted line.
At frequencies below $\sim 4\times 10^{-4}$ Hz the binary GW
background is produced not only by coalescing WD, but by other binaries
as well. So, in this part of the graph we
rely upon numerical calculations \cite{LPP87_gw,PP98_lisa}. 
At the limiting frequency $\sim 10^{-3}$~Hz, individual Galactic 
WD binaries become resolvable in a 1 year observation time,  
and the binary WD noise drops 
below the LISA sensitivity. It then continues as a noise produced
by isotropic distribution of extra-Galactic binaries (see Sec. 5.2). 
Fig. \ref{LPG:fig8} also shows the expected background of 
relic gravitational waves (see Section \ref{sec:secIV}).
Keeping in mind that the real noise caused by merging 
Galactic WD can be smaller than the plotted one and,
in any case, is direction dependent, we conclude that in the
frequency interval $\sim 10^{-3}-10^{-1}$ Hz none of GW backgrounds of
Galactic origin should be higher than the LISA
sensitivity in the frequency bins 
$\Delta f = 3\times 10^{-8} Hz$. If LISA detects a GW background
in this frequency interval, it is expected to be of a primordial 
origin. 
Although the low-frequency part of the binary confusion 
limit is somewhat more model-dependent
(it is determined by the actual number of binary stars
in the Galaxy, their space distribution and details of
binary evolution), the 
calculated noise drops below the LISA sensitivity curve 
at frequencies below $\sim 10^{-4}$~Hz. 
This leaves open for
the search for cosmological backgrounds of primordial origin 
some low-frequency portion of the LISA sensitivity window, in addition to
the already discussed interval $\sim 10^{-3}$--$10^{-1}$~Hz.

\subsection{Gravitational wave noise from extra-Galactic binaries}
\label{sec:NoigeExtraGalBin}

Simple estimates show that the isotropic extra-Galactic background 
is expected to be one and a half order of magnitude smaller than the 
sky-averaged GW noise from Galactic binaries (see
\cite{LP87_gw,LPP87_gw,L&95_gwsky,Hils&90}). Consider a volume of
space with radius $r= 300$~Mpc. This is a large volume, but one can
still neglect effects of curvature and cosmological time-dependence
of star formation rate. According to Eq. (\ref{RV3}), event
rate in this volume $\R_V$ is related with the Galactic event rate $\R_G$
as $\R_V = 3\times 10^6 \R_G$. Formula (\ref{h(f)}), written for
extra-Galactic sources, should now contain
$\R_V$ instead of the galactic rate $\R$, and $r=300 Mpc$ instead of
the characteristic Galactic distance $r= 10 kpc$. Combining the numbers,
one finds the relationship between the extra-Galactic amplitude 
$h_N^{\rm EG}(f)$ and the Galactic amplitude $h_N(f)$: 
\beq{h_N.eg}
h_N^{EG}(f) \approx 5\times 10^{-2} h_N(f). 
\eeq
Thus, the noise amplitude from extra-Galactic binaries is expected
to be a factor 20 smaller than the noise amplitude from binaries 
in our Galaxy.  

More sophisticated calculations
take into account the somewhat larger star formation rate at larger 
red shifts \cite{Kosenko&Postnov98,Schneider&00}. 
According to these studies, unresolved 
extra-Galactic binaries can contribute up to 10\% of the mean 
Galactic noise. This is still smaller than the projected LISA sensitivity
and thus presents no danger of contamination. The contribution
of unresolved extra-Galactic binaries is shown in Fig.~\ref{LPG:fig8} 
to the right of the frequency $\approx 10^{-3}$~Hz, where 
Galactic binaries become resolvable and their noise contribution
sharply drops down.
An additional distinctive feature of GW backgrounds from sources
in distant galaxies is a certain angular anisotropy of the
background caused by inhomogeneities in the distribution of galaxies
over the sky \cite{Kosenko&Postnov99}.


\section{Relic Gravitational Waves and Their Detection}
\label{sec:LPG}
\label{sec:secIV}

\subsection{Introduction}  
\label{sec:secIV:intro}

The existence of relic gravitational waves is a consequence of quite
general assumptions. Essentially, we rely only on the validity of general 
relativity and basic principles of quantum field theory. The strong 
variable gravitational field of the early Universe amplifies the inevitable
zero-point quantum oscillations of the gravitational waves and produces
a stochastic background of relic gravitational waves measurable 
today \cite{LPG:g1:a,LPG:g1:b,LPG:g1:c}.
The detection of relic gravitational waves is 
the only way to learn about the evolution of the very early Universe, 
up to the limits of the Planck era and the big bang.
It is important to appreciate the fundamental and unavoidable nature of this
mechanism. Other physical processes can also generate stochastic backgrounds
of gravitational waves. But those processes either involve many additional
hypotheses, which may turn out to be not true, or produce a gravitational wave 
background (like the one from binary stars in the Galaxy) which should 
be treated as an unwanted noise rather than a useful and interesting signal.
The scientific importance of detecting relic gravitational waves has been 
stressed on several occasions (see, for example, 
\cite{Thorne87_300yr,Thorne95_Next_Mill,LPG:sch}). 
\par
The central notion in the theory of relic gravitons is the phenomenon
of super-adiabatic (parametric) amplification. The roots of this phenomenon
are known in classical physics, and we will remind its basic features. As
every wave-like process, gravitational waves are amenable to the concept of a harmonic
oscillator. The fundamental equation for a free harmonic oscillator is  
\begin{equation}
\label{1}
\ddot{q} + \omega^2 q = 0, 
\end{equation}
where $q$ can be a displacement of a mechanical pendulum or a 
time-dependent amplitude of a mode of the physical field. The energy of the
oscillator can be changed by an external force or, alternatively, by a 
parametric influence, that is,
when a parameter of the oscillator, for instance the length of a pendulum,
is being changed. In the first case, the fundamental equation takes the
form 
\begin{equation}
\label{2}
\ddot{q} + \omega^2 q = f(t),
\end{equation}
whereas in the second case Eq. (\ref{1}) takes the form
\begin{equation}
\label{3}
\ddot{q} + \omega^2 (t) q = 0.  
\end{equation}
Equations (\ref{2}) and (\ref{3}) are profoundly different, both, 
mathematically and physically. 
\par
Let us concentrate on the parametric influence. We consider a pendulum
of length $L$ oscillating in a constant gravitational field $g$. 
The unperturbed pendulum
oscillates with the constant frequency $\omega = \sqrt{g/L}$. 
Fig.~\ref{LPG:fig1}a illustrates the 
variation of the length of the pendulum $L(t)$ by an external agent, 
shown by alternating arrows. Since $L(t)$ varies, the frequency 
of the oscillator does also vary: $\omega(t) = \sqrt{g/L(t)}$. 
The variation in $L(t)$ need not 
be periodic, but it cannot be too slow (i.e., adiabatic) for the result
of the process to be significant. Otherwise, in the adiabatic
regime of slow variations, the energy of the oscillator $E$ and its 
frequency $\omega$ do change slowly, but $E /\omega$ remains constant, so
one can say that the ``number of quanta" 
$E/ \hbar \omega$ in the oscillator remains fixed.  
In other words, for the creation of new ``particles - excitations", 
the characteristic time of the variation should be 
comparable with the period of the oscillator and the adiabatic behaviour 
should be violated. After some duration of the appropriate parametric influence,
the pendulum will oscillate at the original frequency, but will have a 
significantly larger, than before, amplitude and energy. 
This is shown in Fig.~\ref{LPG:fig1}b. Obviously, the energy of the oscillator has 
been increased at the expense of the external agent (pump field). For
simplicity, we have considered a familiar case, when the
length of the pendulum varies, while the gravitational acceleration $g$
remains constant. Variation in $g$ represents a gravitational parametric
influence that would be in an even closer analogy with what we study below.    

\if t\figplace
\begin{figure}
\epsfxsize=0.8\textwidth
\centerline{\epsfbox{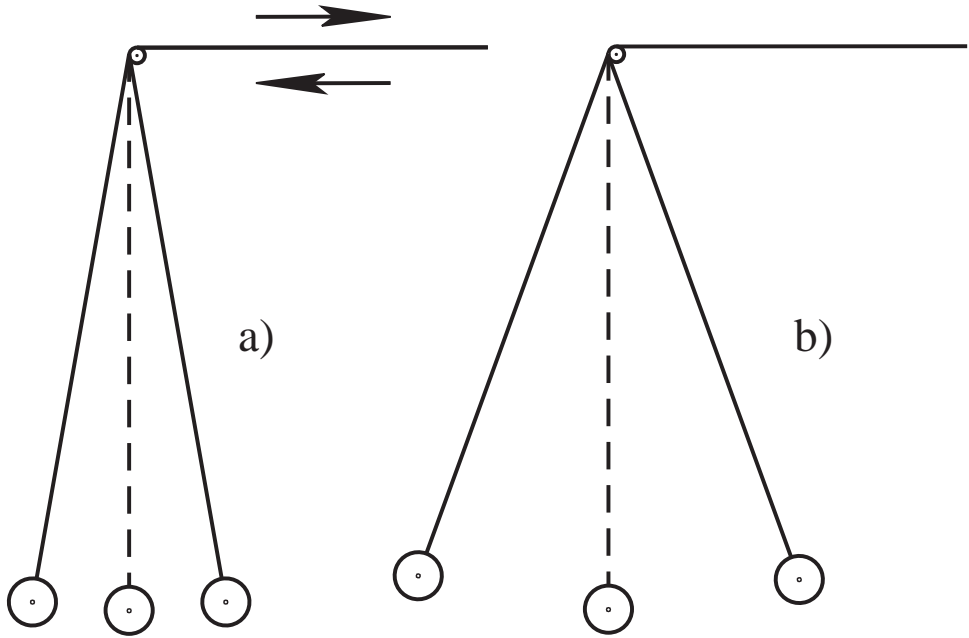}} 
\caption{Parametric amplification. a) variation of the
	length of the pendulum, b) increased amplitude of
        oscillations.}
\label{LPG:fig1}
\end{figure}
\fi
   
A classical oscillator must have a non-zero initial amplitude for the
amplification mechanism to work. Otherwise, if
the initial amplitude is zero, the final amplitude will also be zero. 
Indeed, imagine the pendulum to be strictly at rest, hanging straight down. 
No variation in its length will cause the pendulum 
to oscillate and gain energy. In contrast,
a quantum oscillator does not need to be excited from the very beginning. 
The oscillator can be
initially in its quantum-mechanical vacuum state. The inevitable zero-point
quantum oscillations are associated with the vacuum state energy
$\frac{1}{2} \hbar \omega$. One can imagine a pendulum hanging straight down,
but fluctuating with a tiny amplitude determined by the ``half of the
quantum in the mode". In the classical picture, it is this tiny amplitude 
of quantum-mechanical origin that is being parametrically amplified.

The Schroedinger evolution of a quantum oscillator depends crucially 
on whether the oscillator is being excited parametrically or by a force. 
Consider the phase diagram $(q, p)$, where $q$ is the 
displacement and $p$ is the conjugate momentum. The vacuum state is
described by a circle at the center of the phase-space
(see Fig.~\ref{LPG:fig2}). The mean values of $q$
and $p$ are each zero, but their variances (i.e. the zero-point quantum fluctuations) 
are not zeroes and are equal to
each other. The magnitudes of the variances are represented by the radius of the
circle at the
center. Under the action of a force, the vacuum state evolves into a coherent 
state. The mean values of $p$ and $q$ have increased, but the variances are 
still equal and are described by the circle of the same size as for the
vacuum state. On the other hand, 
under a parametric influence, the vacuum state evolves into a squeezed
vacuum state. (For a recent review of squeezed states see, for example,
\cite{LPG:kn} and references there.) Its variances for the 
conjugate variables $q$ and $p$ 
are significantly unequal and are described by an ellipse. 
As a function of time, the ellipse rotates with respect to the origin 
of the $(q, p)$ diagram, and the numerical values of the variances 
oscillate too. The mean numbers of quanta in the 
two states, one of which is coherent and the other a squeezed vacuum,   
can be equal (similar to the coherent and squeezed states shown in
Fig.~\ref{LPG:fig2}) 
but the statistical properties of these states are significantly different. 
Among other things, the variance
of the phase of the oscillator in a squeezed vacuum state is very 
small (hence the name, squeezed). Graphically,
this is reflected in the fact that the ellipse is very thin, so that that
the uncertainty in the angle between the horizontal axis and the orientation 
of the ellipse is very small. This highly elongated ellipse can be
regarded as a portrait of the gravitational wave quantum state that is being 
inevitably generated by parametric amplification, and which we will be 
dealing with below.    
 
\if t\figplace
\begin{figure}
\epsfxsize=0.8\textwidth
\centerline{\epsfbox{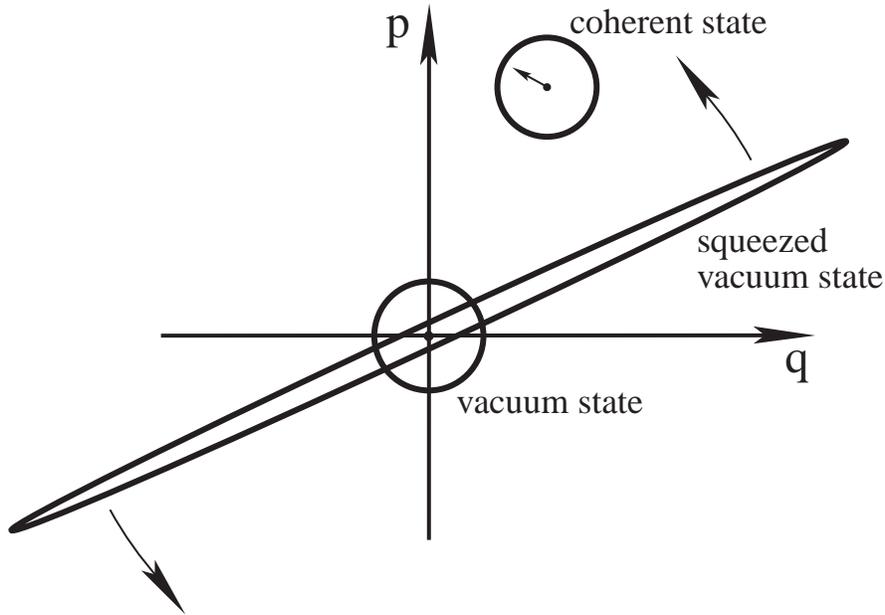}} 
\caption{Some quantum states of a harmonic oscillator.}
\label{LPG:fig2}
\end{figure}
\fi

A wave-field is not a single oscillator, it depends on spatial coordinates 
and time, and may have several independent components (polarization
states). However, the field can be decomposed into a set of 
spatial Fourier harmonics. In this way we represent the gravitational 
wave field as a collection
of many modes, many oscillators. Because of the nonlinear character of
the Einstein equations, each of these oscillators is coupled to the variable
gravitational field of the surrounding Universe. For sufficiently
short gravitational waves of experimental interest, this coupling was
especially effective in the early Universe, when the condition of
the adiabatic behaviour of the oscillator was violated. It is this homogeneous
and isotropic gravitational field of all the matter in the early 
Universe that played the role of an external agent -- the pump field.  
The variable pump field acts parametrically on  
the gravity-wave oscillators and drives them into multi-particle 
states. Concretely, the initial vacuum state of each pair of waves with
oppositely directed momenta evolves into a highly correlated state 
known as the two-mode squeezed vacuum state \cite{LPG:gs:a,LPG:gs:b,LPG:g2:a,LPG:g2:b}. 
The strength and duration of the effective coupling depends 
on the oscillator's frequency. They all start in the vacuum state 
but get excited by various amounts. As a result, a broad 
spectrum of relic gravitational waves is being formed. This spectrum
is accessible to our observations today.

\subsection{Cosmological Gravitational Waves}  

In the framework of general relativity, a homogeneous isotropic gravitational
field is described by the line element
\[
{\rm d}s^2 = c^2{\rm d}t^2 - a^2(t) \delta_{ij} {\rm d}x^i{\rm d}x^j. 
\]
It is more convenient to introduce a new time coordinate $\eta$ and to
write ${\rm d}s^2$ in the form
\begin{equation}
\label{4}
{\rm d}s^2 = a^2({\eta})[{\rm d}\eta^2 - \delta_{ij} {\rm d}x^i{\rm d}x^j].
\end{equation}
In cosmology, the function $a(t)$ (or $a(\eta)$) is called scale factor.
In our discussion, it will represent the gravitational pump field.
\par
Cosmological gravitational waves are small corrections $h_{ij}$ to the metric
tensor. They are defined by the expression 
\begin{equation}
\label{5}
{\rm d}s^2 = a^2({\eta})[{\rm d}\eta^2 - (\delta_{ij} + h_{ij})
{\rm d}x^i{\rm d}x^j].
\end{equation} 
The functions $h_{ij} (\eta ,{\bf x})$ can be expanded over spatial 
Fourier harmonics $e^{i{\bf nx}}$ and $e^{-i{\bf nx}}$, where
${\bf n}$ is a constant wave vector. In this way, we reduce the dynamical
problem to the evolution of time-dependent amplitudes for each 
mode ${\bf n}$. 
Among six functions $h_{ij}$ there are only two independent (polarization) 
components. This decomposition can be made, both, for classical 
and for quantized
field $h_{ij}$. In the quantum version, the functions $h_{ij}$ are 
treated as quantum-mechanical operators. We will use the Heisenberg 
picture, in which the time evolution is carried out by the operators while
the quantum state is fixed. This picture is fully equivalent to the
Schroedinger picture, discussed in the Introduction, in which the vacuum
state evolves into a squeezed vacuum state while the operators are 
time-independent.   

The Heisenberg operator for the quantized real field $h_{ij}$ can 
be written as
\begin{eqnarray}
\label{6}
h_{ij} (\eta ,{\bf x})
= {C\over (2\pi )^{3/2}} \int\limits_{-\infty}^\infty d^3{\bf n}
  \sum_{s=1}^2~{\stackrel{s}{p}}_{ij} ({\bf n})
   {1\over \sqrt{2n}}
\left[ {\stackrel{s}{h}}_n (\eta ) e^{i{\bf nx}}~
                 {\stackrel{s}{c}}_{\bf n}
                +{\stackrel{s}{h}}_n^{\ast}(\eta ) e^{-i{\bf nx}}~
                 {\stackrel{s}{c}}_{\bf n}^{\dag}  \right],
\end{eqnarray}
where $C$ is a constant which will be discussed later.  
The creation and annihilation operators satisfy the conditions 
$[{\stackrel{s'}{c}}_{\bf n},~{\stackrel{s}{c}}_{{\bf m}}^{\dag}]=
\delta_{s's}\delta^3({\bf n}-{\bf m})$, 
${\stackrel{s}{c}}_{\bf n}|0\rangle =0$, where $|0\rangle$ 
(for each ${\bf n}$ and $s$) is the fixed 
initial vacuum state discussed below. 
The wave number $n$ is related with the wave vector ${\bf n}$ by 
$n = (\delta_{ij}n^in^j)^{1/2}$. 
The two polarization tensors ${\stackrel{s}{p}}_{ij}({\bf n})$ $(s = 1, 2)$ 
obey the conditions
\[
 {\stackrel{s}{p}}_{ij}n^j = 0, ~~
 {\stackrel{s}{p}}_{ij}\delta^{ij} = 0, ~~
 {\stackrel{s'}{p}}_{ij}
 {\stackrel{s}{p}}~^{ij} = 2\delta_{ss'}, ~~
 {\stackrel{s}{p}}_{ij}(-{\bf n}) = {\stackrel{s}{p}}_{ij}({\bf n}).
\]
The time evolution, one and the same for all ${\bf n}$ belonging to a 
given $n$, is represented by the complex time-dependent function
${\stackrel{s}{h}}_n(\eta )$. This evolution is dictated by the
Einstein equations. The nonlinear nature of the Einstein equations
leads to the coupling of 
${\stackrel{s}{h}}_n(\eta )$ with the pump field $a(\eta)$.
For every wave number $n$  
and each polarization component $s$,
the functions ${\stackrel{s}{h}}_n(\eta )$ have the form 
\begin{equation}
\label{7}
  {\stackrel{s}{h}}_n(\eta ) = {1\over a(\eta )} 
  [{\stackrel{s}{u}}_n(\eta ) + {\stackrel{s}{v}}_n^{\ast} (\eta )],  
\end{equation}
where ${\stackrel{s}{u}}_n(\eta )$ and ${\stackrel{s}{v}}_n(\eta )$
can be expressed in terms of three
real functions (the polarization index $s$ is omitted): 
$r_n$ -- the squeeze parameter, 
$\phi_n$ -- the squeeze angle, $\theta_n$ -- the rotation angle,   
\begin{equation}
\label{8}
   u_n = e^{i{\theta}_n} \cosh~{r}_n, \qquad
   v_n = e^{-i({\theta}_n - 2{\phi}_n )} \sinh~{r}_n.
\end{equation}
The dynamical equations for $u_n(\eta)$ and $v_n(\eta)$ 
\begin{equation}
\label{9}
i\frac{{\rm d} u_n}{{\rm d}\eta} = n u_n + i\frac{a^\prime}{a} v_n^{*}, \qquad 
i\frac{{\rm d} v_n}{{\rm d}\eta} = n v_n + i\frac{a^\prime}{a} u_n^{*}
\end{equation}
lead to the dynamical equations governing 
the functions $r_n(\eta)$, $\phi_n(\eta)$ and $\theta_n(\eta)$ \cite{LPG:g2:a,LPG:g2:b}: 
\begin{equation}
\label{10}
r_n^{\prime} = \frac{a^{\prime}}{a} \cos{2{\phi}_n}, \quad
\phi_n^{\prime} = -n - \frac{a^{\prime}}{a} \sin{2{\phi}_n}\coth~2{r}_n, \quad 
\theta_n^{\prime} = -n - \frac{a^{\prime}}{a} \sin{2{\phi}_n}\tanh~{r}_n,
\end{equation}
where $^{\prime} = {\rm d}/{\rm d}\eta$, and the evolution begins 
from $r_n = 0$. This value of $r_n$ 
characterizes the initial vacuum state $|0\rangle$ which is defined 
long before the interaction with the pump field
became effective, that is, long before the coupling term 
$a^{\prime}/a$ became comparable with $n$. 
The constant $C$ should be taken as  
$C=\sqrt{16\pi}~l_{Pl}$ where $l_{Pl}=(G\hbar /c^3)^{1/2}$ is 
the Planck length. This particular value of the constant $C$
guarantees the correct quantum normalization of the field: energy
$\frac{1}{2} \hbar \omega$ per each mode in the initial vacuum state.  
The dynamical equations and their
solutions are identical for both polarization components $s$. 
\par
Equations (\ref{9})
can be translated into the more familiar form of the second-order  
differential equation for the function 
${\stackrel{s}{\mu}}_n(\eta ) \equiv 
{\stackrel{s}{u}}_n(\eta ) + {\stackrel{s}{v}}_n^{\ast} (\eta )
\equiv a(\eta) {\stackrel{s}{h}}_n(\eta )$ \cite{LPG:g1:a,LPG:g1:b,LPG:g1:c}:
\begin{equation}
\label{11}
\mu_{n}^{\prime\prime} + \mu_{n} \left[n^2 - 
\frac{a^{\prime\prime}}{a}\right] = 0.    
\end{equation}
\par
Clearly, this is the equation for a parametrically disturbed oscillator
(compare with Eq. (\ref{3})). In absence of the gravitational parametric
influence represented by the term  
$a^{\prime\prime}/a$, the frequency of the oscillator defined in terms
of $\eta$-time would be a constant: $n$. Whenever the term    
$a^{\prime\prime}/a$ can be neglected, the general solution 
to Eq. (\ref{11}) has the usual oscillatory form
\begin{equation}
\label{12}
\mu_n (\eta) = A_n e^{-in\eta} + B_n e^{in\eta},
\end{equation}
where the constants $A_n$, $B_n$ are determined by the initial conditions.   
On the other hand, whenever the term $a^{\prime\prime}/a$ is dominant, 
the general solution to Eq. (\ref{11}) has the form
\begin{equation}
\label{13}
\mu_n (\eta) = C_n a + D_n a \int\limits^{\eta}\frac{{\rm d} \eta}{a^2}.
\end{equation}
In fact, this approximate solution is valid as long as $n$ is small in
comparison with $|a^{\prime}/a|$. This is more clearly seen from the
equivalent form of Eq. (\ref{11}) written in terms of the 
function $h_n (\eta)$ \cite{L&L_v2}:
\begin{equation}
\label{14}
h_n^{\prime\prime} + 2\frac{a^{\prime}}{a} h_n^{\prime} + n^2 h_n = 0.
\end{equation} 
For growing functions $a(\eta)$, that is, in expanding universes, the
second term in Eq.(\ref{13}) is usually smaller than the first one (see below), 
so that, as long as $n \ll a^{\prime}/a$, the dominant solution is the growing
function $\mu_n (\eta) = C_n a(\eta)$, and 
\begin{equation}
\label{d}
h_n = const.
\end{equation} 

Equation (\ref{11}) can also be treated as the Schroedinger 
equation for a particle moving in the presence of an effective 
potential $U(\eta) = a^{\prime\prime}/a$. In those situations that
are normally considered, the potential
$U(\eta)$ has a bell-like shape and forms a barrier (see Fig.~\ref{LPG:fig3}). 
When a given mode $n$ is outside the barrier, its amplitude $h_n$ 
is adiabatically decreasing with time: 
$h_n \propto {e^{{\pm}in\eta}}/{a(\eta)}$. 
This is shown in Fig.~\ref{LPG:fig3} by oscillating lines with decreasing amplitudes 
of oscillations. The modes with sufficiently high frequencies do not 
interact with, and stay above, the barrier. Their amplitudes 
$h_n$ behave adiabatically all the time. For
these high-frequency modes, the initial vacuum state (in the Schroedinger 
picture) remains the vacuum state forever. On the other hand, the modes that 
interact with the barrier are subject to the super-adiabatic amplification.   
Under the barrier and as long as   
$n < a^\prime/a$,  the function $h_n$ stays constant instead of the decreasing
adiabatically. For these modes, the initial vacuum state evolves into a
squeezed vacuum state.

\if t\figplace
\begin{figure}
\epsfxsize=0.8\textwidth
\centerline{\epsfbox{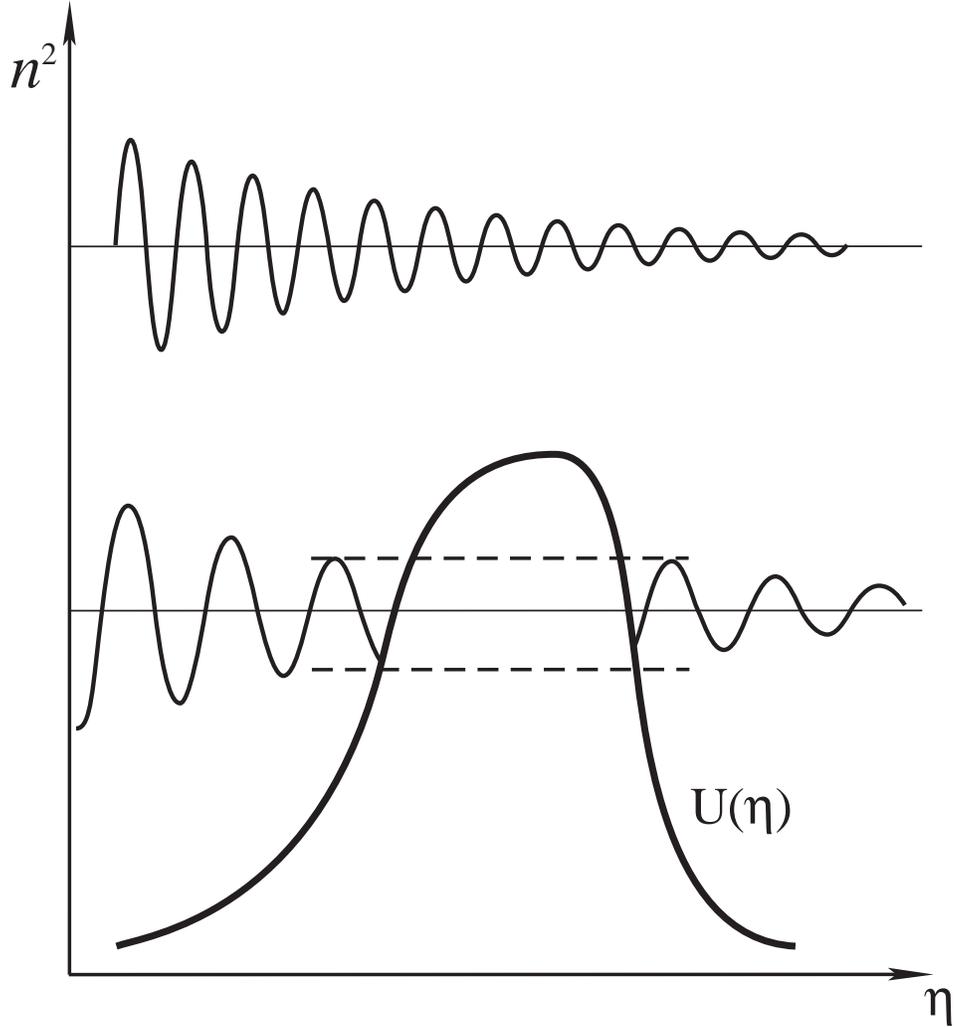}} 
\caption{Effective potential $U(\eta)$.}
\label{LPG:fig3}
\end{figure}
\fi

After having formulated the initial conditions, the present day behaviour 
of $r_n$, $\phi_n$, $\theta_n$ (or, equivalently, the present day behaviour
of $h_n$) is  
essentially all we need to find. The mean number of particles
in a two-mode squeezed state is $2\sinh^2{r_n}$ for each $s$. This number 
determines the 
mean square amplitude of the gravitational wave field. The time behaviour of
the squeeze angle $\phi_n$ determines the time dependence of the correlation 
functions of the field. The amplification (that is, the growth of $r_n$) 
governed 
by Eq. (\ref{10}) is different for different wave numbers $n$. Therefore, 
the present day results depend on the present day frequency $\nu$
($\nu = {cn}/{2 \pi a}$) measured in Hz.  
\par 
In cosmology, the function $H \equiv \dot a/a \equiv c a^{\prime}/a^2$ 
is the time-dependent Hubble parameter. The function $l \equiv c/H$
is the time-dependent Hubble radius. The time-dependent wavelength of the 
mode $n$ is $\lambda = 2 \pi a/n$. The wavelength $\lambda$ has this universal
definition in all regimes. In contrast, the $\nu$ defined 
as $\nu = {cn}/{2 \pi a}$ has the usual meaning of a frequency of an 
oscillating process only in the short-wavelength (high-frequency) regime of
the mode $n$, that is, in the regime where $\lambda \ll l$. 
As we have seen above, the qualitative
behaviour of the solutions to Eqs. (\ref{11}), (\ref{14}), depends 
crucially on the comparative  
values of $n$ and $a^\prime/a$, or, in other words, on the comparative
values of $\lambda(\eta)$ and $l(\eta)$. This relationship is also 
crucial for the solutions to Eq. (\ref{10}), as we shall see now. 
\par
In the short-wavelength regime, that is, during intervals of time
when the wavelength $\lambda(\eta)$ is shorter than the Hubble 
radius $l(\eta) = a^2/a^{\prime}$, the term containing $n$ in (\ref{10}) is dominant. 
The functions $\phi_n(\eta)$ and 
$\theta_n(\eta)$ are, $\phi_n = -n(\eta + \eta_n)$, $\theta_n = \phi_n$
where $\eta_n$ is a constant.  
The factor $\cos 2\phi_n$ is a rapidly oscillating
function of time, so the squeeze parameter $r_n$ stays practically a constant.
This is the adiabatic regime for a given mode. 
\par
In the opposite, long-wavelength regime, the term $n$ can be 
neglected.
The function $\phi_n$ is $\tan \phi_n(\eta) \approx \mbox{const}/a^2(\eta)$,
and the squeeze angle
quickly approaches one of the two values: $\phi_n = 0$ or $\phi_n = \pi$   
(analog of ``phase bifurcation" \cite{LPG:w:a,LPG:w:b}). 
When the long-wavelength regime, for a given $n$, begins 
the squeeze parameter $r_n(\eta)$ grows with time according to   
\begin{equation}
\label{15}
r_n(\eta) \approx ln \left [ \frac{a(\eta)}{a_*} \right ], 
\end{equation} 
where $a_*$ is the value of $a(\eta)$ at $\eta_*$. 
The final value of $r_n$ is
\begin{equation}
\label{16}
r_n \approx ln \left [ \frac{a_{**}}{a_*}\right ], 
\end{equation} 
where $a_{**}$ is the value of $a(\eta)$ at $\eta_{**}$, when the 
long-wavelength regime and 
amplification come to the end. It is important to emphasize that it is not
a ``sudden transition" from one cosmological era to another that is responsible
for amplification, but the entire interval of the long-wavelength 
(non-adiabatic) regime.
\par
After the end of amplification, the accumulated
(and typically large) squeeze parameter $r_n$ stays approximately constant.
The mode is again in the adiabatic regime. In course of the evolution, the 
complex functions 
${\stackrel{s}{u}}_n(\eta ) + {\stackrel{s}{v}}_n^{\ast}(\eta )$
become practically real, and one has 
${\stackrel{s}{h}}_n(\eta ) \approx {\stackrel{s}{h}}_n^{\ast}(\eta ) \approx
\frac{1}{a} e^{r_n} \cos \phi_n(\eta)$.  
Every amplified mode $n$ of the field (\ref{6}) takes the form of 
a product of a function of time and a (random, operator-valued) 
function of spatial coordinates; the mode acquires a 
standing-wave pattern. The periodic dependence $\cos \phi_n(\eta)$
will be further discussed below.  
\par 
It is clearly seen from the fundamental equations (\ref{10}), (\ref{11}), 
(\ref{14}) that 
the final results depend only on $a(\eta)$. Equations do not ask us the 
names of our favorite cosmological prejudices, they ask us about the
pump field $a(\eta)$. Conversely, from the measured relic gravitational waves,
we can deduce the behaviour of $a(\eta)$, which is essentially the purpose
of detecting the relic gravitons.

\subsection{Cosmological Pump Field}

With the chosen initial conditions, the final 
numerical results for relic gravitational waves depend on the concrete 
behaviour of the pump
field represented by the cosmological scale factor $a(\eta)$. We know 
a great deal about $a(\eta)$. We know that $a(\eta)$ behaves as 
$a(\eta) \propto \eta^2$
at the present matter-dominated stage. We know that this stage
was preceded by the radiation-dominated stage in which $a(\eta) \propto \eta$.
At these two stages of evolution the functions $a(\eta)$ are simple 
power-law functions of $\eta$.   
What we do not know is the function $a(\eta)$ describing the initial stage 
of expansion of the very early Universe, that is, before
the era of primordial nucleo-synthesis. It is convenient to
parameterize $a(\eta)$ at this initial stage also by power-law functions
of $\eta$. First, this is a sufficiently broad class of functions, which, 
in addition, allows us to find exact solutions to our fundamental 
equations. Second, it is known \cite{LPG:g1:a,LPG:g1:b,LPG:g1:c} that
the pump fields $a(\eta)$ which have power-law dependence in terms of $\eta$, 
produce gravitational waves with simple power-law spectra in terms of $\nu$.
These spectra are easy to analyze and discuss in the context of detection. 
\par
We model cosmological expansion by several successive eras. Concretely,   
we take $a(\eta)$ at the initial stage of expansion ($i$-stage) as
\begin{equation}
\label{17}
a(\eta) = l_o|\eta|^{1 + \beta},  
\end{equation}
where $\eta$ grows from $- \infty$,  and $1 + \beta < 0$. We will show
later how the available observational data constrain the 
parameters $l_o$ and $\beta$. The $i$-stage lasts up to a certain   
$\eta = \eta_1$, $\eta_1 < 0$. To make our analysis more general,
we assume that the $i$-stage was followed by some interval of the 
$z$-stage ($z$ from Zeldovich). It is known that an
interval of evolution governed by the most ``stiff" matter
(effective equation of state $p = \epsilon$) advocated by Zeldovich, 
leads to a relative increase of gravitational wave amplitudes \cite{LPG:g1:a,LPG:g1:b,LPG:g1:c}.    
It is also known that the requirement of consistency of the graviton production 
with the observational restrictions does not allow the ``stiff" matter interval
to be too long \cite{LPG:g1:a,LPG:g1:b,LPG:g1:c,LPG:zn}. However, we want to 
investigate any interval of
cosmological evolution that can be consistently included. In fact,  
the $z$-stage of expansion that we include is quite general. 
It can be governed by a ``stiffer than radiation" \cite{LPG:gio} matter, 
as well as by a ``softer than
radiation" matter. It can also be simply a part of the radiation-dominated 
era. Concretely, we take $a(\eta)$ at the interval of time from $\eta_1$ 
to some $\eta_s$ ($z$-stage) in the form  
\begin{equation}
\label{18}
a(\eta) = l_o a_z (\eta - \eta_p)^{1 + \beta_s},
\end{equation}  
where $1 + \beta_s > 0$. For the particular choice
$\beta_s = 0$, the $z$-stage reduces to an interval
of expansion governed by the radiation-dominated matter. Starting 
from $\eta_s$ and up to $\eta_2$ the Universe was governed by the 
radiation-dominated matter ($e$-stage). So, in this interval of evolution, 
we take the scale factor in the form   
\begin{equation}
\label{19}
a(\eta) = l_oa_e(\eta - \eta_e). 
\end{equation} 
And, finally, from $\eta =\eta_2$ the expansion switched to the
matter-dominated era ($m$--stage):   
\begin{equation}
\label{20}
a(\eta) = l_oa_m(\eta - \eta_m)^2. 
\end{equation} 
A link between the arbitrary constants participating in 
Eqs. (\ref{17}) - (\ref{20}) is provided
by the conditions of continuous joining of the functions $a(\eta)$ 
and $a^{\prime}(\eta)$ at points of transitions $\eta_1$, $\eta_s$, $\eta_2$.
\par
We denote the present time by $\eta_R$ ($R$ from reception). This time is
defined by the observationally known value of the present-day Hubble
parameter $H(\eta_R)$ and Hubble radius $l_H=c/{H(\eta_R)}$. 
For numerical estimates we will be using  
$l_H  \approx 2 \times 10^{28}~{\rm cm}$. It is convenient to choose   
$\eta_R - \eta_m = 1$, so that $a(\eta_R) = 2l_H$. The ratio 
\[
a(\eta_R)/a(\eta_2) \equiv \zeta_2
\]
is believed to be around $\zeta_2 = 10^4$. We also denote   
\[
a(\eta_2)/a(\eta_s) \equiv \zeta_s~,~~~ a(\eta_s)/a(\eta_1) \equiv \zeta_1~.  
\]
With these definitions, all the constants participating in 
Eqs. (\ref{17}) - (\ref{20}) (except
parameters $\beta$ and $\beta_s$ which should be chosen from
other considerations)
are being expressed in terms of $l_H$, $\zeta_2$, $\zeta_s$,
and $\zeta_1$. For example,
\[
|\eta_1|= \frac{|1+\beta|}{2\zeta_2^{\frac{1}{2}}
\zeta_s{\zeta_1}^{\frac{1}{1+\beta_s}}} ~.
\]
The important constant $l_o$ is expressed as 
\begin{equation}
\label{21}
l_o = b l_H\zeta_2^{\frac{\beta-1}{2}}\zeta_s^{\beta}{\zeta_1}^{\frac{\beta-\beta_s}{1+\beta_s}} ,  
\end{equation}  
where $b \equiv 2^{2+\beta}/{|1+\beta|}^{1+\beta}$. Note that $b=1$ for
$\beta = -2$. 
(This expression for $l_o$ may help to relate formulas written here with 
the equivalent treatment \cite{LPG:g3} which was given in slightly 
different notations.)
The sketch of the entire evolution $a(\eta)$ is given in Fig.~\ref{LPG:fig4}.

\if t\figplace
\begin{figure}
\epsfxsize=0.8\textwidth
\centerline{\epsfbox{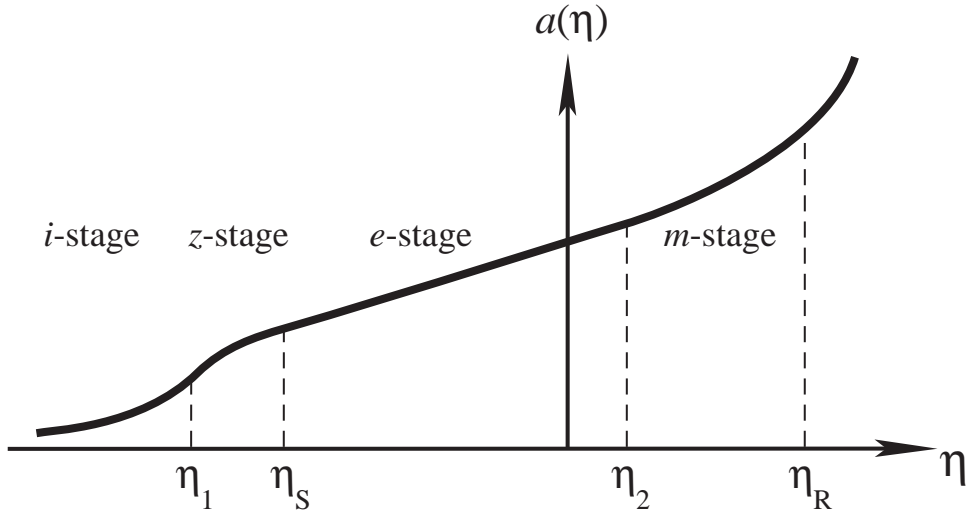}} 
\caption{Scale factor $a(\eta)$.}
\label{LPG:fig4}
\end{figure}
\fi

We work with the spatially-flat models (\ref{4}).
At every instant of time, the energy density $\epsilon(\eta)$ of matter 
driving the evolution is related with the Hubble radius $l(\eta)$ by
\begin{equation}
\label{22}
\kappa \epsilon(\eta) = \frac{3}{l^2(\eta)}, 
\end{equation}
where $\kappa = 8\pi G/ c^4$.
For the case of power-law scale factors $a(\eta) \propto \eta^{1+\beta}$, 
the effective matter pressure $p(\eta)$ is related to the energy density
$\epsilon(\eta)$ by the effective equation of state 
\begin{equation}
\label{23}
p = \frac{1-\beta}{3(1+\beta)} \epsilon.
\end{equation}
For instance, $p=0$ for $\beta =1$, $p=\frac{1}{3} \epsilon$ for $\beta=0$,
$p= -\epsilon$ for $\beta = -2$, and so on. Each interval of
the evolution (\ref{17})-(\ref{20}) is governed by one of these 
equations of state.   
\par
In principle, the function $a(\eta)$ could be even more complicated than 
the one that we consider.  
It could even include an interval of early contraction, 
instead of expansion, leading to a ``bounce" of the scale factor.  
In the case of a decreasing $a(\eta)$ the gravitational-wave equation can 
still be analyzed and the amplification
is still effective \cite{LPG:g1:a,LPG:g1:b,LPG:g1:c}. However, the Einstein equations for
spatially-flat models do not permit a 
regular ``bounce" of $a(\eta)$ (unless $\epsilon$ vanishes at the moment
of ``bounce"). Possibly, a ``bounce" solution can be realized in alternative
theories, such as string-motivated cosmologies 
\cite{LPG:vg:a,LPG:vg:b,LPG:vg:c}. For a  
recent discussion of spectral slopes of gravitational waves produced
in ``bounce" cosmologies, see~\cite{LPG:cr}.

\subsection{Solving Gravitational Wave Equations}

The evolution of the scale factor $a(\eta)$ given by 
Eqs. (\ref{17}) - (\ref{20}) and
sketched in Fig.~\ref{LPG:fig4} allows us to calculate the function $a^\prime/a$. This function
is sketched in Fig.~\ref{LPG:fig5}. In all theoretical generality, the left-hand-side
of the barrier in Fig.~\ref{LPG:fig5} could also consist of several pieces, but we
do not consider this possibility here. The graph also shows the 
important wave numbers
$n_H$, $n_2$, $n_s$, $n_1$: $n_H$ marks the wave whose today's wavelength 
$\lambda(\eta_R)= 2 \pi a(\eta_R)/n_H$ is equal to the Hubble 
radius $l_H$ today. With our parameterization $a(\eta_R) = 2 l_H$, this wave-number
is $n_H = 4 \pi$. $n_2$ marks the wave whose wavelength $\lambda(\eta_2)=
2 \pi a(\eta_2)/n_2$ at $\eta= \eta_2$ is equal to the Hubble 
radius $l(\eta_2)$ at $\eta = \eta_2$. Since
$\lambda(\eta_R)/\lambda(\eta_2) = (n_2/n_H)[a(\eta_R)/a(\eta_2)]$ and 
$l(\eta_R)/l(\eta_2) = [a(\eta_R)/a(\eta_2)][a(\eta_R)/a(\eta_2)]^{1/2}$,
this gives us $n_2/n_H = [a(\eta_R)/a(\eta_2)]^{1/2} = \zeta_2^{1/2}$. 
Working out in a similar fashion other ratios, we find 
\begin{equation}       
\label{24}
\frac{n_2}{n_H} = \zeta_2^{\frac{1}{2}},~~~ \frac{n_s}{n_2} = \zeta_s,
~~\frac{n_1}{n_s} =\zeta_1~^{\frac{1}{1+\beta_s}}. 
\end{equation}

\if t\figplace
\begin{figure}
\epsfxsize=0.8\textwidth
\centerline{\epsfbox{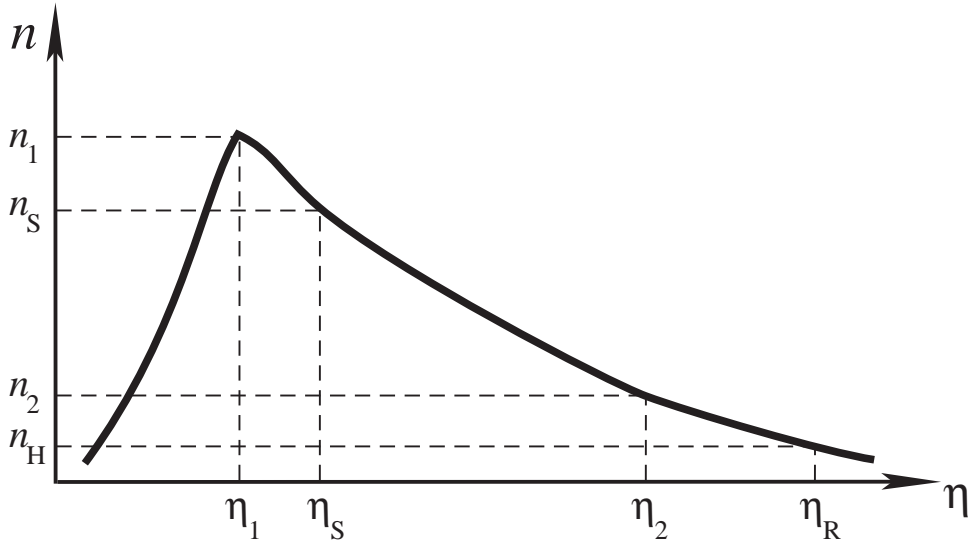}} 
\caption{Function $a^\prime/a$ for the scale factor from
	Fig.~\protect\ref{LPG:fig4}.}
\label{LPG:fig5}
\end{figure}
\fi
   
Solutions to the gravitational wave equations exist for any $a(\eta)$.
At intervals of power-law dependence $a(\eta)$, solutions to Eq. (\ref{11}) 
have simple form of the Bessel functions. We could have found piece-wise 
exact solutions to Eq. (\ref{11}) and join them in the transition points. 
However, we will use a much simpler treatment, which is sufficient for our 
purposes. We know that the
squeeze parameter $r_n$ stays constant in the short-wavelength regimes and
grows according to Eq. (\ref{15}) in the long-wavelength regime. All modes 
start in the vacuum state, that is, $r_n = 0$ initially. After the end of 
amplification, the accumulated value (\ref{16}) stays constant up to today.
To find today's value of $e^{r_n}$ we need to calculate the 
ratio $a_{**}(n)/a_*(n)$. For
every given $n$, the quantity $a_*$ is determined by the condition
$\lambda(\eta_*) = l(\eta_*)$, whereas $a_{**}$ is determined by the
condition $\lambda(\eta_{**}) = l(\eta_{**})$.  

\par
Let us start from the mode $n = n_1$. For this wave number we have 
$a_* = a_{**} = a (\eta_1)$, and therefore, $r_{n_1} = 0$. The higher frequency 
modes, i.e. $n > n_1$ (above the barrier in Fig.~\ref{LPG:fig5}), have never been in the 
amplifying regime, so we can write
\begin{equation}
\label{25} 
e^{r_n} = 1,~~ n\ge n_1.
\end{equation}
Let us now consider the modes $n$ in the interval $n_1 \ge n \ge n_s$. 
For a given $n$ we need to know $a_* (n)$ and $a_{**}(n)$. Using Eq. (\ref{17})
one has $a_*(n)/a_*(n_1) = (n_1/n)^{1 +\beta}$, and using Eq. (\ref{18}) one 
finds $a_{**}(n)/a_{**}(n_s) = (n_s/n)^{1 +\beta_s}$. Therefore, one finds 
\[
\frac{a_{**}(n)}{a_*(n)} = \frac{a_{**}(n_s)}{a_*(n_1)} 
\left(\frac{n_s}{n}\right)^{1+\beta_s}\left(\frac{n}{n_1}\right)^{1+\beta}.   
\]
Since $a_{**}(n_s) = a(\eta_s)$, $a_*(n_1) = a(\eta_1)$, and 
$a(\eta_s)/a(\eta_1) = \zeta_1 =(n_1/n_s)^{1+\beta_s}$, we arrive at
\[
\frac{a_{**}(n)}{a_*(n)} = \left(\frac{n}{n_1}\right)^{\beta - \beta_s}.   
\]
Repeating this analysis for other intervals of the decreasing $n$, 
we come to the conclusion that
\begin{eqnarray}
\label{26} 
e^{r_n}& =& \left(\frac{n}{n_1}\right)^{\beta - \beta_s},~~ 
n_1 \ge n \ge n_s, \nonumber \\ 
e^{r_n}& =& \left(\frac{n}{n_s}\right)^{\beta}
\left(\frac{n_s}{n_1}\right)^{\beta - \beta_s} ,~~ 
n_s \ge n \ge n_2, \nonumber \\
e^{r_n}& =& \left(\frac{n}{n_2}\right)^{\beta - 1}
\left(\frac{n_2}{n_1}\right)^{\beta}\left(\frac{n_s}{n_1}\right)^{-\beta_s},  
 n_2 \ge n \ge n_H. 
\end{eqnarray}
The mnemonic rule of constructing $e^{r_n}$ at successive intervals 
of decreasing $n$ is simple.
If the interval begins at $n_x$, one takes $(n/n_x)^{\beta_{*} - \beta_{**}}$
and multiples with $e^{r_{n_x}}$, that is, with the previous interval's 
value of $e^{r_n}$ calculated at the end of that interval $n_x$. 
For the function $a^\prime/a$ that we are working with, the $\beta_{*}$ is always
$\beta$, whereas the $\beta_{**}$ takes the values $\beta_s$, $0$, $1$, at
the successive intervals.   
\par
The modes with $n < n_H$ are still in the long-wavelength regime. For these
modes, we should take $a(\eta_R)$ instead of $a_{**}(n)$. Combining with
$a_*(n)$, we find
\begin{equation}
\label{27}
e^{r_n} = \left(\frac{n}{n_H}\right)^{\beta +1}
\left(\frac{n_H}{n_2}\right)^{\beta - 1}
\left(\frac{n_2}{n_1}\right)^{\beta}\left(\frac{n_s}{n_1}\right)^{-\beta_s},  
~  n \le n_H.
\end{equation}
Formulas (\ref{25}) - (\ref{27}) give approximate values of $r_n$ 
for all $n$. The
factor $e^{r_n}$ obeys the following inequalities
is $e^{r_n} \ge 1$ for $n \le n_1$, and  
$e^{r_n} \gg 1$ for $n \ll n_1$, and determines the mean square
amplitude of the gravitational waves.  
\par
The mean value of the field $h_{ij}$ is zero at every moment of time $\eta$
and in every spatial point ${\bf x}$:                         
$\langle 0|h_{ij}(\eta, {\bf x})|0\rangle = 0$.
The variance 
\[
\langle 0|h_{ij}(\eta, {\bf x})h^{ij}(\eta, {\bf x})|0\rangle ~\equiv~
\langle h^2 \rangle 
\] 
is not zero, and it determines the mean square amplitude of the generated
field - the quantity of interest for the experiment. Taking the product
of two expressions (\ref{6}) one can show that   	 
\begin{equation}
\label{28}
\langle h^2 \rangle = \frac{C^2}{2\pi^2} \int\limits_0^\infty n \sum_{s=1}^2
\Big| {\stackrel{s}{h}}_n(\eta )\Big|^2 ~{\rm d}n \equiv 
\int\limits_0^\infty h^2(n, \eta) \frac{{\rm d}n}{n}~. 
\end{equation}
Using the representation (\ref{7}), (\ref{8}) in Eq. (\ref{28}) one can 
also write
\begin{equation}
\label{29}
\langle h^2 \rangle = 
\frac{C^2}{{\pi^2}a^2 } \int\limits_0^\infty n {\rm d}n (\cosh2{r_n} + 
\cos2{\phi_n}\sinh2{r_n}). 
\end{equation}
We can now consider the present era and use the fact that $e^{r_n}$ are 
large numbers for all $n$ in the interval of our interest 
$n_1 \ge n \ge n_H$. Then, we can derive  
\begin{equation}
\label{30}
h(n, \eta)\approx \frac{C}{\pi}\frac{1}{a(\eta_R)}n e^{r_n}\cos\phi_n(\eta)=
8 \sqrt{\pi}\left(\frac{l_{Pl}}{l_H}\right)\left(\frac{n}{n_H}\right)
e^{r_n}\cos\phi_n(\eta) ~.
\end{equation}
The quantity $h(n, \eta)$ is the
dimensionless spectral amplitude of the field whose numerical value
is determined by the calculated squeeze parameter $r_n$.  
The oscillatory factor $\cos\phi_n(\eta)$ reflects the squeezing (standing
wave pattern) acquired by modes with $n_1 > n > n_H$. For modes with
$n < n_H$ this factor is approximately $1$. For high-frequency modes 
$n \gg n_H$ one has
$\phi_n(\eta) \approx n(\eta - \eta_n) \gg 1$, so that $h(n, \eta)$ makes 
many oscillations while the scale factor $a(\eta)$ is practically fixed 
at $a(\eta_R)$.
\par 
The integral (\ref{29}) extends formally from $0$ to $\infty$. 
Since $r_n \approx 0$
for $n \ge n_1$, the integral diverges at the upper limit. This is a typical
ultra-violet divergence. It should be discarded (renormalized to zero) 
because it comes from 
the modes which have always been in their vacuum state. At the lower limit,
the integral diverges, if $\beta \le -2$. This is an infra-red divergence
which comes from the assumption that the amplification process has started
from infinitely remote time in the past. One can deal with this 
divergence either by introducing a lower frequency cut-off (equivalent to
the finite duration of the amplification) or by considering only the
parameters $\beta > -2$, in which case the integral is convergent at the lower
limit.
It appears that the available observational data (see below) favour this
second option. The particular case $\beta = -2$ corresponds to  
the de Sitter evolution $a(\eta) \propto |\eta|^{-1}$. In this case, 
the $h(n)$ found in Eqs. (\ref{30}), (\ref{27}) does not depend on $n$. 
This is known as the Harrison-Zeldvich, or scale-invariant, spectrum. 
\par
The spectral amplitudes $h(n)$ can also be derived using the 
approximate solutions (\ref{12}), (\ref{13}) to the wave equation (\ref{11}). 
This method gives exactly the same, as in Eqs. (\ref{30}),
(\ref{25}) - (\ref{27}) numerical values of $h(n)$, but does not 
reproduce the oscillatory factor $\cos\phi_n(\eta)$. 
\par
One begins with the initial spectral amplitude $h_i(n)$ defined 
by quantum normalization:
$h_i(n) = 8 \sqrt{\pi} (l_{Pl}/ \lambda_i)$. This is the amplitude of the mode
$n$ at the moment $\eta_*$ of entering the long wavelength regime, i.e.  
when the mode's wavelength $\lambda_i$ is equal to the Hubble 
radius $l(\eta_*)$. For $\lambda_i$ one derives 
\begin{equation}
\label{31}
\lambda_i = \frac{1}{b} l_o \left(\frac{n_H}{n}\right)^{2+\beta}. 
\end{equation}
Thus, we have
\begin{equation}
\label{32}
h_i(n) = A \left(\frac{n}{n_H}\right)^{2+\beta},
\end{equation}
where $A$ denotes the constant  
\begin{equation}
\label{33}
A = b 8\sqrt{\pi} \frac{l_Pl}{l_o}~. 
\end{equation} 
The numbers $h_i(n)$ are defined at the beginning of the long-wavelength
regime. In other words, they are given along the left-hand-side slope of the 
barrier in Fig.~\ref{LPG:fig5}.
We want to know the final numbers (spectral amplitudes) $h(n)$ 
which describe the field today, at $\eta_R$. 
\par
According to the dominant solution $h_n(\eta) =\mbox{const,}$ of the long-wavelength
regime (see Eq. (\ref{d})), the initial amplitude $h_i(n)$ stays practically 
constant up to the end of the long-wavelength regime at $\eta_{**}$, 
that is, up to the right-hand-side slope of the barrier. 
(The second term in Eq. (\ref{13}) could be
important only at the $z$-stage and only for parameters 
$\beta_s \le -(1/2)$, which
correspond to the effective equations of state $p \ge \epsilon$. 
In order to keep the analysis simple, we do not consider those cases.) 
After the completion of the long-wavelength regime, the amplitudes decrease 
adiabatically in proportion to $1/a(\eta)$, up to the present time. 
Thus, we have 
\begin{equation}
\label{34}  
h(n) = A \left(\frac{n}{n_H}\right)^{2+\beta} \frac{a_{**}(n)}{a(\eta_R)}. 
\end{equation}
\par
Let us start from the lower end of the spectrum, $n \le n_H$, and go 
upward in $n$. The modes $n \le n_H$ have not started yet the adiabatic
decrease of the amplitude, so we have
\begin{equation}
\label{35}  
h(n) = A \left(\frac{n}{n_H}\right)^{2+\beta},~~ n \le n_H. 
\end{equation}
Now consider the interval $n_2 \ge n \ge n_H$. At this interval,
the $a_{**}(n)/a(\eta_R)$ scales as $(n_H/n)^2$, so we have
\begin{equation}
\label{36}  
h(n) = A \left(\frac{n}{n_H}\right)^{\beta},~~ n_2 \ge n \ge n_H. 
\end{equation}
At the interval $n_s \ge n \ge n_2$ the ratio  
$a_{**}(n)/a(\eta_R) = [a_{**}(n)/a(\eta_2)][a(\eta_2)/a(\eta_R)]$ 
scales as $(n_2/n)(n_H/n_2)^2$, so we have
\begin{equation}
\label{37}  
h(n) =A\left(\frac{n}{n_H}\right)^{1+\beta}\frac{n_H}{n_2},~~ n_s\ge n\ge n_2.  
\end{equation}
Repeating the same analysis for the interval $n_1 \ge n \ge n_s$ we find
\begin{equation}
\label{38}  
h(n) =A\left(\frac{n}{n_H}\right)^{1+\beta -\beta_s}
\left(\frac{n_s}{n_H}\right)^{\beta_s}\frac{n_H}{n_2},~~ n_1\ge n\ge n_s.  
\end{equation}
It is seen from
Eq. (\ref{38}) that an interval of the $z$-stage with $\beta_s <0$ 
(the already imposed
restrictions require also $(-1/2) < \beta_s$) bends
the spectrum $h(n)$ upwards, as compared with Eq. (\ref{37}), for larger $n$.  
If one recalls the relationship (\ref{21}) between $l_o$ and $l_H$
and uses (\ref{26}), (\ref{27}) in Eq. (\ref{30}) one 
arrives exactly at Eqs. (\ref{35})-(\ref{38}) up to
the oscillating factor $\cos\phi_n(\eta)$. 
\par 
Different parts of the barrier in Fig.~\ref{LPG:fig5} are responsible for amplitudes and
spectral slopes at different intervals of $n$. The sketch of the generated 
spectrum $h(n)$ in conjunction with the form of the barrier 
is shown in Fig.~\ref{LPG:fig6}.

\if t\figplace
\begin{figure}
\epsfxsize=0.8\textwidth
\centerline{\epsfbox{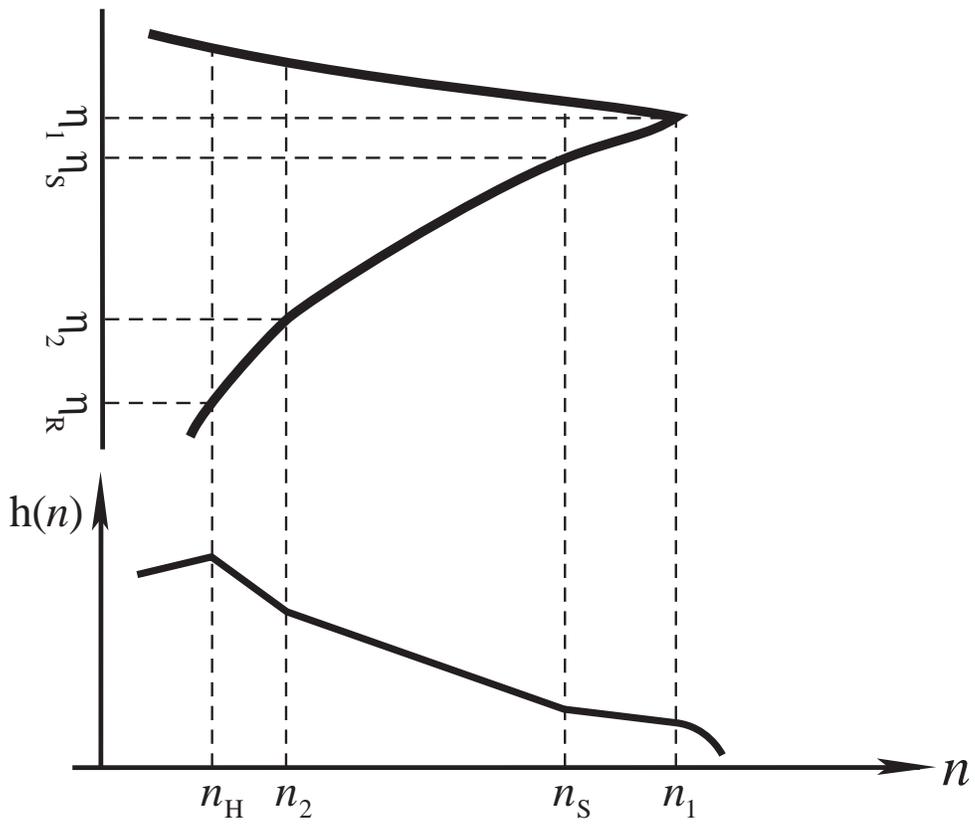}} 
\caption{Amplitudes and spectral slopes of $h(n)$ are
	determined by different parts of the barrier $a^\prime/a$.}
\label{LPG:fig6}
\end{figure}
\fi
   
The present day frequency of the oscillating modes, measured in Hz,
is defined as $\nu = cn/2\pi a(\eta_R)$. The lowest frequency (Hubble
frequency) is $\nu_H = c/l_H$. For numerical estimates we will be using  
$\nu_H \approx 10^{-18}$~Hz. The ratios of $n$ are equal to the ratios of
$\nu$, so that, for example, $n/n_H = \nu/\nu_H$. For high-frequency
modes we will now often use the ratios of $\nu$ instead of ratios of $n$. 
\par
In addition to the spectral amplitudes $h(n)$ the generated field can be
also characterized by the spectral energy density parameter $\Omega_{g}(n)$. 
The energy density $\epsilon_g$ of the gravitational wave field is 
\[
\kappa \epsilon_g = \frac{1}{4} h^{ij}_{~,0} h_{ij,0} =
\frac{1}{4a^2} {h^{ij}}^{\prime} {h_{ij}}^{\prime}.
\]
The mean value $\langle 0|\epsilon_g (\eta, {\bf x})|0\rangle$ is given by
\begin{equation}
\label{39}
\kappa \langle \epsilon_g \rangle = \frac{1}{4a^2} 
\frac{C^2}{2\pi^2} \int\limits_0^\infty n \sum_{s=1}^2
\Big| {{\stackrel{s}{h}}}^{\prime}_n(\eta )\Big|^2 ~{\rm d}n.
\end{equation}
For high-frequency modes, it is only the factor $e^{\pm i n\eta}$ that needs
to be differentiated by $\eta$. After averaging out the oscillating factors, 
one gets  
$\Big| {{\stackrel{s}{h}}}^{\prime}_n\Big|^2 = 
n^2 \Big| {\stackrel{s}{h}}_n\Big|^2$, so that 
\begin{equation}
\label{40}
\kappa \langle \epsilon_g \rangle = \frac{1}{4a^2} 
\int\limits_0^\infty n^2 h^2(n) \frac{{\rm d}n}{n}.  
\end{equation}
In fact, the high-frequency approximation, that has been used, permits 
integration over lower $n$ only up to $n_H$. And the upper limit, as was 
discussed above, is in practice $n_1$, not infinity.    
The parameter $\Omega_g$ is defined as $\Omega_g = 
\langle \epsilon_g \rangle /\epsilon$, where $\epsilon$ is given by
Eq. (\ref{22}) (critical density). So, we derive
\[
\Omega_g =  \int\limits_{n_H}^{n_1} \Omega_g(n) \frac{{\rm d}n}{n} = 
\int\limits_{\nu_H}^{\nu_1} \Omega_g(\nu) \frac{{\rm d}\nu}{\nu} 
\]
and
\begin{equation}
\label{41}
\Omega_g(\nu) = \frac{\pi^2}{3}h^2(\nu)\left(\frac{\nu}{\nu_H}\right)^2.
\end{equation}
\par
The dimensionless quantity $\Omega_g(\nu)$ is useful because it allows us
to quickly evaluate the cosmological importance of the generated field in
a given frequency interval. However, the primary and more universal concept
is $h(\nu)$, not $\Omega_g(\nu)$. It is the field, not its energy density,
that is directly measured by the gravity-wave detector. One should also note
that some authors use quite a misleading definition $\Omega_g (f) =
(1/\rho_c)(d \rho_{gw}/d \ln f)$ which suggests differentiation of the 
gravity-wave energy density by frequency. This would be incorrect and could
cause disagreements in numerical values of $\Omega_g$. Whenever we use
$\Omega_g(\nu)$, we mean relationship (\ref{41}); and for order 
of magnitude estimates one can use \cite{LPG:g1:a,LPG:g1:b,LPG:g1:c}:
\begin{equation}
\label{42}
\Omega_g(\nu) \approx h^2(\nu) \left(\frac{\nu}{\nu_H}\right)^2 .
\end{equation}

\subsection{Theoretical and Observational Constraints}

The entire theoretical approach is based on the assumption that a weak 
quantized gravity-wave field interacts with a classical pump field. We 
should follow the validity of this approximation throughout the analysis. 
The pump field can be treated as a classical gravitational field 
as long as the driving energy
density $\epsilon$ is smaller than the Planck energy density, or, in
other words, as long as the Hubble radius $l(\eta)$ is greater than the
Planck length $l_{Pl}$. This is a restriction on the pump field, but
it can be used as a restriction on the 
wavelength $\lambda_i$ of the gravity-wave mode $n$ at the 
time when it enters the long-wavelength regime. If $l(\eta_*) > l_{Pl}$, then
$\lambda_i > l_{Pl}$. The $\lambda_i$ is given by Eq. (\ref{31}). 
So, we need to ensure that  
\[
b \frac{l_{Pl}}{l_o} \left(\frac{\nu}{\nu_H}\right)^{2+\beta} < 1.
\]
At the lowest-frequency end $\nu = \nu_H$ this inequality
gives $b(l_{Pl}/l_o) < 1$. In fact, the observational constraints 
(see below) give a stronger restriction: 
\begin{equation}
\label{43}
b \frac{l_{Pl}}{l_o} \approx 10^{-6} , 
\end{equation} 
which we accept. Then, at the
highest-frequency end $\nu = \nu_1$ we need to satisfy
\begin{equation}
\label{44}
\left(\frac{\nu_1}{\nu_H}\right)^{2+\beta} < 10^6.
\end{equation}
\par
Let us now turn to the generated spectral amplitudes $h(\nu)$. According
to Eq. (\ref{35}) we have $h(\nu_H) \approx b 8 \sqrt{\pi} (l_{Pl}/l_o)$. The
measured microwave background anisotropies, which we discuss below, 
require this number
to be at the level of $10^{-5}$, which gives the already mentioned 
Eq. (\ref{43}). The quantity $h(\nu_1)$ at the highest 
frequency $\nu_1$ is given by Eq. (\ref{38}):
\[
h(\nu_1) = b 8\sqrt{\pi} \frac{l_{Pl}}{l_o}
\left(\frac{\nu_1}{\nu_H}\right)^{1+\beta -\beta_s}
\left(\frac{\nu_s}{\nu_H}\right)^{\beta_s}\frac{\nu_H}{\nu_2}. 
\]
Using Eq. (\ref{21}) this expression for $h(\nu_1)$ can be rewritten as 
\begin{equation}
\label{45}
h(\nu_1) =8\sqrt{\pi} \frac{l_{Pl}}{l_H}\frac{\nu_1}{\nu_H}= 
8\sqrt{\pi} \frac{l_{Pl}}{\lambda_1}, 
\end{equation}
where $\lambda_1 = c/\nu_1$. This last expression for $h(\nu_1)$ is not 
surprising: the modes with $\nu \ge \nu_1$ are still in the vacuum state,
so the numerical value of $h(\nu_1)$ is determined by quantum normalization.
\par
All the amplified modes have started with small initial amplitudes 
$h_i$, at the level of
zero-point quantum fluctuations. These amplitudes are also small today, since
the $h_i$ could only stay constant or decrease. However, even these relatively
small amplitudes should obey observational constraints. We do not want the
$\Omega_g$ in the high-frequency modes, which might affect the rate of
the primordial nucleosynthesis, to exceed the level of $10^{-5}$. This
means that $\Omega_g(\nu_1)$ cannot exceed the level of $10^{-6}$ or so.   
The use of Eq. (\ref{41}) in combination with 
$\Omega_g(\nu_1) \approx 10^{-6}$ and 
$h(\nu_1)$ from Eq. (\ref{45}), gives us the highest allowed frequency
$\nu_1 \approx 3 \times 10^{10}$~Hz. We will use this value of $\nu_1$
in our numerical estimates.     
Returning with this value of $\nu_1$ to Eq. (\ref{44}) we find that the parameter
$\beta$ can only be $\beta \le - 1.8$. We will be treating $\beta = -1.8$
as the upper limit for the allowed values of $\beta$. 
\par
We can now check whether the accepted parameters leave room for the 
postulated $z$-stage with $\beta_s < 0$. Using Eq. (\ref{21}) we can rewrite
Eq. (\ref{43}) in the form
\begin{equation}
\label{46}
10^{-6} \frac{l_H}{l_{Pl}} =  
\left(\frac{\nu_1}{\nu_H}\right)^{- \beta}
\left(\frac{\nu_1}{\nu_s}\right)^{\beta_s}\frac{\nu_2}{\nu_H}. 
\end{equation}
We know that $\nu_2/ \nu_H = 10^2$ and $\nu_1/\nu_s$ is not smaller 
than $1$. Substituting all the numbers in Eq. (\ref{46}) one can find that
this equation cannot be satisfied for the largest possible $\beta = -1.8$. 
In the case $\beta = -1.9$, 
Eq. (\ref{46}) is only marginally satisfied, in the sense that a significant
deviation from $\beta_s =0$ toward negative $\beta_s$ can only last for 
a relatively short time. For
instance, one can accommodate $\beta_s = -0.4$ and $\nu_s = 10^8$~Hz. 
On the other hand, if one takes  
$\beta = -2$, a somewhat longer interval of the $z$-stage with $\beta_s < 0$ 
can be 
included. For instance, Eq. (\ref{46}) is satisfied if one accepts $\nu_s =
10^{-4}$~Hz and $\beta_s = - 0.3$. This allows us to slightly increase $h(\nu)$
in the interval $\nu_s < \nu < \nu_1$, as compared with the values of
$h(\nu)$ reached in the more traditional
case $\beta = -2$, $\beta_s = 0$. In what follows, we will consider 
consequences of this assumption for the prospects of detection of the
produced gravitational wave signal.    
Finally, let us see what the available information on the microwave background 
anisotropies \cite{LPG:s,LPG:b} allows us to conclude about the 
parameters $\beta$ and $l_o$. 
\par 
Usually, cosmologists operate with the spectral index ${\rm n}$ (not to be
confused with the wave number $n$) of primordial cosmological perturbations.
Taking into account the way in which the spectral index ${\rm n}$ is defined, 
one can relate ${\rm n}$ with the spectral index $\beta + 2$ that shows up
in Eq. (\ref{35}). The relationship between them is ${\rm n} = 2\beta+5$. 
This relationship is valid independently of the nature of cosmological
perturbations. In particular, it is valid for density perturbations,
in which case the $h(n)$ of Eq. (\ref{35}) is the dimensionless 
spectral amplitude
of metric perturbations associated with density perturbations. If primordial
gravitational waves and density perturbations were generated by the 
mechanism that 
we discuss here (an assumption that is likely to be true) then the parameter
$\beta$ that participates in the spectral index is the same as the one that 
participates in the scale factor of Eq. (\ref{17}). 
Primordial gravitational waves and primordial density perturbations 
with the same spectral index $\beta + 2$ produce approximately the same 
lower-order multiple distributions of large-scale anisotropies.       
\par
The evaluation of the spectral 
index ${\rm n}$ of primordial perturbations have resulted 
in ${\rm n} = 1.2 \pm 0.3$ \cite{LPG:b} or even in a somewhat higher 
value. A recent analysis \cite{LPG:melch} of all
available data favors ${\rm n} = 1.2$ and the quadrupole contribution of
gravitational waves twice as large as that of density perturbations. 
One can interpret these evaluations 
as an indication that the true value of ${\rm n}$ lies somewhere near 
${\rm n} = 1.2$ (hopefully, the planned new observational missions will 
determine this index more accurately). This gives us the parameter $\beta$ 
somewhere near $\beta = -1.9$. We will be using $\beta = -1.9$ in our estimates 
below, as the observationally preferred value. The parameter $\beta$ can
be somewhat larger than $\beta = -1.9$. However, as we already 
discussed, the value $\beta = -1.8$ (${\rm n} = 1.4$)      
is the largest one for which the entire approach is well posed.
The Harrison-Zeldovich spectral index ${\rm n} = 1$ corresponds to $\beta = -2$. 
\par
The observed quadrupole anisotropy of the microwave background radiation is
at the level $\delta T/ T \approx 10^{-5}$.   
The quadrupole anisotropy that would be produced by the spectrum 
(\ref{35}) - (\ref{38}) is
mainly accounted for by the wave numbers near $n_H$. Thus, 
the numerical value of the quadrupole anisotropy produced by relic 
gravitational waves is approximately equal to $A$. According to general 
physical considerations and detailed calculations \cite{LPG:gden:a,LPG:gden:b,LPG:gden:c}, the 
metric amplitudes of long-wavelength gravitational waves and density
perturbations generated by the discussed amplification mechanism are of the 
same order of magnitude. Therefore, they contribute roughly equally to the 
anisotropy at lower multipoles. This gives us the estimate $A\approx 10^{-5}$, 
that we have already used in Eq. (\ref{43}). It is not yet proven 
observationally that a significant part of the observed anisotropies
at lower multipoles is indeed provided by 
relic gravitational waves, but we can at least assume this with some degree of
confidence. It is likely that the future measurements of the microwave
background radiation, and especially its polarization, will help us verify 
this theoretical conclusion.   
\par
Combining all the evaluated parameters together, we show
in Fig.~\ref{LPG:fig7} the 
expected spectrum of $h(\nu)$ for the case $\beta = -1.9$. A small allowed
interval of the $z$-stage is also included. The intervals of the spectrum
accessible to space- and ground-based interferometers are indicated
by vertical lines.   

\if t\figplace
\begin{figure}
\epsfxsize=0.8\textwidth
\centerline{\epsfbox{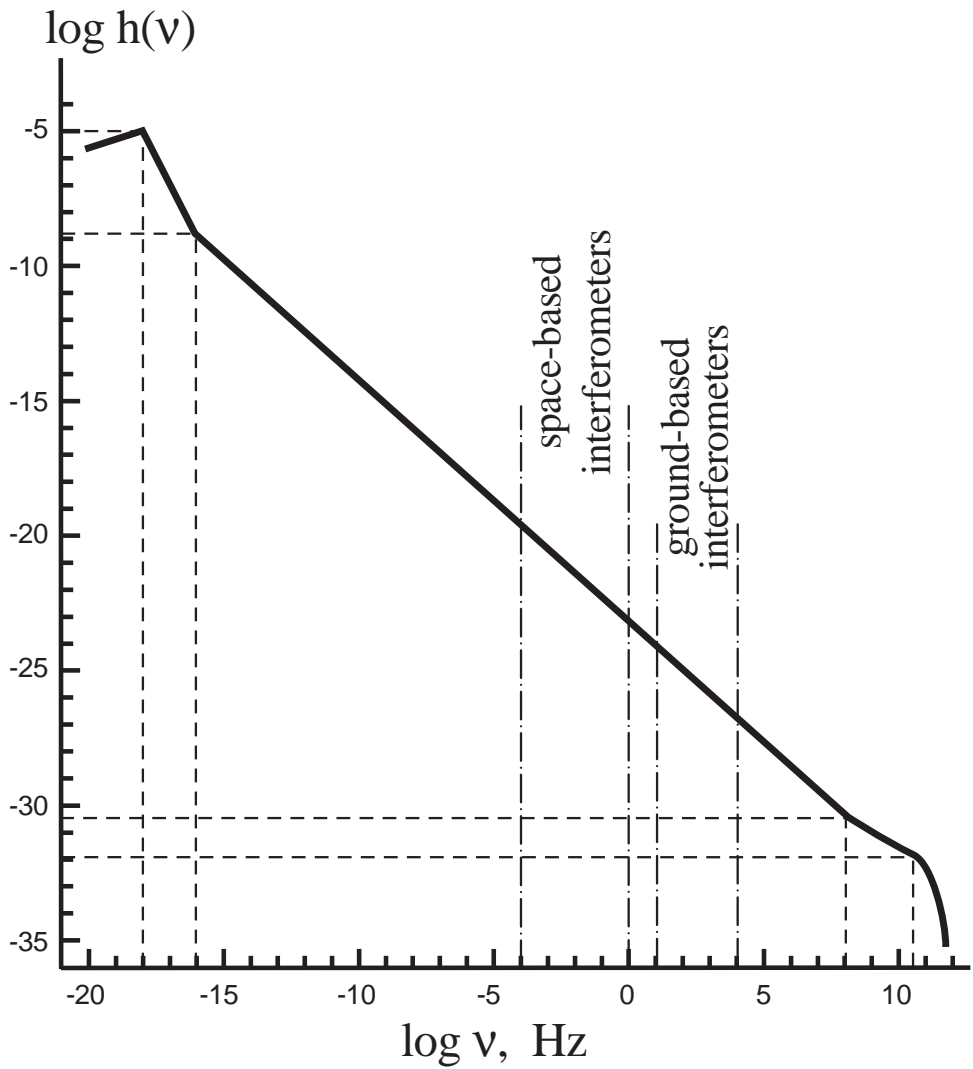}} 
\caption{Expected spectrum $h(\nu)$ for the case $\beta=-1.9$.}
\label{LPG:fig7}
\end{figure}
\fi

It is necessary to note \cite{LPG:gden:a,LPG:gden:b,LPG:gden:c,LPG:g3'} 
that a confirmation of any ${\rm n} > 1$ ($\beta > -2$) will 
mean that the
very early Universe was not driven by a scalar field - the cornerstone of
inflationary considerations. Indeed, the ${\rm n} > 1$ ($\beta > -2$) 
requires the effective equation of state at the initial stage of expansion 
to be $\epsilon + p < 0$ (see Eq. (\ref{23})). But this requirement cannot 
be accommodated by any scalar field $\varphi(t)$ 
whatever the scalar field potential $V(\varphi)$ may be. The energy density 
of the scalar field is $\epsilon = {\dot\varphi}^2/2 + V(\varphi)$, whereas its
pressure is $p = {\dot\varphi}^2/2 - V(\varphi)$, so that $\epsilon + p$
cannot be negative. The available data do not
prove yet that ${\rm n} > 1$, but this possibility seems likely. 

It is also necessary to comment on a certain damage to gravitational 
wave research that was inflicted by the so called ``standard 
inflationary result". The ``standard inflationary
result" predicts infinitely large amplitudes of density perturbations 
$\delta\rho/\rho$ in 
the interval of spectrum with the Harrison-Zeldovich 
slope ${\rm n} =1$ ($\beta = -2$):
$\delta\rho/\rho \propto V^{3/2}/V^{\prime} \propto 1/\sqrt{1 -{\rm n}}$. 
The metric (curvature) 
perturbations $h_S$ accompanying $\delta\rho/\rho$ are
also predicted to be infinitely large, in the same proportion. 
Inflationary literature conceals this predicted infinity of density
perturbations by writing the ratio of the gravitational wave 
amplitude $h_T$ to the scalar metric amplitude $h_S$: 
$h_T/h_S \approx  7 \sqrt{1 - {\rm n}}$, and declaring that the 
contribution of gravitational waves to the cosmic microwave background (CMB) 
anisotropies should be zero, or almost zero, for ${\rm n} \approx 1$. 
Thus, the ``standard'' inflationary theory shifts 
the spectrum of relic gravitational waves, similar in shape to the one shown 
in Fig.~\ref{LPG:fig7}, down by many orders of magnitude.
This claim has incorrectly led to discarding 
the gravitational wave contribution in the analysis of CMB data.
Although one of the recent best fits to the available data indicates the 
presence of the gravitational wave contribution \cite{astro-ph/0002091} 
this fit has been ignored.
For many years, inflationary theorists claimed that their 
arbitrarily large density perturbations were caused by the ``big
amplification during reheating''. It is now universally accepted that
this explanation is false. The scalar metric perturbation, similarly
to the gravitational wave perturbation (see Eq.~(\ref{d})),
remains constant during the long wavelength regime, that is,
its numerical value does not change on the way from the first 
``Hubble-radius-crossing'' to the second ``Hubble-radius-crossing''. 
This fact is also reflected in the constancy of the so-called
``conserved'' gauge-invariant quantity $\zeta$, which is in the center
of inflationary analysis (see, for example, \cite{Kolb&Turner}). 
Since the ``standard'' inflationary theory
predicts arbitrarily large numerical value of $\zeta$, and this
quantity is ``conserved" during the evolution, this arbitrarily large 
number must have been postulated from the very beginning, as quantum
normalization. This happens if one incorrectly assigns the quantum
normalization to the scalar field perturbations alone, as if they were a free
scalar test field in the De-Sitter space-time. Then, the quantity
$\zeta$, calculated from the perturbed Einstein equations, is arbitrarily 
large at the first ``Hubble-radius-crossing", and this number is 
being transmitted to 
the second ``Hubble-radius-crossing". In reality, however, scalar field 
perturbations are always
coupled to metric (gravitational field) perturbations. The correct
quantization of the combined degree of freedom renders the scalar
metric (curvature) perturbation $h_S$ finite and small, and of the same
order of magnitude as gravitational wave perturbation $h_T$. The inflationary
formula for $\delta\rho/\rho$ is incorrect in that it misses the 
dimensionless factor $\sqrt{-\dot H/H^2}$ which
cancels out the zero in the denominator and makes the generated density
perturbations finite and small, even in the interval of spectrum
with the Harrison-Zeldovich slope. The ``standard inflationary result"
is in a severe conflict not only with theory but with observations too: 
when the observers marginalize their data to ${\rm n} = 1$ (enforce this value
of ${\rm n}$ in data analysis) they find finite and small density
perturbations, instead of infinitely large perturbations predicted by
inflationary theorists. 
[For analytical expressions of the ``standard inflationary result" see 
inflationary articles, including recent reviews. For graphical illustration
of the divergent density perturbations and quadrupole 
anisotropies, predicted by inflationary theorists, 
see, for example, \cite{LPG:ms}. For critical analysis and disagreement 
with the ``standard inflationary result'' 
see \cite{LPG:gden:a,LPG:gden:b,LPG:gden:c}].
The most recent articles dealing with perturbations in quasi-deSitter
models rightly emphasize the expected substantial contribution of 
gravitational waves to the large scale CMB anisotropies 
\cite{Hawking&00}. 
In short, general relativity and quantum field theory do not produce
the ``standard inflationary result".

\subsection{Detectability of Relic Gravitational Waves}
\label{sec:secIV:detectability}

We switch now from cosmology to prospects of detecting the predicted relic
gravitational waves. The ground-based 
\cite{S:ligo,S:virgo,LPG:HoughD} 
and space-based \cite{LPG:lisa,LPG:Hel} laser interferometers
(see also \cite{LPG:gw0,LPG:gw1,LPG:gw2}) 
will be in the focus of our attention. We use laboratory frequencies 
$\nu$ and intervals of laboratory time $t$ 
$(c{\rm d}t = a(\eta_R){\rm d}\eta)$. Formulas (\ref{37}) and (\ref{38}), 
with $A = 10^{-5}$, $\nu_2/\nu_H = 10^2$, and the oscillating factor 
restored, can be written as 
\begin{equation}
\label{47}
h(\nu,t) \approx 
10^{-7}\cos[2\pi\nu(t - t_{\nu})]\left(\frac{\nu}{\nu_H}\right)^{\beta+1},
~\nu_2 \le \nu \le \nu_s 
\end{equation} 
and
\begin{equation}
\label{48}
h(\nu,t) \approx 
10^{-7}\cos[2\pi\nu(t - t_{\nu})]
\left(\frac{\nu}{\nu_H}\right)^{1+\beta -\beta_s}
\left(\frac{\nu_s}{\nu_H}\right)^{\beta_s}.~~\nu_s \le \nu \le \nu_1 
\end{equation} 
where the deterministic (not random) constant $t_{\nu}$  
does not vary significantly from one frequency to another at 
the intervals $\Delta\nu \approx \nu$. 
The explicit time dependence of the spectral variance $h^{2}(\nu,t)$ 
of the field, or, in other words, the explicit time dependence of 
the (zero-lag) temporal correlation function of the field at every
given frequency, demonstrates 
that we are dealing with a non-stationary process (a consequence
of squeezing and severe reduction of the phase uncertainty). 
We will first ignore the oscillating factor and will compare the predicted
amplitudes with the sensitivity curves of advanced detectors. The potential
reserve of improving the SNR by exploiting squeezing
will be discussed later.
\par
Let us start from the Laser Interferometer Space Antenna (LISA) \cite{LPG:lisa}.
The instrument will be most sensitive in the interval, roughly, from
$10^{-3}$~Hz to $10^{-1}$~Hz, and will be reasonably sensitive
in a broader range, up to frequencies
$10^{-4}$~Hz and 1\,Hz. The sensitivity graph of LISA to a stochastic
background is usually plotted under the assumption of a 1-year observation
time, that is, the root-mean-square (r.m.s.) instrumental noise is 
being evaluated in frequency 
bins $\Delta \nu = 3\times 10^{-8}$~Hz around each frequency $\nu$.  
We need to rescale our predicted amplitude $h(\nu)$ to these bins. 
\par
The mean square amplitude of the gravitational wave field is given by
the integral (\ref{28}). Thus, the r.m.s. amplitude in the band $\Delta \nu$
centered at a given frequency $\nu$ is given by the expression
\begin{equation}
\label{49}
h(\nu, \Delta\nu) = h(\nu) \sqrt{\frac{\Delta\nu}{\nu}}.
\end{equation}
We use Eqs. (\ref{47}), (\ref{48}) and  calculate expression (\ref{49}) 
assuming $\Delta \nu = 3\times 10^{-8}$~Hz. The results are plotted 
in Fig.~\ref{LPG:fig8}. 
Formula (\ref{47}) has been used 
throughout the covered frequency interval for the realistic
case $\beta = -1.9$ and for the extreme case $\beta = -1.8$. 
The line marked $z$-model describes the signal produced in the 
composite model with $\beta = -2$ up to $\nu_s = 10^{-4}$~Hz 
(formula (\ref{47})) and then followed by formula (\ref{48}) 
with $\beta_s = -0.3$. This
model gives the signal a factor of $3$ higher at $\nu= 10^{-3}$~Hz, 
than the model $\beta = -2$ extrapolated down to this frequency.    

\if t\figplace
\begin{figure}
\epsfxsize=0.8\textwidth
\centerline{\epsfbox{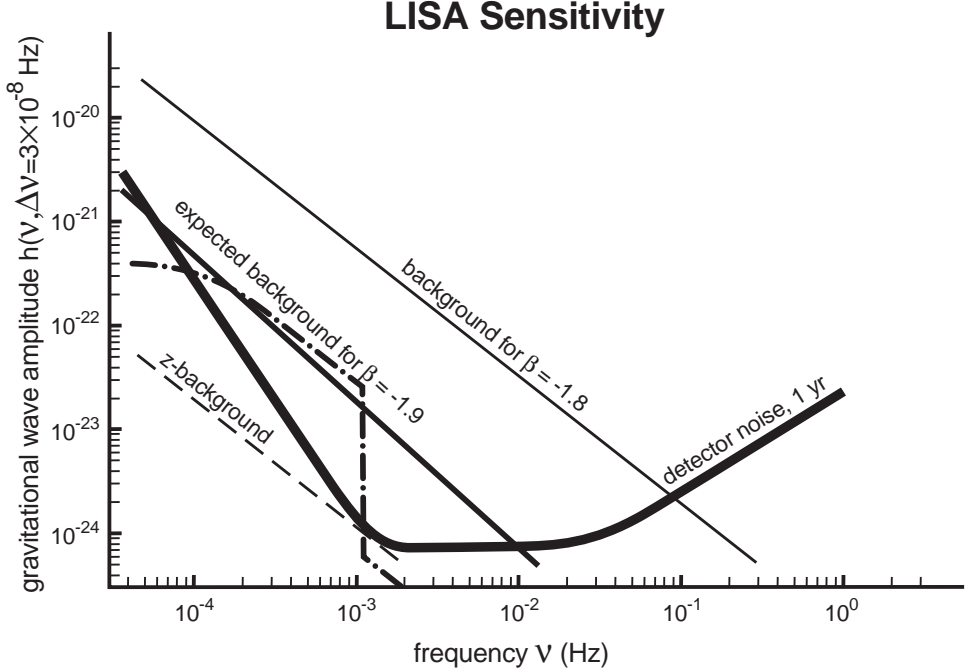}} 
\caption{Expected spectrum $\beta=-1.9$ and other
	possible spectra in comparison with the
        LISA sensitivity.}
\label{LPG:fig8}
\end{figure}
\fi
   
There is no doubt that the signal $\beta = -1.8$ would be easily
detectable even with a single instrument. The signal $\beta = -1.9$
is marginally detectable, with the SNR around $3$ or
so, in a quite narrow frequency interval near and above the frequency 
$3\times 10^{-3}$~Hz. However, at lower frequencies one would
need to be concerned with the possible gravitational wave noise from 
unresolved binary stars in our Galaxy (see Section~\ref{sec:secIII}). 
The further improvement of the
expected LISA sensitivity by a factor of $3$ may prove to be crucial
for a confident detection of the predicted signal with $\beta = -1.9$.   

Let us now turn to the ground-based interferometers operating
in the interval from 10\,Hz to $10^{4}$~Hz. The best sensitivity is
reached in the band around $\nu = 10^2$~Hz. We take this frequency 
as the representative frequency for comparison with the predicted signal. 
We will work directly in terms of the dimensionless quantity $h(\nu)$.
If necessary, the r.m.s. amplitude per Hz$^{1/2}$ at a given $\nu$ can be 
found simply 
as $h(\nu)/{\sqrt\nu}$. The instrumental noise will also be quoted in
terms of the dimensionless quantity $h_{ex}(\nu)$. 

The expected sensitivity of the initial instruments at $\nu = 10^2$~Hz is
$h_{ex} = 10^{-21}$ or better. The theoretical prediction at this
frequency, following from (\ref{47}), (\ref{48}) with $\beta_s = 0$, is 
$h_{th} = 10^{-23}$ for $\beta = -1.8$, 
and $h_{th} = 10^{-25}$ for $\beta = -1.9$. Therefore, the gap between
the signal and noise levels is from 2 to 4 orders of magnitude. The expected
sensitivity of the advanced interferometers, such as 
LIGO-II \cite{LPG:ligoII}, can be as high
as $h_{ex} = 10^{-23}$. In this case, the gap vanishes for the $\beta= -1.8$
signal and reduces to 2 orders of magnitude for the $\beta= -1.9$ signal.
Fig.~\ref{LPG:fig9} illustrates the expected signal in comparison 
with the LIGO-II 
sensitivity. Since the signal lines are plotted in terms of $h(\nu)$, the LISA
sensitivity curve (shown for periodic sources) should be raised and
adjusted in accordance with Fig.~\ref{LPG:fig8}.

\if t\figplace
\begin{figure}
\epsfxsize=0.8\textwidth
\centerline{\epsfbox{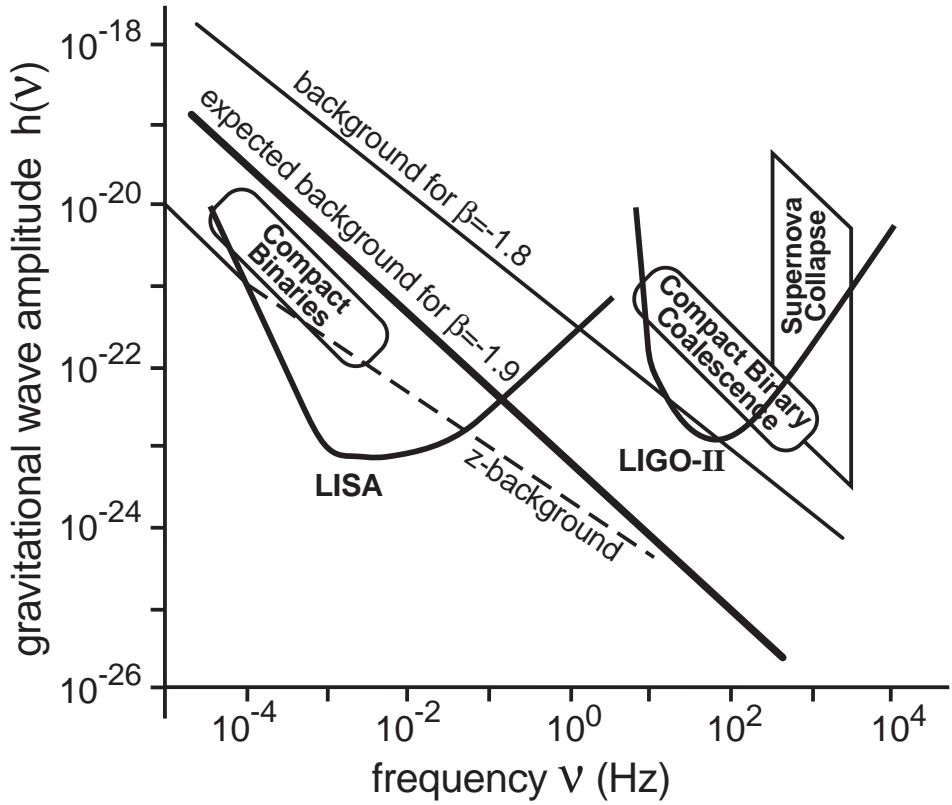}} 
\caption{Full spectrum $h(\nu)$ accessible to laser interferometers.}
\label{LPG:fig9}
\end{figure}
\fi
   
A signal below noise can be detected if the outputs
of two or more detectors can be cross correlated. (For the early estimates 
of detectability of relic gravitational waves see \cite{LPG:g4}.) The cross
correlation will be possible for ground-based 
interferometers, several of which are currently under construction. 
The gap between the signal and the noise levels 
should be covered by a sufficiently long observation time $\tau$.
The duration $\tau$ depends on whether the signal has any temporal 
signature known in advance, or not. We start from the assumption that no
temporal signatures are known in advance. In other words, we first ignore 
the squeezed nature of the relic background and work under the assumption
that the squeezing cannot be exploited to our advantage.  

The response of an instrument to the incoming radiation is 	
$s(t) = F_{ij}h^{ij}$ where $F_{ij}$ depends on the position and 
orientation of the instrument. Since the $h^{ij}$ is a quantum-mechanical
operator (see Eq. (\ref{6})) we need to calculate the mean value of a quadratic
quantity. The mean value of the cross correlation of responses from two
instruments $\langle 0|s_1(t)s_2(t)|0\rangle$ will involve the overlap 
reduction function \cite{LPG:Mich,LPG:Chris,LPG:Flan,LPG:Allen}, 
which we assume to be not much smaller than $1$ \cite{LPG:Flan}. 
The SNR in the measurement of the amplitude 
of a signal with no specific known features increases as
$(\tau \nu)^{1/4}$, where $\nu$ is some characteristic central frequency
(for more detail see Sec.~\ref{sec:data analysis}). 

We apply the guaranteed law $(\tau \nu)^{1/4}$ to initial and advanced
instruments at the representative frequency $\nu = 10^2$~Hz. This law  
requires a reasonably short time $\tau = 10^6~{\rm sec}$ in order
to improve the $S/N$ in initial instruments by two orders of magnitude  
and to reach the level of 
the signal with extreme spectral index $\beta = -1.8$.
The longer integration time or a better sensitivity will make the 
$S/N$ larger than 1. In the case of a realistic spectral index  
$\beta = -1.9$ the remaining gap of 4 orders of magnitude can be
covered by the combination of a significantly better sensitivity and
a longer observation time (not necessarily in one non-interrupted run).   
The sensitivity of the advanced laser interferometers, such as LIGO II, 
at the level $h_{ex} = 10^{-23}$ and the same observation time  
$\tau = 10^6~{\rm sec}$ would be sufficient for reaching the level of the
predicted signal with $\beta = -1.9$.
\par
An additional increase of $S/N$ can be achieved if the statistical
properties of the signal can be properly exploited.
Squeezing is automatically present at all frequencies from $\nu_H$ to
$\nu_1$. The squeeze parameter $r$ is larger in gravitational waves
of cosmological scales, and possibly the periodic structure in 
Eq. (\ref{30}) can be better revealed at those scales. However, we 
are interested here  
in frequencies accessible to ground based interferometers,
say, in the interval 30--100\,Hz. If our intention were to monitor
one given frequency $\nu$ from the beginning of its oscillating regime
and up till now, then, in order to avoid the destructive interference from 
neighbouring modes during all that time,
the frequency resolution of the instrument should have been 
incredibly narrow, of the order of $10^{-18}$~Hz.
Certainly, this is not something what we can, or
intend to do. Although the amplitudes of the waves have adiabatically
decreased and their frequencies redshifted since the beginning of their
oscillating regime, the general statistical properties of the discussed
signal are essentially the same now as they were $10$ years after the 
big bang or will be $1$ million years from now. 
\par 
The periodic structure (\ref{47}) may survive at some level in 
the instrumental window of sensitivity from $\nu_{min}$ (minimal frequency) 
to $\nu_{max}$ (maximal frequency). The mean square value of the field in 
this window is 
\begin{equation}
\label{50}
\int\limits_{\nu_{min}}^{\nu_{max}}h^2(\nu,t)\frac{{\rm d}\nu}{\nu} =
10^{-14}\frac{1}{{\nu_H}^{2\beta+2}}\int\limits_{\nu_{min}}^{\nu_{max}}
\cos^2[2\pi\nu(t - t_{\nu})]\nu^{2\beta+1}{\rm d}\nu~.
\end{equation}
Because of the strong dependence of the integrand on frequency, 
$\nu^{-2.6}$ or $\nu^{-2.8}$, the value of the integral (\ref{50}) 
is determined by 
its lower limit. Apparently, the search through the data should be based on 
the periodic structure that may survive at $\nu = \nu_{min}$. As an illustration,
one can consider such a narrow interval $\Delta \nu = \nu_{max} - \nu_{min}$ 
that the integral (\ref{50}) can be approximated by the formula 
\[
\int\limits_{\nu_{min}}^{\nu_{max}}h^2(\nu,t)\frac{{\rm d}\nu}{\nu} \approx
10^{-14}\left(\frac{\nu_{min}}{{\nu_H}}\right)^{2\beta+2}
\left(\frac{\Delta \nu}{\nu_{min}}\right)\cos^2[2\pi\nu_{min}(t - t_{min})]~.
\]
Clearly, the correlation function is strictly periodic and its structure
is known in advance, in contrast to other possible signals. This is a typical
example of using the a priori information. Ideally, the gain in $S/N$ can
grow as $(\tau \nu_{min})^{1/2}$. This would significantly reduce the 
required observation time $\tau$. For a larger $\Delta \nu$, even an 
intermediate gain between the guaranteed law $(\tau \nu)^{1/4}$ and 
the law $(\tau \nu)^{1/2}$, from the matched filtering 
technique, would help. This could potentially make the signal with
$\beta =-1.9$ measurable even by the initial laser interferometers.  
A straightforward application of (\ref{50}) exploiting squeezing 
may not be possible, as argued in a recent study \cite{LPG:AFP}, but more
sophisticated methods are not excluded. 
\par
For frequency
intervals covered by bar detectors and electromagnetic detectors, the
expected results follow from the same formulas (\ref{47}), (\ref{48}) 
and have been briefly
discussed elsewhere \cite{LPG:g4,LPG:g3}.

\subsection{Conclusion} 

It would be strange, if the predicted signal at the level corresponding
to $\beta =-1.9$ were not seen by the instruments capable of its detection.
There aren't so many cosmological assumptions involved in the derivation, 
that could prove wrong, thus invalidating our predictions. On the other hand, 
it would be even more 
strange (and even more interesting) if the relic gravitational waves
were detected at the level above the $\beta = -1.8$ line. This 
would mean that there is something fundamentally wrong in our 
basic cosmological premises. In the most favorable case, the detection 
of relic gravitational waves can be achieved by the cross-correlation of 
outputs of the initial laser interferometers
in LIGO, VIRGO, GEO600. In the more realistic case, the sensitivity of 
advanced ground-based and space-based laser interferometers will be needed.  
The specific statistical signature of relic gravitational waves, associated
with the phenomenon of squeezing, is a potential reserve for further 
improvement of the SNR.


\section{Gravitational wave detectors and their sensitivity}
\label{sec:detectors}

\subsection{Current status of GW antennas}

Currently, there are a number of  bar detectors in operation: some
of these operate at room-temperature and some others at cryogenic 
temperatures. Bar detectors are resonant, narrow-band detectors.
They can detect
signal amplitudes $h \sim 10^{-20}$ in a band 
width of 10--20~Hz around a central frequency of 1 kHz. 
Asymmetric supernovae in our Galaxy are the best candidate sources
for these detectors. They may also see continuous radiation emitted 
by a neutron star if the frequency happens to lie in their 
sensitivity band.

Interferometric detectors currently under construction will 
increase our ability to directly observe gravitational waves. 
The Japanese have already built a 300 m detector 
in Tokyo, Japan~\cite{S:tama}.  
Several other projects are now nearing completion:
The British-German collaboration is 
constructing a 600 m interferometer (GEO) in Hannover, Germany \cite{S2:geo}, 
the French-Italian collaboration is building a 3 km detector 
(VIRGO) near Pisa, Italy \cite{S2:virgo} and the Americans are
building two 4 km antennas (LIGO), one in Livingston and the other 
in Hanford \cite{S:ligo} in the U.S.A. These detectors 
will start taking data between 2001 and 2003. The larger of these
detectors, LIGO and VIRGO, are likely to  be upgraded in sensitivity
by an order of magnitude with a better low-frequency performance
in 2005.  These ground based interferometers will eventually be
sensitive to sources in the frequency range from 10 Hz to several kHz.
In addition to these ground based antennas, there is a plan to place an 
interferometer in space by the end of this
decade. The Laser Interferometer Space Antenna (LISA) consists 
of three drag-free
satellites, forming an equilateral triangle of size 5 million km,
in a heliocentric orbit, lagging behind the Earth's orbit by $20^\circ$. 
LISA will be sensitive to waves in the low-frequency band of 
$10^{-4}$--$10^{-1}$~Hz.

\subsection{Sensitivity of a GW antenna}

The performance of a GW detector is characterized by the 
{\it power spectral density} (henceforth denoted PSD) of its noise
background.  One can construct the
noise PSD as follows: A GW detector outputs a dimensionless data train, 
say $x(t),$ which in the case of an interferometer is the relative 
strain in the two arms. In the absence of any GW signal the detector 
output is just an instance of noise $n(t),$ that is, $x(t)=n(t).$ 
The noise auto-correlation function $\kappa$ is defined as\footnote{Note that
in earlier Sections we have used angular brackets to denote the
ensemble average; In this Section, however, we shall use an overbar
to denote ensemble average while angular brackets will be reserved
for denoting the scalar product of functions.}:
\begin{equation}
\kappa(t_1,t_2) \equiv \overline {n(t_1) n(t_2)},
\label{eq:noise correlation}
\end{equation}
where an overline indicates the average over an ensemble of noise realizations.
In general, $\kappa$ depends both on $t_1$ and $t_2.$ However, if the detector
output is a stationary noise process, i.e. its performance is, statistically
speaking, independent of time, then $\kappa$ depends only on 
$\tau\equiv t_2 - t_1.$ We shall, furthermore, assume that $\kappa(\tau)
= \kappa(-\tau).$ 
For data from real detectors the above average can be replaced by a time
average under the assumption of ergodicity: 
\begin{equation}
\kappa(\tau) = \frac{1}{T} \int_{-T/2}^{T/2} n(t) n(t-\tau) dt.
\end{equation}

The assumption of stationarity is not strictly valid in the case of real GW
detectors; however, if their performance doesn't vary greatly over 
time scales much larger than typical observation time scales, stationarity
could be used as a working rule.  While this may be good enough in the case
of binary inspiral and coalescence searches, it is a matter of concern for
the observation of continuous and stochastic GW. In this review, for 
simplicity, we shall assume that the detector noise is
stationary. In this case the {\it one-sided} noise PSD, defined only at
positive frequencies, is the Fourier transform of the noise 
auto-correlation function:
\begin{eqnarray}
S_n(f) & \equiv & \frac{1}{2} \int_{-\infty}^{\infty} \kappa(\tau) 
e^{2\pi i f \tau} d\tau,\ \ f\ge 0, \nonumber\\
       & \equiv & 0, \ \ f<0,
\label{eq:psd1}
\end{eqnarray}
where a factor of 1/2 is included by convention. Since we have
assumed that $\kappa(\tau)$ is an even function the above equation
immediately implies that $S_n(f)$ is real. 
It is quite straightforward to show that
\begin {equation}
\overline {\tilde n(f)\tilde n^*(f')} = S_n(f) \delta (f-f'),
\label{eq:psdinfourier}
\end {equation}
where $\tilde n(f)$ is the Fourier transform of $n(t)$
and $\tilde n^*(f)$ denotes the complex conjugate of $\tilde n(f).$ 
The above identity implies that $S_n(f)$ is positive definite.
One derives the above identity by expressing the Fourier transforms
on the left-hand side by their respective time-domain functions,
i.e., $\tilde n(f) \equiv \int_{-\infty}^{\infty} n(t) e^{2\pi i ft} dt,$
and using Eqs.~(\ref{eq:noise correlation}) and (\ref{eq:psd1}).

The autocorrelation function $\kappa(\tau)$ at $\tau=0$ can be expressed
as an integral over $S_n(f).$ Indeed, it is easy to see that
\begin{equation}
\overline {n^2(t)} = 2 \int_0^{\infty} S_n(f) df.
\label{eq:psd3}
\end{equation}
The above equation justifies the name {\it power spectral density} given to
$S_n(f).$ 
It is obvious that $S_n(f)$ has dimensions of time but it is conventional
to use the dimensions of Hz$^{-1}$ since it is a quantity defined in the
frequency domain. The square-root of $S_n(f)$ is 
the noise amplitude, $\sqrt{S_n(f)},$ and has dimensions of {\it per root Hz}.
It is often useful to define the dimensionless quantity 
$h_n^2(f) \equiv fS_n(f),$ called the {\it effective noise.} 
In GW interferometer literature one also comes 
across the {\it displacement noise} or {\it strain noise} defined 
as $h_\ell (f) \equiv \ell h_n(f),$ and the corresponding noise spectrum
$S_\ell(f) \equiv \ell^2 S_n(f),$ where $\ell$ is the arm length of the interferometer. 
The displacement noise gives the smallest strain $\delta \ell/\ell$ in the arms
of an interferometer that can be measured at a given frequency.

\subsection{Source amplitudes vs sensitivity}

One compares the GW amplitudes of astronomical 
sources with the instrumental sensitivity and assesses
what sort of sources will be observable in the following way.
Firstly, as comparisons are almost always made in the 
frequency-domain it is important to note that the Fourier component 
$\tilde h(f) \equiv \int_{-\infty}^\infty dt~h(t) e^{2\pi i ft}$
of a deterministic signal $h(t)$ has dimensions of Hz$^{-1}$ and 
the quantity $f |\tilde h(f)|,$ is dimensionless. It is this last
quantity that should be compared with $h_n(f)$ to deduce the strength of a
source relative to detector noise.  Secondly, it is quite common also to 
compare the amplitude spectrum per logarithmic bin of a source, 
$\sqrt {f} |\tilde h(f)|,$ with the amplitude spectrum of noise, $\sqrt {S_n(f)},$ 
both of which have dimensions of per root Hz. Justification for these
comparisons is given in Sec.~\ref{sec:data analysis}.  Finally, 
for monochromatic sources, one compares the effective noise in a long 
integration period with the expected ``instantaneous'' amplitudes 
in the following way: A monotonic wave of frequency $f_0$ observed 
for a time $T$ is simply a narrow line in a frequency bin
of width $\Delta f\equiv 1/T$ around $f_0.$ The noise in this
bin is $S_n(f) \Delta f=S_n(f)/T.$ 
Thus the SNR $\rho$ after a period of 
observation $T$ is 
\begin {equation}
\rho = \frac {h_0} {\sqrt {S_n(f_0)/T}}.
\label{eq:snrcw1}
\end {equation}
One therefore computes this dimensionless noise spectrum for
a given duration of observation, $S_n(f)/T,$  to assess the 
detectability of  a continuous GW. 

Surely, if the observation time is $T$ then
the total {\it energy} (that is, the integrated 
power spectrum) of both the signal and noise must increase in 
proportion to $T^2.$ Then how does the SNR for a continuous 
wave improve with the duration of observation? 
The point is that while the signal energy is all concentrated in
one bin, the noise is distributed over the entire frequency band. 
As $T$ increases, the frequency resolution improves as $1/T$ and
the number of frequency bins increase in proportion to $T.$
Consequently, the noise intensity {\it per frequency bin} increases only as $T.$
Now, the signal intensity is concentrated in just one bin since the signal
is assumed to be monochromatic. Therefore, the power SNR
increases as $T,$ or, the amplitude SNR increases as $\sqrt{T}.$


\subsection{Noise power spectral density in first interferometers}
\label{sec:noise psd}

As mentioned in the beginning of this Section the performance of a GW detector 
is characterised by the one-sided noise power spectral density (PSD).
The noise PSD plays an important role in signal analysis.  We shall
only discuss the PSDs of interferometric gravitational wave
detectors since the two prime candidate sources discussed in this Review
are both detectable conclusively only in a broadband detector. Interferometers
have a very broad band sensitivity and are, therefore, ideal for 
the detection of these sources. Since our aim is to foresee the
first possible detections we shall mainly concentrate on the initial 
interferometers, GEO, LIGO-I, TAMA and VIRGO mentioning future
ground- and space-based detectors where appropriate.
The sensitivity of ground based detectors is limited at
frequencies less than a Hertz by the time-varying local gravitational 
field caused by a variety of different noise sources, e.g. low frequency 
seismic vibrations, density variation in the atmosphere due to winds, etc.
Thus, for data analysis purposes, the noise PSD is assumed to be essentially
infinite below a certain lower cutoff $f_s.$ Above this cutoff,  i.e. 
for $f\ge f_s,$ Table \ref{table:psd} lists the noise PSD $S_n(f)$ for
various interferometric detectors.
The effective noise $h_n(f)=\sqrt{fS_n(f)}$ expected in these detectors 
is summarized in Fig.~\ref{fig:noise curves}.
The GEO noise curve is somewhat different from others as it uses 
a signal enhancement technique called 'signal recycling' \cite{S:meers}.
It is also for this reason that its sensitivity is close to LIGO and VIRGO
though it is only a sixth/fifth in size as compared to larger interferometers.

\if t\figplace
\begin{figure}
\epsfxsize=4.5in
\centerline{\epsfbox{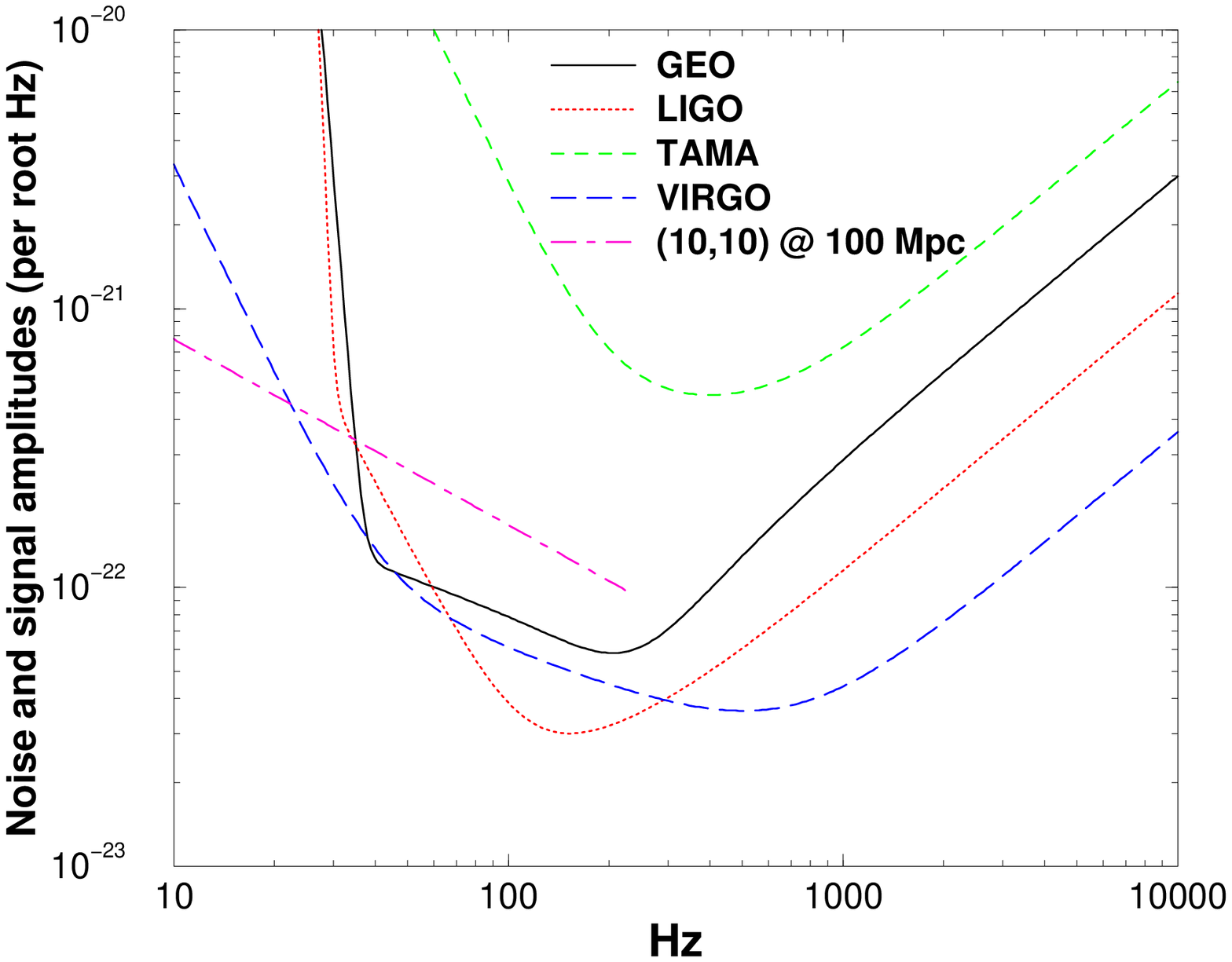}} 
\caption{This figure shows the amplitude noise spectral density,
$\sqrt{S_n(f)},$ in initial interferometers.
On the same graph we also plot the signal 
amplitude, $ \sqrt{f\,}|\tilde h(f)|$, of a binary black hole
inspiral occurring at a distance of 100 Mpc. Each black hole
is taken to be of mass equal to 10M$_\odot.$
(See text in the Section~\ref{sec:inspiral snrs} for a discussion.)}
\label{fig:noise curves}
\end{figure}
\fi
   
\if t\figplace
\begin{figure}
\epsfxsize=4.5in
\centerline{\epsfbox{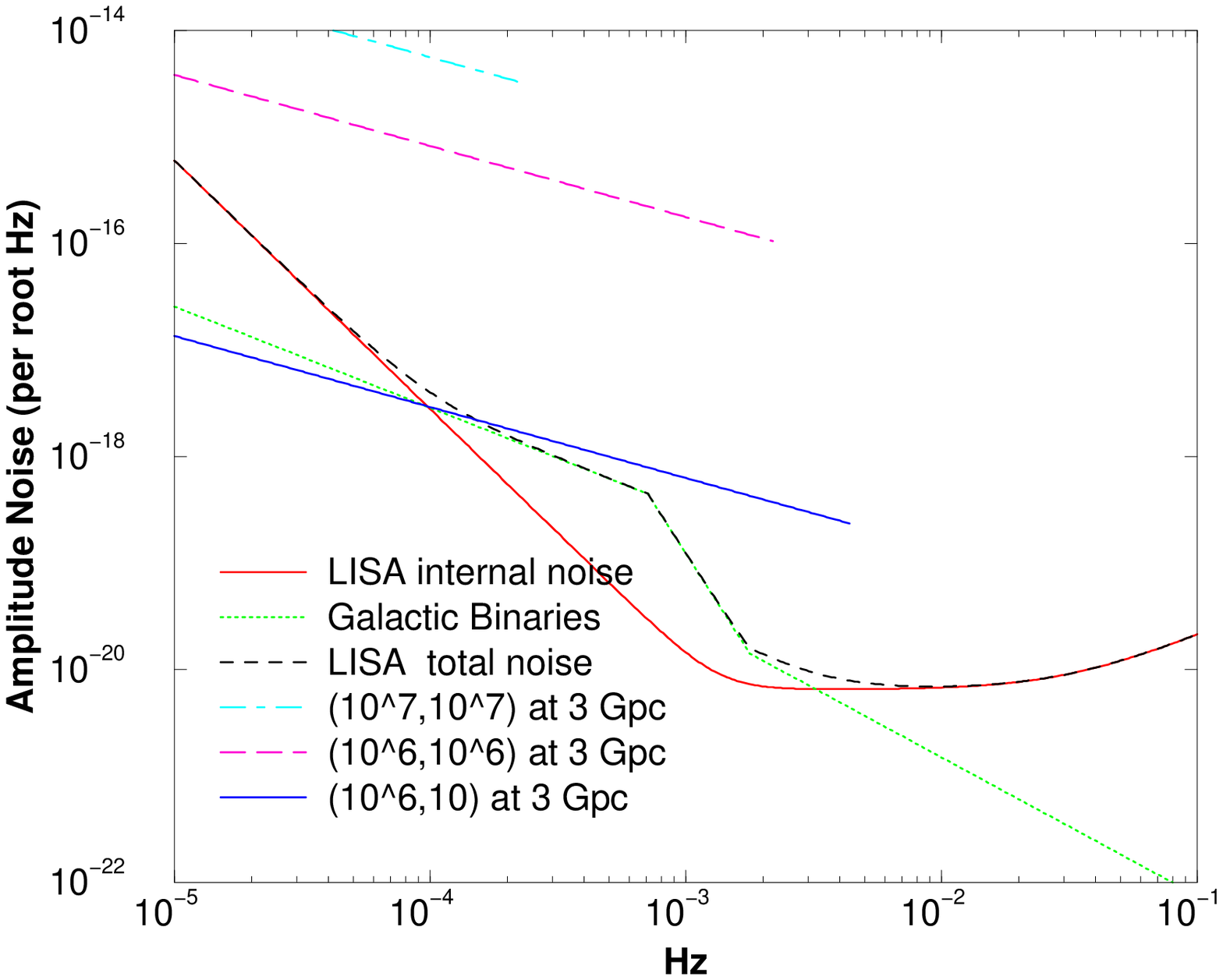}} 
\caption{Same as Fig.~\protect{\ref{fig:noise curves}} but
for the LISA detector. Note that this figure, in contrast to
Fig.~\protect\ref{LPG:fig8}, uses amplitudes per $\sqrt{\hbox{\rm Hz}\,}$. 
We also
plot signals from supermassive black holes. The supermassive BH 
sources are
assumed to lie at a red-shift of $z=1$ but LISA can detect these
sources with a good SNR practically anywhere in the Universe.
}
\label{fig:lisa noise curve}
\end{figure}
\fi

\begin {table}
\caption {Noise power spectral densities of initial interferometers,
$S_n(f)$. For each detector the noise PSD is given in 
terms of a dimensionless frequency $x=f/f_0$ and raises steeply above a 
lower cutoff $f_s.$
(Also see Figs.~\protect{\ref{fig:noise curves}} 
and~\protect{\ref{fig:lisa noise curve}}.).)
}
\centering
\begin {tabular}{cccc}
\hline
\hline
Detector & $f_s$/Hz & $f_0$/Hz & $10^{46} \times S_n(x)$/Hz$^{-1}$ \\
\hline
GEO      & 40       &  150     & $\left [(3.4 x)^{-30} + 34 x^{-1} + 20(1-x^2+x^4/2)/(1+x^2/2) \right ]$  \\   
LIGO-I   & 40       &  150     & $9.00 \left [ (4.49x)^{-56} + 0.16 x^{-4.52} + 0.52 + 0.32 x^2\right ]$\\
TAMA     & 75       &  400     & $7.50 \left [x^{-5} + 13x^{-1}+9(1+x^2)\right ]$ \\
VIRGO    & 20       &  500     & $3.24 \left [ (6.23 x)^{-5} + 2 x^{-1} + 1 + x^2 \right]$\\  
LISA    & $10^{-5}$ &  $10^{-3}$ & $420 \left [ (x/5.62 x)^{-14/3} + 10^3 + x^{2} \right]$\\  
\hline
\hline
\end {tabular}
\label{table:psd}
\end {table}

For LISA the Table gives the internal instrumental noise only. However, 
in the frequency range $10^{-4}$--$3\times10^{-3}$~Hz, LISA will
be limited in its sensitivity by the background produced by several populations
of Galactic binary systems,  such as closed white-dwarf binaries, 
binaries consisting of cataclysmic variables, etc. 
This binary {\it confusion noise} has been well-modeled 
(see~\cite{LPG:lisa} and Section~\ref{sec:secIII}) and gives the following 
`external' LISA noise $S_{\rm ext}(x)$:
\begin {eqnarray}
10^{-46} S_e(x) & = & (x/5.56 \times 10^4)^{-1.9},\ \  x < 10^{-0.15}\nonumber \\
                & = & (x/12.3)^{-7.5},\ \  10^{-0.15}<x<10^{0.25}\nonumber \\
                & = & (x/471)^{-2.6},\ \  10^{0.25}<x,
\label{eq:lisa3}
\end{eqnarray}
where $x=f/f_0$ and $f_0=10^{-3}$~Hz.


\subsection{False alarms and detection threshold}

Gravitational wave event rates in initial interferometers are 
expected to be rather low: about a few per year. Therefore, one has to 
set a high
threshold so that the noise generated false alarms mimicking an event is 
negligible. 
For a detector output sampled at 1 kHz and processed through 
a large number of filters, say $10^3,$ one has  
$\sim 3 \times 10^{13}$ instants of noise in a year. If the noise is Gaussian
then demanding that no more than one false alarm occurs in a year's observation
sets the threshold to be about 7.5 times the standard deviation of noise.
Therefore, a source is detectable  only if its amplitude is 
significantly larger than the effective noise amplitude: i.e. 
$f\tilde h(f) \gg h_n(f).$ 
The reason for accepting only such high-sigma events is because 
the event rate of a transient source, i.e. a source lasting for a few seconds to
mins, such as a binary inspiral, could be as low as a few per year and
the noise generated false alarms, at low SNRs $\sim 3$-4, over a period of an 
year, tend to be quite large. Setting higher thresholds for detection helps in
removing spurious, noise generated events.
However, signal enhancement techniques 
(cf. Sec.~\ref{sec:data analysis}) make it possible to detect a signal of
relatively lower amplitude provided there are a large number of
wave cycles and the shape of the wave is known accurately.

\subsection{Beam pattern functions}

Gravitational wave detectors are sensitive to waves coming from almost
any direction in the sky although the degree of their sensitivity depends
on the actual direction. The sensitivity of a detector to the direction of
the wave is described by what is called the {\it beam pattern function,} also
referred to as the {\it antenna pattern} \cite{Thorne87_300yr}. A GW
antenna responds best if the waves are incident at right angles to the
principal direction, as in the case of a cylindrical bar antenna, 
or principal plane, as in the case of an interferometric detector.
For waves incident in any other direction the response will alter
by trigonometric factors that will be different for the two
polarisations, $h_+$ and $h_\times.$ Denoting the two beam
pattern functions by $F_+$ and $F_\times,$ 
the response $C$ of an antenna to a GW of polarisation amplitudes
$h_+$ and $h_\times$ is given by
\begin{equation}
C = \left [ F_+^2(\theta, \varphi, \psi) h_+^2 + F_\times^2 (\theta, \varphi, \psi) h_\times^2 \right ]^{1/2}. 
\end{equation}
where $(\theta,$ $\varphi)$ denotes the direction to the source
and $\psi$ is the polarisation angle.
For a resonant bar detector, with its longitudinal axis
aligned along the $z$-axis, the response is,
\begin{equation}
F_+=\sin^2\theta \cos 2\psi, \ \ \ \ F_\times = \sin^2\theta \sin 2\psi, 
\end{equation}
and for an interferometer with its arms in the $(x,y)$ plane and
at right angles to each other and the $x$--axis bisecting the two arms
\cite{Thorne87_300yr,S:dt88}
\begin {eqnarray}
F_+(\theta, \varphi, \psi) & = & \frac{1}{2} (1+\cos^2\theta) \cos 2\varphi \cos 2\psi
                           - \cos \theta \sin 2 \varphi \sin 2\psi,\nonumber\\
F_\times (\theta, \varphi, \psi) & = & \frac{1}{2} (1+\cos^2\theta) \cos 2\varphi \sin 2\psi
                           + \cos \theta \sin 2 \varphi \cos 2\psi.
\label{eq:beam pattern functions}
\end {eqnarray}

If a gravitational wave lasts long
enough, then the detector's motion relative to the source will cause
the detector to see different polarisations at different
times (i.e. $C$ is a function of time). Thus, long duration observation 
will help in resolving the two polarisations. However, this will not 
be possible for inspirals observed by ground-based antennas as the waves
will only last for a few mins, during which $C$ essentially remains
a constant.  In this case an interferometric antenna will observe 
only a certain combination of the two polarisations and information of
the source direction and wave polarisation can only be extracted if
several widely separated antennas observe the same signal. 

\section{Data Analysis}
\label{sec:data analysis}
\label{sec:DataAn}

Observing GWs requires a data analysis strategy which
is in many ways different from conventional astronomical data analysis.
There are several reasons why this is so:

\begin {itemize}
\item Gravitational wave antennas are essentially omni-directional with
their response better than 50\% of the average over 75\% of the sky.
Hence our data analysis systems will have to carry out all-sky searches
for sources. 

\item Interferometers are typically broad-band covering 3 to 4
orders of magnitude in frequency. While this is obviously to our
advantage as it helps to track sources whose frequency 
may change rapidly, it calls for searches to be carried over 
a wide-band of frequencies. 

\item In contrast to EM radiation, most astrophysical GWs
are tracked in phase and the SNR 
is built up by coherent superposition of many wave cycles emitted 
by a source. Consequently, the SNR is proportional to the 
amplitude and only falls off, with the distance to
the source $r,$ as $1/r.$ Therefore,
the number of sources of a limiting SNR increases as $r^{3}$ 
for a homogeneous distribution of sources in a flat Universe, as opposed
to EM sources that increase only as $r^{3/2}.$ 

\item Finally, GW antennas
acquire data continuously for many years at the rate of several
mega-bytes per second. It is expected that about a hundredth of
this data will have to pass through our search analysis systems. Unless
on-line processing can be done we cannot hope to make our searches.
This places huge demands on the speed of our data analysis hardware
and a careful study of our search algorithms with a view to making
them as optimal (maximum SNR) and efficient (least search times) 
as one possibly can.
\end {itemize}

Let us first clarify our notation in this Section
and recall two important theorems.  We shall use $x(t)$ to denote
the detector output which is assumed to consist of a background
noise $n(t)$ and a useful gravitational wave signal $h(t).$
The Fourier transform of a function $x(t)$ will be denoted $\tilde x(f)$ 
and is defined as
\begin {equation}
\tilde x(f) \equiv \int_{-\infty}^\infty x(t) e^{2\pi i f t} dt.
\end {equation}
With this definition the inverse Fourier transform is
$x(t) \equiv \int_{-\infty}^\infty \tilde x(f) e^{-2\pi i f t} df.$
The Fourier transform of a real function $x(t)$ obeys
$\tilde x(-f) = \tilde x^*(f).$ A shift in the time-domain 
simply appears as a constant phase shift in the Fourier domain, i.e.
if $\tilde x(f)$ is the Fourier transform of $x(t)$ then
the Fourier transform of $x(t-t_a)$ is $e^{2\pi i f t_a} \tilde x(f).$

In this Section we use a system 
of units in which $G=c=1.$ Thus,
for instance, $1M_\odot = 4.925 \times 10^{-5}$~s and 
1~Mpc$=1.08 \times 10^{14}$~s.
We would like to caution the reader that
$h(f)$ is a dimensionless gravitational
wave amplitude, while in this Section $\tilde h(f)$ is the Fourier 
transform of $h(t)$ and has physical dimensions of time.

\subsection{Matched filtering and Optimal Signal-to-Noise Ratio}
\label{sec:matched filtering}

Matched filtering is a data analysis
technique that efficiently searches for a signal of known shape 
buried in noisy data \cite{S:cwh68}.
The technique consists in correlating the 
output of a detector with a waveform, variously known as a
template or a filter. Given a signal $h(t)$ buried in noise
$n(t),$ the task is to find an `optimal' template $q(t)$ that 
would produce, on the average, the best possible
SNR. In this review we shall
treat the problem of matched
filtering as an operational exercise. However, this intuitive
picture has a solid basis in the theory of hypothesis testing.
The interested reader may consult any standard text
book on signal analysis, for example Helstrom \cite{S:cwh68},
for details. 

\subsubsection{Optimal filter}

Let $x(t)$ denote the detector output. If no signal is present then
$x(t)$ is just a realisation of noise $n(t),$ i.e. $x(t)=n(t),$ while in the
presence of a deterministic signal $h(t)$ it takes the form,
\begin {equation}
x(t) = h(t-t_a) + n(t),
\end {equation}
where $t_a$ is a signal that simply shifts the signal relative to
the origin of time. In the case of signals $h(t)$ whose frequency changes
with time different values of $t_a$ correspond to different frequencies
at $t=0.$ Since a wave detector can only observe signals
in a certain frequency band $t_a$ could mark the time at which the
signal enters the detector's sensitivity band. For this reason $t_a$
is called the {\it time-of-arrival.} Given the time-of-arrival the time 
at which a signal leaves the detector is determined by the parameters 
characterising the signal, such as the masses of the component stars in the
case of a binary inspiral signal. Time-of-arrival is an important parameter
in data analysis as it will usually be unknown and must
be determined by observation. Measuring arrival times in a network of
detectors also helps in determining the direction to a source.

The correlation $c$ of a template $q(t)$ with the detector output is 
defined as
\begin {equation}
c \equiv \int_{-\infty}^{\infty} x(t) q(t) dt.
\label{eq:correlation}
\end {equation}
The purpose of the above correlation integral is to improve the 
visibility of the signal. The following analysis reveals how
this is achieved wherein we shall work out the {\it optimal}
filter $q(t)$ that maximises the statistical average of the correlation
$c$ when a signal $h(t)$ is present in the detector output.
To do this let us first write the correlation integral in the Fourier
domain by substituting for 
$x(t)$ and $q(t),$ in the above integral, 
their Fourier transforms $\tilde x(f)$ and $\tilde q(f),$ i.e.,
$x(t) \equiv \int_{-\infty}^\infty \tilde x(f) 
\exp \left (-2 \pi i ft \right ) df$ and $q(t) \equiv 
\int_{-\infty}^\infty \tilde q(f) \exp \left (-2 \pi i ft \right ) df,$ 
respectively.  After some straightforward algebra one obtains
\begin {equation}
c = \int_{-\infty}^{\infty} \tilde x(f) \tilde q^*(f) df
\end {equation}
where $\tilde q^*(f)$ denotes the complex conjugate of $\tilde q(f).$
In general, $c$ consists of a sum of a two terms, a filtered signal
$S$ and filtered noise $N$:
\begin{equation}
c = S + N,
\end{equation}
where $S\equiv \int_{-\infty}^\infty \tilde h(f) q^*(f) e^{2\pi i f t_a}df$ 
and $N\equiv \int_{-\infty}^\infty \tilde n(f) q^*(f) df.$

Since $n$ is a real random process, $c$ is also a real random process. 
If $n$ is specified by a Gaussian random process with zero mean then 
$c$ will also be described by a Gaussian distribution function, although 
its mean and variance will, in general, differ from those of $n$. 
The mean value of $c$ is, clearly, $S$ -- the correlation of the template $q$ 
with the signal $h,$ since the mean value of $n$ is zero:
\begin {equation}
\overline{c} = S = \int_{-\infty}^{\infty} \tilde h(f) 
         \tilde q^*(f) e^{2\pi i f t_a} df. 
\label{eq:signal}
\end{equation}
The variance of $c,$ that is $\overline {(c- \overline c)^2},$ 
turns out to be,
\begin {equation}
\overline{ (c-\overline c)^2} = \overline {N^2} =
\int_{0}^{\infty} S_n(f) \left | \tilde q(f) \right |^2 df,
\label{eq:noise}
\end {equation}
where $S_n(f)$ is the power noise spectral density defined 
in Eq.~(\ref{eq:psd1}). 
Now the power SNR is defined as 
$\rho^2 \equiv S^2/\overline{N^2}$ and the amplitude SNR is 
$\rho.$ The form of integrals in Eqs. (\ref{eq:signal}) 
and (\ref{eq:noise}) motivates the definition 
of the scalar product of functions, which could either be templates
or waveforms. Given two functions $a(t)$ and $b(t)$ we define their 
scalar product $\left <a, b\right >$ to be
\begin {equation}
\left < a, b \right > \equiv 2 \int_{0}^\infty
{df \over S_n(f)} \left [\tilde a(f) \tilde b^*(f) + \tilde a^*(f) \tilde b(f)\right].
\label{eq:scalar product}
\end {equation}
Recall that $S_n(f)$ is real and positive definite,
[cf. Eq.~(\ref{eq:psdinfourier})],
consequently, the above scalar product defines a positive definite
norm: The norm of $a,$ denoted $||a||,$ is given by
\begin{equation} 
||a|| = 2 \left [ \int_0^\infty  \frac{df}{S_n(f)} |\tilde a(f) |^2 \right ]^{1/2}.
\end{equation} 

Using the reality of the time-domain function $h(t)$ we can write 
down the SNR in terms of the above scalar product as
\begin {equation}
\rho^2 = \frac {\left <he^{2\pi i f t_a}, S_n q\right >^2 }
               {\left <S_n q, S_n q\right >}.
\label{eq:snr1}
\end {equation}
Now, the scalar product of two functions $\langle a, b \rangle$
acquires its maximum value when $a=b.$ Applying this to the
above equation one finds that the template $q$ that 
maximises $\rho,$ called the {\it optimal template,} denoted
$\tilde {q}_{\rm opt}(f),$ is simply
\begin {equation}
\tilde {q}_{\rm opt}(f) = \gamma \frac {\tilde {h}(f)e^{2\pi ift_a}} {S_n(f)},
\label{eq:optimal}
\end {equation}
where $\gamma$ is an arbitrary constant. The inverse Fourier transform
of Eq.(\ref{eq:optimal}) gives the optimal template $q_{\rm opt}(t)$
in the time-domain.  One can see that $q_{\rm opt}(t)$ is the convolution
of the {\it time-translated} signal $h(t)$ with the inverse Fourier 
transform of $1/S_n(f).$ Note that to achieve the maximum of the SNR the optimal
template has to not only match the shape of the signal but also its
time-of-arrival $t_a$ (cf: the factor $e^{2\pi i ft_a}$ in the expression for
the optimal template). Since one would not know the time-of-arrival of the
signal before hand one will have to construct the correlation of the
detector output for several different relative lags of the template
with respect to the detector output. In other words, one constructs the
correlation function $c(t'),$
\begin {equation}
c(t') \equiv \int_{-\infty}^{\infty} x(t) q(t-t') dt
= \int_{-\infty}^{\infty} \tilde x(f) \tilde q^*(f) e^{-2\pi i f t'} df
\label{eq:correlation2}
\end {equation}
where $t'$ is called the lag parameter.

\subsubsection{Optimal signal-to-noise ratio}
We can now work out the maximum, or optimal, SNR by substituting 
Eq.~(\ref{eq:optimal}) for the optimal template in Eq.~(\ref{eq:snr1}),
\begin {equation}
\rho_{\rm opt} = \left < h, h \right >^{1/2} = 2\left [ \int_{0}^\infty
df\frac{ \left |\tilde h(f)\right|^2 }{S_n(f)} \right]^{1/2}.
\label{eq:snr}
\end {equation}
We note that the optimal SNR is {\it not} proportional to
the total energy of the signal,
which is $4\int_{0}^\infty df |\tilde h(f) |^2$,
but rather the integrated signal power weighted down by the noise
PSD. This is in accordance with what we would guess intuitively:
the contribution to the SNR from a frequency
bin where the noise PSD is high, and hence less reliable,
should be smaller than from a bin where the noise PSD is low. 
Thus, an optimal filter automatically 
takes into account the nature of the noise PSD. In this final
expression for the optimal SNR the parameter $t_a$ does not
appear because the time-of-arrival optimal template matches
with that of the signal and hence cancels out in the scalar
product.

The expression for the optimal SNR Eq.~(\ref{eq:snr}) suggests how
one may compare signal strengths with the noise performance of a
detector.  Note that one cannot directly compare $\tilde |h(f)|^2$
with $S_n(f)$ as they have different physical dimensions. 
In GW literature one writes the optimal SNR in one of the following
equivalent ways
\begin {equation}
\rho_{\rm opt} = 
2 \left [ \int_{0}^\infty
\frac {df}{f} \frac {\left |\sqrt{f} \tilde h(f)\right|^2}{S_n(f)} 
\right ]^{1/2}  =
2 \left [ \int_{0}^\infty
\frac {df}{f} \frac {\left |f \tilde h(f)\right|^2}{f S_n(f)} 
\right ]^{1/2},
\label{eq:snr2}
\end{equation}
which facilitates the comparison of signal strengths
with noise performance. One can compare the dimensionless quantities
$f|\tilde h(f)|$ and $\sqrt{f S_n(f)}$ or dimensionful quantities
$\sqrt{f} |\tilde h(f)|$ and $\sqrt{S_n(f)}.$

Signals of interest to us are characterised by several (a priori unknown)
parameters, such as the masses of the component stars in a binary, their 
intrinsic spins, etc., and an optimal filter must agree with 
both the signal shape and its parameters.
A filter whose parameters are slightly mis-matched
with that of a signal can greatly degrade the SNR.
For example, even a mis-match of one cycle in $10^4$ cycles
can degrade the SNR by a factor 2. 

When the parameters of a filter and its shape are precisely 
matched with that of a signal what is the improvement brought 
about as opposed to the case when no knowledge of the signal 
is available?  Matched filtering helps in enhancing the SNR in 
proportion to the square-root of the number of signal cycles covered in 
the detector band, as opposed to the case when the signal shape 
is not known and all that can be done is to Fourier transform the 
detector output and compare the signal energy in a frequency bin 
to noise energy in that bin (see, e.g., \cite{S2:schutz91} for a 
proof).

\subsubsection{Matched filtering of continuous GW}
We will now apply the matched filtering
theorem to observations of sources emitting continuous gravitational
waves (CW) at a single frequency. If we observe a monochromatic source of
frequency $f_0,$ that is $h(t) = h_0 \cos(2\pi f_0t),$ for a 
duration $T$ then its Fourier transform is a sinc function
(${\rm sinc}\,x \equiv \sin x/x$):
\begin{equation}
\tilde h(f) = (h_0T/2) {\rm sinc} [2\pi (f-f_0) T].
\end{equation}
In the above expression we have ignored the sinc function that
occurs at $f=-f_0,$ as it would not contribute to the
SNR since the integral in Eq.(\ref{eq:snr}) runs only over
positive frequencies.  Since the
sinc funciton is strongly peaked around $f=f_0$ and also
because the noise PSD is slowly varying in the frequency range
100-1000 Hz, one can treat the power spectrum appearing
in the SNR integrand Eq.(\ref{eq:snr}) as a constant and write
the optimal SNR as:
\begin{equation}
\rho_{\rm opt}^{\rm CW} =  \frac{2}{\sqrt{S_n(f_0)}} 
\left [\int_0^\infty |\tilde h(f)|^2 df\right ]^{1/2} = 
\frac{h_0}{\sqrt{S_n(f_0)/T}}
\label{eq:snrcw}
\end{equation}
where we have used the identity
$\int_0^\infty {\rm sinc}^2(2\pi fT) df = 1/4T.$
Equation (\ref{eq:snrcw}) justifies Eq.(\ref{eq:snrcw1}) which
was derived heuristically.

\subsection{Matched Filtering Inspiral Waves from Compact Binaries}

As pointed in earlier Secs., the last few mins in the evolution of a
compact binary (NS+NS, NS+BH, BH+BH) is a promising source for 
interferometers that are currently being built. The waveform from
these sources is known very accurately and therefore matched filtering
is the best choice for detecting these sources. 
Matched filtering is very sensitive to the phasing of 
the waves. It is important, therefore, to keep accurate
phase information in our search templates; their amplitudes can be 
taken to be that given by the lowest order post-Newtonian theory.
Such an approximation which works only with phase corrections and
neglects amplitude corrections is called the {\it restricted 
post-Newtonian approximation} \cite{S:last3min}. 

\subsubsection{Accurate templates for inspiral search}

To compute the waveform one must know the evolution of its phase 
and amplitude, which involves the computation of the relative
velocity $v_A(t)$ of the two stars and the phase evolution 
$\varphi_A(t)$ of the binary. The subscript $A$ (which will be
dropped, for convenience, from $v_A$) denotes the
fact that we know these quantities only in a certain approximation.
In the restricted post-Newtonian approximation the binary inspiral
waveform is given by (see, e.g., Ref. \cite{S:dis98})
\begin {equation}
h^A(t) = h_0 v_A^2(t) \cos \phi_A[v_A(t)].
\label{eq:wave}
\end {equation}
The amplitude $h_0$ depends on the masses $M_1$ and $M_2$ of the 
component stars, the distance to the binary $r$
and the orientation of the source relative to the antenna.
More precisely,
\begin {equation}
h_0 = \frac{4\eta M}{r} C(i,\theta, \varphi, \psi). 
\end {equation}
Here, $M=M_1+M_2$ is the total mass of the binary. $\eta=M_1M_2/M^2$ is a
dimensionless (symmetric) mass ratio, which takes a maximum value of 1/4
when $M_1=M_2$. $\eta$ can also be thought of as a measure of how strongly 
the geometry is different from Schwarzschild geometry of a single
body and is sometimes refereed to as the
{\it deformation parameter}. $C$ is a function of the various angles 
describing the polarisation of the wave and the orientation of the 
source relative to the antenna. For
a binary whose orbit is inclined at an angle $i$ to the line of sight,
$C$ is given by
\begin {equation}
C(i,\theta, \varphi, \psi) = \sqrt{A^2 + B^2}, \ \  
A=\frac{1}{2} (1+\cos^2 i)F_+(\theta, \varphi, \psi),\ \ 
B=\cos i\  F_\times(\theta, \varphi, \psi).
\end {equation}
In the above equation the angles $\theta,$ $\varphi$ and $\psi$ parameterize
both the propagation direction and the polarisation of the GW with respect
to the detector and $F_+$ and $F_\times$ are
detector beam pattern functions given in Eq.(\ref{eq:beam pattern functions}). 

For the purpose of computing the SNR
from candidate binaries we can either assume the source to be {\it ideally}
oriented, that is 
\begin{equation}
i=\theta=\varphi=\psi=0,\ \ F_+=1, \ \ F_\times = 0,\ \  C=1, 
\label{eq:ideal C}
\end{equation}
giving us the best 
possible SNR $\rho_{\rm ideal},$ or we can aim at computing the {\it rms}
(root-mean-square) SNR $\rho_{\rm rms}$ by averaging over all angles. 
In the latter case we have following rms values
\begin{equation}
{\left < F_+^2 \right >_{\theta, \varphi, \psi}^{1/2}} = 
{\left < F_\times^2 \right >_{\theta, \varphi, \psi}^{1/2} }= \frac{1}{\sqrt 5}, \ \ 
{\left < C^2 \right >_{i, \theta, \varphi, \psi}^{1/2}} = \frac{2}{5},
\label{eq:rms C}
\end{equation}
which will be used in the next Section to compute the various SNR.

The relative velocity $v_A(t)$ and the GW phase 
$\phi_A(t)\equiv 2 \varphi_A(t)$ in Eq.(\ref{eq:wave})
are given in terms of the following ordinary differential equations
\cite{S2:dis3}:
\begin {equation}
\frac{dv}{dt} + \frac{{\cal F}_A(v)}{ME'_A(v)} = 0,\ \ \ \ 
\frac{d\phi_A}{dt} - \frac{2v^3}{M} = 0,
\label{eq:ode representation}
\end {equation}
where $E_A(v)$ is the relativistic total energy per unit mass,
i.e., $E=(E_{\rm total}-M)/M$, $E'_A(v)\equiv
dE_A/dv$ being its $v$-derivative, and ${\cal F}_A \equiv -dE_A/dt$ is the
GW flux escaping the system, at the given 
approximation\footnote{See \protect\ref{sec:appA} and \protect\ref{sec:appC}
for the post-Newtonian expansions of the energy and flux functions.}. 
The above differential equations can be formally solved to obtain
\begin {equation}
t_A = t_0 - M \int_{v_0}^{v} d\overline {v} \frac {E_A'(\overline v)}{{\cal F}_A(\overline {v})}, \ \ 
\phi_A = \phi_0 - 2 \int_{v_0}^v d\overline {v}\ \overline {v}^3\frac {E_A'(\overline v)}{{\cal F}_A(\overline {v})}.
\label{eq:integral representation}
\end {equation}
In general, there is no closed-form
solution to the above integrals. However, while working with
post-Newtonian theory one essentially has Taylor series expansions 
of GW energy and flux functions and the above integrals, therefore, 
can be solved to obtain a solution of the form $t_A=t_A(v)$ and 
$\phi_A=\phi_A(v),$ which can in turn be inverted to obtain 
$v_A=v_A(t)$ and $\phi_A=\phi_A(t).$ However, 
there is much debate in the literature on what is the best 
representation of an exact inspiral wave emitted by a binary system:
Some authors work with post-Newtonian expansions of flux and energy
and the closed form solution mentioned above \cite{S2:grasp}, others with 
the ODEs Eq.~(\ref{eq:ode representation}),
or, equivalently, with the integrals Eq.~(\ref{eq:integral representation})
\cite{S:poisson.95,S2:dis3}, and yet others with a more accurate representation 
of the energy and flux functions, called P-approximants \cite{S:dis98,S2:dis2}. 
These latter representations have been
shown to be extremely accurate in the test mass limit (i.e.
$\eta \rightarrow 0$) and are also expected to be well-behaved
when $\eta \ne 0$ \cite{S:dis98,S2:dis2,S2:dis3}.

At the lowest post-Newtonian order,
that is at the quadrupole approximation, and for circular orbits we have
\begin {equation}
E_N(v) = -\eta v^2/2,\ \ {\cal F}_N(v) = -32 \eta^2 v^{10}/5,
\label{eq:newtonian energy and flux}
\end {equation}
where a subscript $N$ is used to denote that the quantity is given
at the lowest, i.e. Newtonian, order in post-Newtonian theory.
Substituting these expressions in Eq.(\ref{eq:integral representation})
and inverting the resulting equations gives the following waveform:
\begin{eqnarray}
t(v) & = & t_0 + \frac {5 M}{256 \eta} \left (v_0^{-8} - v^{-8} \right )\\ 
\phi(t) & = & \phi_0 + \frac{1}{16\eta} \left(v_0^{-5} - v^{-5} \right ).\\
h(t) & = & h_0 v^2(t) \cos \phi(t). 
\end{eqnarray}
where we have dropped subscript $N$ for brevity, $v_0$ and $\phi_0$ are the 
relative velocity of the two stars and the phase of the wave at time 
$t_0$ and $v$ in the last two equations is that 
which is obtained by inverting the first equation.

\subsubsection{Signal-to-noise ratios for binary inspiral signals}
\label{sec:inspiral snrs}

The optimal SNR Eq.~(\ref{eq:snr}) is given in terms of the
Fourier transform of the signal. For compact binary inspiral events 
an estimate of the optimal SNR can be made by using the stationary 
phase approximation to the Fourier transform of the signal \cite{S:bssd91}.
In this approximation, the modulus of the Fourier transform 
of the signal in Eq.(\ref{eq:wave}) is found to be \cite{S2:dis2}
\begin{equation}
|\tilde h(f)| = h_0 \left ( \frac{5}{384\eta} \right )^{1/2} \frac{f^{-7/6}}{\pi^{2/3} M^{1/6}}.
\end{equation}
Substituting for $h_0$ in terms of the quantity $C$ and using the resulting expression in
Eq.~(\ref{eq:snr}) we have the following optimal SNR
\begin{equation}
\rho_{\rm opt} = \frac {CM^{5/6}}{r\pi^{2/3}} \left ( \frac{5 \eta}{6} \right)^{1/2} 
\left [ \int_{f_s}^{f_{\rm LSO}} df \frac{f^{-7/3}}{S_n(f)} \right ]^{1/2},
\label{eq:snr3}
\end{equation}
where we have now introduced a specific lower and upper limit in the 
integral. The lower cutoff is dictated by the noise PSD that rises steeply
below a certain frequency $f_s.$ There is hardly any SNR to be gained
by extending our templates to frequencies below this lower cutoff.
For ground-based interferometers $f_s$ is in the range 1-50 Hz 
while for the space-based LISA $f_s$ is slightly smaller than 0.1 mHz. 
The upper cutoff $f_{\rm LSO}$ is determined by the location of
the last stable orbit of the binary.  In the test mass limit 
(i.e., $\eta \rightarrow 0$) the inspiral signal would 
terminate when the test mass is at a distance
$R_{\rm LSO}=6M$ from the central body.
This corresponds to a GW frequency of $f_{\rm LSO}$ given by
\begin {equation}
f_{\rm LSO}^2=\frac{M}{\pi^2 R_{\rm LSO}^3} = \frac {1}{6^3 \pi^2 M^2},
\end {equation}
where the first equality is just a statement of Kepler's law and in the
second equality we have used $R_{\rm LSO}=6M.$
This upper limit\footnote{The justification for using an upper limit 
in the frequency domain based on an upper limit that occurs in the 
time-domain is somewhat  technical; the interested reader is 
referred to \cite{S2:dis3} and references therein for a discussion.}
is important since it is this which limits the kind of inspiral
signals that an interferometer can observe. 
For instance, a ground-based interferometer
will not be able to observe the inspiral of massive binary black holes of 
mass $\sim 10^6$-$10^9 M_\odot,$
since this signal would terminate at frequencies in the milli-Hertz
region, where ground-based interferometers are too noisy. However, a
space-based detector, such as LISA, is sensitive
to the inspiral of massive black holes occurring anywhere in the Universe.

We are finally in a position to write down the rms and ideal SNRs. 
For a binary at a distance $r$ from Earth consisting of stars of 
individual masses $m_1$ and $m_2,$ total mass $m$ and symmetric mass 
ratio $\eta$ the rms and ideal SNRs are obtained by using the rms 
and ideal values of $C$ from Eq.~(\ref{eq:rms C}) and
Eq.~(\ref{eq:ideal C}), respectively, in Eq.~(\ref{eq:snr3}): 
\begin{equation}
\rho_{\rm rms} = \frac {m^{5/6}}{r\pi^{2/3}} \left ( \frac{2 \eta}{15} \right)^{1/2} 
\left [ \int_{f_s}^{f_{\rm LSO}} df \frac{f^{-7/3}}{S_n(f)} \right]^{1/2},\ \ 
\rho_{\rm ideal} = \frac{5}{2} \rho_{\rm rms}.
\label{eq:final snr}
\end{equation}
Note that the SNR depends only the combination ${\cal M}=m\eta^{3/5}$ of 
the two masses called
the chirp mass \cite{S:schutz.nature} and on the integral whose value
crucially depends on the upper limit. In ground-based detectors
for $f_{\rm LSO}$ larger than
about 300~Hz the upper limit is not too important while for lower values
of the LSO frequency the integrals begins to degrade.

Though the SNR depends only on the chirp mass it does not mean that we
cannot measure the two masses by GW observations. Recall that the
optimal SNRs are obtained only when the phases of the signal and the
template are finely matched. In matching the phase one resolves the
degeneracy in the two masses. Indeed, post-Newtonian theory offers
the opportunity to measure more than two parameters although, in general
relativity, each of those parameters, in the point particle limit
(i.e., neglecting intrinsic spins) depend only on the two component masses.
This can be potentially used to test general relativity in the strongly
non-linear regime. (See Ref.~\cite{S:lbbs94,S:lbbs95} for a discussion.) 

A detector's sensitivity is sometimes measured by the
distance $r_{5}^{(1.4,1.4)}$ to an ideally oriented NS-NS source that 
would produces an SNR of 5.  We find that 
$r_{5}^{(1.4,1.4)} =$ 30~Mpc for GEO,  45~Mpc for LIGO, 1.8 Mpc for TAMA
and 50~Mpc for VIRGO.

\begin{table}
\caption {Signal-to-noise ratios for some archetypal binaries consisting
of neutron stars (NS) of 1.4~$M_\odot$ and/or black holes (BH) of 
10 and 15~$M_\odot$ at 100 Mpc and for the hypothetical MACHO BH binaries of 
$0.5 M_\odot$ at 10 Mpc.}
\centering
\begin{tabular}{cccccc}
\hline
Detector$\backslash$ Binary  & (0.5,0.5)&   (1.4,1.4)&  (1.4,10) & (10,10) & (15,15)\\
\hline
GEO                        & 2.6        &   0.58     &   1.2      & 2.8 & 3.6\\
LIGO                       & 3.8        &   0.95     &   2.0      & 4.6 & 5.5\\
VIRGO                      & 3.5        &   0.78     &   1.5      & 3.4 & 4.3\\
\hline
\end{tabular}
\label{table:snrs}
\end{table}

The integral in the expression for the SNR Eq.~(\ref{eq:final snr})
depends on the detector
in question via the noise PSD $S_n(f)$ and can be computed (numerically) 
using the expression for noise PSDs given in Table~\ref{table:psd}. 
The SNRs so obtained are plotted in Fig.\ref{fig:snr estimates} as a function of
the total mass of a binary consisting of equal masses (i.e. $\eta =1/4$) and 
listed in Table~\ref{table:snrs} 
for three archetypal binaries at 100 Mpc and hypothetical
MACHO binaries at 1 Mpc. {\it We note that an inspiral event from a BH-BH binary
at 100 Mpc will have a reasonably good SNR in all the three larger 
interferometers, namely GEO, LIGO and VIRGO.  A coincidence search 
for these sources can, therefore, unambiguously pick out these signals
and hence we believe that the first signals to be detected are 
binary black hole inspirals.}

\if t\figplace
\begin{figure}
\epsfxsize=4.5in
\centerline{\epsfbox{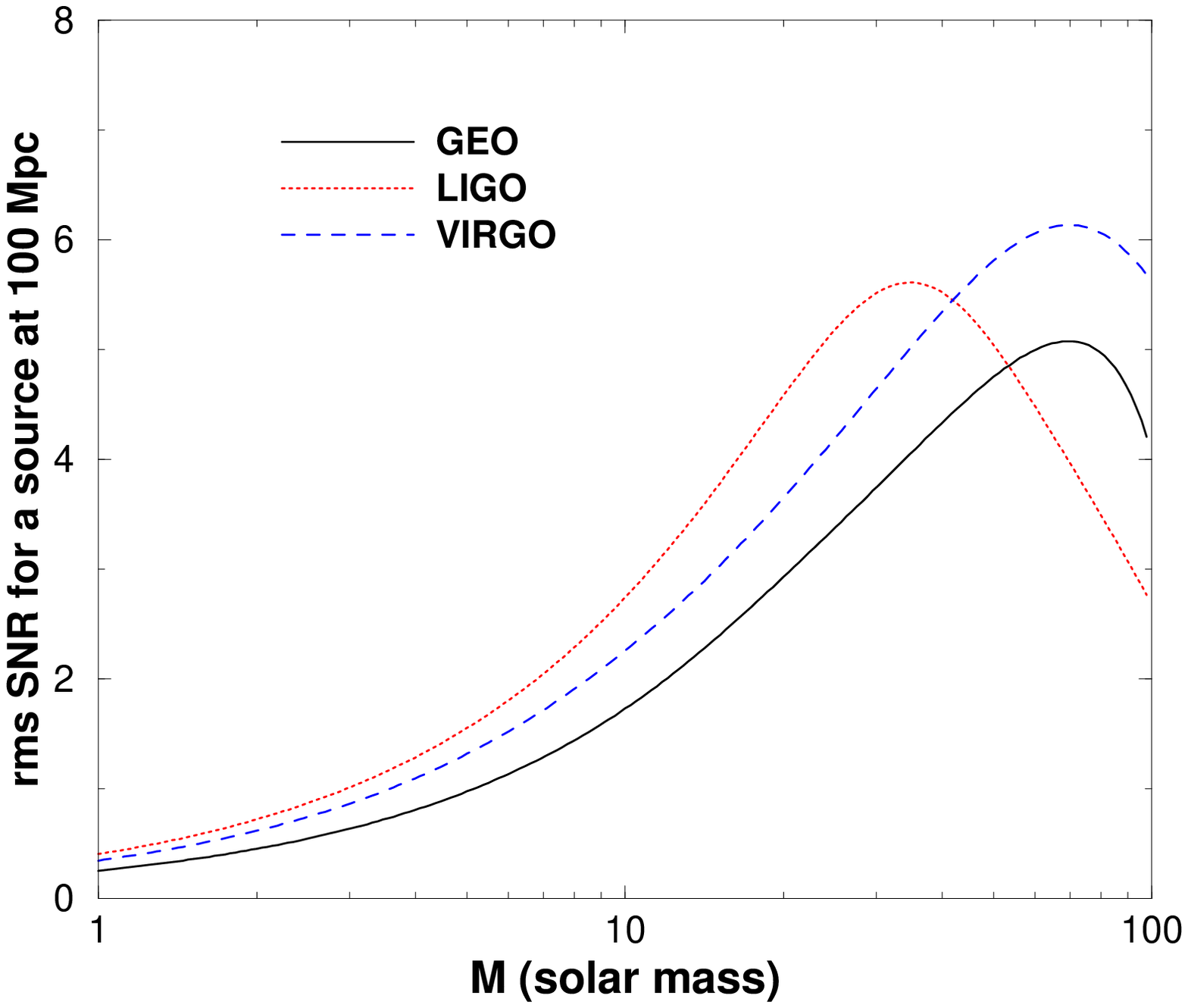}} 
\caption {Signal-to-noise ratio, in initial interferometers, 
as a function of total mass, for inspiral signals from binaries
of equal masses at 100 Mpc
and averaged over source inclination and location. 
(TAMA has been left out as the SNRs in that case are too 
low for these sources.)}
\label{fig:snr estimates}
\end{figure}
\fi

For intermediate mass black hole binaries of mass in
the range of 100 solar masses, VIRGO and GEO can achieve excellent SNRs. 
There is
no convincing evidence for the existence of BHs of such masses, and 
their binaries,
but future GW observations should shed light on these systems.

As mentioned earlier, enhanced interferometers will
have ten times better sensitivity and bandwidth and will therefore be able 
to achieve similar
SNRs at 1 Gpc. As we have seen in earlier Sections at such distances 
the event rate
builds up to several per week. Construction of these advanced 
interferometers will, therefore,
make routine detection of compact binary inspirals and 
coalescences possible and thereby
open up the new gravitational window for observation.

The SNRs achievable for the LISA detector is phenomenal and is listed 
in Table \ref{table:lisa.snrs}. 
LISA should be able to observe $\sim$ million solar mass BH inspiral 
almost anywhere in the Universe.
The time spent by a binary in the LISA band width can be larger than
the life time of the detector. For this reason we have expressed the
SNR in the above equation in terms of the frequency range $(f_1, f_2)$
in which the signal is extracted during a given observational period. 
For a given duration of observation
the binary signal will be the strongest if the observation starts
closest to the instant of coalescence and
therefore we shall take $f_2=f_{\rm LSO}.$ 

\begin{table}[h]
\caption{Signal-to-noise ratios in LISA for supermassive binary black holes and
stellar mass compact objects falling into them, at a cosmological distance
of 3 Gpc.}
\centering
\begin{tabular} {ccccr}
\hline
$ m_1 (M_\odot)$& $m_2 (M_\odot)$ & $ f_1 $ (Hz) & $f_{\rm lso}$  (Hz)& $\rho$ \\
\hline
$10^7$ & $10^7$& $1.08 \times 10^{-5}$  & $ 2.2 \times 10^{-4}$  & $ 1900$\\
$10^6$ & $10^6$& $4.54 \times 10^{-5}$  & $ 2.2 \times 10^{-3}$  & $ 4900$\\
$10^5$ & $10^5$& $1.92 \times 10^{-4}$  & $ 2.2 \times 10^{-2}$  & $ 1700$\\[0.5 cm]
 
$10^6$ & $10$  & $2.75 \times 10^{-3}$  & $ 4.4 \times 10^{-3}$  & $ 21 $\\
$10^5$ & $10$  & $5.55 \times 10^{-3}$  & $ 4.4 \times 10^{-2}$  & $ 11 $\\[0.5 cm]
  
$10^6$ & $1.4$ & $3.93 \times 10^{-3}$  & $ 4.4 \times 10^{-3}$  & $ 3.8 $\\
$10^5$ & $1.4$ & $1.15 \times 10^{-2}$  & $ 4.4 \times 10^{-2}$  & $ 2.2 $\\
\hline
\end{tabular}
\label{table:lisa.snrs}
\end{table}

\subsection{Sensitivity to stochastic gravitational waves}
\label{sec:filtering stochastic background}

In Secs. \ref{sec:secIII} and \ref{sec:secIV} we have seen how astrophysical processes can generate
a stochastic background of GW and what physical interactions can produce
a background in the early Universe, respectively. A stochastic background
is easily detected if the instrumental and environmental background noise
levels in a detector are much smaller than the GW background and especially
if the statistical properties of the GW background or its spectral
characteristics, are different from what is expected for the background
noise. 

Even when the GW background is far too small compared to the
instrumental noise backgrounds, the theory of matched 
filtering developed in the previous Sections, for
the detection of deterministic signals, can be generalised to the case of 
a stochastic background of gravitational radiation. As opposed to 
deterministic signals we will not have the advantage of using a 
template that we can compute before hand. Rather, the idea is to use
the (noisy) data in one detector, in which a stochastic background 
may be present, as a template to detect the background in (noisy)
data from another detector. In other words, we can use the `noisy
template' from one detector to match filter the background in another.
Because our template is noisy, the enhancement in SNR
will not be as good as in the case of matched filtering a deterministic
signal. However, as the background will always be present, if we can
correlate the outputs of two detectors for a long enough duration, 
in principle, the background will show up above other noise sources.

The idea that one detector output can be used as a template
to enhance the visibility of the background in another essentially 
assumes that (1) the two detectors record identical GW signals and (2) 
the instrumental noise in the two detectors are uncorrelated. In reality
neither of these assumptions will hold good perfectly. If a pair of detectors
happen to be close together on Earth, so that they record identical
stochastic signals, then they will also be affected by similar
environmental disturbances such as seismic activity, wind, storm, etc.,
so that the background noise might have a large non-GW correlation.
If the two detectors are far apart then their environmental disturbances
are unlikely to be correlated but they may be registering 
different polarisations
and phases of the background GW so that the cross-correlation might
be insignificant. Clearly, a compromise is in order: Two widely separated
detectors are good for cancellation of the noise background, while 
two nearby detectors are  favoured for enhancing the signal background.
In what follows we will derive the
SNR enhancement achieved on cross-correlating data from two identical
detectors placed at the same location. 
The following derivation is a heuristic one
to indicate how the cross-correlation works; 
see Ref.~\cite{LPG:Flan,LPG:Allen} 
for a more rigorous derivation involving two detectors of different
orientations.

Let $x_1(t)$ and $x_2(t)$ denote the outputs of two interferometric
antennas located nearby to each other and having the same orientation.
They will both have the same stochastic signal $h(t)$ and we will 
assume that the internal noises $n_1(t)$ and $n_2(t)$ are uncorrelated.
We will also assume that the rms value of
the internal noise is much larger than the rms value of the stochastic
signal we wish to detect. Thus, we can write
\begin{equation}
x_1(t) = n_1(t) + h(t), \ \ \ \ x_2(t) = n_2(t) + h(t), \ \ \ \ 
\overline{n_1^2} = \overline{n_2^2} \gg \overline {h^2},
\label{eq:two detector outputs}
\end{equation}
where an overline indicates the ensemble average and we have
assumed that the ensemble averages of the background noise and
stochastic signal are both zero.
In order to extract the stochastic background we construct the 
cross-correlation integral $C.$ The cross-correlation of the
two detector outputs each lasting a time $T$ is given by
\begin{equation}
C^2 \equiv \int_{-T/2}^{T/2} x_1(t) x_2(t) dt \equiv \left (x_1,x_2 \right ),
\end{equation}
where we have introduced the bracket notation to denote the cross-correlation
of two data sets $x_1$ and $x_2.$ The cross-correlation is denoted as $C^2$
since it is quadratic in the useful signal $h.$ 
Using Eq.~(\ref{eq:two detector outputs})
$C^2$ can be written as 
\begin{equation}
C^2  = \left (n_1,n_2 \right )+\left (n_1,h \right )+ 
\left (h,n_2 \right )+\left (h,h \right ).
\end{equation}
$C^2$ is a random process whose average, under the assumption that
the internal detector noises are uncorrelated, gives the useful
signal $S^2$
\begin{equation}
S^2 = \overline {C^2} = \overline {\left (h,h \right )}.
\label{eq:stochastic signal}
\end{equation}
From the above equation we conclude that the signal 
component of $C^2$ grows in proportion to
the integration time $T,$ i.e., $S^2 \propto T.$
This, of course, is only the statistical average of $C^2$ and it must
be compared with the fluctuation in $C^2$ to deduce its significance.
Let us define the noise component of $C^2$ by 
\begin{equation}
N^2 \equiv C^2 - \overline{C^2} = \left(n_1, n_2\right).
\end{equation}
$N^2$ is also a random variable and its rms value will involve the 
noise spectral densities [cf. Eqs.~(\ref{eq:psd1}) and (\ref{eq:psd3})]
$S_{n_1}(f)$ and $S_{n_2}(f)$ of the two
detectors. Assuming that the two detectors have identical noise spectra,
say $S_n(f)$, the rms value of $N^2$ turns out to be 
\begin{equation}
\overline {N^4} \propto T \int_{-\infty}^{\infty} df S_n^2(f).
\label{eq:stochastic noise}
\end{equation}
We see from Eqs.(\ref{eq:stochastic signal}) and (\ref{eq:stochastic noise})
that the SNR for the stochastic background grows as
$S/N \propto T^{1/4}.$ Therefore, in principle, a stochastic background
that is below the internal noise can be extracted by integrating for a 
sufficiently long duration. Conversely, given that we have data sets of
a certain duration $T$ with internal noise spectrum $S_n(f),$
we can estimate the minimum level of the stochastic background we can
extract. These estimates show that initial interferometers will be able
to detect a stochastic background at the level of 
$h(f) \approx 2 \times 10^{-23}$, or 
$\Omega_{\rm GW} \sim 5 \times 10^{-6},$ 
(cf. Sec.\ref{sec:secIV}),
 with a 90 \% confidence level,
after three months of integration \cite{LPG:Allen}. Advanced 
ground-based detectors should be able to reach a level that is five 
orders of magnitude lower (in terms of $\Omega_{\rm GW}$). 
The space-based LISA will have sensitivity
to primordial background similar to the advanced ground-based detectors
but in a frequency range $\sim 10^{-4}$-$10^{-1}$Hz where the
primordial background has larger spectral amplitudes $h(f)$. 
Since LISA is likely to
be in place around the same time as the third generation advanced
ground-based detectors, there is an exciting possibility to detect
primordial background, of the kind discussed in Sec.~\ref{sec:secIV},
within the next decade. They will also help us to understand
populations of astrophysical sources that generate a stochastic GW
background of the sort discussed in Sec.~\ref{sec:secIII}.

\subsection{Computational Costs}
\label{sec:compute costs}

Matched filtering places stringent demands on
the knowledge of the signal's phase evolution which depends on two things:
(1) our modeling of the signal and (2) the parameters characterising
the signal. If our signal model is inaccurate or if the signal's parameters
are unknown, there could be a loss in the SNR extracted.
For instance, in the case of inspiral signals, a mismatch of one cycle 
in $10^4$ cycles leads to a drop in the SNR by more than a factor 
of two, losing a factor of eight in the number of potentially detectable 
events. (Recall that the SNR is inversely proportional to the distance to
a source; thus an SNR loss by a factor $a$ will reduce the
span of a detector by the same factor, resulting in a decrease in 
the volume of observation, and hence the number of events, by a factor $a^3.$)
Moreover, since the parameters of a signal will not be known in advance,
it is necessary to filter the data with a family of templates located
at various points in the parameter space---e.g., placed on a lattice---such that
any signal will lie close enough to at least one of the templates
to have a good cross-correlation with that template.
The number of such templates is typically very large. This places
a great demand on the computational resources needed
to make an on-line search, that is to search for signals in the detector
output at the same rate at which data is acquired. 
We shall discuss below the method of
finding the number of templates to filter any known signal
and the computational resources required to analyse the data on-line.
In the next Section we will discuss the tools needed in 
parameter estimation. 

\subsubsection{Ambiguity function}

Ambiguity function, well known in statistical theory of
signal detection \cite{S:cwh68}, is a very powerful tool in signal analysis:
It helps in making estimates of variances and covariances involved
in the measurement of various parameters, in computing biases introduced in
using a family of templates whose shape is not the same as that of a family of
signals intended to be detected, in assessing the number of
templates required to span the parameter space of the signal, etc. We 
discuss parameter estimation in this Section and computational costs in 
the next.

We begin by defining a {\it normalised} waveform. 
A waveform $a$ is said to be {\it normalised} if it has unit norm: 
$||a|| \equiv \left < a, a \right >^{1/2}  = 1.$ The norm $h_0$ of a 
signal $h$ is also its optimal SNR: $\rho_{\rm opt} = \left<h, h\right >^{1/2}
= ||h|| = h_0.$ For this reason the norm of a signal is 
also referred to as the signal strength.

The {\it ambiguity function} is defined as the scalar product of
two normalised waveforms maximised over the initial phase of one of the
waveforms. More precisely, it is the absolute value of the 
scalar product of two normalised waveforms\footnote {Working 
with analytic signals 
$h(t)=a(t) e^{i(\phi(t)+ \phi_0)},$ where $a(t)$ and $\phi(t)$
are the time-varying amplitude and phase of the signal, respectively,
we observe that the initial phase $\phi_0$ of the signal simply factors
out as a constant phase in the Fourier domain and we can maximise
over this initial phase by taking the absolute value of the
scalar product of a template with a signal.}. 

Let $a(t; {\alpha})$ be a normalised waveform. 
Note that we use the symbol $a$ to denote
a family of waveforms all having the same functional form but
differing from one another in the parameter values. Indeed,
$ {\alpha} =\{\alpha^A | A=0,\ldots,p\}$ denotes the 
parameter vector comprising of $p+1$ parameters.
It is conventional to choose the parameter
$\alpha^0$ to be the lag $t',$ which simply corresponds to a
coordinate time when an event occurs and is, therefore, called 
an {\it extrinsic} parameter, while the rest of the $p$ parameters,
$\alpha^k,\ k=1,\ldots,p,$ are called the {\it intrinsic } parameters and 
characterise the GW source.  Given two normalised waveforms, $a(t; {\alpha_1})$
and $a(t; {\alpha_2}),$ whose parameter vectors are not necessarily
the same, the ambiguity ${\cal A}$ is defined as
\begin {equation}
{\cal A} ({ \alpha_1}, { \alpha_2}) \equiv 
\max_{\alpha^0}\left | \left <a({ \alpha_1}), a({ \alpha_2})\right > \right |.
\label{defcorr}
\end {equation}
Since the waveforms are normalised it follows that
${\cal A} ({\alpha_1}, {\alpha_1}) = 1,$ and, if $\alpha_1$ is not
equal to $\alpha_2,$ ${\cal A} ({\alpha_1}, {\alpha_2}) \le 1.$ 

It is important to
note that in the definition of the ambiguity function there is
no need for the functional forms of the template and signal to be
the same; the definition holds good for any signal-template
pair of waveforms. Moreover, the number of template parameters need 
not be identical (and usually aren't) to the number of parameters 
characterising the signal. For instance, a binary can be characterised
by a large number of parameters, such as the masses, spins, eccentricity 
of the orbit, etc., while we may take as a model waveform 
the one involving only the masses. 
In the context of inspiral waves $a(t;{\alpha_2})$ is the exact
general relativistic waveform emitted by a binary, whose form we do not
know, while the template family is a post-Newtonian, or some other,
approximation to it, that will be used in detecting the true waveform. 
Another example would be signals emitted by spinning neutron stars, 
isolated or in binaries, whose 
time evolution is unknown, either because we cannot anticipate all the
physical effects that affect their spin, or because the parameter
space is so large that we cannot possibly take into account all of them
in a realistic search.  
Of course, in such cases we cannot compute the ambiguity
function since one of the arguments to the ambiguity function would be
unknown. These are indeed issues where substantial work is called for. 
For this Review it suffices to assume that the signal and template 
wave forms are of identical shape and the number of parameters in the
two cases is the same.

In the definition of the ambiguity function Eq.~(\ref{defcorr})
${\alpha_1}$ can be thought of
as the parameters of a (normalised) template, while ${\alpha_2}$ those of
a signal. With the template parameters ${\alpha_1}$ fixed,
the ambiguity function is a function of $p$ signal parameters $\alpha^k_2,\
k=1,\ldots,p,$ giving the SNR obtained by the template for different signals. 
However, one can equally well interpret the ambiguity function as
the SNR obtained for a given signal by filters of different parameter values.
Now, the region in the signal parameter space for which a 
template obtains SNRs larger than a chosen value,
called the {\it minimal match} \cite{S:owen96}, is the {\it span}
of that template. Templates should be chosen so that together
they span the entire signal parameter space of interest with
the least overlap of one other's spans. 

\subsubsection{Metric on the space of waveforms}
\label{sec:metric}

The computational costs of our searches and the estimation of
parameters of a signal, afford a lucid geometrical picture
\cite{S:rbbsssd96,S:owen96}. To develop this picture we
begin with the space of signal waveforms.
A waveform $e(t;{\alpha}),$ with a given
set of values of its parameters, can be thought of as a unit 
{\it vector.} (In much of the discussion below we will deal
with unit signal and template vectors.) 
The set of sample values $h=\{h_k | k=0,\ldots, N-1\}$ 
of a waveform, sampled at times $t_k,$ $k=0, \ldots, N-1,$ can
be thought of as an $N$-dimensional signal vector but not necessarily of
unit norm.  Indeed, the output of a detector sampled in the same way can also
be regarded as an $N$-dimensional vector.  The set of all detector output
vectors forms a vector space. Returning to the 
signal vector, as its parameters are varied the signal vector spans a
space in the underlying N-dimensional vector space of detector outputs.
The dimension of this sub-space is equal to the number of parameters 
of the signal and is called the {\it signal space}. 

The signal space, which is a sub-space of the full vector space,
has a manifold structure, the parameters of the wave
constituting a coordinate system and the dimensionality of the space
being equal to the number of parameters. Having defined a manifold we
can ask if it is possible to define a meaningful metric on this
manifold. Indeed, we already have the necessary tool to define the
metric, namely the ambiguity function. 

Let us fix the template parameters $\alpha_1$ of the template in 
which case the ambiguity function is a function of $p+1$ signal parameters $\alpha_2^k,$
$k=0,\ldots, p.$ Expanding ${\cal A}(\alpha_1,\alpha_2)$ about its maximum 
$\alpha_2=\alpha_1$ and retaining only quadratic terms we get:
\begin{equation}
{\cal A}({\alpha_1},{\alpha_2}) = 1 -
\gamma_{AB} \Delta\alpha^A \Delta \alpha^B + O[(\Delta {\alpha})^3],
\label{eq:match}
\end{equation}
where $\Delta{\alpha}\equiv {\alpha_2}-{\alpha_1}$ and
$\gamma_{AB}$ is the metric on the space of waveforms:
\begin{equation}
\label{def:metric}
\gamma_{AB}({\alpha_1})\equiv
-\frac{1}{2}\left.\frac{\partial^2 {\cal A}({\alpha_1},{\alpha_2})}
{\partial\alpha_2^A\partial\alpha_2^B}\right|
{\atop\scriptstyle\alpha_2=\alpha_1}.
\label{eq:metric}
\end{equation}
This is the metric at the point $\alpha_1$ on the manifold. From now
on we shall drop the suffix $1$ on the parameter $\alpha$ since it 
really represents an arbitrary point on the manifold where a template
resides.  Since one can easily maximise over the lag parameter $\alpha^0$ it is
desirable to work with the metric $g_{ij}$that is projected orthogonal to $\alpha^0,$ 
namely 
\begin{equation}
g_{ij} \equiv \gamma_{ij} - \frac{\gamma_{i0} \gamma_{j0}}{\gamma{ij}}\ \ \ \ i,j=1,
\ldots,p.
\end{equation}
Let us suppose that it is required to make a choice of a template bank.
By a template bank we mean a discrete family of signal waveforms chosen 
in a given region of the signal parameter space. In such a template
bank no template waveform will perfectly match an incoming signal
but if the density of templates is large enough then it may be
possible to extract every signal with an SNR larger than a certain
fraction of the optimal SNR. Most of the early GW data analysis 
literature concerned itself in studying efficient algorithms to
set up a template bank so as to minimise the computational costs
of a search.  The density of templates
in the bank depends on what is the largest fraction of the optimal SNR 
one is prepared to lose in a search: Smaller this fraction greater is
the density of templates. One is normally interested in setting up a
bank such that each possible signal will have its {\it maximised overlap}
larger than a certain {\it minimal match} $(MM)$ with at least some 
member of the bank. By overlap we mean the scalar product of a 
(normalised) signal waveform with a (normalised) template waveform
and maximisation is over all the template parameters.
The maximised overlap, sometimes referred to as the {\it match}, 
is always smaller than or equal to one.
Demanding that the proper distance between templates, namely
$g_{ij}\Delta \alpha^i \Delta\alpha^j$ 
be as large as possible, for a given minimal match $MM,$ we can 
obtain the following formula for a spacing of templates
using Eq.~(\ref{eq:match}):
\begin{equation}
\Delta\alpha^k = \sqrt {\frac{2(1-MM)}{g_{kk}}}
\label{eq:spacings}
\end{equation}
where a factor 2 in the numerator arises because a proper distance of
$(1-MM)$ between the least matched signal and a template implies that
the proper distance between templates be twice that value, assuming
that the templates are placed on a square lattice.

The distance between templates can also be computed more precisely by
employing numerical methods, as was done in \cite{S:bssd91} (also see
\cite{S:bjobss}). 
A comparison of the metric-based method discussed above and
the numerical method shows that the quadratic approximation~(\ref{eq:spacings})
is good typically for ${\cal A} \geq 0.95$. Thus, in the limit of 
close template spacing, Eq.~(\ref{eq:spacings}) can be used to make
a choice of templates.

We now turn our attention to the use of the metric
in calculating the number of templates needed for a search.
If the number ${\cal N}$ of templates needed
to cover a region of interest is large,
${\cal N}$ is well approximated by dividing the proper volume
of the region of interest on the signal manifold space by the proper volume 
per template.  The proper volume per template, $\Delta V$, depends on the packing
algorithm used, which in turn depends on $p$ the dimension of the
parameter space. For instance, we used a square lattice above.
For $p=2$, the optimal packing is a hexagonal lattice, 
and thus
\begin{equation}
\Delta V=\frac{3\sqrt{3}}{2}(1-MM).
\end{equation}
There is no packing scheme which is optimal for all $p$,
but it is always possible (though inefficient) to use a hypercubic lattice,
for which
\begin{equation}
\Delta V=(2\sqrt{(1-MM)/p})^p,
\end{equation}
as a starting point.
Once the span of a template is known the total number of templates
is straightforward to compute using
\begin{equation}
{\cal N}=\frac{\int d^p{\alpha}\,\sqrt{\det\|g_{ij}\|}}{\Delta V},
\end{equation}
where $p$ is, as before, the dimension of the parameter space.

\subsubsection{Computational costs for binary inspiral search}

The number of templates for binary inspiral searches has been
computed for the post-Newtonian signals and ground-based interferometers
discussed earlier \cite{S:bjobss}. The numbers required at a high minimal
match $MM=0.97$ are several hundreds of thousands 
(cf. Table~\ref{table:volumes}) and imply huge
computational costs. For instance, to search for inspiral waves on-line
(i.e. to search at the same rate as the rate at which data is recorded)
we would require a dedicated computer that can carry out $10^{10}$
floating point operations per second (i.e., 10 GFLOPS). Building economical 
computers of such speed is the prime concern of various data analysis groups at
the moment. 
\begin{table}[t]
\caption{Number of templates required to search for inspiral waves
from a binary consisting of stars of mass $m_1,m_2 \ge 0.2M_\odot$
and minimal match of 0.97. The numbers are given 
for different post-Newtonian families of waveforms.}
\begin {center}
\begin{tabular}{lccc}
\hline
Interferometer & 1PN & 1.5PN & 2PN\\
\hline
LIGO-I & $2.5\times10^5$ & $5.3\times10^5$ & $4.7\times10^5$\\
VIRGO & $1.4\times10^7$ & $1.4\times10^7$ & $1.3\times10^7$\\
GEO600 \ & $4.3\times10^5$ & $8.5\times10^5$ & $7.5\times10^5$\\
\hline
\end{tabular}
\end {center}
\label{table:volumes}
\end{table}

\subsubsection{Computational costs to search for continuous waves}
\label{sec:periodic}

The search problem for continuous waves from spinning neutron stars
is the most compute-intensive job in gravitational wave data analysis.
Today, there is little hope that all-sky searches lasting
for a year or more, can be made. It is easy to see why this is
such an intensive job: Firstly, the data has to be collected continuously
for months together and at a good sensitivity. No one has run
interferometers for periods as long as that and we do not yet
know if this would be possible.
Secondly, though a neutron star emits a periodic signal
in its rest frame, save for the neutron star spin-down which indeed
induces some modulation in the waveform, because of Earth's acceleration
relative to the source, the detector does not see a periodic wave. The
wave is both frequency- and amplitude-modulated. One can, fortunately,
de-modulate these effects since Earth's motion is known quite accurately,
and hence recover the original periodic signal.
{\em But} de-modulation requires a knowledge of the source's direction
and its frequency, which are unknown in a blind search. The angular
resolution one obtains in a year's integration is
$\Delta \theta = \lambda/D,$ where $\lambda$ is the wave length of 
radiation and $D$ is the baseline of the detector in a year's integration,
namely 1 A.U. Thus, for $f=100$~Hz we have $\Delta\theta=10^{-5}$~rad or 
about two arcsec. Now, assuming that the source may be in any one
of the 4 arcsec$^2$ patches on the sky we get the number of patches 
in the sky for which we will have to try out a de-modulation correction
to be $4 \pi/(\Delta \theta)^2 = 4 \pi 10^{10}.$ It is quite an
impossible task to apply Doppler de-modulation to the detector output
for each of these $\sim 10^{11}$ patches and compute as many Fourier
transforms. 

One, therefore, asks the question given a compute power what is the
best possible search one can do? Is there any advantage in going from
a one-step search to a two- or multi-step hierarchical search? What
about directional searches? These are some of the problems for which
we have some answer; but a great deal of work is needed and
is currently under progress, to improve and optimise search algorithms.
In the following we will provide a summary of the current status.

The differential geometric formalism discussed above
has been used \cite{BradyEtal1998} to compute the number of days of data that a 
TFLOPS-class computer can analyse on-line (that is, analyse $T$-hours of 
data in $T$ hours) and carry out a blind (that is, unknown direction, 
frequency and spin-down rate) search. Unfortunately, the longest
data we can integrate on-line, for neutron stars with spin frequencies
$f\le 100$~Hz and spin-down rates less than 1000 years, is about 18 days.
This yields a SNR lower by a 
factor of 5 as compared to a year's worth of observing. 
On-line searches for neutron stars with $f\le 500$~Hz (largest observed
frequencies of millisecond pulsars) and spin-down rates
of 40 years (shortest observed spin-down rates), can only be made for
a data set lasting for a duration of 20 hours or less. If the source's 
position is
known in advance, but not its frequency, then one can carry out an on-line
search, again with a TFLOPS-class computer,  for the frequency of the source
in a data set that is worth 3 months long. This is good news since there
are many known pulsars and X-ray binary systems that are potential sources
of radiation. In addition, the obvious targeted search locations are
the centre of the Galaxy and globular clusters.

There have been efforts \cite{S2:brady.creighton,S2:papa.schutz} to study
the effectualness of a two-step hierarchical method for a blind search. Here  
the basic idea is to construct Fourier transforms of data sets of duration
smaller than the period in which Doppler modulations will be important
and to stack spectral densities obtained in this way and to add them all up. 
This is an incoherent way of building the signal since one adds spectral
densities that have no phase information. Therefore, 
one gains in SNR less than what an optimal matched filtering method
is able to achieve.  However, this does not matter since
(i) the targeted SNR's are quite high $\sim 10$ and (ii) candidate
events can always be followed-up using coherent integration
methods. These methods afford an on-line all-sky blind search (i.e., a search
in which no assumptions are made about the parameters of the source)
for continuous gravitational waves for a period of 4 months or less using
a 20 GFLOPS computer. Detector groups are planning to build computers
of this kind to aid in their search for continuous GW.

\subsection{Covariance Matrix and Parameter Estimation}
\label{sec:estimation}

After a detection has been made, say because a high SNR has been recorded
that could not be vetoed out, the next step in data analysis is estimation
of parameters characterising the event and provide error bounds
on the measured values. The first thing to note is that one can never
be absolutely certain that a signal is present in a data train; one can
only give confidence levels about its presence which could be close
to 100\% at high values of the SNR. Confidence level is a measure of
the probability that the observation of an event, such as a large
peak in the correlated output, is generated by a gravitational wave
signal as opposed to a random, non-gravitational wave process. 
The next thing to note is that
howsoever high the SNR may be one can't be absolutely certain about the
true parameters of the signal: At best one can compute
a range of values in which the true parameters of the signal are
most likely to lie. The width of the range
depends on the confidence level demanded, being larger for 
higher confidence levels.

In our search for a  signal in the output of a
detector we use a discrete, rather than a continuous, family of
templates. Each template has a particular set of values of the parameters
and the templates together span an interesting region of the parameter
space.  The spacing between templates in the parameter space will,
in general, be quite small. A common estimate of the signal parameters
is given by the parameters of the template that obtains the 
maximum SNR. Such an estimate is called the {\it maximum likelihood 
estimate} -- so-named because the parameters of this template 
maximise what is called the {\it likelihood ratio} \cite{S:cwh68}.
Maximum likelihood estimates are not always the 
minimum uncertainty estimates, as has been particularly demonstrated for
the case of binary inspiral signals \cite{S:rbbsssd96}.
{\it Bayesian} estimates, which take into account any prior 
knowledge that may be available about the distribution of the source
parameters, as well as the knowledge from the output of a whole bank of 
templates rather than a single template, often give a much better estimate 
\cite{S:nv98}. The reason for the better performance of Bayesian estimates
is that they make a quantitative use of the information at hand.

In a measurement process any estimation of parameters, howsoever efficient, 
robust and accurate, is unlikely to give the actual parameters of the signal 
since, at any finite SNR, the presence of noise alters the input signal. 
In geometric terms, the signal vector is being altered
by the noise vector resulting in a vector that lies outside the signal
manifold. Techniques such as matched filtering aim at computing the
best projection of this altered vector onto the signal space. 
The true parameters of the signal are expected to lie within an ellipsoid of 
$p$ dimensions at a certain level of confidence -- the volume of 
the ellipsoid increasing with the level of confidence. The axes of the 
ellipsoid are the 1--$\sigma$ uncertainties in the estimation of parameters 
and the confidence level corresponding to a 1--$\sigma$ uncertainty is 
$0.67^p$, confidence level corresponding to a 2--$\sigma$ uncertainty
is $0.95^p,$ and so on. 

The topic of parameter estimation deserves a much wider discussion
than given here. However, our goal here is only to assemble the necessary tools 
from estimation theory for ready use. Interested reader can consult
the ever-growing literature for further details. (See, Ref. \cite{S:cwh68} for 
estimation theory and \cite{S:rbbsssd96,S:rbbsssd96a,S:cf94,S:epcw95} for applications in 
GW observations.)

\subsubsection{Covariance matrix}
The scalar product Eq.~(\ref{eq:scalar product}) induces the following metric
on the signal manifold \cite{S:rbbsssd96}:
\begin {equation}
G_{ij} = \left < {\partial a({ \alpha}) \over \partial \alpha^i},
{\partial a({ \alpha}) \over \partial \alpha^j} \right >,
\label{eq:full metric}
\end {equation}
where $a(\alpha)$ is a signal vector of unit norm and $\alpha^i$ are
the coordinates. The metric $G_{ij}$ is defined on the manifold
of all the parameters including the lag $t'$ and the initial
phase $\phi_0$ of the waveform. It is easy to show that 
$G$ projected orthogonal to the initial phase $\phi_0$  yields
the metric $\gamma$ in Eq.(\ref{def:metric}).

The metric on the space of waveforms introduced above is also called
the Fisher information matrix \cite{S:cwh68}. Indeed, it contains
the `information' about how similar or dissimilar are the waveforms
in a small neighbourhood around the parameter $\alpha$ of the signal.
Large values of the metric imply that even small changes in signal
and template parameters can greatly affect their overlap while the
opposite is true when the metric coefficients are small.

The inverse of the information matrix is the {\it covariance matrix} 
$C_{ij},$ whose diagonal 
and off-diagonal elements are, in the limit of large SNR,
the variances in the measured values of the parameters and 
correlation coefficients among different parameters, respectively
\cite{S:cwh68}:
\begin {equation}
C_{ij} = \left (G \right )^{-1}_{ij}.
\end {equation}

Covariance matrix based errors in the estimation of the total mass,
reduced mass and the instant of coalescence $t_C$ have been computed
by Poisson \& Will \cite{S:epcw95} for the second post-Newtonian 
inspiral waveforms.  They are listed in Table~\ref{table:errors}
for three archetypal binaries. The relative errors are smaller, for a given
SNR, in the case of lighter binaries. This is because lighter
binaries last for a longer duration and have a larger number of
cycles in the detector, making it relatively easier, as compared
to higher mass binaries, to discriminate waveforms of different parameters.

\begin {table}
\caption {Errors in the estimation of instant of coalescence 
$\Delta t_C,$ phase at coalescence $\Delta \phi_C,$
chirp mass $\Delta{\cal M}/{\cal M},$ and symmetric mass ratio
$\Delta \eta/\eta,$ and percentage bias in the estimation
of the total mass ${\cal B}_m=100(1-m^T/m^X)$ and percentage bias in
the estimation of the mass ratio ${\cal B}_\eta=100(1-\eta^T/\eta^X),$ 
where $T$ stands for the parameter of the template and
$X$ for the parameter of exact wave form,
for the second-post-Newtonian corrected
inspiral waveform neglecting the effect of spins. Values are
quoted for three archetypal binaries consisting of two 1.4~$M_\odot$
NSs (system NS-NS), a 1.4~$M_\odot$ NS and a 10~$M_\odot$ black hole 
(system NS-BH), and two 10~$M_\odot$ black holes (system BH-BH).}
\begin {center}
\begin {tabular}{lcccccc}
\hline
System & $\Delta t_C$ (ms) & $\Delta \phi_C$ (rad) & 
$\Delta {\cal M}/{\cal M}$ ($10^{-3}$) & $\Delta \eta/\eta$ 
& ${\cal B}^T_m$ & ${\cal B}^T_\eta$ \\
\hline
NS-NS & 1.07 & 2.94 & 0.36 & 0.28 & $ 0.214$ & $0.211$  \\
NS-BH & 1.72 & 2.27 & 2.20 & 0.50 & $-6.96$  & $ 12.2$  \\
BH-BH & 1.50 & 2.19 & 5.40 & 1.50 & $ 1.40$  & $ 0.282$ \\
\hline
\end {tabular}
\end {center}
\label{table:errors}
\end{table}

\subsubsection{Biases in estimation}
\label{sec:biases}

There are two ways in which an error can occur in the estimation
of signal parameters. Firstly, an error in the
measurement of a parameter occurs because of internal noise which
alters the input signal and hence, in the process of maximising the
SNR, we err in the estimation of parameter by an amount that depends
on the SNR. This type of error is a {\it random error} and
normally goes down in inverse proportion
to the SNR. In the limit of an infinitely large ensemble of measurements
the estimated values converge to the true values of the signal parameters.
Secondly, our estimator may be a {\it biased} estimator in the sense
that the average over an ensemble of measurements might 
converge to values different from the true values of the parameters. 
This can happen because of one or both of the following two reasons: 
The search templates
we use in our detection algorithms might only be an approximation
to the true signal, as in the case of inspiral wave searches where
we use post-Newtonian search templates to look for the fully general
relativistic signal. As we shall show below, such a search
would induce a bias in the estimation of parameters, which we shall
refer to as the bias of the {\it first kind}. 
Alternatively, the estimator might be inherently a biased estimator and
may give erroneous values even if the search templates are not faulty 
which we shall refer to as the bias of the {\it second kind.}

An example of the bias of the second kind is the following. In the
maximum likelihood method we always aim for the largest value of the
likelihood ratio. This has the effect of making a higher estimation of
the amplitude parameter of the signal as shown in~\cite{S:rbbsssd96}. 
For inspiral signals, this would mean that the distance
to the source would be under-estimated. 

Biases in parameter estimation for inspiral signals have been discussed
in \cite{S:sathya:lh}. Biases in the estimation
of the total mass and the symmetric mass ratio on using standard post-Newtonian
approximation are quoted in Table~\ref{table:errors} and discussed
in detail in \cite{S:dis98}.
They find that while using standard post-Newtonian approximation to the
waveform the bias could be quite large, whereas $P$--approximants to
inspiral waves greatly reduce the bias in the estimation of parameters.

\subsection{Conclusions}
Gravitational waves in ground-based interferometers are expected to be
below the noise levels of first and second generation instruments. 
We have seen in this Section how a prior knowledge of the signal's
shape can be used in enhancing the visibility of the signal. With the
aid of matched filtering a network of initial interferometers, consisting of
GEO, LIGO and VIRGO, should be able to 
survey a volume of $3 \times 10^6$ Mpc$^3$ for binary black holes at a 
minimum SNR of 3. In this
volume conservative estimates of coalescence rates predict about a few per
year. Thus, the binary black holes may be the first events to be
registered in our detectors.

Initial interferometers should also be able detect
primordial gravitational wave background at the level of $\Omega_{\rm GW} 
\sim 5 \times 10^{-6}$ by cross-correlating data in nearby interferometer pairs
(LIGO-LIGO or GEO-VIRGO). However, this is at a level rather too small
from a theoretical point of view. New data analysis algorithms that
exploit the specific signature of the primordial background may aid
in detecting the background at a much lower level but, at the moment, 
we do not know how to achieve this.

Initial interferometers will also be able to detect continuous gravitational
waves from newly born or rapidly spinning  
non-spherical neutron stars provided that the amplitude is $h\ge 10^{-26}$
and the signal lasts for a few months.  This, of course,  assumes
that we know the evolution of the phase of the signals emitted in the
process which, for old neutron stars is a simple sinusoid. However, in the
case of waves emitted due to one of the instabilities we do not have
the right templates. This is also true of transient waves emitted during
a supernova explosion or the merger of black holes. Thus, much work is
needed in understanding sources of gravitational waves. Indeed, it may very
well be that the new generation of gravitational wave antennas will open
up a new window for observing the Universe which will aid in our 
understanding of strong and non-linear gravity.

\subsection*{Acknowlegements}

We appreciate useful discussions with K. Thorne, V. Braginsky,
and B. Schutz. 
This work was supported in part by a joint research grant from the
Royal Society.


\def\theequation{A\arabic{equation}}
\def\thesubsection{A\arabic{subsection}}
\def\thesubsubsection{\thesubsection.\arabic{subsubsection}}
\setcounter{equation}{0}
\setcounter{section}{0}
\setcounter{subsection}{0}

\section{Appendix}
\label{sec:app}

Here we summarize some basic formulae relevant for the 
description of massive binary evolution.

\subsection{Keplerian binary system and radiation back reaction}
\label{sec:appA}

Binary stars is one of the main topics
of the present paper, so it is necessary to remind the reader
some basic facts about Keplerian motion in a binary system.
The stars are highly concentrated objects, so their
treatment as point masses is usually adequate for the description 
of their interaction in the binary.
Further, the Newtonian gravitation theory is sufficient 
for this purpose as long as the orbital velocities
are small in comparison with the speed of light $c$. The systematic
change of the orbit caused by the emission of gravitational waves will 
be considered in a separate paragraph below.

\subsubsection{Keplerian motion}

Let us consider two point masses $M_1$ and  $M_2$ orbiting each other 
under the force of gravity.
It is well known (see~\cite{L&L_v1}) that this problem
is equivalent to the problem of a single body with mass $\mu$ moving 
in an external gravitational potential. The value of the external
potential is determined by the total mass $M$. 
The total mass $M$ of the system is
\beq{B:M}
	M = M_1 + M_2\,,
\eeq
and the reduced mass $\mu$ is
\beq{B:mu}
	\mu = \frac{M_1M_2}{M}\,.
\eeq
The body $\mu$ moves along an elliptic orbit with eccentricity $e$
and major semiaxis $a$. 
The orbital period $P$ and orbital frequency $\Omega=2\pi/P$
are related with $M$ and $a$ by the 3d Kepler's law
\beq{B:3Kepl}
	\Omega^2 = \myfrac{2\pi}{P}^2 = \frac{GM}{a^3}\,.
\eeq
This relationship is true for any eccentricity $e$.

Individual bodies $M_1$ and $M_2$ move around the 
barycentre of the system in elliptic orbits with the same
eccentricity $e$. Major semiaxes $a_i$ of the two ellipses
are inversely proportional to the masses:
\beq{B:a1}
	\frac{a_1}{a_2} = \frac{M_2}{M_1}\,
\eeq
and satisfy the relationship $a=a_1+a_2$. 
The position vectors of the bodies from the system's barycenter are 
$\vec r_1 = M_2 \vec r /(M_1+M_2)$ and 
$\vec r_2 = - M_1 \vec r /(M_1+M_2)$, where $\vec r = \vec r_1 - \vec r_2$
is the relative position vector. Therefore,  
the velocities of the bodies with respect to the system's
barycentre are related by 
\beq{B:v1}
	-\frac{\vec V_1}{\vec V_2} = \frac{M_2}{M_1}\,,
\eeq
and the relative velocity is $\vec{V}= \vec{V}_1-\vec{V}_2$.

The total conserved energy of the binary system is 
\beq{B:E}
	E
        = \frac{M_1 \vec{V_1}^2}{2} + \frac{M_2 \vec{V_2}^2}{2} - 
\frac{GM_1M_2}{r}
        = \frac{\mu \vec{V}^2}{2} - \frac{GM_1M_2}{r}
        = -\frac{GM_1M_2}{2a}\,
\eeq
where $r$ is the distance between the bodies.
The orbital angular momentum vector is perpendicular to the orbital plane
and can be written as
\beq{B:vecJ}
\vec{J_{orb}} = M_1\vec{V}_1\times\vec{r}_1 +
                 M_2\vec{V}_2\times\vec{r}_2
        = \mu\vec{V}\times\vec{r}. 
\eeq
The absolute value of the orbital angular momentum is
\beq{B:Je}
        |\vec{J_{orb}}| = \mu\sqrt{GMa(1-e^2)\,}.
\eeq

For circular binaries with $e=0$
the distance between orbiting bodies does not depend on time
$$
	r(t,e=0) = a
$$
and is usually referred to as orbital separation.
In this case, velocities of the bodies, as well as their 
relative velocity, are also time-independent: 
\beq{B:Vorb}
      V\equiv  |\vec  V| = \Omega a= \sqrt{GM/a}\,,
\eeq
and the orbital angular momentum becomes
\beq{B:J}
      |\vec J_{orb}|  = \mu Va = \mu \Omega a^2. 
\eeq

\subsubsection{Gravitational radiation from a binary}

The plane of the orbit is determined by the orbital
angular momentum vector $\vec{J}_{orb}$. The
line of sight is defined by a unit vector $\vec{n}$. The
binary inclination angle $i$  is defined by the relation
$\cos i = (\vec{n},\vec{J}_{orb}/J_{orb})$
such that $i=90^\circ$ corresponds to
the system visible edge-on.

Let us start from two point masses $M_1$ and $M_2$
in a circular orbit. 
In the quadrupole approximation
\cite{L&L_v2}, the two polarization amplitudes of GW at a 
distance $r$ from the source are given by
\beq{A:h_+}
h_+=\frac{G^{5/3}}{c^4}
\frac{1}{r}\,2(1+\cos^2i)(\pi f M )^{2/3}\mu\cos(2\pi f t)
\eeq
\beq{A:h_x}
h_\times=\pm \frac{G^{5/3}}{c^4}
\frac{1}{r}\,4 \cos i (\pi f M )^{2/3}\mu\sin(2\pi f t)
\eeq
Here $M$ is the total mass, $\mu$ is the reduced mass,
and $f=\Omega/\pi$ is frequency of the emitted GW
(twice the orbital frequency).
Note that for a fixed distance $r$ and a given frequency $f$, 
the GW amplitudes are fully determined by $\mu M^{2/3} = {\M}^{5/3}$, 
where  the combination
$$
\M\equiv\mu^{3/5}M^{2/5}
$$
is called the ``chirp mass" of the binary. 
After averaging over orbital period (so that
the squares of periodic functions are replaced by 1/2)
and orientations of binary orbital plane, 
one arrives at the averaged (characteristic) GW amplitude
\beq{A:meanh}
h(f,\M,r)=(\langle h_+^2 \rangle+ \langle h_\times^2 \rangle)^{1/2}
= \myfrac{32}{5}^{1/2}\frac{G^{5/3}}{c^4}
\frac{\M^{5/3}}{r}(\pi f)^{2/3}
\label{A:h1}
\eeq

\subsubsection{Energy and angular momentum loss}

In the approximation and under the choice of coordinates that we are 
working with, it is sufficient to use the Landau-Lifshitz gravitational 
pseudotensor \cite{L&L_v2} when calculating the gravitational wave's 
energy and flux. (This calculation can be justified with the help of a 
fully satisfactory gravitational energy-momentum
tensor that can be derived in the field-theoretical formulation of general
relativity \cite{Babak&Grishchuk99}). Energy $dE$ carried by a gravitational wave along its direction
of propagation per area $dA$ per time $dt$ is given by 
\beq{A:flux}
\frac{dE}{dAdt}\equiv F=\frac{c^3}{16\pi G}
	\left[ \myfrac{\partial h_+}{\partial t}^2 +
	\myfrac{\partial h_\times}{\partial t}^2\right]\,. 
\eeq
The energy output $dE/dt$ from a localized source in all directions is given by
the integral
\beq{A:loss}
\frac{dE}{dt}= \int F(\theta,\phi) r^2 d\Omega.
\eeq
Replacing
$$
\myfrac{\partial h_+}{\partial t}^2 + \myfrac{\partial h_\times}{\partial t}^2 = 
4\pi^2 f^2 h^2(\theta, \phi), 
$$
and introducing
$$
h^2 = \frac{1}{4 \pi} \int h^2(\theta, \phi) d \Omega,
$$
we write Eq. (\ref{A:loss}) in the form
\beq{A:loss2}
\frac{dE}{dt}= \frac{c^3}{G} (\pi f)^2 h^2 r^2.
\eeq

Specifically for a binary system in a circular orbit, one finds
the energy loss from the system (sign minus) with the help of 
Eqs.~(\ref{A:loss2}) and (\ref{A:meanh}) :
\beq{A:dEdt}
	\frac{dE}{dt}=-\myfrac{32}{5}\frac{G^{7/3}}{c^5}
        (\M\pi f)^{10/3}\,.  
\eeq
This expression is exactly the same one that can be obtained directly from
the quadrupole formula~\cite{L&L_v2}
\beq{A:GW:dEdt}
\frac{dE}{dt} =
-\frac{32}{5}\frac{G^4}{c^5} \frac{M_1^2M_2^2M}{a^5}\,
\eeq
rewritten using the definition of the chirp mass and the 
Kepler's law. Since energy and angular momentum are continuously
carried away by gravitational radiation, two masses in orbit spiral
toward each other, thus increasing their orbital frequency $\Omega$. 
The GW frequency $f=\Omega/\pi$ and the GW amplitude $h$ 
are also increasing functions of time. The rate of the frequency 
change is \footnote{A signal with such an increasing frequency is 
reminiscent of a chirp of a bird. This explains the
origin of the term ``chirp mass'' for the 
parameter $\M$ which fully determines the GW frequency and 
amplitude behaviour.}
\beq{A:dotf}
\dot f=\myfrac{96}{5}\frac{G^{5/3}}{c^5}\pi^{8/3}\M^{5/3}f^{11/3}. 
\eeq

In spectral representation, the flux of energy  per unit area
per unit frequency interval is given by the right-hand-side of the
expression 
\beq{A:S_h}
\frac{dE}{dA\,df}=\frac{c^3}{G}\frac{\pi f^2}{2}(|\tilde h(f)_+|^2+
|\tilde h(f)_\times|^2)
\equiv \frac{c^3}{G}\frac{\pi f^2}{2} S_h^2(f),
\eeq
where we have introduced the spectral density $S_h^2(f)$ of the gravitational
wave field $h$.
In case of a binary system, the quantity $S_h$ is calculable from
Eqs. (\ref{A:h_+}) and (\ref{A:h_x}): 
\beq{A:S_h:2}
S_h=\frac{G^{5/3}}{c^3}\frac{\pi}{12}\frac{\M^{5/3}}{r^2}
\frac{1}{(\pi f)^{7/3}}\,.
\eeq

\subsubsection{Binary coalescence time}
\label{A:GW_evol}

A binary system in a circular orbit loses energy according to Eq. (\ref{A:dEdt}).
For orbits with non-zero eccentricity $e$, the right-hand-side of 
this formula should be multiplied by the factor
$$
f(e)=(1+\frac{73}{24}e^2+\frac{37}{96}e^4)(1-e^2)^{-7/2}\,
$$
(see \cite{Peters64}).
The initial binary separation $a_0$ decreases and, 
assuming Eq. (\ref{A:GW:dEdt}) is 
always valid, it should vanish in a time
\beq{A:GW:t_0}
t_0=\frac{c^5}{G^3}\frac{5 a_0^4}{256 M^2\mu}=\frac{5c^5}{256}
\frac{(P_0/2\pi)^{8/3}}{(G\M)^{5/3}}
\approx (9.8\times10^6\,\hbox{years})
\myfrac{P_0}{1\,\hbox{h}}^{8/3}\myfrac{\M}{M_\odot}^{-5/3}\,. 
\eeq
As we noted above, gravitational radiation from
the binary depends on the chirp mass $\M$, which can also
be written as $\M\equiv M\eta^{3/5}$, where $\eta$ is the dimensionless 
ratio $\eta=\mu/M$. 
Since $\eta\le 1/4$, one has $\M\simlt 0.435 M$. For example, for
two NS with  equal masses $M_1=M_2=1.4 M_\odot$
the chirp mass is $\M\approx 1.22 M_\odot$. This explains the 
choice of normalization in Eq. (\ref{A:GW:t_0}).

The coalescence time for
an initially eccentric orbit with $e_0 \ne 0$ and separation
$a_0$ is shorter than the coalescence time for a circular orbit 
with the same initial separation $a_0$ \cite{Peters64}:  
\beq{A:GW:t_c}
	t_c(e_0) = t_0~ f(e_0)
\eeq
where the correction factor $f(e_0)$ is 
\beq{A:GW:f(e)}
f(e_0)=\frac{48}{19}\frac{(1-e_0^2)^4}
{e_0^{48/19}\left(1+\frac{121}{304}e_0^2\right)^{3480/2299}}
\int\limits_0^{e_0}\frac{\left(1+\frac{121}{304}e^2\right)^{1181/2299}}
{(1-e^2)^{3/2}}\,e^{29/19}\,de\;.
\eeq
To merge in a time interval shorter than the
Hubble time ($t_H\approx 10$~Gyr),
the binary should have a small enough initial orbital period
$P_0\le P_{cr}(e_0,\M)$ and, accordingly, a
small enough initial semimajor axis $a_0\le a_{cr}(e_0,\M)$.
These critical orbital periods and semi-major axes are shown
as functions of the initial eccentricity $e_0$ in
Fig.~\ref{A:GW:p-e} and \ref{A:GW:a-e}, respectively.
The lines are plotted for three typical sets of masses:
two neutron stars with equal masses (1.4M$_\odot$+1.4M$_\odot$),
a black hole and a  neutron star (10M$_\odot$+1.4M$_\odot$), and
two black holes with equal masses (10M$_\odot$+10M$_\odot$).
Note that in order to get a significantly shorter coalescence time,
the initial binary eccentricity should be $e_0\ga 0.6$.

\if t\figplace
\begin{figure}
\epsfxsize=0.7\textwidth
\centerline{\epsfbox{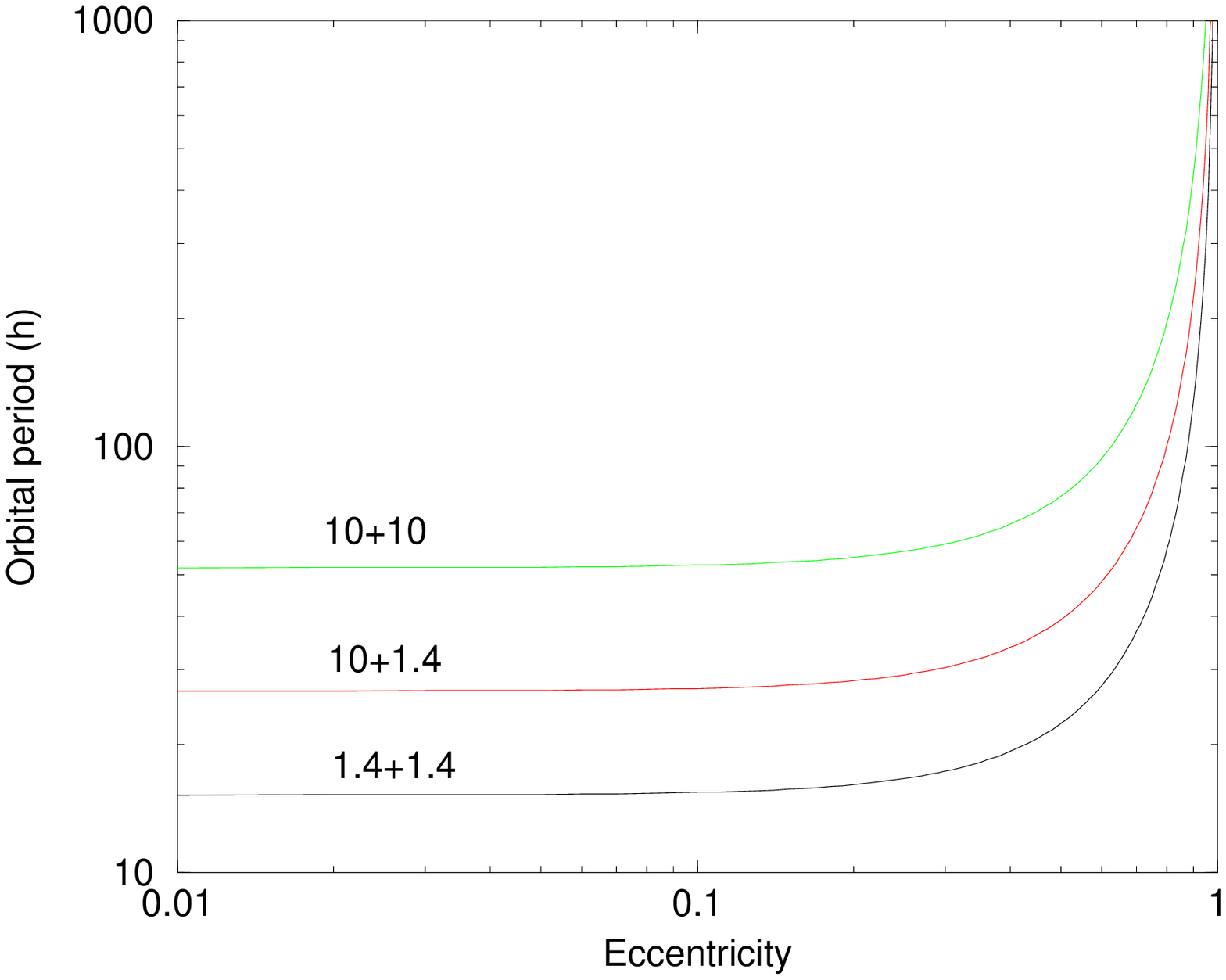}} 
\caption{The maximal initial orbital period (in hours)
of two point masses which will coalesce
due to gravitational wave emission in a time interval shorter
than $10^{10}$ years, as a function of the initial eccentricity $e_0$.
The lines are calculated for 
10M$_\odot$+10M$_\odot$ (BH+BH), 10M$_\odot$+1.4M$_\odot$ (BH+NS),
and 1.4M$_\odot$+1.4M$_\odot$ (NS+NS).}
\label{A:GW:p-e}
\bigskip\bigskip
\epsfxsize=0.7\textwidth
\centerline{\epsfbox{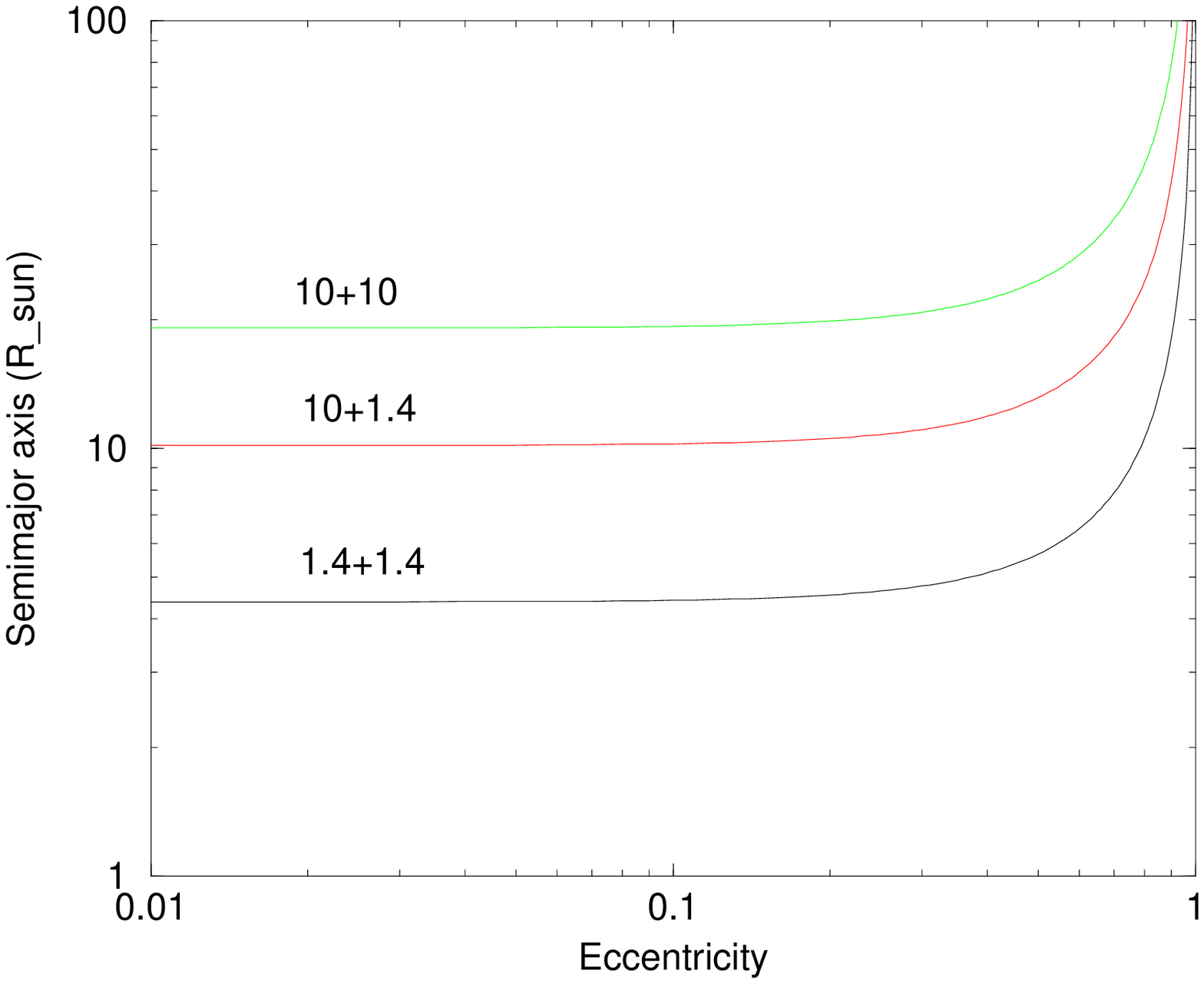}} 
\caption{The maximal initial semimajor axis (in R$_\odot$)
of two point masses which will coalesce
due to gravitational wave emission in a time interval shorter
than $10^{10}$ years, as a function of the initial eccentricity $e_0$.
The lines are calculated for 
10M$_\odot$+10M$_\odot$ (BH+BH), 10M$_\odot$+1.4M$_\odot$ (BH+NS),
and 1.4M$_\odot$+1.4M$_\odot$ (NS+NS).}
\label{A:GW:a-e}
\end{figure}
\fi

\subsection{Mass transfer modes and mass loss in binary systems}
\label{A:mass_transf}
\label{sec:appB}

The gravitational wave emission is the sole factor responsible for 
the change of orbital parameters of a pair of compact (degenerate) stars. 
However, at the early stages of binary evolution, it is the 
mass transfer between the components and the loss of matter and its 
orbital momentum that play dominant dynamical role. 
Strictly speaking, these processes should be treated hydrodynamically
and they require complicated numerical calculations. However,
binary evolution can also be described semi-qualitatively,
using a simplified description in terms of point-like bodies. 
The change of their integrated physical quantities, such as masses,
orbital angular momentum, etc. governs the evolution of the orbit.
This description turns out to be successful in reproducing the
results of more rigorous numerical calculations
(see e.g. \cite{Heuvel94_int_bin} for a review). In this approach, 
the key role is allocated to the total orbital angular momentum $J_{orb}$
of the binary.

Let star 2 lose matter at a rate $\dot M_2<0$ and let $\beta$
$(0\le\beta\le1)$ be a fraction of the ejected matter which leaves
the system (the rest falls on the first star),
i.e. $\dot M_1=-(1-\beta)\dot M_2\ge 0$.
Consider circular orbits with orbital angular momentum given by
(\ref{B:J}). Differentiate both parts of Eq. (\ref{B:J}) by time $t$ 
and exclude $d\Omega/dt$ with the help of the third Kepler's
law (\ref{B:3Kepl}). This gives us the rate of change of the
orbital separation:
\beq{A:dotaa}
\frac{\dot a}{a}=-2\left[1+(\beta-1)\frac{M_2}{M_1}-\frac{\beta}{2}
\frac{M_2}{M}\right]\frac{\dot M_2}{M_2}+2\frac{\dot J_{orb}}{J_{orb}}
\eeq
One defines the mass transfer as conservative if both  
$\beta=0$ and $\dot J_{orb}=0$. The mass transfer is called 
non-conservative if at least one of these conditions is violated.

For massive binaries, which we are mostly interested in, it is important
to distinguish some specific cases (modes) of mass transfer.
They amount to: (1) conservative accretion mode,
(2) non-conservative Jeans's mode (or fast wind mode),
(3) non-conservative isotropic re-emission,
(4) sudden mass loss from one of the components
during supernova explosion, and 
(5) common-envelope stage. Separately we consider
orbit evolution due to (6) gravitational wave emission,
which becomes the main factor for short-period compact binaries.
For non-conservative
modes, one can also introduce some subcases, such, for example, as 
a ring-like mode in which a circumbinary ring of expelled matter is 
being formed (e.g. \cite{Soberm&97}). Here, we will not go into the details
of such subcases.

\subsubsection{Conservative accretion.}

In the case of conservative accretion, matter from $M_2$ is fully 
deposited to $M_1$. The transfer process preserves the
total mass $M=\const$ ($\beta=0$)
and the orbital angular momentum $J_{orb}=\const$ of the system. 
It follows from Eq. (\ref{A:dotaa}) that 
$$
M_1M_2\sqrt{a}=\const\,,
$$
so that the initial and final binary separations are related as
\beq{A:conserv}
\frac{a_f}{a_i}=\myfrac{M_{1i}\,M_{2i}}{M_{1f}\,M_{2f}}^2\,.
\eeq            
The well-known ``rule of thumb'' for this case says that
the orbit shrinks when the more massive component
loses matter, and the orbit widens in the opposite situation. 
During such a mass exchange,
the orbital separation passes through a minimum, 
if the masses become equal in course of the mass transfer.

\subsubsection{The Jeans (fast wind) mode.}

In this mode the ejected matter completely escapes from the system,
that is, $\beta =1$.
The escape of matter can take place either in a
spherically symmetric way or in the form of bipolar jets moving
from the system at high velocity.
In both cases, matter carries away some amount of the total 
orbital momentum
proportional to the orbital angular momentum $J_2 = (M_1/M) J_{orb}$ 
of the mass loosing star (we
neglect a possible proper rotation of the star,
see \cite{Heuvel93_in_accr_driv_303}). 
For the loss of orbital momentum $\dot{J}_{orb}$ it is reasonable to take 
\beq{A:jspecif}
\dot J_{orb} = \frac{\dot M_2}{M_2} J_2\,.
\eeq
In the case $\beta =1$, Eq. (\ref{A:dotaa}) can be written as
\beq{A:dotaa2} 
\frac{(\Omega a^2)\,\dot{}}{\Omega a^2} =\frac{\dot{J}_{orb}}{J_{orb}} -
\frac{M_1\dot{M}_2}{M M_2}\,.
\eeq
Then Eq. (\ref{A:dotaa2}) in conjunction with Eq. (\ref{A:jspecif}) 
give $\Omega a^2 = \const$, that is, 
$\sqrt{GaM}=\const$. Thus, as a result of such a mass loss,
the change in orbital separation is 
\beq{A:Jeans}
\frac{a_f}{a_i}=\frac{M_{i}}{M_{f}}\,.
\eeq
Since the total mass decreases, the orbit always widens.

\subsubsection{Isotropic re-emission.}

The matter lost by star 2 can first accrete to star 1, and then,
a fraction $\beta$ of the accreted matter, can be expelled from the system. 
This happens when a massive star transfers matter to a compact star on 
the thermal timescale ($<10^6$ years). The accretion luminosity 
may exceed the Eddington luminosity limit, and 
the radiation pressure pushes the infalling matter away from the system,
in a manner similar to the spectacular example of the SS~433 
binary system. In this mode of the mass transfer, the binary orbital 
momentum carried away 
by the expelled matter is determined by the orbital momentum of
the accreting star $M_1$, rather than by the orbital momentum of the 
mass-loosing star $M_2$. The orbital momentum loss can be written as  
\beq{A:jspecif_1}
\dot{J}_{orb} = \beta \frac{\dot M_2}{M_1} J_1\,,
\eeq
where $J_1=(M_2/M)J_{orb}$ is the orbital momentum of the star $M_1$.
In the limiting case when all the mass attracted
by $M_1$ is fully pushed away, $\beta=1$, Eq. (\ref{A:jspecif_1})
simplifies to 
\beq{A:jreemiss}
\frac{\dot J_{orb}}{J_{orb}} = \frac{\dot {M_2} M_2}{M_1 M}\,.
\eeq
After substitution of this formula into Eq.~(\ref{A:dotaa})
and integration over time, one arrives at
\beq{A:re-em}
\frac{a_f}{a_i}= \frac{M_i}{M_f}\myfrac{M_{2i}}{M_{2f}}^2
\exp{\left(-2\,\frac{M_{2i}-M_{2f}}{M_1}\right)}\,.
\eeq
The exponential term makes this mode of the mass transfer
very sensitive to the components mass ratio.
If $M_1/M_2\ll 1$, the separation $a$ between the stars
may decrease so greatly that the approximation of point masses becomes
invalid. The tidal orbital instability
(Darwin instability) may set in, and the compact star
may start spiraling toward the companion star center
(the common envelope stage; see section~\ref{A:CE} below). 

\subsubsection{Supernova explosion.}
\label{A:SN}

Supernova explosion in a binary system occurs on a timescale
much shorter than the orbital period, so
the loss of mass is practically instantaneous.
This case can be treated analytically
(e.g. \cite{Blaauw61,Flann&Heuv75,Yamaoka&93}). 
In general, the loss of matter and radiation is aspherical, so that  
the remnant of the supernova
explosion (neutron star or black hole) acquires 
some recoil velocity called kick velocity $\vec{w}$.
In a binary, kick velocity should be added to
the orbital velocity of the pre-supernova star.

The usual treatment proceeds as follows.
Let us consider a pre--SN binary with initial masses $M_1$
and $M_2$. The stars move in a circular orbit with orbital separation
$a_i$ and relative velocity $\vec{V}_i$. The star $M_1$ explodes 
leaving a compact remnant of mass $M_c$. The total mass of the binary
decreases by the amount $\Delta M = M_1 - M_c$. The compact star acquires 
some kick velocity $\vec{w}$. Unless the binary is disrupted, it will
end up in a new orbit with eccentricity $e$, major semiaxis $a_f$,
and the angle $\theta$ between the orbital planes before and 
after the explosion. In general, the new barycenter will also 
receive some velocity, but we neglect this motion. 
The goal is to evaluate the parameters $a_f$, $e$, and $\theta$.
   
It is convenient to work in an instantaneous 
reference frame centered on $M_2$ right at the time of explosion. 
The x--axis is the line from $M_2$ to $M_1$, the y--axis points in
the direction of $\vec{V}_i$, and the z--axis is perpendicular to the 
orbital plane. In this frame, the pre--SN relative velocity is
$\vec{V}_i = (0, V_i, 0)$, where  
$V_i=\sqrt{G(M_1+M_2)/a_i}$ (Eq. (\ref{B:Vorb}).
The initial total orbital momentum is $\vec{J}_i = \mu_i a_i (0, 0, -V_i)$.
The explosion is considered to be instantaneous. 
Right after the explosion, the position vector of the exploded 
star $M_1$ has not changed: $\vec{r}= (a_i, 0, 0)$. However,
other quantities has changed: $\vec{V}_f= (w_x, V_i+w_y, w_z)$
and $\vec{J}_f = \mu_f a_i (0, w_z, -(V_i+w_y))$, where 
$\vec{w} = (w_x, w_y, w_z)$ is the kick velocity and
$\mu_f= M_cM_2/(M_c+M_2)$ is the reduced mass of the system after explosion.
The parameters $a_f$ and $e$ are being found from equating the total
energy and the absolute value of orbital momentum at the initial
circular orbit to their expressions at the resulting elliptical orbit
(see Eqs. (\ref{B:E}), (\ref{B:J}), and (\ref{B:Je})):   
\beq{A:SN:E}
\mu_f\frac{V_f^2}{2}-\frac{GM_cM_2}{a_i}=-\frac{GM_cM_2}{2a_f},
\eeq
\beq{A:SN:J}
\mu_f a_i \sqrt{w_z^2 +(V_i+w_y)^2}=\mu_f \sqrt{G(M_c+M_2)a_f(1-e^2)}.
\eeq
For the resulting $a_f$ and $e$ one finds 
\beq{A:SN:afai}
\frac{a_f}{a_i} = \left(2-\chi \left[\frac{w_x^2+w_z^2+(V_i+w_y)^2}
{V_i^2}\right]\right)^{-1}
\eeq
and
\beq{A:SN:ecc}
1-e^2 = \chi \frac{a_i}{a_f}\left[\frac{w_z^2+(V_i+w_y)^2}{V_i^2}\right]
\eeq
where $\chi \equiv (M_1+M_2)/(M_c+M_2) \ge 1$. The angle $\theta$ is 
defined by
$$
\cos \theta = \frac{\vec{J}_f \cdot \vec{J}_i}{|\vec{J}_f|~|\vec{J}_i|},
$$
which results in
\beq{A:SN:theta}
\cos\theta = \frac{V_i+w_y}{\sqrt{w_z^2+(V_i+w_y)^2}}.  
\eeq

The condition of disruption of the binary system depends on the absolute
value $V_f$ of the final velocity, and on the parameter $\chi$.
The binary disrupts if its total energy defined by the left-hand-side
of Eq. (\ref{A:SN:E}) becomes non-negative or, equivalently, if its eccentricity
defined by Eq. (\ref{A:SN:ecc}) becomes $e \ge 1$. From either of these
requirements one derives the condition of disruption:
\beq{A:disrupt}
\frac{V_f}{V_i} \ge \sqrt{\frac{2}{\chi}}.
\eeq
The system remains bound if the opposite inequality is satisfied. 
Eq. (\ref{A:disrupt}) can also be written in terms of the escape (parabolic)
velocity $V_e$ defined by the requirement
$$
\mu_f\frac{V_e^2}{2}-\frac{GM_cM_2}{a_i}= 0. 
$$
Since $\chi = M/(M - \Delta M)$ and $V_e^2 = 2 G(M - \Delta M)/ a_i =
2V_i^2 / \chi$, one
can write Eq. (\ref{A:disrupt}) in the form
\beq{A:disrupt2}
V_f \ge V_e.
\eeq
The condition of disruption simplifies 
in the case of a spherically symmetric SN explosion, that is, when there is
no kick velocity, $\vec{w}= 0$, and, therefore, $V_f = V_i$. In this case,
Eq. (\ref{A:disrupt}) reads $\chi \ge 2$, which is equivalent to
$\Delta M \ge M/2$. Thus, the system unbinds if more
than a half of mass of the binary is lost. In other words, 
the resulting eccentricity 
\beq{A:SN:symm-ecc}
e=\frac{M_1-M_c}{M_c+M_2}\,
\eeq
following from (\ref{A:SN:afai}), (\ref{A:SN:ecc}), and $\vec{w} = 0$
becomes larger than 1, if $\Delta M > M/2$. 
So far, we have considered
an originally circular orbit. If the pre--SN star moves in an 
originally eccentric orbit, the condition of disruption of the system 
under symmetric explosion reads
$$
	\Delta M = M_1-M_c > \frac{1}{2} \frac{r}{a_i}\,,
$$
where $r$ is the distance between the components at the moment of explosion.

\subsubsection{Common envelope stage.}
\label{A:CE}

This is a very important stage in
binary evolution. A possibility of this stage was
first suggested in \cite{Pacz76}. Generally, 
it occurs in binary systems where the mass transfer from the
mass-losing star is high, and the companion
cannot accrete all the matter. The common envelope stage appears 
unavoidable on the observational grounds.
The evidence for a dramatic orbital
angular momentum decrease at some preceding
evolutionary stage follows from observations
of certain types of close binary stars. They include
cataclysmic variables, in which a white dwarf
accretes matter from a small red dwarf
main-sequence companion, planetary nebulae with double cores,
low-mass X-ray binaries and X-ray transients (neutron stars and
black holes accreting matter from low-mass main-sequence dwarfs).
The radii of progenitors of compact stars in these binaries
typically should have been
100--1000 solar radii, that is, much larger than the observed
binary separations. This testifies to some
dramatic reduction of the orbital momentum at earlier stages
of evolution and eventual removal of the common envelope.

There is no exact criterion for the formation of a  
common envelope. However, a high mass overflow onto a compact star 
from a normal star is always expected when the normal star 
leaves the main-sequence and develops a convective envelope. 
The critical mass ratio for the unstable Roche lobe overflow
depends on specifics of the stars, but is close to 1.
Another way for the formation of a common envelope is 
direct penetration of a compact star into the dense outer
layers of the companion. This can happen as a result of the 
Darwin tidal orbital instability in
binaries \cite{Counsel73,Bagot96}, 
or when a compact remnant of supernova explosion with appropriately
directed kick velocity finds itself in an elliptic orbit whose  
minimum periastron distance $a_f(1-e)$ is  
smaller than the stellar radius of the companion. 

A simplified treatment of the common envelope stage is usually 
done as follows \cite{Webbink84}. The orbital evolution of
the compact star $M_c$ inside the envelope of the normal star $M_1$ is
driven by the dynamical friction drag. This leads to a gradual
spiral-in process of the compact star. The released orbital energy
$\Delta E_{orb}$, or a fraction of it, can become numerically
equal to the binding energy $E_{bind}$ of the envelope with the rest of
the binary system. It is usually assumed that this equality 
provides a condition for expulsion of the common envelope. 
What remains of the normal star $M_1$ is its stellar core $M_{sc}$. 
The final orbital parameters are being derived from the condition 
$E_{bind}=\alpha_{CE}\Delta E_{orb}$, where $\alpha_{CE}$
is the efficiency parameters less or equal to one. This condition reads
\beq{A:CE:eq}
\frac{GM_1(M_1-M_{sc})}{\lambda R_L}=\alpha_{CE}\left(
\frac{GM_cM_{sc}}{2a_f}-\frac{GM_1M_c}{2a_i}\right), 
\eeq
where $a_i$ and $a_f$ are the initial and the final
orbital separations, 
$\lambda$ is a numerical coefficient of order 1.
$R_L$ is the Roche lobe radius of the
normal star approximated by \cite{Eggl83}
\beq{A:Roche}
\frac{R_L}{a_i}= \frac{0.49}{0.6+q^{2/3}\ln (1+q^{-1/3})}
\eeq
and $q\equiv M_1/M_2$. From Eq. (\ref{A:CE:eq})  
one derives 
\beq{A:CE:afai}
\frac{a_f}{a_i}=\frac{M_{sc}}{M_1}\left(1+\frac{2a_i}
{\lambda\alpha_{CE}R_L} \frac{M_1-M_{sc}}{M_c}\right)^{-1}
\la \frac{M_{sc}}{M_1}\frac{M_c}{\Delta M}\,,
\eeq
where $\Delta M=M_1-M_{sc}$.
Recent studies (e.g. \cite{Heuvel94_CE,Rasio&Liv96})
show that $\alpha_{CE}\lambda$ falls within the range from 0.5 to 2.
The mass $M_{sc}$ of a helium core of a massive star 
is (see \cite{Iben&Tut85}) 
\beq{A:MHe}
M_{\rm He}\approx 0.073 (M_1/M_\odot)^{1.42}, 
\eeq
so the orbital separation during the common envelope stage may 
decrease as much as by factor 30--60. 


\subsection{Post-Newtonian expansions of GW flux and energy}
\label{appx:pn}
\label{sec:appC}

The gravitational wave flux escaping a system of two compact stars 
in quasi-circular orbit when their orbital frequency is 
$f_{\rm orb}$ (the dominant GW frequency being $f=2f_{\rm orb}$), is 
given by \cite{S:bdiww,S2:bdi,S2:ww,S2:biww}
\begin{eqnarray}
{\cal F}(v)
& = & \frac{32\eta^2 v^{10}}{5} \Biggl [ 1
      - \left(\frac{1247}{336} + \frac{35\eta}{12}\right) v^2
      + 4\pi v^3 \nonumber \\
      & - & \left( \frac{44711}{9072} + \frac{9271\eta}{504} + \frac{65\eta^2}{18}
        \right ) v^4 
    - \left(\frac{8191}{672} + \frac{535\eta}{24}\right) \pi v^5 \Biggr ]
\end{eqnarray}
where $v= (\pi M f)^{1/3}$ is the relative velocity of the two stars and
$\eta=M_1M_2/M^2$ is the symmetric mass ratio.
The $\eta$-parameter takes a maximum values of 1/4 when the two
masses are equal. It characterises
the extent to which two-body effects in the system are important.
The relativistic energy $E(v)$ of the system is given by 
\begin{eqnarray}
E(v) = -\frac{\eta v^2}{2} \left [ 1 - \left ( \frac{9+\eta}{12} \right ) v^2 
       - \left ( \frac{81-57\eta+\eta^2}{24} \right ) v^4 \right ].
\end{eqnarray}

Solving the differential equations in Eq.~(\ref{eq:ode representation}) for
time and phase in terms of $v$ gives the following equations: 
\begin{eqnarray}
t(v) & = & -\frac{5M}{256 \eta v^8} \Biggl [ 1 + 
      \left ( \frac{743}{252} + \frac{11\eta}{3} \right ) v^2 
      - \frac{32\pi}{5} v^3 \nonumber \\
      & + & \left ( \frac{3058673}{508032} + \frac{5429\eta}{504} + \frac{617\eta^2}{72} \right ) v^4 
      - \left(\frac{7729}{252}+ \eta\right)\pi v^5 \Biggr ],
\end{eqnarray}
\begin{eqnarray}
\phi(v) & = & -\frac{1}{16\eta v^5} \Biggl [ 1 + 
      \left ( \frac{3715}{1008}+\frac{55\eta}{12} \right ) v^2
      - 10 \pi v^3 
      + \left ( \frac{15293365}{1016064} + \frac{27145\eta}{1008 } + 
      \frac{3085\eta^2}{144} \right ) v^4 \nonumber \\
      & + & \left (\frac{38645}{672} + \frac{15\eta}{8 } \right ) \pi 
\ln \left ( \frac{v}{v_{\rm lso}} \right ) v^5 \Biggr ].
\end{eqnarray}
One can invert the first of the equations above to
express $v$ in terms of a post-Newtonian expansion in $t$ and then 
use the resulting expression in the second equation to arrive at
an explicit phasing formula. Introducing a new time parameter $\theta$
defined by $\theta=[\eta (t_{\rm lso}-t)/(5M)]^{-1/8},$ 
where $t_{\rm lso}$ is a reference time taken to be 
the time at which the GW frequency is equal to
twice the orbital frequency at the last stable circular orbit, we find
\begin{eqnarray}
\phi(\theta) & = & -\frac{2}{\eta \theta^5} \Biggl [ 1 + 
      \left ( \frac{3715}{8064}+\frac{55\eta}{96} \right ) \theta^2 
       - \frac{3\pi}{4} \theta^3
       + \left ( \frac{9275495}{14450688}+\frac{284875\eta}{258048 } + 
\frac{1855\eta^2}{2048 } \right ) \theta^4 \nonumber \\
       & + & \left (\frac {38645}{21504} + \frac{15\eta}{256 } \right ) \pi 
\ln \left ( \frac {\theta}{\theta_{\rm lso}} \right ) \theta^5 \Biggr ].
\end{eqnarray}

\newpage



%
\if t\figplace\relax\else

\clearpage      
\begin{figure}[H]
\begin{center}
\epsfxsize=0.5\hsize
\fbox{\epsfbox{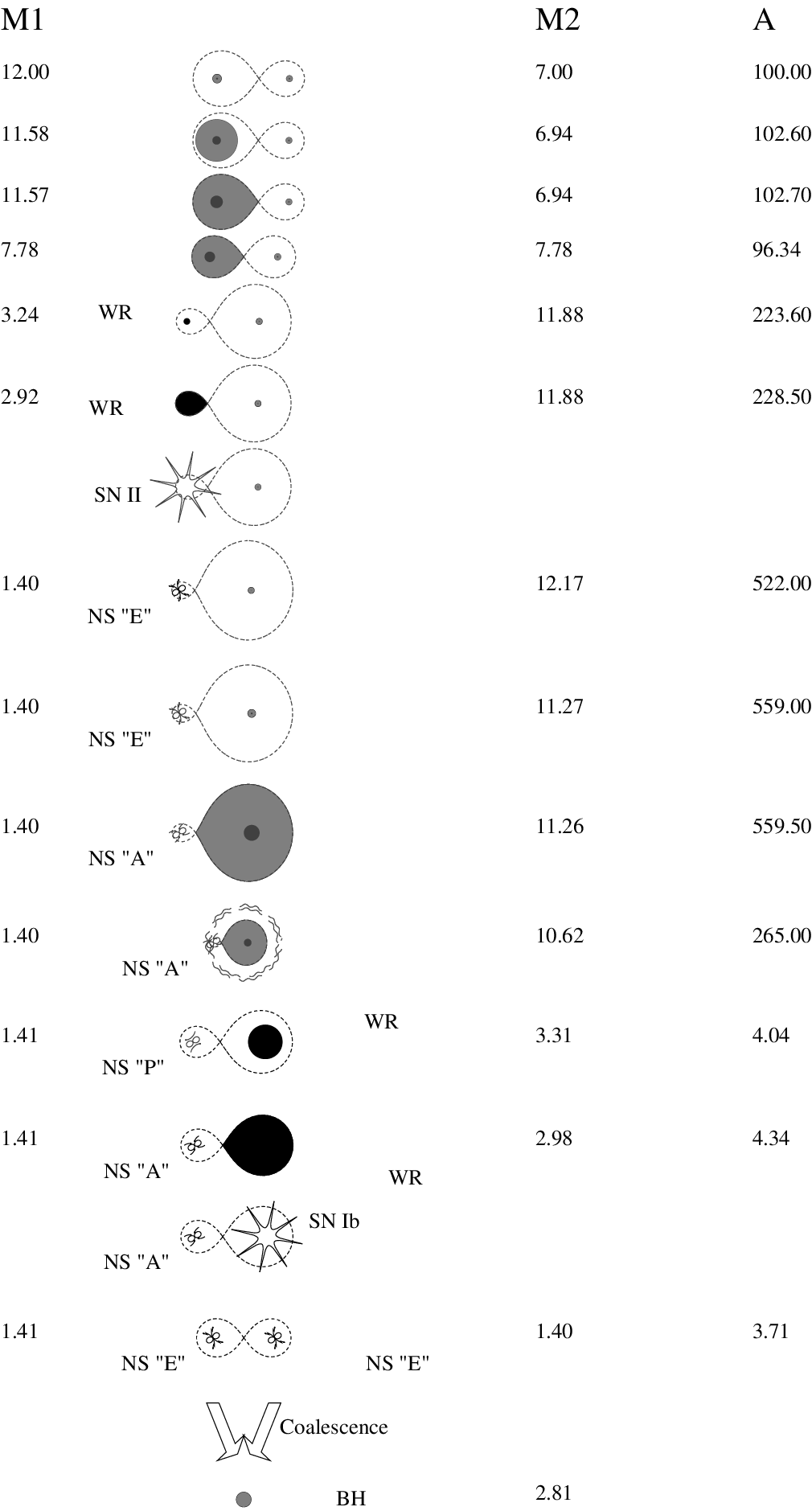}} 
\caption{Evolutionary track of a massive binary star leading
to the formation and coalescence of two NS.}
\label{f:track}
\end{center}
\end{figure}

\clearpage      
\begin{figure}[H]
\begin{center}
\epsfxsize=0.5\hsize
\fbox{\epsfbox{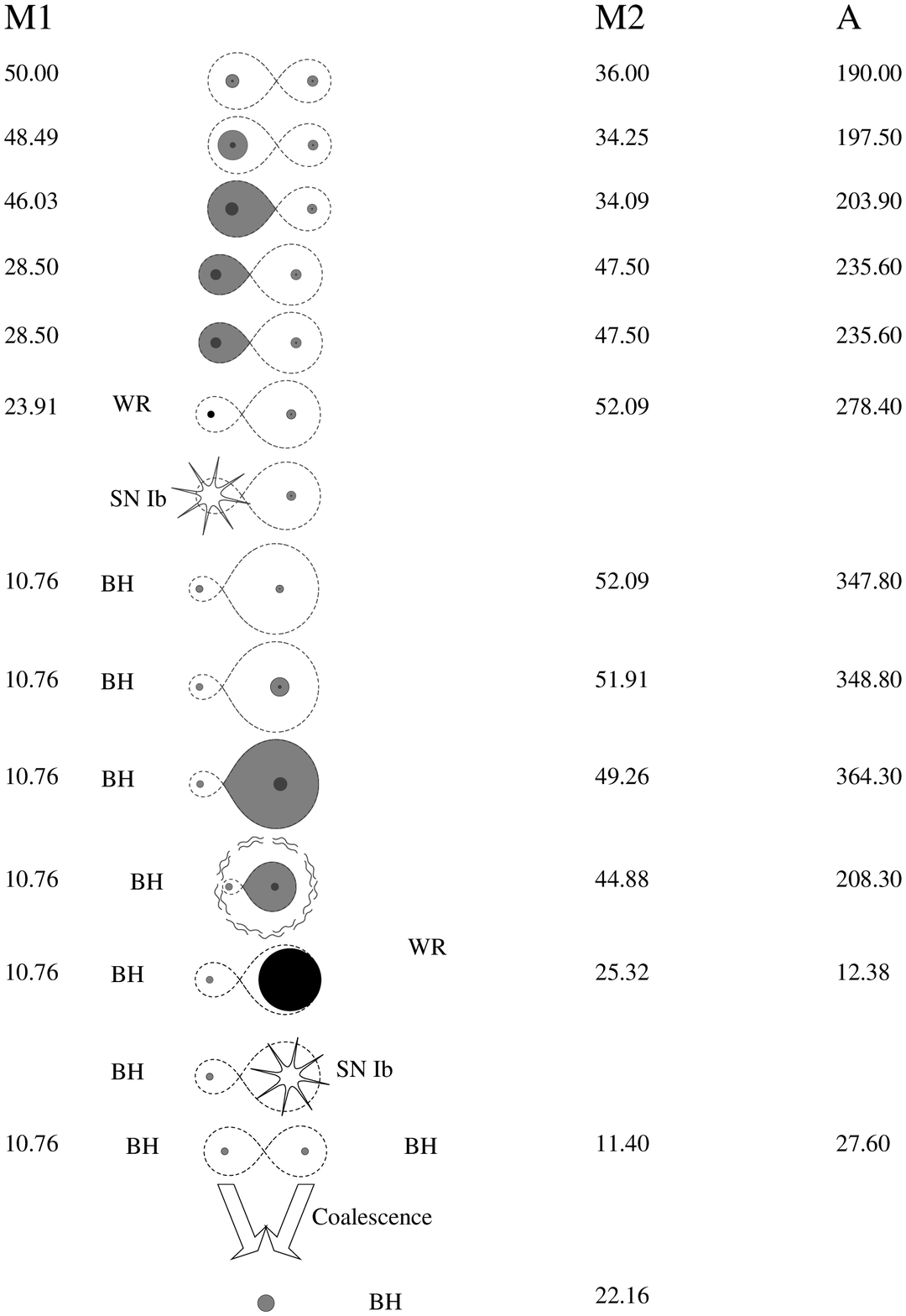}} 
\caption{Evolutionary track of a massive binary star leading
to the formation and coalescence of two BH.
The low stellar wind mass loss scenario is used.}
\end{center}\label{f:bhbh}
\end{figure}

\clearpage      
\begin{figure}[H]
\begin{center}
\epsfxsize=0.5\hsize
{\epsfbox{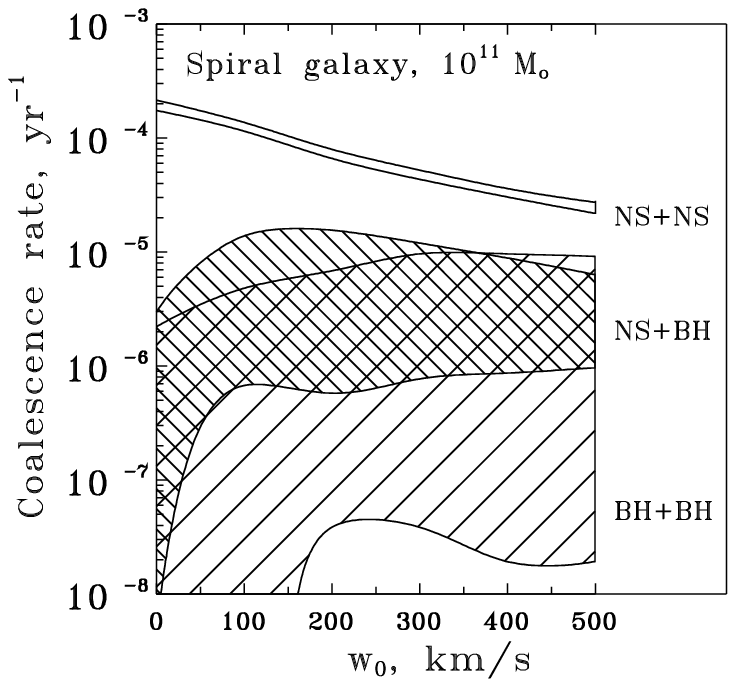}} 
\end{center}
\caption{NS+NS, BH+NS, and BH+BH merging rates in a
$10^{11}$~M$_\odot$ galaxy 
as functions of the kick velocity parameter $w_0$ for
Lyne-Lorimer kick velocity distribution (\protect\ref{bin:LL}).
Star formation rate in the galaxy is assumed constant.
BH formation parameters are $M_*=15$--50~M$_\odot$, $k_{BH}=0.25$.}
\label{f:grate}
\end{figure}

\clearpage      
\begin{figure}[H]
\begin{center}
\epsfxsize=0.7\hsize
{\epsfbox{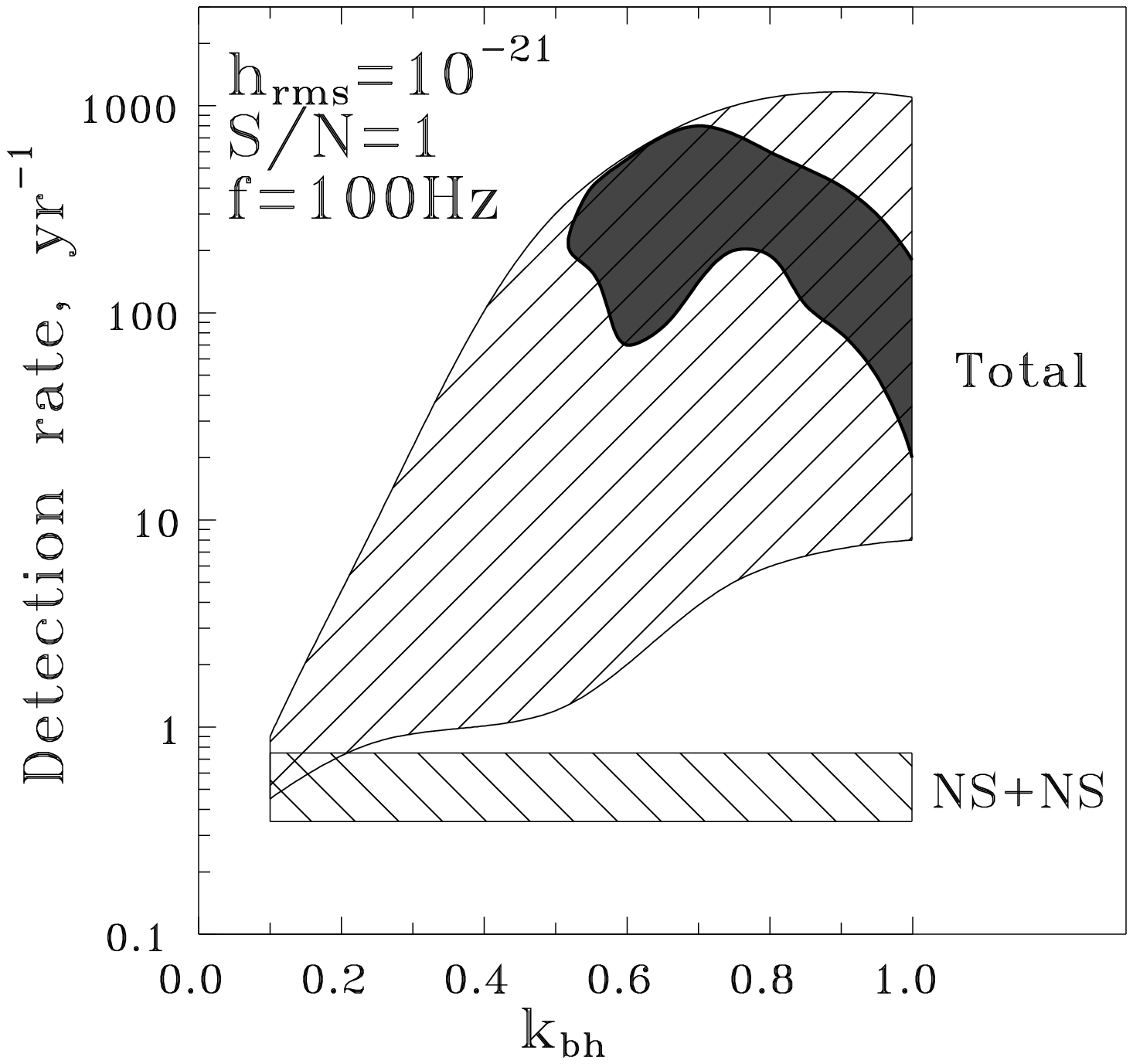}} 
\end{center}
\caption{The detection rate $\D$ of GW events in a detector
with the sensitivity $h_{\rm rms}=10^{-21}$ at frequency 100 Hz and the
signal-to-noise level $S/N=1$, as a function of BH formation parameter
$k_{BH}$. The calculations were performed for the Lyne--Lorimer kick velocity
distribution with $w_0=400$~km/s. The spread of $\D$ at fixed $k_{BH}$
is due to variation of the parameter $M_{cr}$ 
from 15 $M_\odot$ to 50 $M_\odot$. The bottom rectangular area is drawn
for binary NS coalescences.
Their rate is independent 
of $k_{BH}$ (as it should be) and predicts a couple of events per 1--3 years
at this level.   
The total detection rate can be 2--3 orders of magnitude higher 
then the NS+NS rate
at the expense of 
BH+BH and BH+NS coalescences. The hatched area
shows the region of the most probable parameters for the low stellar wind
mass loss scenario. Inside this region
the outcomes of calculations are in agreement with 
the upper limit on the galactic number of binary 
BH with radiopulsars (less than 1 per 700 single pulsars)
and the galactic number of BH candidates similar to Cyg~X--1 (from 1 to 10).}
\label{f:sheja}
\end{figure}

\clearpage      
\begin{figure}[H]
\hbox to \textwidth{ 
\hbox to 0.5\textwidth{
\epsfxsize=0.5\textwidth
\epsfbox{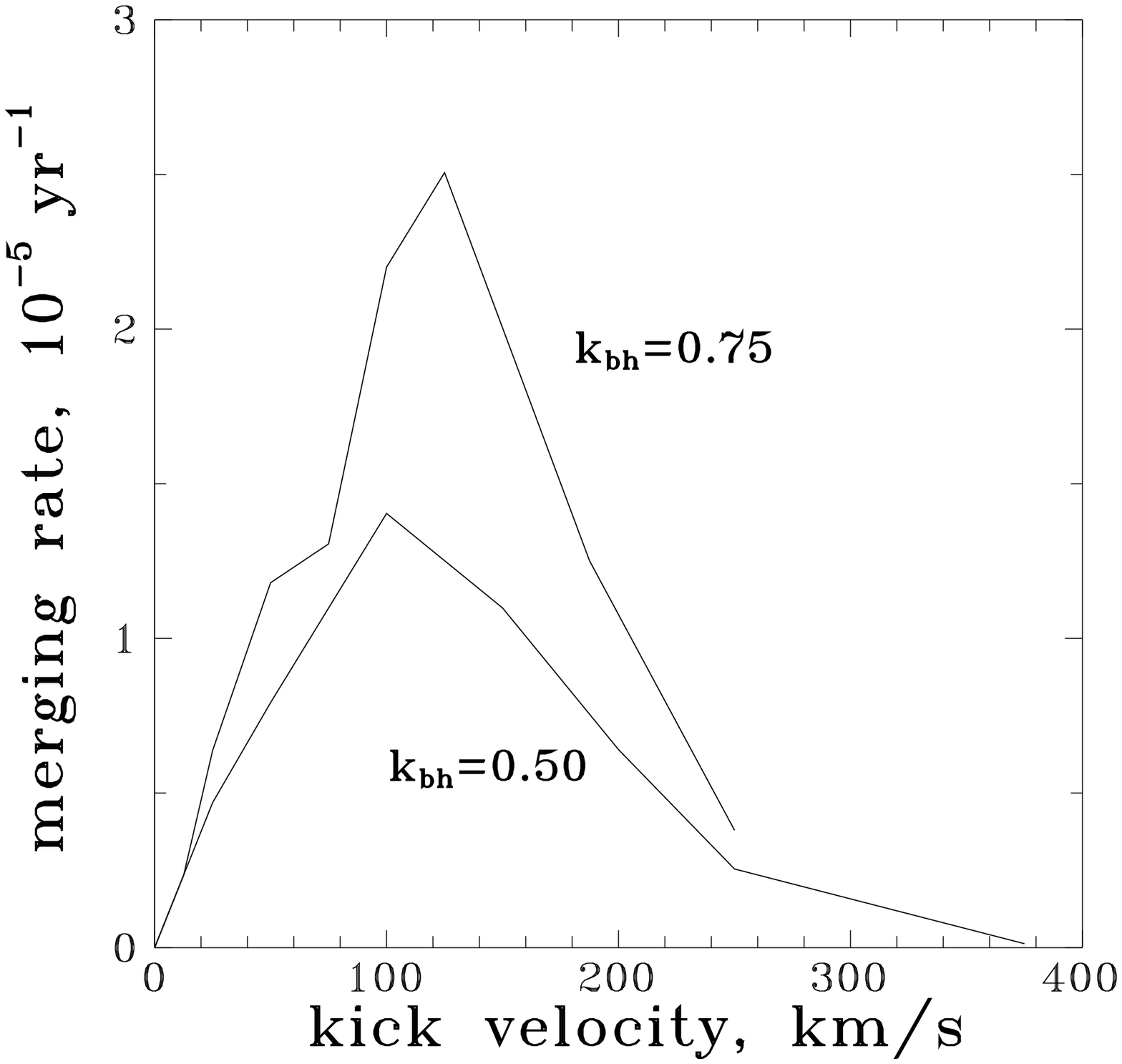} 
}
\hss
\hbox to 0.5\textwidth{
\epsfxsize=0.5\textwidth
\epsfbox{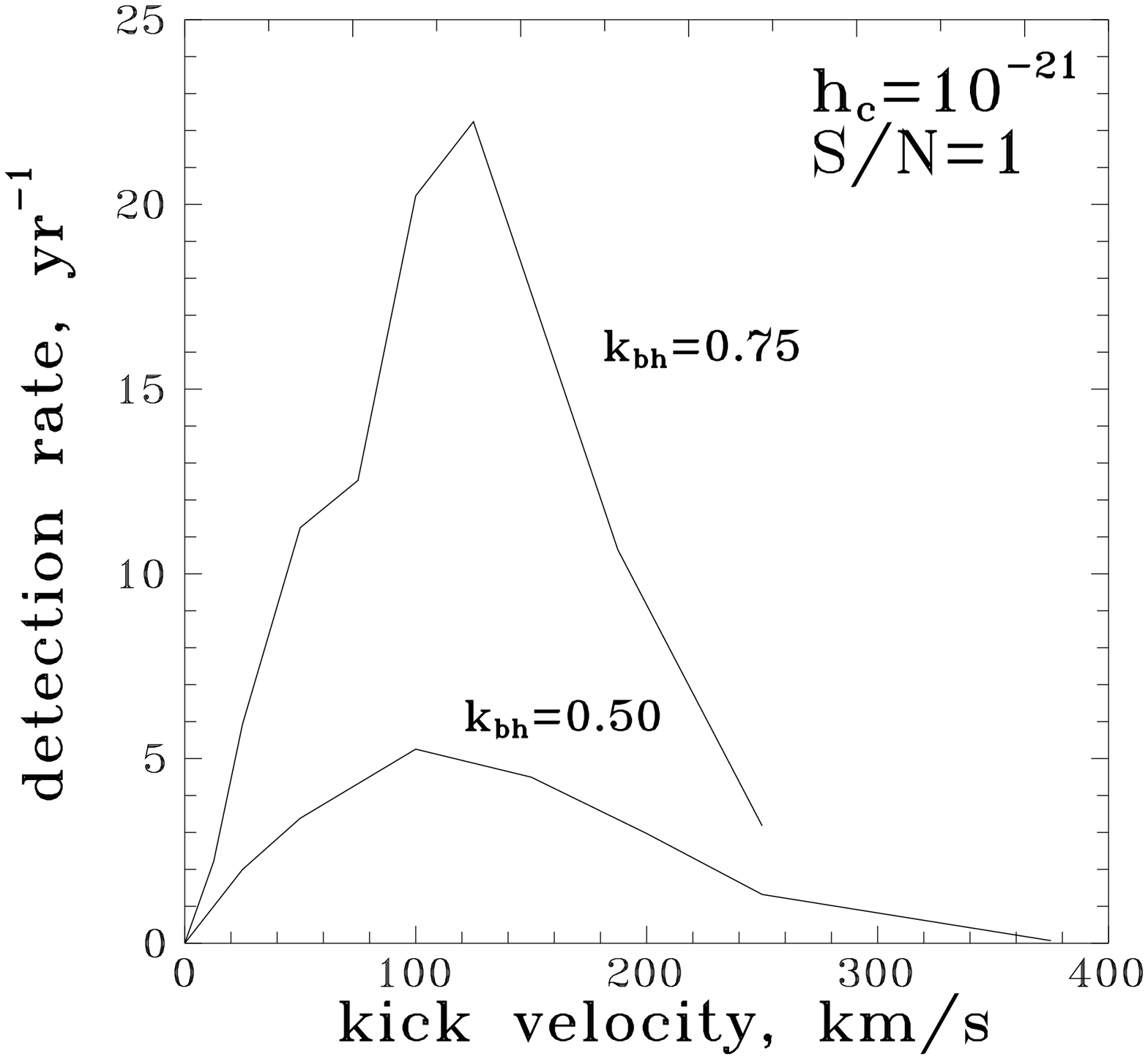} 
}
}
\caption{New scenario --- superhigh wind.
Left: BH+BH merging rate calculated for a $10^{11} M_\odot$
galaxy with a constant star formation rate, as a function of the kick 
velocity during BH formation with $M_{cr}=35 M_\odot$,
for $k_{BH}=0.5$ and $0.75$. 
Right: Detection rate of 
BH+BH mergings by the initial laser
interferometers ($h_{\rm rms}=10^{-21}$ at $f=100$~Hz), as a function of 
the kick velocity during BH formation.}
\label{f:cls-dlt}
\label{f:det-dlt}
\end{figure}

\clearpage
\begin{figure}[H]
\epsfxsize=0.8\textwidth
\centerline{\epsfbox{f06.ai}} 
\caption{Parametric amplification. a) variation of the
	length of the pendulum, b) increased amplitude of
        oscillations.}
\label{LPG:fig1}
\end{figure}

\clearpage
\begin{figure}[H]
\epsfxsize=0.8\textwidth
\centerline{\epsfbox{f07.ai}} 
\caption{Some quantum states of a harmonic oscillator.}
\label{LPG:fig2}
\end{figure}

\clearpage
\begin{figure}[H]
\epsfxsize=0.8\textwidth
\centerline{\epsfbox{f08.ai}} 
\caption{Effective potential $U(\eta)$.}
\label{LPG:fig3}
\end{figure}

\clearpage
\begin{figure}[H]
\epsfxsize=0.8\textwidth
\centerline{\epsfbox{f09.ai}} 
\caption{Scale factor $a(\eta)$.}
\label{LPG:fig4}
\end{figure}

\clearpage      
\begin{figure}[H]
\epsfxsize=0.8\textwidth
\centerline{\epsfbox{f10.ai}} 
\caption{Function $a^\prime/a$ for the scale factor from
	Fig.~\protect\ref{LPG:fig4}.}
\label{LPG:fig5}
\end{figure}

\clearpage      
\begin{figure}[H]
\epsfxsize=0.8\textwidth
\centerline{\epsfbox{f11.ai}} 
\caption{Amplitudes and spectral slopes of $h(n)$ are
	determined by different parts of the barrier $a^\prime/a$.}
\label{LPG:fig6}
\end{figure}

\clearpage      
\begin{figure}[H]
\epsfxsize=0.8\textwidth
\centerline{\epsfbox{f12.ai}} 
\caption{Expected spectrum $h(\nu)$ for the case $\beta=-1.9$.}
\label{LPG:fig7}
\end{figure}

\clearpage      
\begin{figure}[H]
\epsfxsize=0.8\textwidth
\centerline{\epsfbox{f13.ai}} 
\caption{Expected spectrum $\beta=-1.9$ and other
	possible spectra in comparison with the
        LISA sensitivity.}
\label{LPG:fig8}
\end{figure}

\clearpage      
\begin{figure}[H]
\epsfxsize=0.8\textwidth
\centerline{\epsfbox{f14.ai}} 
\caption{Full spectrum $h(\nu)$ and its intervals 
accessible to ground-based and cosmic interferometers.}
\label{LPG:fig9}
\end{figure}

\clearpage      
\begin{figure}
\epsfxsize=4.5in
\centerline{\epsfbox{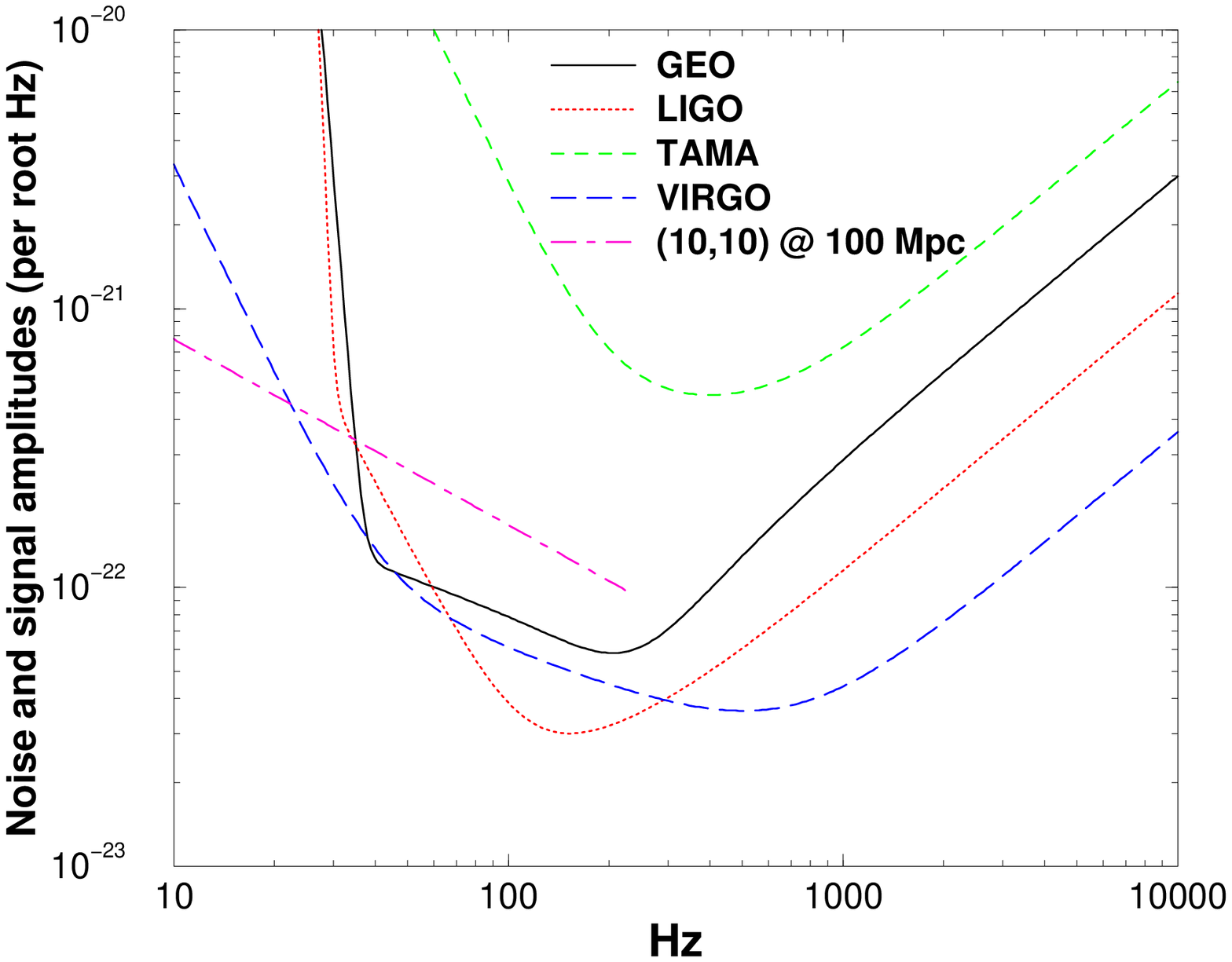}} 
\caption{This figure shows the amplitude noise spectral density,
$\sqrt{S_n(f)},$ in initial interferometers.
On the same graph we also plot the signal 
amplitude, $ \sqrt{f\,}|\tilde h(f)|$, of a binary black hole
inspiral occurring at a distance of 100 Mpc. Each black hole
is taken to be of mass equal to 10M$_\odot.$
(See text in the Section~\ref{sec:inspiral snrs} for a discussion.)}
\label{fig:noise curves}
\end{figure}

\clearpage      
\begin{figure}
\epsfxsize=4.5in
\centerline{\epsfbox{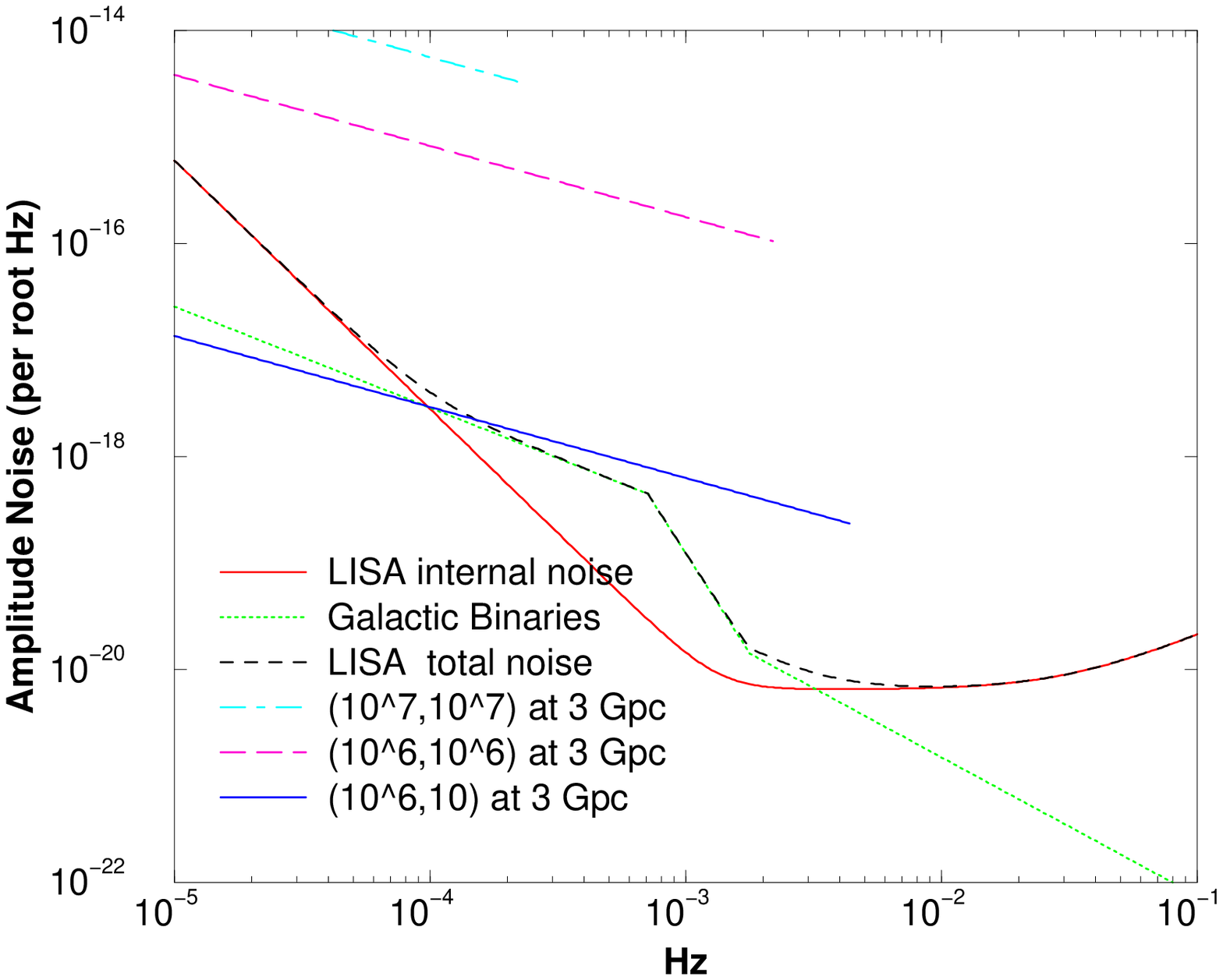}} 
\caption{Same as Fig.~\protect{\ref{fig:noise curves}} but
for the LISA detector. Note that this figure, in contrast to
Fig.~\protect\ref{LPG:fig8}, uses amplitudes per $\sqrt{\hbox{\rm Hz}\,}$. We also
plot signals from supermassive black holes. The supermassive BH 
sources are
assumed to lie at a red-shift of $z=1$ but LISA can detect these
sources with a good SNR practically anywhere in the Universe.
}
\label{fig:lisa noise curve}
\end{figure}

\clearpage
\begin{figure}
\epsfxsize=4.5in
\centerline{\epsfbox{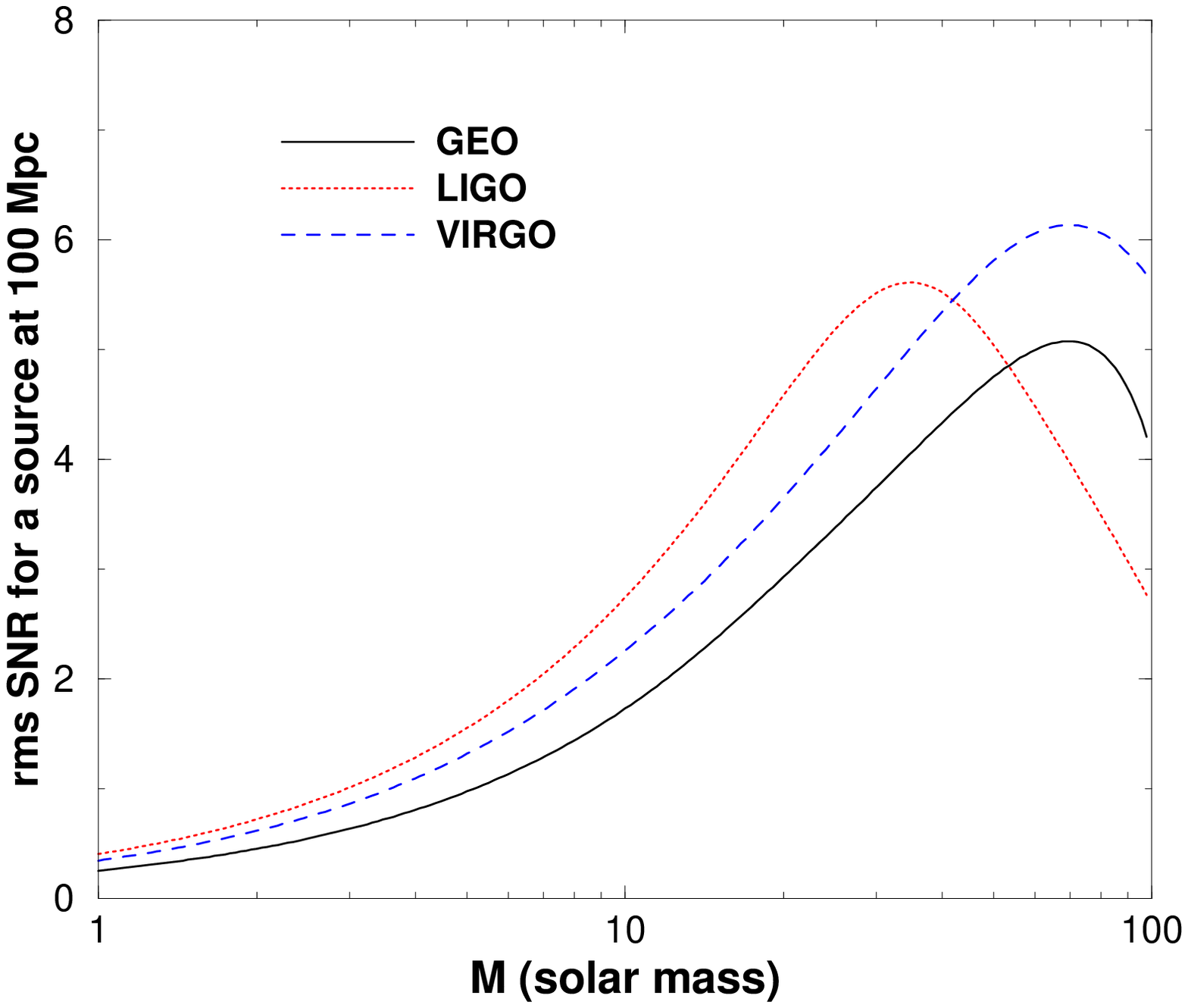}} 
\caption {Signal-to-noise ratio, in initial interferometers, 
as a function of total mass, for inspiral signals from binaries
of equal masses at 100 Mpc
and averaged over source inclination and location. 
(TAMA has been left out as the SNRs in that case are too 
low for these sources.)}
\label{fig:snr estimates}
\end{figure}

\clearpage
\begin{figure}[H]
\epsfxsize=\textwidth
\centerline{\epsfbox{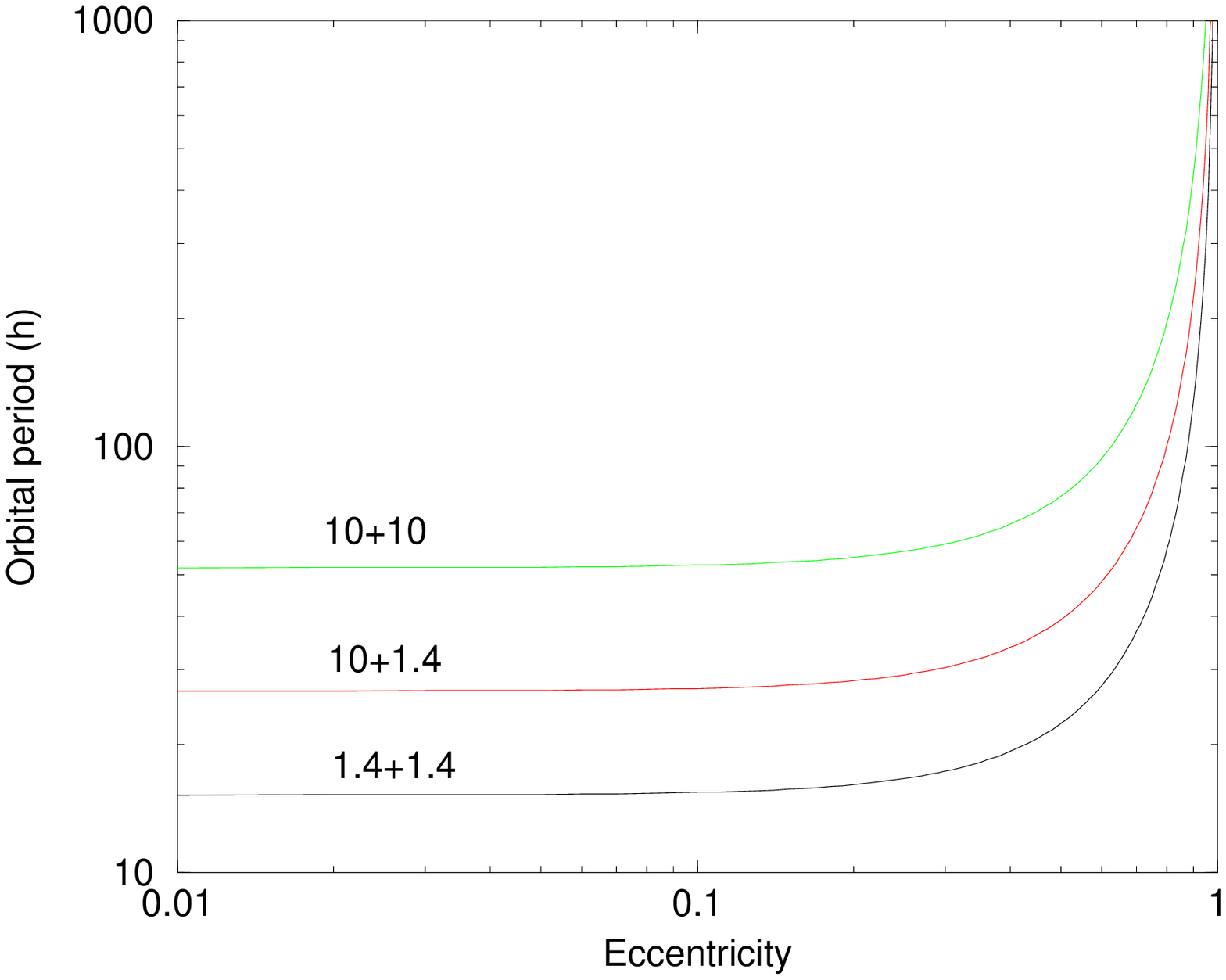}} 
\caption{The initial orbital period (in hours)
of two point masses which will coalesce
due to gravitational wave emission in a time interval 
$10^{10}$ years, as a function of the initial eccentricity $e_0$.
The lines correspond to binaries with  
10M$_\odot$+10M$_\odot$ (BH+BH), 10M$_\odot$+1.4M$_\odot$ (BH+NS),
and 1.4M$_\odot$+1.4M$_\odot$ (NS+NS). Binary with 
a shorter (longer) orbital period will coalesce in 
a shorter (longer) time.}
\label{A:GW:p-e}
\end{figure}

\clearpage
\begin{figure}[H]
\epsfxsize=\textwidth
\centerline{\epsfbox{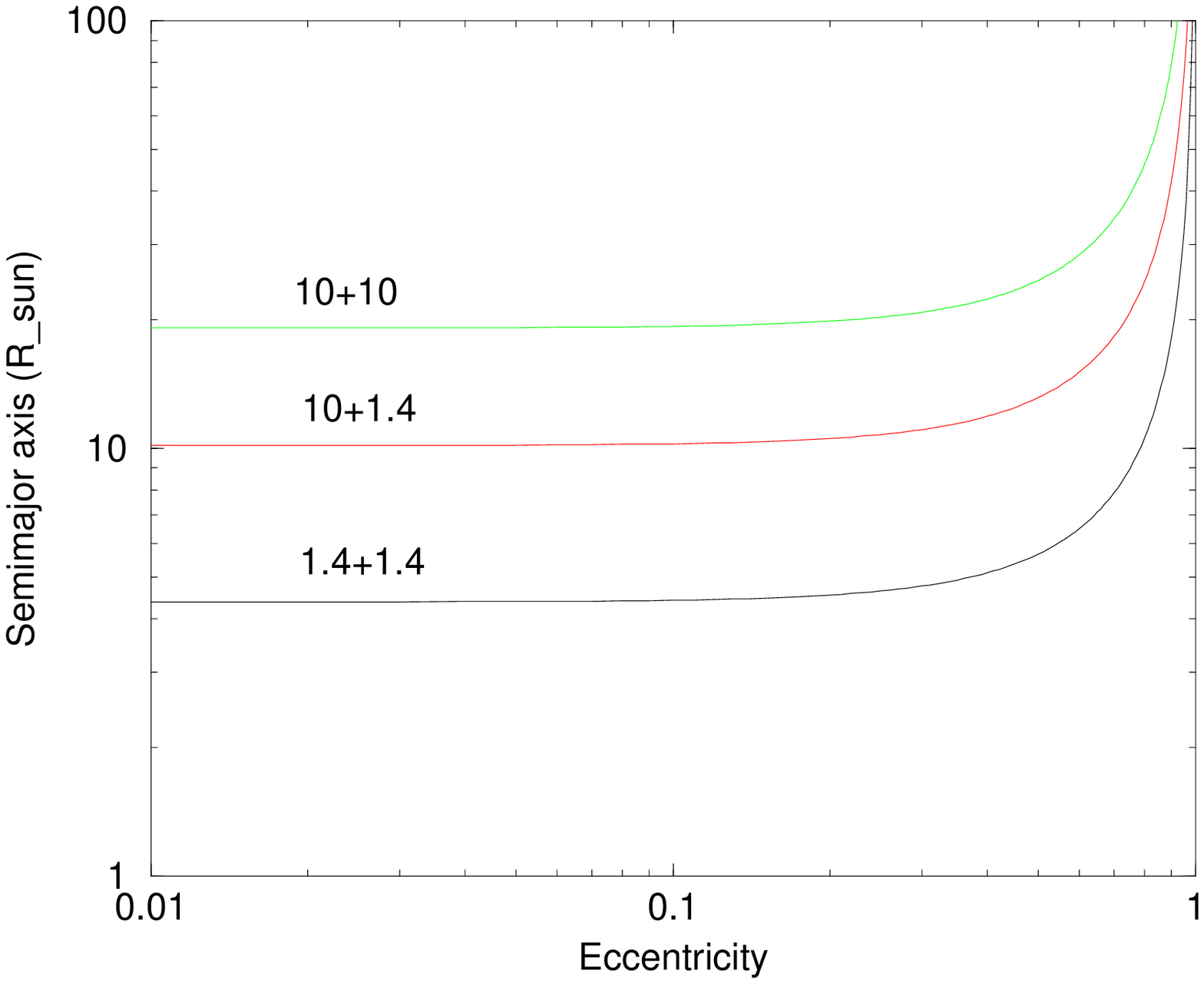}} 
\caption{The initial semimajor axis (in R$_\odot$)
of two point masses which will coalesce
due to gravitational wave emission in a time interval 
$10^{10}$ years, as a function of the initial eccentricity $e_0$.
The lines correspond to binaries with
10M$_\odot$+10M$_\odot$ (BH+BH), 10M$_\odot$+1.4M$_\odot$ (BH+NS),
and 1.4M$_\odot$+1.4M$_\odot$ (NS+NS). Binary with a smaller (larger)
semimajor axis will coalesce in a shorter (longer) time.}
\label{A:GW:a-e}
\end{figure}

\fi

\end{document}